\documentclass[longauth,traditabstract]{aa}
\pdfmapfile{+txfonts.map}

\usepackage[utf8]{inputenc}
\usepackage[nonamebreak]{natbib}
\usepackage[stable]{footmisc}
\usepackage{amsmath}
\usepackage{txfonts}
\usepackage{placeins}
\usepackage{natbib}
\bibpunct{(}{)}{;}{a}{}{,} %
\usepackage{graphicx}
\usepackage{epstopdf}
\usepackage{ifthen}
\usepackage[breaklinks, colorlinks, citecolor=blue, pdfencoding=auto,linkcolor=black]{hyperref}
\usepackage{overpic}
\usepackage{fixltx2e}
\usepackage{multirow}
\usepackage{rotating}
\usepackage[table,usenames,dvipsnames]{xcolor}
\usepackage{multirow}
\usepackage{afterpage}
\usepackage{float}

\def\setsymbol#1#2{\expandafter\def\csname #1\endcsname{#2}}
\def\getsymbol#1{\csname #1\endcsname}

\def\Planck{\textit{Planck}}

\newbox\tablebox    \newdimen\tablewidth
\def\leaderfil{\leaders\hbox to 5pt{\hss.\hss}\hfil}
\def\endPlancktable{\tablewidth=\columnwidth 
    $$\hss\copy\tablebox\hss$$
    \vskip-\lastskip\vskip -2pt}

\def\tablenote#1 #2\par{\begingroup \parindent=0.8em
    \abovedisplayshortskip=0pt\belowdisplayshortskip=0pt
    \noindent
    $$\hss\vbox{\hsize\tablewidth \hangindent=\parindent \hangafter=1 \noindent
    \hbox to \parindent{$^#1$\hss}\strut#2\strut\par}\hss$$
    \endgroup}
\def\doubleline{\vskip 3pt\hrule \vskip 1.5pt \hrule \vskip 5pt}

\def\L2{\ifmmode L_2\else $L_2$\fi}

\def\DeltaT{\ifmmode \Delta T\else $\Delta T$\fi}
\def\deltat{\ifmmode \Delta t\else $\Delta t$\fi}
\def\fknee{\ifmmode f_{\rm knee}\else $f_{\rm knee}$\fi}
\def\Fmax{\ifmmode F_{\rm max}\else $F_{\rm max}$\fi}
\def\solar{\ifmmode{\rm M}_{\mathord\odot}\else${\rm M}_{\mathord\odot}$\fi}
\def\Msolar{\ifmmode{\rm M}_{\mathord\odot}\else${\rm M}_{\mathord\odot}$\fi}
\def\Lsolar{\ifmmode{\rm L}_{\mathord\odot}\else${\rm L}_{\mathord\odot}$\fi}
\def\inv{\ifmmode^{-1}\else$^{-1}$\fi}
\def\mo{\ifmmode^{-1}\else$^{-1}$\fi}
\def\sup#1{\ifmmode ^{\rm #1}\else $^{\rm #1}$\fi}
\def\expo#1{\ifmmode \times 10^{#1}\else $\times 10^{#1}$\fi}
\def\,{\thinspace}
\def\lsim{\mathrel{\raise .4ex\hbox{\rlap{$<$}\lower 1.2ex\hbox{$\sim$}}}}
\def\gsim{\mathrel{\raise .4ex\hbox{\rlap{$>$}\lower 1.2ex\hbox{$\sim$}}}}

\def\simprop{\mathrel{\raise .4ex\hbox{\rlap{$\propto$}\lower 1.2ex\hbox{$\sim$}}}}
\def\deg{\ifmmode^\circ\else$^\circ$\fi}
\def\pdeg{\ifmmode $\setbox0=\hbox{$^{\circ}$}\rlap{\hskip.11\wd0 .}$^{\circ}
          \else \setbox0=\hbox{$^{\circ}$}\rlap{\hskip.11\wd0 .}$^{\circ}$\fi}
\def\arcs{\ifmmode {^{\scriptstyle\prime\prime}}
          \else $^{\scriptstyle\prime\prime}$\fi}
\def\arcm{\ifmmode {^{\scriptstyle\prime}}
          \else $^{\scriptstyle\prime}$\fi}
\newdimen\sa  \newdimen\sb
\def\parcs{\sa=.07em \sb=.03em
     \ifmmode \hbox{\rlap{.}}^{\scriptstyle\prime\kern -\sb\prime}\hbox{\kern -\sa}
     \else \rlap{.}$^{\scriptstyle\prime\kern -\sb\prime}$\kern -\sa\fi}
\def\parcm{\sa=.08em \sb=.03em
     \ifmmode \hbox{\rlap{.}\kern\sa}^{\scriptstyle\prime}\hbox{\kern-\sb}
     \else \rlap{.}\kern\sa$^{\scriptstyle\prime}$\kern-\sb\fi}
\def\ra[#1 #2 #3.#4]{#1\sup{h}#2\sup{m}#3\sup{s}\llap.#4}
\def\dec[#1 #2 #3.#4]{#1\deg#2\arcm#3\arcs\llap.#4}
\def\deco[#1 #2 #3]{#1\deg#2\arcm#3\arcs}
\def\rra[#1 #2]{#1\sup{h}#2\sup{m}}

\def\dots{\relax\ifmmode \ldots\else $\ldots$\fi}
\def\WHzsr{\ifmmode $W\,Hz\mo\,sr\mo$\else W\,Hz\mo\,sr\mo\fi}
\def\mHz{\ifmmode $\,mHz$\else \,mHz\fi}
\def\GHz{\ifmmode $\,GHz$\else \,GHz\fi}
\def\mKs{\ifmmode $\,mK\,s$^{1/2}\else \,mK\,s$^{1/2}$\fi}
\def\muKs{\ifmmode \,\mu$K\,s$^{1/2}\else \,$\mu$K\,s$^{1/2}$\fi}
\def\muKRJs{\ifmmode \,\mu$K$_{\rm RJ}$\,s$^{1/2}\else \,$\mu$K$_{\rm RJ}$\,s$^{1/2}$\fi}
\def\muKHz{\ifmmode \,\mu$K\,Hz$^{-1/2}\else \,$\mu$K\,Hz$^{-1/2}$\fi}
\def\MJysr{\ifmmode \,$MJy\,sr\mo$\else \,MJy\,sr\mo\fi}
\def\MJysrmK{\ifmmode \,$MJy\,sr\mo$\,mK$_{\rm CMB}\mo\else \,MJy\,sr\mo\,mK$_{\rm CMB}\mo$\fi}
\def\microns{\ifmmode \,\mu$m$\else \,$\mu$m\fi}

\def\muK{\ifmmode \,\mu$K$\else \,$\mu$\hbox{K}\fi}
\def\microK{\ifmmode \,\mu$K$\else \,$\mu$\hbox{K}\fi}
\def\muW{\ifmmode \,\mu$W$\else \,$\mu$\hbox{W}\fi}
\def\kms{\ifmmode $\,km\,s$^{-1}\else \,km\,s$^{-1}$\fi}
\def\kmsMpc{\ifmmode $\,\kms\,Mpc\mo$\else \,\kms\,Mpc\mo\fi}
\providecommand{\sorthelp}[1]{}

\defcitealias{planck2013-p11}{PCP13}
\defcitealias{planck2013-p08}{PPL13}
\defcitealias{planck2013-p17}{PCI13}
\defcitealias{pb2015}{BKP}
\defcitealias{Array:2015xqh}{BK14}
\defcitealias{planck2014-a24}{PCI15}
\defcitealias{planck2014-a15}{PCP15}
\defcitealias{planck2014-a13}{PPL15}
\defcitealias{planck2014-a10}{POD16}
\defcitealias{planck2014-a25}{PRE16}
\defcitealias{planck2016-l05}{PPL18}
\defcitealias{planck2016-l06}{PCP18}
\defcitealias{planck2016-l08}{PPLe18}
\defcitealias{bk15}{BK15}

\def\redcomment#1{}

\def\LCDM{$\Lambda$CDM}
\def\NHUNIT{\ifmmode {\rm \,cm^{-2}} \else $\rm \,cm^{-2}$ \fi} 

\def\aap{{A\&A}}
\def\apj{{ApJ}}
\def\mnras{{MNRAS}}
\def\apjs{{ApJ Supp.}}
\def\muKcmb{\ifmmode \,\mu$K$_{\rm CMB}$\else \,$\mu$K$_{\rm CMB}$\fi}
\newcommand{\OmegaM}{\ifmmode\Omega_{\rm M}\else $\Omega_{\rm M}$\fi}

\def\email#1{{\textcolor{magenta}{\tt#1}}}

    \renewcommand{\dbltopfraction}{0.9} 
    \renewcommand{\textfraction}{0.00}  

\begin{document}

\title{\vglue -10mm\Planck\ 2018 results. X. Constraints on inflation}

\author{\small
Planck Collaboration: Y.~Akrami\inst{55, 57}
\and
F.~Arroja\inst{59}
\and
M.~Ashdown\inst{65, 5}
\and
J.~Aumont\inst{94}
\and
C.~Baccigalupi\inst{78}
\and
M.~Ballardini\inst{21, 39}
\and
A.~J.~Banday\inst{94, 8}
\and
R.~B.~Barreiro\inst{60}
\and
N.~Bartolo\inst{28, 61}
\and
S.~Basak\inst{85}
% ...
}
\institute{\small
AIM, CEA, CNRS, Universit\'{e} Paris-Saclay, F-91191 Gif sur Yvette, France. AIM, Universit\'{e} Paris Diderot, Sorbonne Paris Cit\'{e}, F-91191 Gif sur Yvette, France.\goodbreak
\and
APC, AstroParticule et Cosmologie, Universit\'{e} Paris Diderot, CNRS/IN2P3, CEA/lrfu, Observatoire de Paris, Sorbonne Paris Cit\'{e}, 10, rue Alice Domon et L\'{e}onie Duquet, 75205 Paris Cedex 13, France\goodbreak
\and
African Institute for Mathematical Sciences, 6-8 Melrose Road, Muizenberg, Cape Town, South Africa\goodbreak
\and
Aix Marseille Univ, CNRS, CNES, LAM, Marseille, France\goodbreak
\and
Astrophysics Group, Cavendish Laboratory, University of Cambridge, J J Thomson Avenue, Cambridge CB3 0HE, U.K.\goodbreak
\and
Astrophysics \& Cosmology Research Unit, School of Mathematics, Statistics \& Computer Science, University of KwaZulu-Natal, Westville Campus, Private Bag X54001, Durban 4000, South Africa\goodbreak
\and
CITA, University of Toronto, 60 St. George St., Toronto, ON M5S 3H8, Canada\goodbreak
\and
CNRS, IRAP, 9 Av. colonel Roche, BP 44346, F-31028 Toulouse cedex 4, France\goodbreak
\and
Cahill Center for Astronomy and Astrophysics, California Institute of Technology, Pasadena CA,  91125, USA\goodbreak
\and
California Institute of Technology, Pasadena, California, U.S.A.\goodbreak
\and
Centre for Theoretical Cosmology, DAMTP, University of Cambridge, Wilberforce Road, Cambridge CB3 0WA, U.K.\goodbreak
\and
Computational Cosmology Center, Lawrence Berkeley National Laboratory, Berkeley, California, U.S.A.\goodbreak
\and
DTU Space, National Space Institute, Technical University of Denmark, Elektrovej 327, DK-2800 Kgs. Lyngby, Denmark\goodbreak
\and
D\'{e}partement de Physique Th\'{e}orique, Universit\'{e} de Gen\`{e}ve, 24, Quai E. Ansermet,1211 Gen\`{e}ve 4, Switzerland\goodbreak
\and
Departamento de Astrof\'{i}sica, Universidad de La Laguna (ULL), E-38206 La Laguna, Tenerife, Spain\goodbreak
\and
Departamento de F\'{\i}sica, Universidad de Oviedo, C/ Federico Garc\'{\i}a Lorca, 18 , Oviedo, Spain\goodbreak
\and
Department of Astrophysics/IMAPP, Radboud University, P.O. Box 9010, 6500 GL Nijmegen, The Netherlands\goodbreak
\and
Department of General Education, National Institute of Technology, Kagawa College, 355 Chokushi-cho, Takamatsu, Kagawa 761-8058, Japan\goodbreak
\and
Department of Mathematics, University of Stellenbosch, Stellenbosch 7602, South Africa\goodbreak
\and
Department of Physics \& Astronomy, University of British Columbia, 6224 Agricultural Road, Vancouver, British Columbia, Canada\goodbreak
\and
Department of Physics \& Astronomy, University of the Western Cape, Cape Town 7535, South Africa\goodbreak
\and
Department of Physics and Astronomy, University College London, London WC1E 6BT, U.K.\goodbreak
\and
Department of Physics and Astronomy, University of Sussex, Brighton BN1 9QH, U.K.\goodbreak
\and
Department of Physics, Gustaf H\"{a}llstr\"{o}min katu 2a, University of Helsinki, Helsinki, Finland\goodbreak
\and
Department of Physics, Princeton University, Princeton, New Jersey, U.S.A.\goodbreak
\and
Department of Physics, University of California, Santa Barbara, California, U.S.A.\goodbreak
\and
Department of Physics, University of Illinois at Urbana-Champaign, 1110 West Green Street, Urbana, Illinois, U.S.A.\goodbreak
\and
Dipartimento di Fisica e Astronomia G. Galilei, Universit\`{a} degli Studi di Padova, via Marzolo 8, 35131 Padova, Italy\goodbreak
\and
Dipartimento di Fisica e Scienze della Terra, Universit\`{a} di Ferrara, Via Saragat 1, 44122 Ferrara, Italy\goodbreak
\and
Dipartimento di Fisica, Universit\`{a} La Sapienza, P. le A. Moro 2, Roma, Italy\goodbreak
\and
Dipartimento di Fisica, Universit\`{a} degli Studi di Milano, Via Celoria, 16, Milano, Italy\goodbreak
\and
Dipartimento di Fisica, Universit\`{a} degli Studi di Trieste, via A. Valerio 2, Trieste, Italy\goodbreak
\and
Dipartimento di Fisica, Universit\`{a} di Roma Tor Vergata, Via della Ricerca Scientifica, 1, Roma, Italy\goodbreak
\and
European Space Agency, ESAC, Planck Science Office, Camino bajo del Castillo, s/n, Urbanizaci\'{o}n Villafranca del Castillo, Villanueva de la Ca\~{n}ada, Madrid, Spain\goodbreak
\and
European Space Agency, ESTEC, Keplerlaan 1, 2201 AZ Noordwijk, The Netherlands\goodbreak
\and
Gran Sasso Science Institute, INFN, viale F. Crispi 7, 67100 L'Aquila, Italy\goodbreak
\and
Haverford College Astronomy Department, 370 Lancaster Avenue, Haverford, Pennsylvania, U.S.A.\goodbreak
\and
Helsinki Institute of Physics, Gustaf H\"{a}llstr\"{o}min katu 2, University of Helsinki, Helsinki, Finland\goodbreak
\and
INAF - OAS Bologna, Istituto Nazionale di Astrofisica - Osservatorio di Astrofisica e Scienza dello Spazio di Bologna, Area della Ricerca del CNR, Via Gobetti 101, 40129, Bologna, Italy\goodbreak
\and
INAF - Osservatorio Astronomico di Padova, Vicolo dell'Osservatorio 5, Padova, Italy\goodbreak
\and
INAF - Osservatorio Astronomico di Trieste, Via G.B. Tiepolo 11, Trieste, Italy\goodbreak
\and
INAF, Istituto di Radioastronomia, Via Piero Gobetti 101, I-40129 Bologna, Italy\goodbreak
\and
INAF/IASF Milano, Via E. Bassini 15, Milano, Italy\goodbreak
\and
INFN - CNAF, viale Berti Pichat 6/2, 40127 Bologna, Italy\goodbreak
\and
INFN, Sezione di Bologna, viale Berti Pichat 6/2, 40127 Bologna, Italy\goodbreak
\and
INFN, Sezione di Ferrara, Via Saragat 1, 44122 Ferrara, Italy\goodbreak
\and
INFN, Sezione di Milano, Via Celoria 16, Milano, Italy\goodbreak
\and
INFN, Sezione di Roma 1, Universit\`{a} di Roma Sapienza, Piazzale Aldo Moro 2, 00185, Roma, Italy\goodbreak
\and
INFN, Sezione di Roma 2, Universit\`{a} di Roma Tor Vergata, Via della Ricerca Scientifica, 1, Roma, Italy\goodbreak
\and
IUCAA, Post Bag 4, Ganeshkhind, Pune University Campus, Pune 411 007, India\goodbreak
\and
Imperial College London, Astrophysics group, Blackett Laboratory, Prince Consort Road, London, SW7 2AZ, U.K.\goodbreak
\and
Institut d'Astrophysique Spatiale, CNRS, Univ. Paris-Sud, Universit\'{e} Paris-Saclay, B\^{a}t. 121, 91405 Orsay cedex, France\goodbreak
\and
Institut d'Astrophysique de Paris, CNRS (UMR7095), 98 bis Boulevard Arago, F-75014, Paris, France\goodbreak
\and
Institut f\"{u}r Theoretische Teilchenphysik und Kosmologie, RWTH Aachen University, D-52056 Aachen, Germany\goodbreak
\and
Institute Lorentz, Leiden University, PO Box 9506, Leiden 2300 RA, The Netherlands\goodbreak
\and
Institute of Astronomy, University of Cambridge, Madingley Road, Cambridge CB3 0HA, U.K.\goodbreak
\and
Institute of Theoretical Astrophysics, University of Oslo, Blindern, Oslo, Norway\goodbreak
\and
Instituto de Astrof\'{\i}sica de Canarias, C/V\'{\i}a L\'{a}ctea s/n, La Laguna, Tenerife, Spain\goodbreak
\and
Instituto de Astrof\'{\i}sica e Ci\^{e}ncias do Espa\c{c}o, Faculdade de Ci\^{e}ncias da Universidade de Lisboa, Campo Grande, PT1749-016 Lisboa, Portugal\goodbreak
\and
Instituto de F\'{\i}sica de Cantabria (CSIC-Universidad de Cantabria), Avda. de los Castros s/n, Santander, Spain\goodbreak
\and
Istituto Nazionale di Fisica Nucleare, Sezione di Padova, via Marzolo 8, I-35131 Padova, Italy\goodbreak
\and
Jet Propulsion Laboratory, California Institute of Technology, 4800 Oak Grove Drive, Pasadena, California, U.S.A.\goodbreak
\and
Jodrell Bank Centre for Astrophysics, Alan Turing Building, School of Physics and Astronomy, The University of Manchester, Oxford Road, Manchester, M13 9PL, U.K.\goodbreak
\and
Kavli Institute for Cosmological Physics, University of Chicago, Chicago, IL 60637, USA\goodbreak
\and
Kavli Institute for Cosmology Cambridge, Madingley Road, Cambridge, CB3 0HA, U.K.\goodbreak
\and
LERMA, CNRS, Observatoire de Paris, 61 Avenue de l'Observatoire, Paris, France\goodbreak
\and
LERMA/LRA, Observatoire de Paris, PSL Research University, CNRS, Ecole Normale Sup\'erieure, 75005 Paris, France\goodbreak
\and
Laboratoire AIM, CEA - Universit\'{e} Paris-Saclay, 91191 Gif-sur-Yvette, France\goodbreak
\and
Laboratoire de Physique Subatomique et Cosmologie, Universit\'{e} Grenoble-Alpes, CNRS/IN2P3, 53, rue des Martyrs, 38026 Grenoble Cedex, France\goodbreak
\and
Laboratoire de Physique Th\'{e}orique, Universit\'{e} Paris-Sud 11 \& CNRS, B\^{a}timent 210, 91405 Orsay, France\goodbreak
\and
Lawrence Berkeley National Laboratory, Berkeley, California, U.S.A.\goodbreak
\and
Leung Center for Cosmology and Particle Astrophysics, National Taiwan University, Taipei 10617, Taiwan\goodbreak
\and
Low Temperature Laboratory, Department of Applied Physics, Aalto University, Espoo, FI-00076 AALTO, Finland\goodbreak
\and
Max-Planck-Institut f\"{u}r Astrophysik, Karl-Schwarzschild-Str. 1, 85741 Garching, Germany\goodbreak
\and
Mullard Space Science Laboratory, University College London, Surrey RH5 6NT, U.K.\goodbreak
\and
NAOC-UKZN Computational Astrophysics Centre (NUCAC), University of KwaZulu-Natal, Durban 4000, South Africa\goodbreak
\and
Nicolaus Copernicus Astronomical Center, Polish Academy of Sciences, Bartycka 18, 00-716 Warsaw, Poland\goodbreak
\and
SISSA, Astrophysics Sector, via Bonomea 265, 34136, Trieste, Italy\goodbreak
\and
San Diego Supercomputer Center, University of California, San Diego, 9500 Gilman Drive, La Jolla, CA 92093, USA\goodbreak
\and
School of Chemistry and Physics, University of KwaZulu-Natal, Westville Campus, Private Bag X54001, Durban, 4000, South Africa\goodbreak
\and
School of Physical Sciences, National Institute of Science Education and Research, HBNI, Jatni-752050, Odissa, India\goodbreak
\and
School of Physics and Astronomy, Cardiff University, Queens Buildings, The Parade, Cardiff, CF24 3AA, U.K.\goodbreak
\and
School of Physics and Astronomy, Sun Yat-sen University, 2 Daxue Rd, Tangjia, Zhuhai, China\goodbreak
\and
School of Physics and Astronomy, University of Nottingham, Nottingham NG7 2RD, U.K.\goodbreak
\and
School of Physics, Indian Institute of Science Education and Research Thiruvananthapuram, Maruthamala PO, Vithura, Thiruvananthapuram 695551, Kerala, India\goodbreak
\and
School of Physics, The University of New South Wales, Sydney NSW 2052, Australia\goodbreak
\and
Simon Fraser University, Department of Physics, 8888 University Drive, Burnaby BC, Canada\goodbreak
\and
Sorbonne Universit\'{e}-UPMC, UMR7095, Institut d'Astrophysique de Paris, 98 bis Boulevard Arago, F-75014, Paris, France\goodbreak
\and
Space Research Institute (IKI), Russian Academy of Sciences, Profsoyuznaya Str, 84/32, Moscow, 117997, Russia\goodbreak
\and
Space Science Data Center - Agenzia Spaziale Italiana, Via del Politecnico snc, 00133, Roma, Italy\goodbreak
\and
Space Sciences Laboratory, University of California, Berkeley, California, U.S.A.\goodbreak
\and
The Oskar Klein Centre for Cosmoparticle Physics, Department of Physics, Stockholm University, AlbaNova, SE-106 91 Stockholm, Sweden\goodbreak
\and
UPMC Univ Paris 06, UMR7095, 98 bis Boulevard Arago, F-75014, Paris, France\goodbreak
\and
Universit\'{e} de Toulouse, UPS-OMP, IRAP, F-31028 Toulouse cedex 4, France\goodbreak
% ...
}


\date{\vglue -1.5mm 10 June 2019\vglue -5mm}
\abstract{
We report on the implications for cosmic inflation of the 2018
release of the \Planck\ cosmic microwave background (CMB) anisotropy 
measurements. The results are fully consistent with those reported 
using the data from the two previous \Planck\ cosmological releases, 
but have smaller uncertainties thanks to improvements in the 
characterization of polarization at low and high multipoles. \Planck\ 
temperature, polarization, and lensing data determine the spectral 
index of scalar perturbations to be $n_\mathrm{s}=0.9649\pm 0.0042$ 
at 68\,\% CL. We find no evidence for a scale dependence of 
$n_\mathrm{s}$, either as a running or as a running of the running. 
The Universe is found to be consistent with spatial flatness with a 
precision of $0.4\,\%$ at 95\,\% CL by combining \Planck\ with a compilation of BAO data. 
The \Planck\ 95\,\% CL upper limit on the tensor-to-scalar ratio, $r_{0.002}<0.10$, is further tightened 
by combining with the BICEP2/Keck Array BK15 data to obtain 
$r_{0.002}<0.056$. In the framework of standard single-field inflationary 
models with Einstein gravity, these results imply that: 
(a) the predictions of slow-roll models with a concave potential, $V'' (\phi) < 0$, are 
increasingly favoured by the data; and (b) based on two different 
methods for reconstructing the inflaton potential, we find no 
evidence for dynamics beyond slow roll. Three different methods for 
the non-parametric reconstruction of the primordial power spectrum 
consistently confirm a pure power law in the range of comoving scales 
$0.005\,\mathrm{Mpc}^{-1} \lesssim k \lesssim 0.2\,\mathrm{Mpc}^{-1}$. 
A complementary analysis also finds no evidence for theoretically 
motivated parameterized features in the \Planck\ power spectra. For 
the case of oscillatory features that are logarithmic or linear in $k,$ this 
result is further strengthened by a new combined analysis including 
the \Planck\ bispectrum data. The new \Planck\ polarization data 
provide a stringent test of the adiabaticity of the initial 
conditions for the cosmological fluctuations. In correlated, mixed 
adiabatic and isocurvature models, the non-adiabatic contribution to 
the observed CMB temperature variance is constrained to 1.3\,\%, 
1.7\,\%, and 1.7\,\% at 95\,\% CL for cold dark matter, neutrino density, 
and neutrino velocity, respectively. \Planck\ power spectra plus lensing set constraints 
on the amplitude of compensated cold dark matter-baryon 
isocurvature perturbations that are consistent with current complementary measurements.
The polarization data 
also provide improved constraints on inflationary models that predict  
a small statistically anisotropic quadupolar modulation of the 
primordial fluctuations.  However, the polarization data do not 
support physical models for a scale-dependent dipolar modulation. 
All these findings support the key predictions of the standard single-field
inflationary models, which will be further tested by future 
cosmological observations.

}
   

\authorrunning{Planck Collaboration}
\titlerunning{Constraints on Inflation}
   \maketitle




\clearpage 
\newpage 

\setcounter{tocdepth}{3}
\tableofcontents
\hypersetup{linkcolor=red}

\redcomment{RED COMMENTS ON}
\newpage

\section{Introduction \label{sec:intro}}

\begin{figure*}[!tbp] 
\includegraphics[width=\columnwidth]{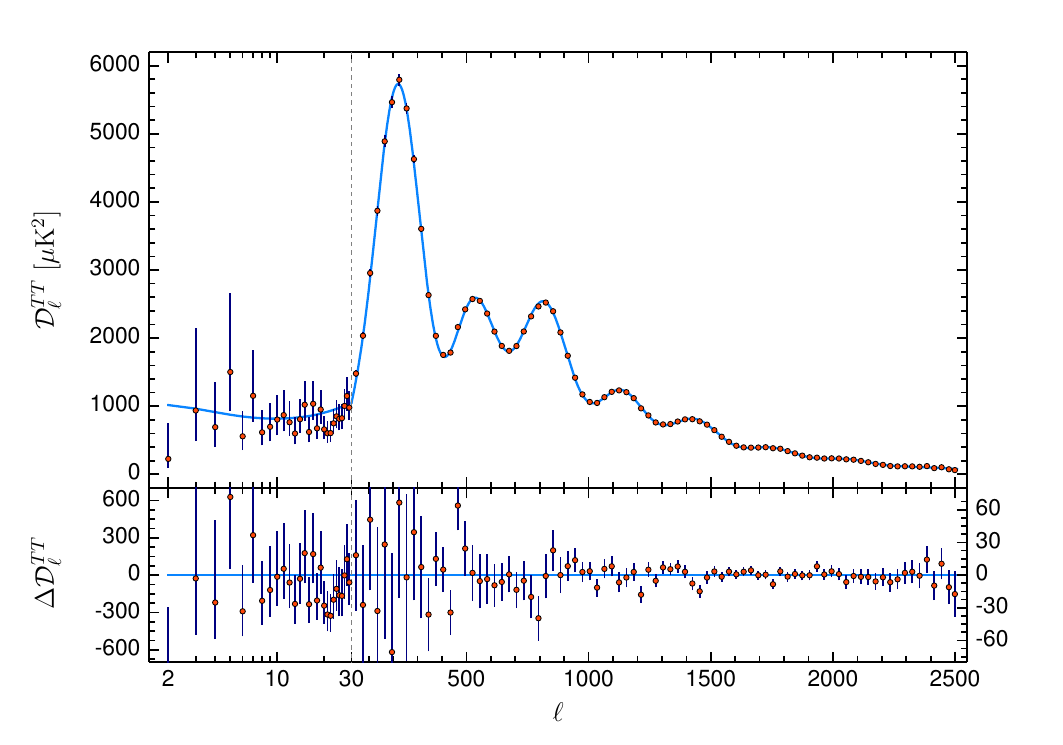}
\includegraphics[width=\columnwidth]{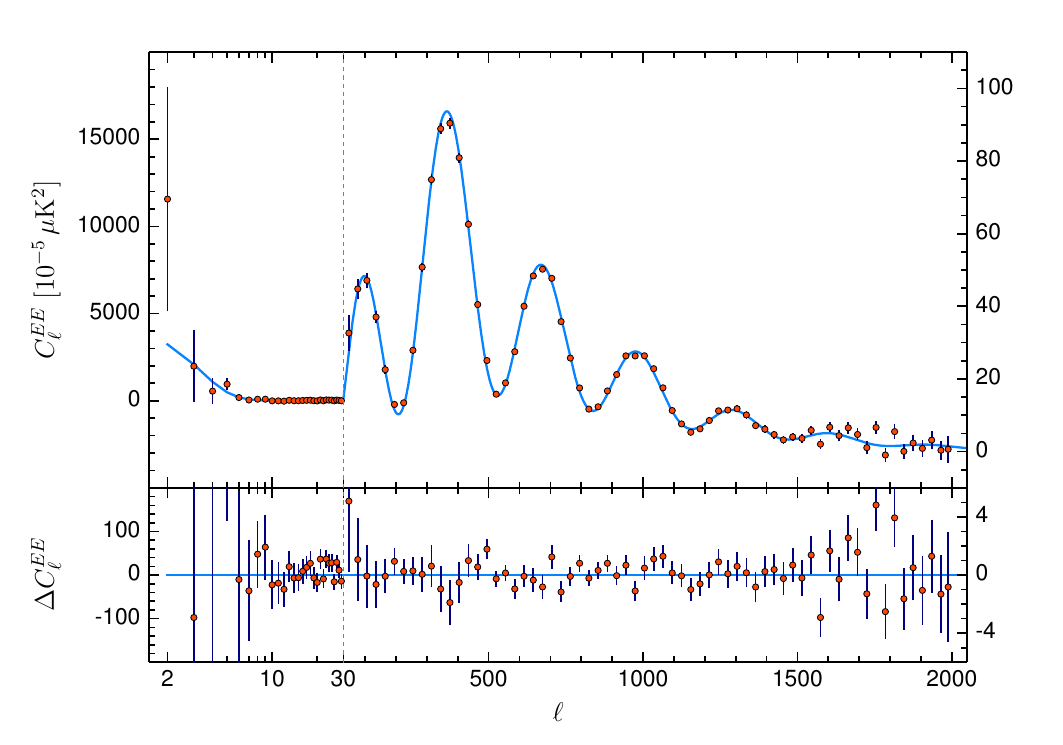}
\\
\includegraphics[width=\columnwidth]{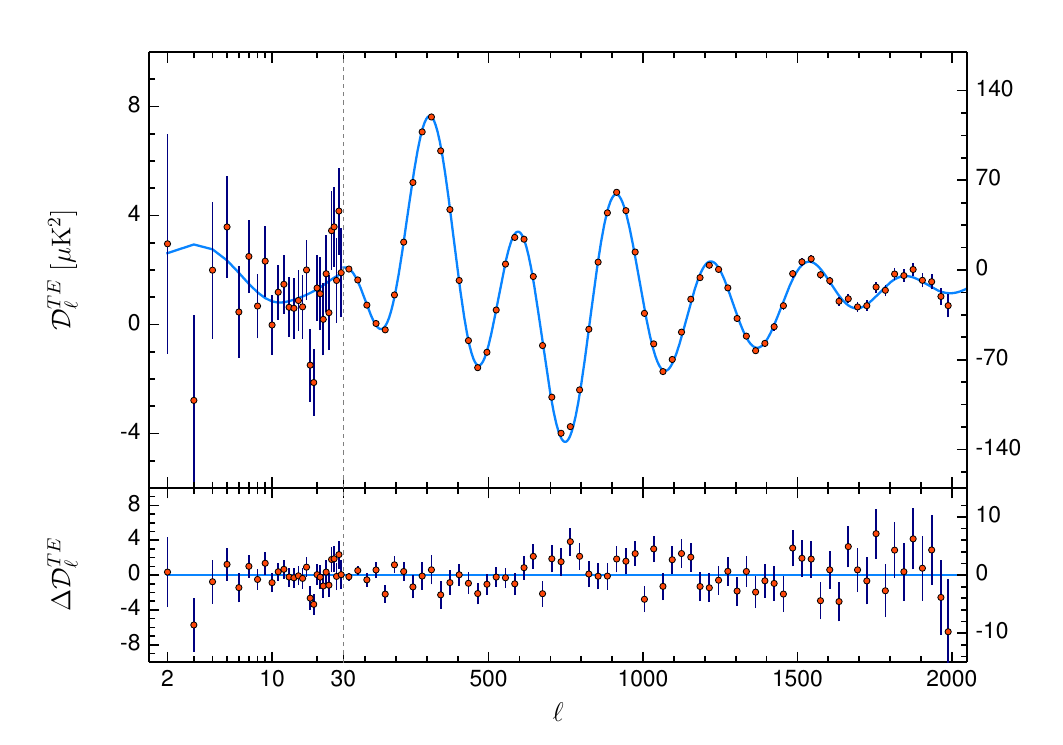}
\includegraphics[width=\columnwidth]{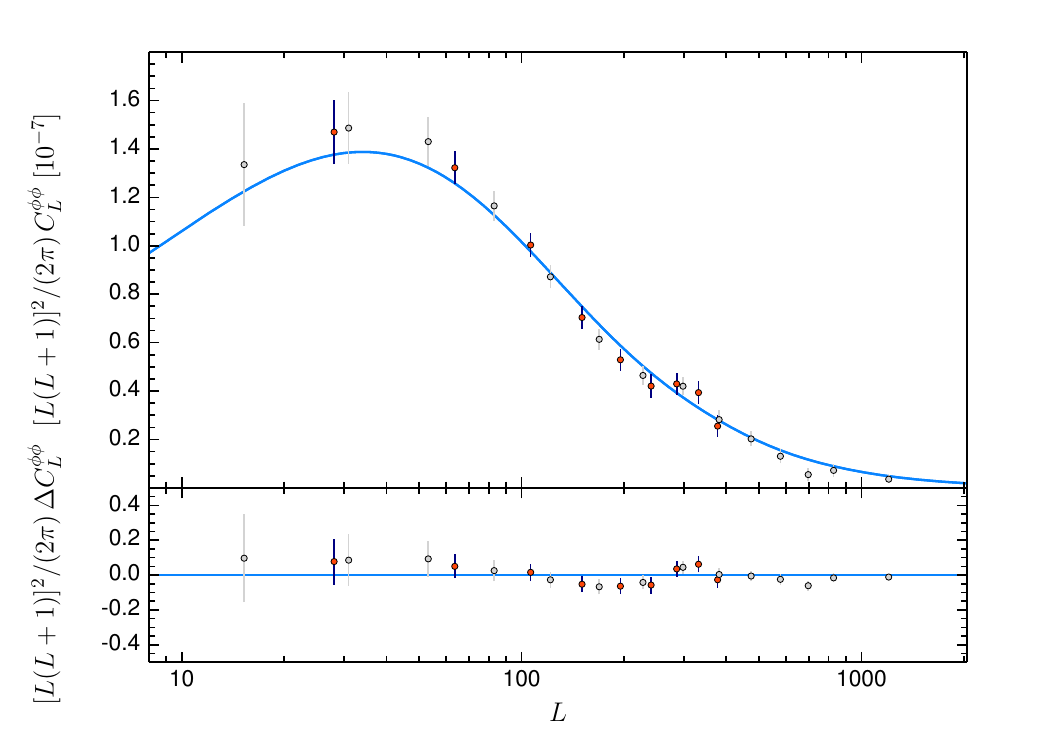}
\caption{\label{fig:Planck_Cl_2018} \Planck\ 2018 CMB angular power spectra, compared with
the base-$\Lambda$CDM best fit to the \Planck\ TT,TE,EE+lowE+lensing data (blue curves). For
each panel we also show the residuals with respect to this baseline best fit. Plotted are
${\cal D}_\ell=\ell(\ell+1)C_\ell/(2\pi)$ for $TT$ and $TE$, $C_\ell$ for $EE$, and 
$L^2 (L+1)^2 C_L^{\phi \phi}/(2\pi)$ for lensing. For $TT$, $TE$, and $EE$, the multipole
range $2 \le \ell \le 29$ shows the power spectra
from {\tt Commander} ($TT$) and {\tt SimAll} ($TE$, $EE$), while at $\ell \ge 30$ we display the
co-added frequency spectra computed from the \texttt{Plik} cross-half-mission likelihood,
with foreground and other nuisance parameters fixed to their best-fit values in the
base-$\Lambda$CDM cosmology.  For the \Planck\ lensing potential angular power spectrum, we show
the conservative (orange dots; used in the likelihood) and aggressive (grey dots) cases.
Note some of the different horizontal and vertical scales on either side of $\ell = 30$ for the
temperature and polarization spectra and residuals.}
\end{figure*}

This paper, one of a set associated with the 2018 release of data from the 
\Planck\footnote{\Planck\ (\url{http://www.esa.int/Planck}) is a project of 
the European Space Agency  (ESA) with instruments provided by two scientific 
consortia funded by ESA member states and led by Principal Investigators from 
France and Italy, telescope reflectors provided through a collaboration between 
ESA and a scientific consortium led and funded by Denmark, and additional contributions from NASA (USA).} 
mission, presents the implications for cosmic inflation 
of the 2018 \Planck\ measurements of the cosmic microwave background (CMB) 
anisotropies. In terms of data, this paper updates \cite{planck2013-p17} 
(henceforth \citetalias{planck2013-p17}), which was based on the temperature data of the nominal \Planck\ 
mission (``PR1''), including the first 14 months of observations, 
and \cite{planck2014-a24} (henceforth \citetalias{planck2014-a24}), 
which used temperature data and a first set of polarization data from the full \Planck\ mission (``PR2''), 
comprising 29 and 52 months 
of observations for the high frequency instrument (HFI) and low frequency instrument (LFI), respectively. 

The ideas underlying cosmic inflation were developed 
during the late 1970s and early 1980s in order to remedy a number of defects
of the hot big-bang cosmological model (e.g., the 
horizon, smoothness, flatness, and monopole problems) 
\citep{Brout:1977ix, Starobinsky:1980te, Kazanas:1980tx, Sato1981,
Guth:1980zm, Linde:1981mu, Albrecht:1982wi, Linde:1983gd}.
Subsequently, it was realized 
that, on account of quantum vacuum fluctuations, cosmic inflation also provides a 
means to generate the primordial cosmological perturbations
\citep{Mukhanov:1981xt, Mukhanov:1982nu,Hawking:1982cz,Guth:1982ec,Starobinsky:1982ee,Bardeen:1983qw,Mukhanov:1985rz}.
The development of cosmic inflation is one of the major success stories
of modern cosmology, and in this paper we explore how the latest 
2018 release of the \Planck\ data constrains inflationary models.

\Planck\ data currently provide 
the best constraints on the CMB anisotropies, 
except on very small angular scales beyond the resolution limit of \Planck . 
The \Planck\ data set has enabled a precision characterization of the primordial 
cosmological perturbations and has allowed cosmological parameters 
to be constrained at the sub-percent level. 
One of the main data products, described in more detail in the following
section, is the \Planck\ $TT$, $TE$, $EE$, and lensing power spectra, which are
shown in Fig.~\ref{fig:Planck_Cl_2018}, together with the residuals from the six-parameter
concordance $\Lambda$ cold dark matter ($\Lambda$CDM) model using the best-fit parameter values.

In order to provide a quantitative estimate of the improvement 
achieved by \Planck, as well as to show where \Planck\
stands compared to an ultimate cosmic-variance-limited survey,
we consider an idealistic estimator for the number of modes 
[i.e., the effective number of $a_{\ell m}$'s measured \citep{planck2014-a01}]:
\begin{equation}
N_\mathrm{modes}^{X Y} (\ell) \equiv 2 \sum_{\ell'=2}^{\ell} \left( \frac{C_{\ell'}^{X Y}}{\Delta C_{\ell'}^{X Y}} \right)^2 \,, \label{numbermodes_per_multipole}
\end{equation}
where $C_\ell^{X Y}$ ($\Delta C_\ell^{X Y}$) is the (error on the) 
angular power spectrum of the $XY$ channel \citep{planck2014-a01,Scott:2016fad}.
The number of modes measured by \Planck\ is $1\,430\,000$ and $109\,000$
for temperature ($XY=TT$ up to $\ell=2500$) and polarization ($XY=EE$ 
up to $\ell=2000$), respectively \citep{planck2016-l01}.
The number of modes measured is increased by approximately a 
factor of 7 (570) for temperature (polarization)
with respect to the WMAP 9-year measurement, but there is still a 
factor of 3 (40) to gain for a cosmic-variance-limited experiment 
up to $\ell=2500$ accessing
70\,\% of the sky. The additional
modes measured by \Planck\ play a key role in improving the 
constraints on
the initial conditions for the cosmological perturbations and on 
models of inflation with respect to previous measurements of CMB anisotropies.

\Planck\ data have also greatly improved the constraints on bispectral 
non-Gaussianity, both for the ``local'' pattern,
as predicted by many inflationary models, and for other templates 
such as the equilateral one, as analysed
and reported in detail in the dedicated \Planck\ non-Gaussianity papers 
\citep{planck2013-p09a,planck2014-a19,planck2016-l09}. The constraints on
the non-Gaussianity parameter $f_\mathrm{NL}$
are limited by a combination of cosmic variance and instrumental noise. 
An order-of-magnitude estimate for the signal-to-noise
ratio for the local pattern (with $f_\mathrm{NL}^\mathrm{loc}=1$) 
is given by
\begin{equation}
\left( \frac{S}{N} \right)^2 \propto \Omega_\mathrm{sky} \ell_\mathrm{max}^2
\ln \left( \frac{\ell_\mathrm{max}}{\ell_\mathrm{min}}\right) .
\end{equation}
For the local shape, the logarithm enters because most of the 
signal derives
from detecting the modulation of the small-scale power by the 
large-scale CMB anisotropy,
highlighting the importance of full-sky maps for this kind of 
analysis. For other shapes such as equilateral,
one instead has $(S/N)^2\sim \Omega_\mathrm{sky}\ell_\mathrm{max}^2$. 
\Planck\ has significantly sharpened the
constraints on $f_{\rm NL}$, largely on account of its measurement 
of high multipoles with higher signal-to-noise ratio 
compared to past surveys. Some improvement 
has also been obtained from
including polarization.

The \Planck\ measurements have significantly constrained the 
physics of inflation.
The hypothesis of adiabatic Gaussian scalar
fluctuations with a power spectrum described by a simple power law,
which is the key prediction of the standard single-field 
slow-roll inflationary models,
has been tested to unprecedent accuracy 
(\citetalias{planck2013-p17,planck2014-a24};\citealt{planck2013-p09a,planck2014-a19}).  
\Planck\ has set tight constraints on the amount of 
inflationary gravitational 
waves by exploiting the shape of the CMB temperature 
spectrum \citepalias{planck2013-p17}.
These results have inspired a resurgence of activity 
in inflationary model building. 
For more details, see, for example, the following 
review articles and references therein:
\citet{Linde:2014nna}; \citet{Martin:2013nzq}; \citet{Guth:2013sya}; and \citet{Burgess:2013sla}. 
\Planck\ analysis and interpretation
have also sparked a debate on the likelihood of initial conditions 
for some inflationary models                            
\citep{Ijjas:2013vea,Ijjas:2015hcc,Linde:2017pwt}, which is primarily 
of theoretical interest and is not addressed in this paper.
In combination with more sensitive $B$-mode ground-based polarization 
measurements, as from BICEP-Keck Array \citepalias{pb2015}, \Planck\ 
has convincingly ruled out the slow-roll inflationary model with 
a quadratic potential \citepalias{planck2014-a24}.
In terms of physics beyond
the simplest slow-roll inflationary models, the pre-\Planck\ 
hints of a running spectral
index \citep{Hou:2012xq} or of large non-Gaussianities \citep{Bennett:2012fp} 
have disappeared as a result of the \Planck\ measurements.
How to interpret anomalies on the largest angular scales and at high multipoles is a question 
motivating the search for new 
non-standard inflationary models. We 
discuss how the \Planck\ 2018 release data further test these ideas.

This paper is organized as follows. In Sect.~\ref{sec:data} we 
describe the statistical methodology, 
the \Planck\ likelihoods, and the complementary data sets used 
in this paper. In Sect.~\ref{sec:like_updates} 
we discuss the updated constraints on the spectral index of the scalar 
perturbations, on spatial curvature, and on the tensor-to-scalar ratio. 
Section~\ref{sec:single} is devoted to constraining slow-roll 
parameters and to a Bayesian model comparison of 
inflationary models, 
taking into account the uncertainties 
in connecting the inflationary expansion to the subsequent 
big-bang thermalized era. 
In Sect.~\ref{sec:taylor} the potential for standard single-field inflation is reconstructed 
using two different methodologies. 
Section~\ref{sec:reconstruction} describes the primordial power 
spectrum reconstruction using three different approaches.
In Sect.~\ref{sec:parametrized}, the parametric search for 
features in the primordial scalar power spectrum is described, 
including a dedicated study of the axion monodromy model. In 
Sect.~\ref{sec:combined}, the \Planck\ power spectrum data are
combined with information from the \Planck\ bispectrum in a search 
for oscillations in the primordial spectra. The constraints 
on isocurvature modes are summarized in Sect.~\ref{sec:isocurvature}. 
Section~\ref{sec:anisotropic} updates and extends the constraints 
on anisotropic inflationary models of inflation. We summarize our 
conclusions in Sect.~\ref{sec:conclusions}, highlighting 
the key results and the legacy of \Planck\ for inflation.


\section{Methodology and data \label{sec:data}}

The general theoretical background and analysis methods applied in this paper closely 
match those of the previous \Planck\ inflation papers~\citepalias{planck2013-p17,planck2014-a24}. 
Consequently, in this section we provide only a brief summary of the methodology and 
focus on changes in the \Planck\ likelihood relative to previous releases.

\subsection{Cosmological models and inference}

For well over a decade, the base-$\Lambda$CDM model has been established as the simplest 
viable cosmological model. Its six free parameters can be
divided into primordial and late-time parameters. The former describe the state of 
perturbations on observable scales (corresponding to a wavenumber range of 
$10^{-4}\,\mathrm{Mpc}^{-1} \lesssim k \lesssim 10^{-1}\,\mathrm{Mpc}^{-1}$ today) prior 
to re-entering the Hubble radius around recombination.  In base $\Lambda$CDM, the initial 
state of perturbations is assumed to be purely adiabatic and scalar, with the spectrum of 
curvature perturbations given by the power law
\begin{equation}
\ln \mathcal{P_R}(k) = \ln A_\mathrm{s}  + (n_\mathrm{s} - 1) \ln(k/k_*) \equiv \ln \mathcal{P}_0(k),
\end{equation}
where $k_*$ denotes an arbitrary pivot scale. The late-time parameters, on the other hand, 
determine the linear evolution of perturbations after re-entering the Hubble radius. By 
default we use the basis 
$(\omega_\mathrm{b},\omega_\mathrm{c},\theta_\mathrm{MC},\tau)$\footnote{Refer to 
Table~\ref{table:CPDefinitions} for definitions.} for the late-time parameters, but 
occasionally also consider non-minimal late-time cosmologies.
Because of the inflationary perspective of this paper, we are mainly interested
in exploring modifications of the primordial sector and their interpretation in terms of
the physics of the inflationary epoch.

{
\begin{table*}[h]
\begingroup
\newdimen\tblskip \tblskip=5pt
\caption{Baseline and optional late-time parameters, primordial power spectrum parameters, and slow-roll parameters. All primordial quantities are evaluated at a pivot scale of $k_*=0.05\,{\mathrm{Mpc}}^{-1}$, unless otherwise stated.}
\label{table:CPDefinitions}
\nointerlineskip
\vskip 3mm
\footnotesize
\setbox\tablebox=\vbox{
      \newdimen\digitwidth
      \setbox0=\hbox{\rm 0}
      \digitwidth=\wd0
      \catcode`"=\active
      \def"{\kern\digitwidth}
      \newdimen\signwidth
      \setbox0=\hbox{+}
      \signwidth=\wd0
      \catcode`!=\active
      \def!{\kern\signwidth}
\halign{\tabskip=0pt\hbox to 1.5in{$#$\leaderfil}\tabskip=3em&
 #\hfil\tabskip=0pt\cr
\noalign{\doubleline}
\omit\hfil Parameter\hfil&\omit\hfil Definition\hfil\cr
\noalign{\vskip 3pt\hrule\vskip 5pt}
\omega_\mathrm{b} \equiv \Omega_\mathrm{b} \, h^2 & Baryon density today\cr
\omega_\mathrm{c} \equiv \Omega_\mathrm{c} \, h^2  & Cold dark matter density today\cr
\theta_\mathrm{MC}& Approximation to the angular size of sound horizon at
 last scattering\cr
\tau& Optical depth to reionization\cr
\noalign{\hrule\vskip 2pt}
N_\mathrm{eff}& Effective number of neutrino species \cr
\Sigma m_\nu& Sum of neutrino masses\cr
\Omega_K& Spatial curvature parameter\cr
w_0 & Dark energy equation of state parameter \cr
\noalign{\hrule\vskip 2pt}
A_{\mathrm{s}}& Scalar power spectrum amplitude \cr
n_\mathrm{s}& Scalar spectral index \cr
d n_\mathrm{s}/d \ln k& Running of scalar spectral index \cr
d^2 n_\mathrm{s}/d \ln k^2& Running of running of scalar
 spectral index \cr
r& Tensor-to-scalar power ratio \cr
n_\mathrm{t}& Tensor spectral index \cr
\noalign{\hrule\vskip 2pt}
\epsilon_1 = - {\dot H}/{H^2}& First Hubble slow-roll parameter\cr
\epsilon_{n+1}={\dot \epsilon_n}/(H \epsilon_n)& $(n+1)$st Hubble slow-roll parameter ($n \ge 1$)\cr
\epsilon _V={M_\mathrm{pl}^2 V_\phi^2}/(2 V^2)& First potential slow-roll parameter, where $\vphantom{X}_\phi \equiv \mathrm{d}/\mathrm{d}\phi$\cr
\eta _V={M_\mathrm{pl}^2 V_{\phi \phi}}/V& Second potential slow-roll parameter\cr
\xi^2_V={M_\mathrm{pl}^4 V_{\phi} V_{\phi \phi \phi}}/{V^2}& Third potential slow-roll
 parameter\cr
\varpi^3_V={M_\mathrm{pl}^6 V_{\phi}^2 V_{\phi \phi \phi \phi}}/{V^3}&
 Fourth potential slow-roll parameter\cr
\noalign{\vskip 2pt\hrule\vskip 3pt}}}
\endPlancktable
\endgroup
\end{table*}

}

Perturbations produced by generic single-field slow-roll models of inflation are typically 
well approximated by the following form of the adiabatic scalar and tensor components:
\begin{align}
\label{eq:powerlaw} \ln \mathcal{P_R}(k) =& \ln \mathcal{P}_0(k)
+ \frac{1}{2} \frac{d \ln n_\mathrm{s}}{d \ln k} \ln(k/k_*)^2\nonumber\\
&\quad + \frac{1}{6} \frac{d^2 \ln n_\mathrm{s}}{d \ln k^2} \ln(k/k_*)^3  + \ldots ,\\
\ln \mathcal{P}_\mathrm{t}(k) =& \ln (r A_\mathrm{s}) + n_\mathrm{t} \ln(k/k_*) + \ldots  ,
\end{align}
which allows for a weak scale dependence of the scalar spectral index, $n_\mathrm{s}$, 
modelled by a running, ${d \ln n_\mathrm{s}}/{d \ln k}$, or a running 
of the running, ${d^2 \ln n_\mathrm{s}}/{d (\ln k)^2}$.\footnote{Unless 
explicitly stated otherwise, we adopt a default pivot scale $k_* = 0.05\,$Mpc$^{-1}$ in 
this work. As in previous \Planck\ releases, we will also quote the tensor-to-scalar 
ratio $r_{0.002}$ at $k_* = 0.002\,$Mpc$^{-1}$ in order to facilitate comparison with 
earlier primordial tensor-mode constraints.}  The power spectrum parameterization in 
Eq.~\ref{eq:powerlaw} can be extended to address wider classes of inflation-related 
questions, (e.g., the search for isocurvature perturbations, specific primordial features 
in the spectra, etc.),
as described in subsequent sections.
We also go beyond simple functions to parameterize the primordial power spectrum. In the 
spirit of reconstructing the primordial spectrum from the data,
we consider some general parameterizations (e.g., taking the power spectrum as an interpolation 
between knots of freely varied amplitudes at fixed or varying wavenumbers).

One could argue that the primordial power spectra are merely intermediate quantities and 
assess theories directly from more fundamental parameters.
By using the slow-roll approximation, or by evolving the mode equations to obtain exact 
numerical predictions for the spectra without
resorting to the slow-roll approximation, we can relate the primordial perturbations to 
the dynamics of the Hubble parameter during inflation or to the inflaton potential and its derivatives,
thus constraining these quantities directly.

For any given model, theoretical predictions of CMB-related and other cosmological observables are calculated
using appropriately modified versions of the Boltzmann codes \texttt{CAMB}~\citep{CAMB} or \texttt{CLASS}~\citep{Blas:2011rf}.
As in \citepalias{planck2013-p17,planck2014-a24}, we compare models 
$\mathcal{M}_1$ and $\mathcal{M}_2$ by the difference in the logarithm of the likelihood of their best fits,
or {\it effective} $\Delta \chi ^2 \equiv 2 \left[\ln \mathcal{L}_{\text{max}}(\mathcal{M}_1) - \ln \mathcal{L}_{\text{max}}(\mathcal{M}_2)\right]$.
We apply Bayesian statistical methods to infer the posterior probability distributions of the model parameters and select
between competing models~\citep{Trotta:2008qt}, using either the Metropolis-Hastings Markov-chain Monte Carlo (MCMC) sampling
algorithm, as implemented in \texttt{CosmoMC}~\citep{cosmomc} and \texttt{Monte Python}~\citep{Audren:2012wb}, or software based on
nested sampling~\citep{2004AIPC..735..395S},
such as \texttt{MultiNest}~\citep{Feroz:2008xx,Feroz:2013hea} or \texttt{PolyChord}~\citep{Handley2015a,Handley2015b}.
The latter can simultaneously estimate the Bayesian evidence $\mathcal{E}_i$
of a model $\mathcal{M}_i$, allowing the comparison between different models via
the Bayes factor, \mbox{$B = \mathcal{E}_2/\mathcal{E}_1$}, where $\left| \ln B \right| > 5$ is commonly considered ``strong'' evidence
for or against the respective model~\citep{Jeffreys1998BK,Trotta:2005ar}.

\subsection{Data}

\subsubsection{\Planck\ data}

The \Planck\ data processing has
improved in a number of key aspects with respect to the
previous 2015 cosmological release. We briefly summarize the main points here,
referring the interested reader to \cite{planck2016-l02} and \cite{planck2016-l03} for details.

The flagging procedure in the LFI~2018 pipeline has been made more
aggressive, in particular for the first 200 operational days. However, the most important 
improvement in the LFI pipeline is in the calibration approach. 
Whereas in the 2015 release, the main calibration source for LFI was the \Planck\ orbital dipole 
(i.e., the amplitude modulation of the CMB dipole induced by the satellite orbit) of each 
single radiometer model, the 2018 procedure also includes the Galactic emission along with 
the orbital dipole in the calibration model and becomes iterative~\citep{planck2016-l02}.

The HFI data for the 2018 release have also been made more conservative, cutting 
approximately 22 days of observations under non-stationary conditions with respect  to 2015. 
The main change in the HFI data processing is the use of a new map making and calibration 
algorithm called {\tt SRoll}, whose first version was introduced in \cite{planck2014-a10} for 
the initial analysis of HFI polarization on large angular scales. This algorithm employs 
a generalized polarization destriper which uses the redundancy in the data to
extract
several instrumental systematic-effect parameters directly from the data \citep{planck2016-l03}.

These improvements have a
minor impact on \Planck\ temperature maps, but are much more important for polarization, 
particularly on large angular scales, allowing, for instance, the removal of the high-pass 
filtering adopted in the 2015 study of isotropy and statistics~\cite{planck2014-a18}.

In the following, we summarize the essentials of the \Planck\ inputs used in this paper 
(i.e., the \Planck\ likelihood approach to the information contained in the 2-point 
statistics of the temperature and polarization maps
and the \Planck\ CMB lensing likelihood). As for previous cosmological releases, the 
\Planck\ likelihood approach is hybridized between low- and high-multipole regions, which 
therefore are summarized separately below. We refer the interested reader to the relevant 
papers \citet{planck2016-l05} (henceforth \citetalias{planck2016-l05}) and 
\citet{planck2016-l08} (henceforth \citetalias{planck2016-l08}) for a more complete 
description of these inputs.

\subsubsection*{\Planck\ low-$\ell$ likelihood}

As in the \Planck~2015 release, several options are available for evaluating the 
temperature likelihood on large angular scales, each with its own computational complexity 
and approximations. One option is based on the {\tt Commander} framework and implements 
full Bayesian sampling of an explicit parametric model that includes both the cosmological 
CMB signal and non-cosmological astrophysical signals, such as thermal dust, CO, and 
low-frequency foregrounds. This framework is described in earlier papers [see 
\citet{planck2014-a13} and \citet{planck2014-a10} and references therein for details]. 
The only changes since the 2015 implementation concern the data and model selection. As 
described in \citet{planck2016-l04}, we only use the \Planck~2018 data in the current data 
release, whereas the previous 2015 version additionally included WMAP \citep{bennett2012} 
and Haslam \citep{haslam1982} observations. With fewer frequencies available, this requires 
a simpler model, and in particular we now fit for only a single low-frequency foreground 
component, rather than individual synchrotron, free-free, and spinning dust emission 
components, and we only fit a single CO component, rather than for individual CO~line 
components at 100, 217, and 353\,GHz. On the one hand, this results in fewer internal 
foreground degeneracies compared to the 2015 version, and a likelihood that only depends 
on \Planck\ data, but at the same time the simpler foreground modelling also requires a 
slightly larger Galactic mask. Overall, the two versions are very compatible in terms of 
the recovered CMB power spectra, as discussed in \citetalias{planck2016-l05}. For additional 
details on the {\tt Commander} temperature analysis, see \citet{planck2016-l04}.




\vspace{.3cm}

The HFI low-$\ell$ polarization likelihood is based on the
full-mission HFI 100-GHz and 143-GHz $Q$ and $U$ low-resolution maps
cleaned through a template-fitting procedure with LFI 30\,GHz and HFI
353\,GHz information\footnote{The polarized synchrotron
component is fitted only at 100\,GHz, being negligible at 143\,GHz.
For the polarized dust component, following the
prescription in \citet{planck2016-l03}, the low-$\ell$ HFI
polarization likelihood used the 353-GHz map constructed
only from polarization-sensitive bolometers.} used
as tracers of polarized synchrotron and thermal dust, respectively
(see \citetalias{planck2016-l05} for details
about the cleaning procedure). The likelihood method, called {\tt
SimAll}, represents a follow-up of the {\tt SimBaL}
algorithm presented in \citet{planck2014-a10} and uses the FFP10
simulations to construct empirically the probability
for the $EE$ and $BB$ spectra. The method is based on the quadratic
maximum likelihood estimation of the cross-spectrum
between 100 and 143\,GHz, and its multipole range spans from $\ell=2$
to $\ell=29$. We only built the likelihood for $EE$ and $BB$ and not
for $TE$, due to the poor statistical consistency of the $TE$ spectrum
for $\ell>10$, and due to the difficulty of describing accurately the
correlation with $TT$ and $EE$, given the limited number of simulations available; 
see discussions in section 2.2.6 of \citetalias{planck2016-l05}. Further
details about the method and consistency tests are presented in
\citetalias{planck2016-l05}. When combined with the low-$\ell$
temperature likelihood (based on the {\tt Commander} CMB solution),
the low-$\ell$ polarization likelihood implies
$\tau=0.051\pm0.009$ and $r_{0.002}<0.41$ at 95\,\% CL.



\vspace{.3cm}

As an alternative to
the {\tt Commander} and {\tt SimAll} low-$\ell$ likelihood, an update of the joint 
temperature and polarization pixel-based low-$\ell$ LFI likelihood used in 2015 is 
part of this \Planck\ data release. Its methodology (see \citetalias{planck2016-l05} 
for details) is similar to that of 2015 \citep{planck2014-a13}, i.e., a pixel-based 
approach in $TQU$ at $N_\mathrm{side}=16$, and employs the {\tt Commander} solution in 
temperature along  with the LFI 70-GHz linear polarization maps, foreground cleaned 
using the \Planck\ 30-GHz and 353-GHz channels as tracer templates for synchrotron 
and dust, respectively. This 2018 version allows for a larger sky fraction in 
polarization (66.4\,\%, compared to the previous 46\,\%) and retains the sky surveys 2 and 
4 that were excluded in 2015. By performing a two-parameter estimate for $A_\mathrm{s}$ 
and $\tau$ restricted to $\ell < 30$, we find using this likelihood that 
$\tau = 0.063 \pm 0.020$ and $\mathrm{ln(10^{10}A_{\rm s})}= 2.975 \pm 0.056$ at 
68\,\% CL. The latter value has been derived by varying the $TT$, $EE$, and $TE$ CMB spectra.

\subsubsection*{\Planck\ high-$\ell$ likelihood}

The 2018 baseline high-$\ell $ likelihood ({\tt Plik}) is an update of the 2015 baseline version. The 
{\tt CamSpec} likelihood is also used to explore alternative data cuts and modelling of the 
data and is described below. Both approaches implement a Gaussian likelihood approximation using cross-spectra 
between the 100-, 143-, and 217-GHz maps.  {\tt Plik} covers the multipoles $30\leq \ell \leq 2509$ in temperature and 
$30\leq \ell \leq 1997$ in polarization (i.e., for $TE$ and $EE$). In order to avoid noise bias, the high-$\ell$ 
likelihood relies only on half-mission map cross-spectra, which have been demonstrated to be largely free of 
correlated noise. The spectra are computed on masked maps in order to reduce the anisotropic Galactic 
contamination (dominated by dust emission), and in the case of $TT$ also strong point sources and CO 
emission. The {\tt Plik} masks, identical to the 2015 masks, are tailored to each frequency channel and 
differ in temperature and polarization to take into account differing foreground behaviour and channel beams. 
The {\tt Plik} intensity (polarization) masks effectively retain 66, 57, and 47\,\% (70, 50, and 41\,\%) of the sky after apodization 
for the 100, 143, and 217 GHz channels, respectively.
Unlike in 2015, when the map beams were computed for an average sky fraction, they 
are now computed for the exact sky fraction used at each frequency. 
The data vector used in the  
likelihood approximation discards multipoles 
that are 
highly 
contaminated by foregrounds
or have low signal-to-noise ratios. 

The {\tt Plik} power spectra are binned using the same scheme 
as in 2015. Unbinned likelihoods are also available. When forming the data vector, individual 
cross-frequency spectra are not co-added. This allows for independent exploration
of the calibration, nuisance, and foreground parameter 
space for each cross-spectrum using dedicated templates in the theory vector. 

The {\tt Plik} (and {\tt CamSpec}) covariance matrices are computed for a fixed fiducial CMB including 
the latest estimate of the foreground and systematics, which are all assumed to be Gaussian. As verified in 2015, 
after the masks have been applied this is a reasonable assumption.
The covariance matrix computation uses an approximation to account for 
mask-induced correlations.  {\tt Plik} uses only the large Galactic mask in the analytic computation and then takes extra correlations 
due to the point-source mask into account using a Monte Carlo estimate 
of the extra variance induced. Missing pixels are ignored in the covariance.  In 2015 its was shown 
that this approach (i.e., Gaussian approximation and approximate covariance) induced only a less than $0.1\sigma$ bias on 
$n_{\rm s}$ (from the $30<\ell <100$ modes).

The {\tt Plik} noise model has been re-estimated on the latest HFI maps 
using the same methodology as in 2015, based on a comparison between noise-biased 
auto-spectra and cross-spectra. This procedure avoids correlated glitch residuals, which had biased 
previous noise estimates \citep{planck2014-a13}, particularly in polarization at $\ell \lesssim 500$. 

The 2018 HFI data processing pipeline has refined the maps used in the likelihood relative to 2015.  For example, an improved destriping 
procedure reduced the residual scatter in the polarization 
maps, in particular at 143\,GHz (yielding about $12\,\%$ lower noise on the half-mission cross-spectrum). More stringent selection cuts resulted in the 
discarding of the last 1000 rings of data, increasing the noise in the temperature half-mission spectrum by about $3\,\%$.  Also, a higher 
threshold was imposed on the conditioning of the $TQU$ intra-pixel noise covariance matrix for a pixel to be considered
well-measured, resulting in more missing pixels relative to 2015.

The data modelling has also significantly improved, in particular for 
polarization, making cosmological constraints from polarization more reliable. 
In 2015, the polarization spectra 
($TE$ and $EE$) displayed relatively large inter-frequency disagreements. A plausible explanation (at least for $TE$) 
was the temperature-to-polarization leakage induced by beam and calibration 
differences (so-called ``beam leakage''). The beam-leakage modelling 
has improved substantially in 
2018 \citep{Hivon:2016qyw}
so that we can now propagate the beams, gain differences, polarization angles, etc. to 
compute a reliable template for the beam leakage and thus remove these leakage effects. These improvements
substantially 
reduce the 
$TE$ 
inter-frequency disagreements.

We also reassessed the estimates of the polarization efficiency for the 
polarized channels. Comparing different data-based estimates demonstrates that the ground-based 
polarization efficiency uncertainty estimates (of the order of a fraction of a percent) were too optimistic by 
a factor of 5 to 10. Correcting for the observed polarization efficiency errors (at the percent level) very 
significantly reduces the $EE$ inter-frequency disagreements. This calibration correction relies on cosmological 
priors (using the $TT$ best-fit cosmology).  Calibrating using either the $TE$ or the $EE$ 
spectra yields generally consistent results, except at 143\,Ghz where there is disagreement at more than $2\sigma$. 
At this level, this discrepancy can be caused either by a statistical fluctuation, or by an unknown residual.

The {\tt Plik} baseline likelihood implements a map-based calibration. The $TE$ calibration is deduced from the $TT$ and $EE$ calibrations,
including at 143\,GHz. 
Other improvements over the 2015 version are the following.
The dust model has been 
improved in temperature and polarization, using also the latest version of the 353-GHz maps. The level of 
synchrotron contamination in the 100-GHz and 143-GHz maps has been estimated and shown to be negligible. 
Sub-pixel noise has been included in $TT$ and $EE$ (and demonstrated to have a negligible effect on the cosmological 
parameters). Finally, a correlated component of the noise has been observed in the end-to-end HFI simulation, affecting the 
large scales and very small scales of the $EE$ auto-frequency spectra. The large-scale contribution 
affects the dust correction and the $n_{\rm s}$ constraints. We constructed an empirical model of this correlated noise 
from our simulations, which is included in the {\tt Plik} likelihood.

\vspace{.5cm}

{\tt CamSpec} was the baseline for the 2013 release and was described in detail in \citet{planck2013-p08}, and used cross-spectra formed 
from detector-set temperature maps using data from the nominal mission period.  It was extended for the 2015 release 
to include both polarization and temperature-polarization 
cross-spectra and to use the data from the full mission period.  Similarly to {\tt Plik},
{\tt CamSpec} switched from detector-set cross-spectra to cross-spectra formed from frequency maps constructed from separate halves of the full 
mission data in order to mitigate the effects of noise correlated between 
detectors. In 2015, the foreground modelling was also modified and the sky fraction retained at each frequency was 
increased, using common masks with {\tt Plik} in temperature.   {\tt CamSpec} used a more conservative mask in polarization than {\tt Plik}.

Differently from {\tt Plik}, {\tt CamSpec} corrects each $TE$ and $EE$ cross-frequency spectrum with a fixed dust and temperature-to-polarization
leakage template before co-adding them to form the $EE$ and $TE$ components of its data vector and bases 
its noise estimate on differences between maps constructed using alternating pointing periods.
Note also that {\tt CamSpec} uses an individual-spectrum-based calibration scheme, where the $TE$ calibrations 
are not fixed to be those inferred from the $TT$ and $EE$ ones. 

In the 2018 release further improvements in the {\tt CamSpec} foreground modelling have been implemented. The dust model in temperature 
has been updated in a way similar to {\tt Plik}.  {\tt CamSpec}
now uses a richer model of the cosmic 
infrared background, allowing for the exploration of any impact on the cosmological parameters. As explained above, the 
noise modelling was also modified. 
Further modifications of the masking have been made for polarization, still using the same masks for each frequency channel, 
but different from the {\tt Plik} mask. 
As we discussed above, beam-leakage and polarization-efficiency corrections are applied to the individual 
polarization spectra before their addition for inclusion in the 
likelihood. More details on the {\tt Plik} and {\tt CamSpec} likelihoods can be found in \citetalias{planck2016-l05}.

\vspace{.5cm}

As in 2015, the high-$\ell$ {\tt Plik} and {\tt CamSpec} likelihoods are in excellent agreement for 
temperature. The different assumptions  for the 
polarization-efficiency parameters and the masks \citep{planck2014-a13} propagate into differences in 
cosmological parameter estimates. For the baseline $\Lambda$CDM model, the difference in cosmological 
parameters between the {\tt Plik} likelihood and the {\tt CamSpec} 
likelihood (using joint TT,TE,EE in combination with {\tt Commander}, {\tt 
SimAll}, and lensing) is at most $0.5\sigma$ (for $\Omega_b h^2$) (\citealt{planck2014-a15}, henceforth \citetalias{planck2014-a15}). Similar differences 
in cosmological parameters occur in extended cosmological models. The differences between the {\tt 
Plik} and {\tt CamSpec} parameters are dominated by calibration-model differences for the joint TT,TE,EE 
and TE-only cases, and by mask differences for the EE-only case.  To a large extent, the 
{\tt CamSpec} results can be reproduced within the {\tt Plik} framework simply by changing in {\tt Plik} the 
calibration model (for $TE$) and the polarization mask (for $EE$). Below
we use {\tt Plik} as the \Planck\ baseline high-$\ell$ likelihood. {\tt CamSpec} results are used 
to assess the residual uncertainty from modelling and mask choices.
We quote values obtained with {\tt CamSpec} only for a few cases.

\vspace{.5cm}

We use the following conventions for naming the \Planck\ likelihoods: 
(i)~\Planck~TT+lowE denotes the combination of the high-$\ell$ TT likelihood at 
multipoles $\ell \ge 30$ and the low-$\ell$ temperature-only {\tt Commander} likelihood, 
plus the {\tt SimAll} low-$\ell$ EE-only likelihood in the range $2 \le \ell \le 29$;
(ii)~\Planck~TE and \Planck~EE denote the TE and EE likelihood at $\ell \ge 30$, respectively; 
(iii)~\Planck~TT,TE,EE+lowE denotes the combination of the combined likelihood using $TT$, $TE$, 
and $EE$ spectra at $\ell \ge 30$, the low-$\ell$ temperature {\tt Commander} likelihood, and 
the low-$\ell$ {\tt SimAll} EE likelihood; and (iv)~\Planck~TT,TE,EE+lowP denotes the combination 
of the likelihood using $TT$, $TE$, and $EE$ spectra at $\ell \ge 30$ and the alternative joint temperature-polarization 
likelihood at $2 \le \ell \le 29$, based on the temperature {\tt Commander} map and the 70-GHz polarization map.
Unless otherwise stated, high-$\ell$ results are based on the \texttt{Plik} likelihood and 
low-$\ell$ polarization information is based on {\tt SimAll}.

\subsubsection*{\Planck\ CMB lensing likelihood}

The \Planck\ 2018 lensing likelihood, presented in \citetalias{planck2016-l08}, uses the 
lensing trispectrum to estimate the power spectrum of the lensing potential $C_L ^{\phi\phi}$. 
This signal is extracted using a minimum-variance combination of a full set of temperature- 
and polarization-based quadratic lensing estimators \citep{Okamoto:2003zw} applied to the 
{\tt SMICA} CMB map over approximately $70\,\%$ of the sky using CMB multipoles 
$100\le \ell \le 2048$, as described in \citetalias{planck2016-l08}. We use the
lensing bandpower likelihood, with bins spanning lensing multipoles $8 \le L \le 400,$ which 
has been validated with numerous consistency tests.
Because its multipole range has been extended down to $L=8$ (compared to $L=40$ for the 
\Planck\ 2015 analysis), the statistical power of the lensing likelihood used here is slightly greater.


\subsubsection{Non-\Planck\ data}

While the data derived exclusively from \Planck\ observations are by themselves already 
extremely powerful at constraining cosmology, external data sets can provide helpful 
additional information.  The question of consistency between \Planck\ 
and external data sets is discussed in detail in \citet{planck2016-l06} (henceforth \citetalias{planck2016-l06}).
Here we focus on two data sets that are particularly useful for breaking degeneracies and whose 
errors can be assessed reliably. We consider the measurement of the CMB $B$-mode 
polarization angular power spectrum by the BICEP2/Keck~Array collaboration and 
measurements of the baryon acoustic oscillation (BAO) scale.  The supplementary $B$-mode 
data provide independent constraints on the tensor sector, which are better than those 
that can be derived from the \Planck\ data alone (based on the shape of the scalar power 
spectrum).
The BAO data, on the other hand, do not directly constrain the primordial perturbations.  
These data, however, provide invaluable low-redshift information that better constrain the 
late-time cosmology, especially in extensions of $\Lambda$CDM, and thus allow degeneracies 
to be broken.

\subsubsection*{BICEP2/Keck Array 2015 $B$-mode polarization data}

Although \Planck\ measured the CMB polarization over the
full sky, its polarization sensitivity in the cosmological frequency channels is not sufficient
to compete with current suborbital experiments surveying small, particularly low-foreground patches of the
sky very deeply using many detectors.
In \citetalias{planck2014-a24}, constraints on $r$ using the joint BICEP2/Keck Array and
\Planck\ (BKP) analysis \citep{pb2015} were reported.
Here we make use of the most recent $B$-mode polarization data available from the
analysis of the BICEP2/Keck field (\citealt{bk15},
henceforth \citetalias{bk15}), unless otherwise stated.
The BK15 likelihood draws on data from the new Keck array
at 220\,GHz in addition to those already in use for the \citetalias{Array:2015xqh} \citep{Array:2015xqh}
likelihood, i.e., the 95 and 150-GHz channel,
as well as from \Planck\ and WMAP to remove foreground contamination.
The BK15 observations measure $B$-mode polarization
using 12 auto- and 56 cross-spectra between
the BICEP2/Keck maps at 95, 150, and 220\,GHz, the
WMAP maps at 23 and 33\,GHz, and the \Planck\ maps at 30, 44, 70, 100, 143, 217, and 353\,GHz,
using nine bins in multipole number. By using $B$-mode information only within the BK15 likelihood,
a 95\,\% upper limit of $r < 0.07$ is found \citepalias{bk15}, 
which improves on the corresponding 95\,\% CL $r < 0.09$ \citepalias{Array:2015xqh} based on the BK14 likelihood.

\subsubsection*{Baryon acoustic oscillations}

Acoustic oscillations of the baryon-photon fluid prior to recombination are responsible for the
acoustic peak structure of the CMB angular power spectra.  The counterpart to the CMB acoustic 
peaks in the baryon distribution are the BAOs, which remain imprinted into the matter distribution 
to this day.  In the position-space picture, the BAOs of the power spectrum correspond to a peak 
in the correlation function, defining a characteristic, cosmology-dependent length scale that 
serves as a standard ruler and can be extracted (e.g., from galaxy redshift surveys). The transverse 
information of a survey constrains the ratio of the comoving angular diameter distance and the 
sound horizon at the drag epoch (i.e., when the baryon evolution becomes unaffected by coupling 
to the photons), $D_M/r_\mathrm{d}$, whereas the line-of-sight information yields a measurement 
of $H(z) r_\mathrm{d}$.  Sometimes, these two observables are combined to form the direction-averaged 
quantity $D_V/r_\mathrm{d} \equiv \left[ cz D_M^2(z) H^{-1}(z) \right]^{1/3}/r_\mathrm{d}$.

For our BAO data compilation, we use the measurements of $D_V/r_\mathrm{d}$ from the 6dF survey at an 
effective redshift \mbox{$z_\mathrm{eff} = 0.106$}~\citep{Beutler:2011hx}, and the SDSS Main 
Galaxy Sample at \mbox{$z_\mathrm{eff} = 0.15$}~\citep{Ross:2014qpa}, plus the final interpretation 
of the SDSS III DR12 data~\citep{Alam:2016hwk}, with separate constraints on $H(z) r_\mathrm{d}$ 
and $D_M/r_\mathrm{d}$ in three correlated redshift bins at $z_\mathrm{eff} = 0.38$, $0.51$, and 
$0.61$.  In~\cite{Addison:2017fdm}, the same set of BAO data combined with either non-\textit{Planck} 
CMB data or measurements of the primordial deuterium fraction was shown to favour a cosmology 
fully consistent with, but independent of, \textit{Planck} data.


\section{\Planck\ 2018 results for the main inflationary observables \label{sec:like_updates}}

As in \citetalias{planck2013-p17} and \citetalias{planck2014-a24}, we start by 
describing \Planck\ measurements of the key inflationary parameters. Some of the 
results reported in this section can be found in the Planck Legacy 
Archive.\footnote{\url{http://www.cosmos.esa.int/web/planck/pla}}

\subsection{Results for the scalar spectral index}

\begin{table*}[!h]
\begin{center}
\begin{tabular}{cccccc}
\noalign{\hrule\vskip 2pt} \noalign{\hrule\vskip 3pt}
Parameter & TT+lowE & EE+lowE & TE+lowE & TT,TE,EE+lowE & TT,TE,EE+lowE+lensing \\
\noalign{\vskip 1pt\hrule\vskip 2pt} $\Omega_{\mathrm{b}}h^2$& $ 0.02212 \pm 0.00022$ 
& $0.0240 \pm 0.0012$ & $0.02249 \pm 0.00025$ &$0.02236 \pm 0.00015$ & $0.02237 \pm 
0.00015$
\\
$\Omega_{\mathrm{c}}h^2$& $ 0.1206 \pm 0.0021$ & $ 0.1158 \pm 0.0046$ & $0.1177 \pm 
0.0020$ & $0.1202 \pm 0.0014$ & $0.1200 \pm 0.0012$
\\
$100\theta_{\mathrm{MC}}$& $1.04077 \pm 0.00047$ & $1.03999 \pm 0.00089$ & $1.04139 
\pm 0.00049$ & $1.04090 \pm 0.00031$ & $1.04092 \pm 0.00031$
\\
$\tau$ & $0.0522 \pm 0.0080$& $ 0.0527 \pm 0.0090$ & $0.0496 \pm 0.0085$ & 
$0.0544^{+0.0070}_{-0.0081}$ & $0.0544 \pm 0.0073$
\\
$\ln(10^{10} A_\mathrm{s})$& $3.040 \pm 0.016$& $3.052 \pm 0.022$ & 
$3.018^{+0.0020}_{-0.0018}$ &$3.045 \pm 0.016$ & $3.044 \pm 0.014$
\\
$n_\mathrm{s}$&$0.9626 \pm 0.0057$& $0.980 \pm 0.015$ & $ 0.967 \pm 0.011$ & $0.9649 
\pm 0.0044$ & $0.9649 \pm 0.0042$
\\
\noalign{\hrule\vskip 3pt} $H_0$&$ 66.88 \pm 0.92 \phantom{0}$ & 
$69.9 \pm 2.7 \phantom{0}$ & $68.44 \pm 0.91\phantom{0}$ & $ 67.27 \pm 0.60 
\phantom{0}$ & $67.36 \pm 0.54 \phantom{0}$
\\
$\Omega_{\mathrm{m}}$& $0.321 \pm 0.013$& $0.289^{+0.026}_{-0.033}$ & $0.301 \pm 
0.012$ & $0.3166 \pm 0.0084$ & $0.3153 \pm 0.0073$
\\
$\sigma_8$& $0.8118 \pm 0.0089$& $0.796 \pm 0.018$&
$ 0.793 \pm 0.011$ & $0.8120 \pm 0.0073$ & $0.8111 \pm 0.0060$ \\
\hline
\end{tabular}
\end{center}
\caption{Confidence limits for the cosmological parameters in the 
base-$\Lambda$CDM model from \Planck\ temperature, polarization, and 
temperature-polarization cross-correlation separately and combined, in combination 
with the EE measurement at low multipoles.} \label{tab:lambdaCDMTvsP}
\end{table*}

\Planck\ temperature data in combination with the $EE$ measurement at low multipoles 
determine the scalar spectral tilt in the $\Lambda$CDM model as
\begin{equation}
n_{\mathrm s} = 0.9626 \pm 0.0057 \quad (68\,\%\ \text{CL, \Planck\ TT+lowE}) \,. 
\label{eq:TT_ns}
\end{equation}
This result for $n_{\mathrm s}$ is compatible with the \Planck\ 2015 68\,\% CL value 
$n_{\mathrm s} = 0.9655 \pm 0.0062$ for \Planck\ TT+lowP \citepalias{planck2014-a15}. 
The slightly lower value for $n_{\mathrm s}$ is mainly driven by a corresponding shift in the 
average optical depth $\tau$, now determined as
\begin{equation}
\tau = 0.052 \pm 0.008 \quad (68\,\%\ \text{CL, \Planck\ TT+lowE}) \,, 
\label{eq:TT_tau}
\end{equation}
which is to be compared with the \Planck\ 2015 value \mbox{$\tau=0.078 \pm 0.022$} 
\citepalias{planck2014-a15}. This more precise determination of $\tau$ is due to 
better noise sensitivity of the HFI 100- and 143-GHz channels employed in the 
low-$\ell$ {\tt SimAll} polarization likelihood, compared to the joint 
temperature-polarization likelihood based on the LFI 70-GHz channel in 2015. Because 
of the degeneracy between the average optical depth and the amplitude of the 
primordial power spectrum, $A_{\mathrm s}$ and $\sigma_8$ are also lower than in the 
\Planck\ 2015 release. These shifts from the \Planck\ 2015 values for the 
cosmological parameters have been anticipated with the first results from the HFI 
large-angular polarization pattern \citep{planck2014-a10,planck2014-a25}.

The trend toward smaller values for ($n_{\mathrm s}$,$\tau$) with respect to the 
\Planck\ 2015 release also occurs for different choices for the low-$\ell$ 
likelihood. By substituting {\tt Commander} and {\tt SimAll} with the updated joint 
temperature-polarization pixel likelihood coming from the LFI 70-GHz channel, we obtain 
in combination with high-$\ell$ temperature data:
\begin{align}
n_{\mathrm s} & = 0.9650 \pm 0.0061  &(68\,\%\ \text{CL, \Planck\ TT+lowP}) \,; \\
\tau & = 0.072 \pm 0.016  &(68\,\%\ \text{CL, \Planck\ TT+lowP}) \,. 
\label{eq:TTplusbflike}
\end{align}
Although with larger errors, these latter results are consistent with the shifts 
induced by a determination of a lower optical depth than in the \Planck\ 2015 
release,\footnote{As in 2015, the combination with high-$\ell$ data pulls 
$\tau$ to larger values than the low-$\ell$ pixel likelihood alone,  
i.e. $\tau = 0.063 \pm 0.020$ at 68\,\% CL; see Sect.~\ref{sec:data}. This effect is less pronounced 
for the {\tt SimAll} likelihood. 
} as found in 
Eqs.~(\ref{eq:TT_ns}) and (\ref{eq:TT_tau}). Given this broad agreement and the consistency in 
the values of $\tau$ derived with {\tt SimAll} separately from the three 
cross-spectra \mbox{$70 \times 100$}, \mbox{$70 \times 143$}, and \mbox{$100 \times 
143$} \citepalias{planck2016-l05}, we will mainly use the baseline low-$\ell$ 
likelihood in the rest of the paper.

\begin{figure*}[!ht]
\includegraphics[width=18cm]{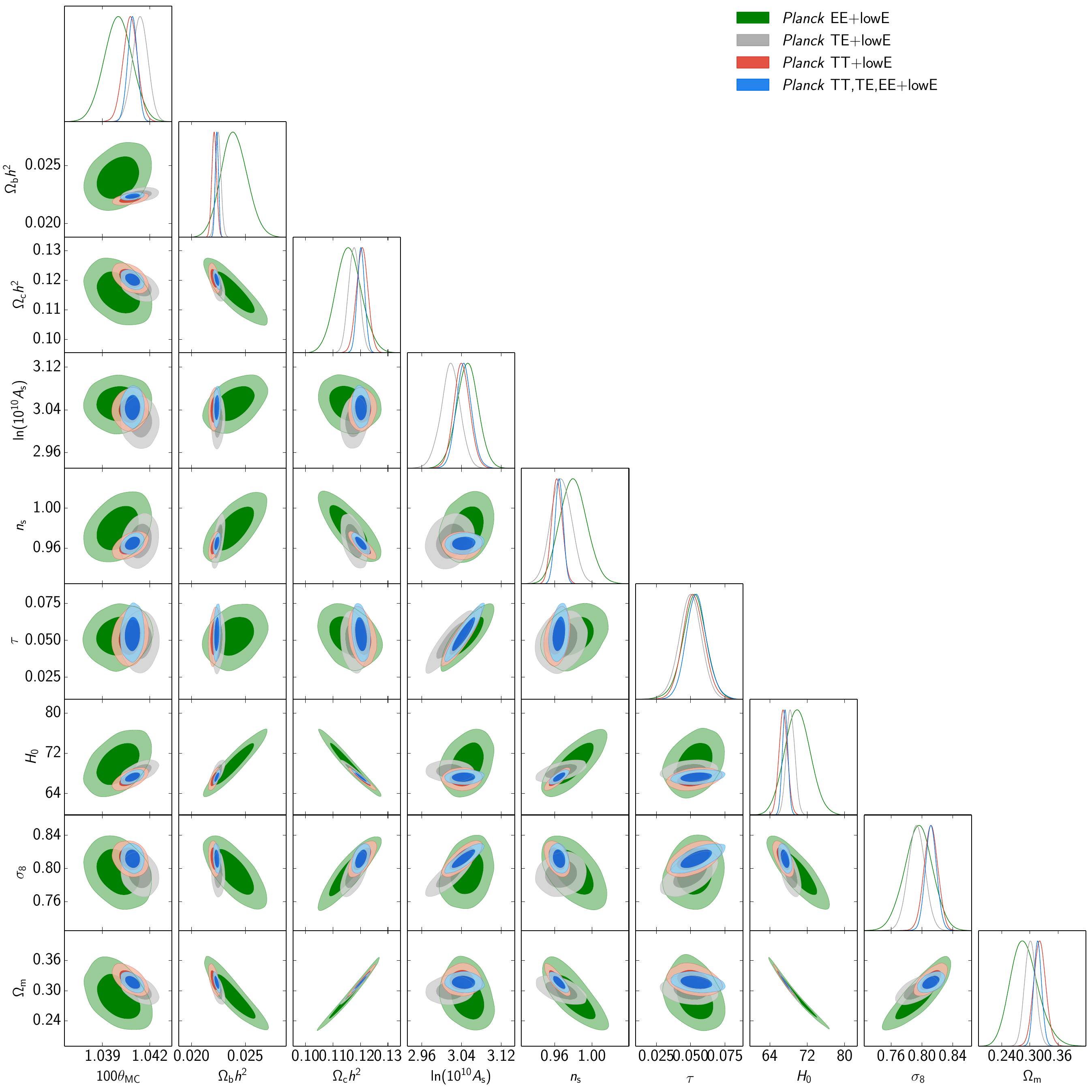}
\caption{Marginalized joint 68\,\% and 95\,\%~CL regions for the cosmological parameters 
in $\Lambda$CDM with \Planck\ TT, EE, TE, and joint TT,TE,EE, all in combination with the 
EE likelihood at low multipoles.} \label{fig:TvsP}
\end{figure*}

As anticipated in 2015, the information in the high-$\ell$ polarization \Planck\ data is 
powerful for breaking degeneracies in the parameters and to further decrease 
parameter uncertainties compared to temperature data alone. The addition of 
high-$\ell$ polarization leads to a tighter constraint on $n_{\mathrm s}$:
\begin{align}
n_{\mathrm s} &= 0.9649 \pm 0.0044 \nonumber \\
&\quad (68\,\%\ \text{CL, \Planck\ TT,TE,EE+lowE}). \label{eq:TTTEEE_ns}
\end{align}
This is in good agreement with the \Planck\ 2015 TT,TE,EE+lowP 68\,\% CL result, 
$n_{\mathrm s} = 0.9645 \pm 0.0049$. In this 2018 release the mean value of $n_{\mathrm 
s}$ is approximately  $0.5\sigma$ larger than the temperature result in 
Eq.~(\ref{eq:TT_ns}). This pull is mainly due to a higher value for the scalar tilt 
preferred by \Planck\ 2018 polarization and temperature-polarization 
cross-correlation data only:
\begin{equation}
n_{\mathrm s} = 0.969 \pm 0.009 \quad (68\,\%\ \text{CL, \Planck\ TE,EE+lowE}) \,.  
\label{eq:TEEE_ns}
\end{equation}
This pull is then mitigated in combination with temperature due to the 
larger uncertainty in the determination by TE,EE only. 
Similar considerations hold for the alternative {\tt CamSpec} high-$\ell$ likelihood, 
which leads to a 68\,\% CL result $n_{\mathrm s} = 0.9658 \pm 0.0045$, consistent with the 
baseline {\tt Plik} reported in Eq. (\ref{eq:TTTEEE_ns}).
Overall, the cosmological 
parameters from \Planck\ baseline temperature, polarization, and temperature-polarization 
cross-correlation separately and combined are very consistent, as can be seen from 
Table~\ref{tab:lambdaCDMTvsP} and Fig.~\ref{fig:TvsP} for the $\Lambda$CDM model.

\begin{figure*}[!ht]
\includegraphics[width=18cm]{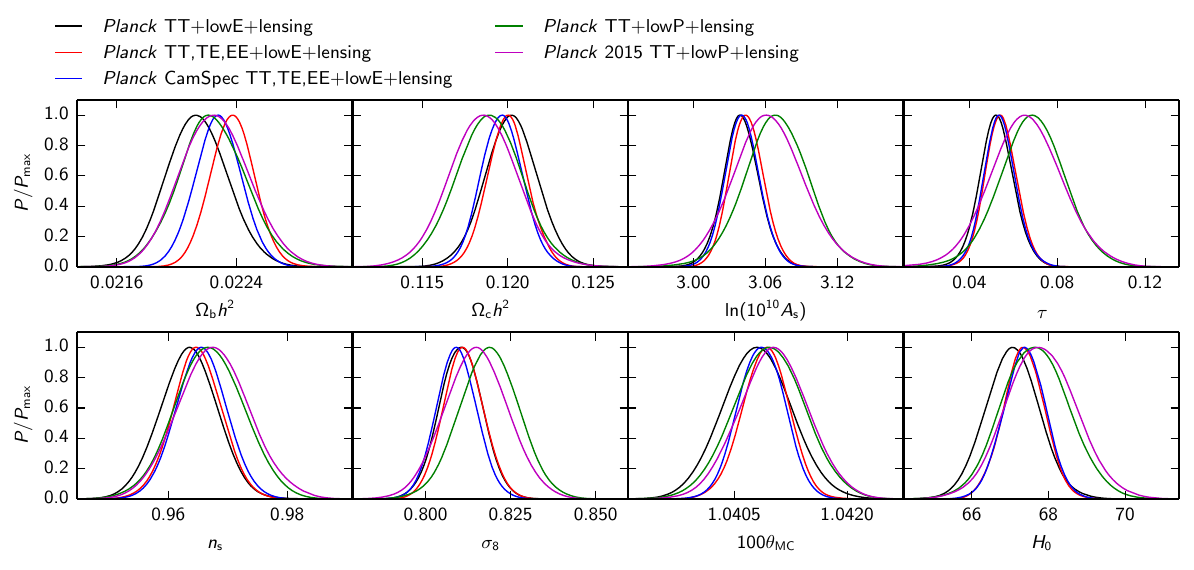}
\caption{Comparison of the marginalized probability density of the primary parameters and $(\sigma_8 \,, H_0)$ for the 
baseline cosmological model from \Planck\ TT+lowE+lensing (black curves), TT,TE,EE+lowE+lensing (red curves), and 
the alternative likelihood {\tt Camspec}. 
For comparison we also display the \Planck\ 2018 TT+lowP+lensing (blue curves) and the corresponding \Planck\ 2015 TT+lowP+lensing (green curves) results.}
\label{fig:2015vs2018} 
\end{figure*}

After combining with \Planck\ lensing, we obtain
\begin{align}
n_{\mathrm s} &= 0.9634 \pm 0.0048 \, \nonumber \\
&\quad (68\,\%\ \text{CL, \Planck\ TT+lowE+lensing}),\\
n_{\mathrm s} &= 0.9649 \pm 0.0042 \nonumber \\
&\quad (68\,\%\ \text{CL, \Planck\ TT,TE,EE+lowE+lensing}).
\end{align}
The shift in $n_{\mathrm s}$ (and, more generally, in the cosmological 
parameters of the base-$\Lambda$CDM model) obtained when \Planck\ lensing is combined with TT,TE,EE+lowE
is smaller than in 2015 because of the improved 
polarization likelihoods. The combination with lensing is, however, powerful for breaking 
parameter degeneracies in extended cosmological models, and, therefore, for this 2018 
release we will consider the full information contained in temperature, polarization,  
and lensing, i.e., TT,TE,EE+lowE+lensing, as the baseline \Planck\ data set.  
Figure~\ref{fig:2015vs2018} shows a comparison of the \Planck\ 2018 baseline results with 
those from alternative likelihoods and from the 2015 baseline for the $\Lambda$CDM cosmological parameters.

As in 2013 and 2015, BAO measurements from galaxy surveys are consistent with 
\Planck.  When BAO data are combined, we obtain for the base-$\Lambda$CDM 
cosmology:
\begin{align}
n_{\mathrm s} &= 0.9665 \pm 0.0038 \, \\
&\quad (68\,\%\ \text{CL, \Planck\ TT,TE,EE+lowE+lensing+BAO}) \,. \nonumber
\label{eq:Planck_BAO_ns}
\end{align}
The combination with BAO data decreases (increases) the marginalized value of $\Omega_{\rm c} 
h^2$ ($\Omega_{\rm b} h^2$) obtained by \Planck, and this effect is compensated for by a shift 
in $n_{\mathrm s}$ towards slightly larger values.

\subsection{Ruling out $n_\mathrm{s} = 1$}

One of the main findings drawn from previous \Planck\ releases was that the 
scale-independent Harrison-Zeldovich (HZ) spectrum \citep{Harrison:1969fb,Zeldovich:1972zz,1970ApJ...162..815P} is decisively ruled out. This 
conclusion is reinforced in this release: in standard $\Lambda$CDM late-time 
cosmology, the scalar spectral index from Table~\ref{tab:lambdaCDMTvsP} lies $6.6$, 
$8.0$, and $8.4\sigma$ away from $n_\mathrm{s} = 1$, for \Planck\ TT+lowE, \Planck\ TT,TE,EE+lowE, and 
\Planck\ TT,TE,EE+lowE+lensing, respectively. The corresponding effective $\Delta \chi^2$ between 
the power-law spectrum and the best-fit HZ model are $\Delta \chi^2 = 43.9$, 
$66.9$, and $72.4$.

Simple one-parameter modifications of the cosmological model are not sufficient to 
reconcile a scale-invariant power spectrum with \Planck\ data. For instance, when the 
effective number of neutrino species $N_\mathrm{eff}$ is allowed to float for a 
cosmology with a scale-invariant spectrum, the effective $\Delta \chi^2$ with respect 
to the power-law spectrum are $\Delta \chi^2 = 12.9$, $27.5$, and $30.2$, 
respectively.

When instead the assumption of flat spatial sections is relaxed,\footnote{For non-flat 
models the power spectra encode the eigenvalues of the corresponding Laplacian 
operator of the spatial sections and scale invariance holds at scales much smaller 
than the curvature radius.} we obtain effective $\Delta \chi^2$ values of 
$\Delta \chi^2 = 11.8$, $28.8$, and $40.9$, 
respectively, for the same data sets. Therefore, the corresponding closed 
cosmological models fitting \Planck\ TT+lowE ($\Omega_{K} = - 
0.122^{+0.039}_{-0.029}$, $H_0 = 44.2^{+3.1}_{-4.3}$\,km\,s$^{-1}$\,Mpc$^{-1}$ at 68\,\% 
CL),  \Planck\ TT,TE,EE+lowE ($\Omega_{K} = - 0.095^{+0.029}_{-0.019}$, $H_0 
= 47.1 \pm 3.2$\,km\,s$^{-1}$\,Mpc$^{-1}$ at 68\,\% CL), and \Planck\ 
TT,TE,EE+lowE+lensing ($\Omega_{K} = - 0.032^{+0.006}_{-0.007}$, $H_0 = 58.9 
\pm 2.0$\,km\,s$^{-1}$\,Mpc$^{-1}$ at 68\,\% CL) provide a worse fit compared to the 
tilted flat $\Lambda$CDM model.\footnote{This is not a new result based on the 
\Planck\ 2018 release,  but just an update of a similar conclusion also reached with 
the \Planck\ 2015 data. Compared to the flat $\Lambda$CDM tilted model, we 
obtain $\Delta \chi^2 = 12.3$, $34.8$, and $45$ with \Planck\ 2015 TT+lowP, \Planck\ 
2015 TT,TE,EE+ lowP, and \Planck\ 2015 TT,TE,EE+lowP+lensing, respectively. 
Therefore, even with \Planck\ 2015 data, a closed model with $n_\mathrm{s} = 1$ 
provides a worse fit than tilted $\Lambda$CDM and is not compelling as 
claimed in \cite{Ooba:2017ukj}.}

\subsection{Constraints on the scale dependence of the scalar spectral index}

The \Planck\ 2018 data are consistent with a vanishing running of the scalar 
spectral index. Using \Planck\ TT,TE,EE+lowE+lensing we obtain
\begin{equation}
\frac{d n_\mathrm{s}}{d \ln k} = -0.0045 \pm 0.0067 \quad (68\,\%\ \text{CL}) \,.
\end{equation}
These results are consistent with, and improve on, the \Planck\ 2015 result, $d 
n_\mathrm{s} / d \ln k = -0.008 \pm 0.008$ \citepalias{planck2014-a15}.

As discussed in \citetalias{planck2013-p17} and \citetalias{planck2014-a24},
a better fit to the temperature low-$\ell$ deficit was found in 2015, thanks to a 
combination of non-negative values for the running and the running of the running. 
The \Planck\ 2018 release has significantly reduced the parameter volume of this 
extension of the base-\LCDM\ model.
The \Planck\ 2018 TT(TT,TE,EE)+lowE+lensing constraints for the model including 
running of running are
\begin{align}
n_\mathrm{s} &= 0.9587 \pm 0.0056  \,\,(0.9625 \pm 0.0048) \,, \\
d n_\mathrm{s}/d \ln k &= 0.013 \pm 0.012 \,\,(0.002 \pm 0.010) \,, \\
d^2 n_\mathrm{s}/d \ln k^2 &= 0.022 \pm 0.012 \,\,(0.010 \pm 0.013) \,,
\end{align}
all at 68\,\%~CL.  It is interesting to note that the high-$\ell$ temperature data still allow a sizable 
value for the running of the running, although slightly decreased with respect to the \Planck\ 
2015 results \citepalias{planck2014-a24}.
However, when high-$\ell$ \Planck\ 2018 polarization data are also included, $d n_\mathrm{s}/d 
\ln k$ and $d^2 n_\mathrm{s}/d \ln k^2$ are tightly constrained.

The model including a scale-dependent running can produce a better fit to the 
low-$\ell$ deficit at the cost of an increase of power at small scales; this latter 
effect is constrained in this release. As an example of a model with suppression 
only on large scales, we also reconsider the phenomenological model with an 
exponential cutoff:
\begin{equation}
\mathcal{P_R}(k) = \mathcal{P}_{0}(k) \left\{ 1 - \exp \left[- \left( 
\frac{k}{k_\mathrm{c}} \right)^{\lambda_\mathrm{c}} \right] \right\},
\label{eq:cutoff}
\end{equation}
which can be motivated by a short stage of inflation \citep{Contaldi:2003zv, 
Cline:2003ve} (see also \citet{Kuhnel:2010pp}, \citet{Hazra:2014jka}, and \citet{Gruppuso:2015xqa} for other types of large-scale suppressions). We do not find any statistically significant detection of 
$k_\mathrm{c}$ using either logarithmic or linear priors and for different values of 
$\lambda_\mathrm{c}$, with any combination of \Planck\ baseline likelihoods. 
We have also checked that these 
results depend only weakly on the exclusion of the $EE$ quadrupole in {\tt SimAll} and 
are stable to the substitution of {\tt Commander} and {\tt SimAll} with the 
joint temperature-polarization likelihood based on the 70-GHz channel.

\subsection{Constraints on spatial curvature}
\label{curvSubsection}

\begin{table*}[!ht]
\begin{center}
\begin{tabular}{ccccc}
\noalign{\hrule\vskip 2pt}
\noalign{\hrule\vskip 3pt}
Cosmological model & Parameter & \Planck\ TT,TE,EE & \Planck\ TT,TE,EE & \Planck\ TT,TE,EE \\
$\Lambda$CDM+$r$ & & +lowEB+lensing & +lowE+lensing+BK15 & +lowE+lensing+BK15+BAO \\
\noalign{\hrule\vskip 2pt}

\multirow{2}{*}{} & $r$ & $< 0.11$ & $< 0.061$ & $< 0.063$ \\
& $r_{0.002}$ &  $<0.10$ & $< 0.056$ &$ < 0.058$ \\
& $n_{\mathrm{s}}$ & $0.9659 \pm 0.0041$ & $0.9651 \pm 0.0041$ & $0.9668 \pm 0.0037$ \\

\noalign{\hrule\vskip 2pt}
& $r$ & $< 0.16$  & $< 0.067 $ & $< 0.068$\\
& $r_{0.002}$ &  $ < 0.16$ & $< 0.065$  & $< 0.066$ \\
+$d n_{\mathrm{s}}/d \ln k$ & $n_{\mathrm{s}}$ & $ 0.9647 \pm 0.0044$
& $ 0.9639 \pm 0.0044$ & $0.9658 \pm 0.0040$ \\
& $d n_{\mathrm{s}}/d \ln k$ & $ - 0.0085 \pm 0.0073$
& $ -0.0069\pm 0.0069$ & $-0.0066 \pm 0.0070$ \\

\noalign{\hrule\vskip 2pt}
& $r$ &  $< 0.092$ & $< 0.061$ & $< 0.064$ \\
& $r_{0.002}$ &  $< 0.085$ & $< 0.055$ & $< 0.059$ \\
+$N_\mathrm{eff}$ & $n_{\mathrm{s}}$
& $0.9607_{-0.0084}^{+0.0086}$ & $ 0.9604 \pm 0.0085$ & $0.9660\pm 0.0070 $\\[0.2ex]
& $N_\mathrm{eff}$  & $2.92 \pm 0.19$ & $ 2.93\pm 0.19$ & $3.02 \pm 0.17 $ \\[0.5ex]

\noalign{\hrule\vskip 2pt}
& $r$  &  $< 0.097$ & $< 0.061$ & $< 0.061$ \\
& $r_{0.002}$ &  $< 0.091$ & $< 0.056$ & $< 0.056$ \\
+$m_\nu$ & $n_\mathrm{s}$ & $0.9654 \pm 0.0044$ & $ 0.9649 \pm 0.0044$ & $0.9668\pm 0.0036$\\
& $\sum m_\nu$ [eV] & $< 0.24$ & $< 0.23$ &  $< 0.11$ \\

\noalign{\hrule\vskip 2pt}
& $r$ & $< 0.12$  & $< 0.066$ & $< 0.062$ \\
& $r_{0.002}$ &  $< 0.12$ & $<0.062$ & $< 0.057$ \\
+$\Omega_K$ & $n_{\mathrm{s}}$ & $0.9703_{-0.0046}^{+0.0045}$ & $ 0.9697 \pm 0.0046$ & $0.9663 \pm 0.0044$\\[0.3ex]
& $\Omega_K$ & $-0.012_{-0.006}^{+0.007}$ & $ -0.012_{-0.007}^{+0.006}$ & $0.0006\pm 0.0019 $\\[0.5ex]

\noalign{\hrule\vskip 2pt}
& $r$ &  $< 0.11$ & $< 0.064$ & $< 0.062$ \\
& $r_{0.002}$ &  $< 0.10$ & $<0.059$ & $< 0.057$ \\
+$w_0$ & $n_{\mathrm{s}}$ & $0.9675 \pm 0.0042$ & $ 0.9669 \pm 0.0042$ & $0.9659 \pm 0.0040$\\[0.3ex]
& $w_0$ & $-1.58_{-0.34}^{+0.14}$ & $ -1.58_{-0.34}^{+0.14}$ & $-1.04 \pm 0.05 $\\[0.5ex]
\hline
\end{tabular}
\end{center}
\caption{Constraints on the tensor-to-scalar ratio $r$ and scalar tilt $n_\mathrm{s}$ 
for the $\Lambda$CDM+$r$ model and some important extensions and different data sets. 
For each model we quote 68\,\% confidence limits on measured parameters and 
95\,\% upper bounds on other parameters.}
\label{tab:r_extended}
\end{table*}

Since the vast majority of inflation models predict that the Universe has been driven 
towards spatial flatness, constraints on the spatial curvature provide an important 
test of the standard scenario.  Therefore in this subsection we extend the 
base-$\Lambda$CDM model with the addition of the spatial curvature parameter, $\Omega_K$.  
For the case of \Planck\ TT,TE,EE+lowE+lensing, we find a constraint of
\begin{equation}
\Omega_K = -0.011^{+0.013}_{-0.012} \quad (95\,\%\ \text{CL}) \,.
\label{eq:OmegaK_TElens}
\end{equation}
The inclusion of \Planck\ lensing information only weakly breaks the geometrical 
degeneracy \citep{ebdegen99} which results in the same primary fluctuations while 
varying the total matter density parameter, $\Lambda$, and $H_0$.  The degeneracy can 
be effectively broken with the addition of BAO data, in which case \Planck\ 
TT,TE,EE+lowE+lensing+BAO gives
\begin{equation}
\Omega_K = 0.0007 \pm 0.0037 \quad (95\,\%\ \text{CL}) \,. \label{eq:OmegaK_TElensBAO}
\end{equation}
Although $\Omega_K$ is one of the cosmological parameters exhibiting some differences 
between {\tt Plik} and {\tt Camspec}, the constraints in 
Eqs.~(\ref{eq:OmegaK_TElens}) and (\ref{eq:OmegaK_TElensBAO}) are quite robust due to 
the inclusion of lensing (and BAO) information.

   A constraint on the curvature parameter can be translated into a constraint on the 
radius of curvature, $R_K$, via
\begin{equation}
R_K = \left(a_0H_0\sqrt{|\Omega_K|}\right)^{-1},
\end{equation}
in units such that $c = 1$.  For the case of \Planck\ TT,TE,EE+lowE+lensing+BAO we find
\begin{align}
R_K & > 67\,{\rm Gpc} \quad {\rm (open)},\\
R_K & > 81\,{\rm Gpc} \quad {\rm (closed)},
\end{align}
both at $95\,\%$ confidence.  These lengths are considerably greater than our current 
(post-inflation) particle horizon, at $13.9\,$Gpc.

Our tightest constraint, Eq.~(\ref{eq:OmegaK_TElensBAO}), tells us that our 
observations are consistent with spatial flatness, with a precision of about 0.4\,\%.  
However, even if inflation has driven the background curvature extremely close to 
zero, the presence of fluctuations implies a fundamental ``cosmic variance'' for 
measurements of curvature confined to our observable volume.  In particular, the 
known amplitude of fluctuations implies a standard deviation for $\Omega_K$ of 
roughly $2\times10^{-5}$ \citep{wz08}.  Therefore our best constraint is still a 
factor of roughly $10^2$ above the cosmic variance limit for a flat universe.  A 
future measurement of negative curvature above the cosmic variance floor would point 
to open inflation
\citep{Gott:1982zf,Gott:1984ps,Bucher:1994gb,Yamamoto:1995sw,Ratra:1994vw,Lyth:1990dh}, 
while a measurement of positive curvature could pose a problem for 
the inflationary paradigm due to the difficulty of producing closed inflationary 
models \citep{Kleban:2012ph}.

Alternatively, excess spatial curvature might be evidence for the intriguing
possibility that there was ``just enough'' inflation to produce structure on the 
largest observable scales.  Indeed an upper limit on spatial curvature implies a 
lower limit on the total number of $e$-folds of inflation (see, e.g., 
\citealt{Komatsu:2008hk}).  We can relate these limits to the number of $e$-folds of 
inflation, $N_* = N(k_*)$, after scale $k_*$ left the Hubble radius during inflation, 
to be given explicitly in Eq.~(\ref{eq:nefolds}).  We define the (constant) curvature 
scale, $k_K$, as the inverse of the comoving radius of curvature, i.e.,
\begin{equation}
k_K \equiv aH\sqrt{|\Omega_K|}. \label{eq:curvscale}
\end{equation}
In the absence of special initial conditions, inflation will begin with a curvature 
parameter of order unity.  Equation~(\ref{eq:curvscale}) then implies that $k_K \sim 
aH$ at the start of inflation, i.e., the curvature scale is ``exiting the horizon'' 
at that time.  Then the lower limit on the number of $e$-folds of inflation will 
simply be $N_K \equiv N(k_K)$, i.e., the number of $e$-folds after scale $k_K$ left 
the Hubble radius during inflation.  With Eq.~(\ref{eq:nefolds}) this 
gives\footnote{This expression ignores a negligible correction, $(1/2)\ln 
V_{k_K}/V_{k_*}$, due to the different inflationary potential scales at $k_*$ and $k_K$ 
exit.}
\begin{equation}
N_K = \ln\frac{k_*}{a_0H_0} - \frac{1}{2}\ln|\Omega_K| + N_*.
\end{equation}
With the pivot scale of $k_* = 0.002$\,Mpc$^{-1}$ (for comparison with the values in 
Sect.~\ref{sec:implns_sel_slow_roll}) and our tightest upper limit on 
$\Omega_K$ from Eq.~(\ref{eq:OmegaK_TElensBAO}), this becomes
\begin{equation}
N_K \gtrsim 4.9 + N_*. \label{eq:NKlimit}
\end{equation}
That is, our constraint on spatial curvature implies that inflation must have lasted 
at least about 5 $e$-folds longer than required to produce the pivot scale $k_*$.  
Equation~(\ref{eq:NKlimit}) provides a model-independent comparison between the 
$e$-folds required to solve the flatness problem (to current precision) and to 
produce large-scale fluctuations (at scale $k_*$).  We stress that this limit assumes 
a unity curvature parameter at the start of inflation (although the dependence on 
this assumption, being logarithmic, is weak).

   For comparison with the result of \citet{Komatsu:2008hk}, we can simplify to
the case of instantaneous thermalization and constant energy density during 
inflation.  Then we find
\begin{equation}
N_K \gtrsim 34.2 + \ln\frac{T_{\rm end}}{1\,{\rm TeV}},
\end{equation}
where $T_{\rm end}$ is the reheating temperature.

\subsection{Constraints on the tensor-to-scalar ratio}
\label{sec:rconstraints}

This subsection updates constraints on the tensor-to-scalar ratio $r$
assuming that the tensor tilt satisfies the 
consistency relation, $n_{\mathrm t} = - r/8$, which is the case for slow-roll 
inflation driven by a single scalar field with a canonical kinetic term.

By combining \Planck\ temperature, low-$\ell$ polarization, and lensing we obtain
\begin{equation}
r_{0.002} < 0.10 \quad  (95\,\%\ \text{CL, \Planck\ TT+lowE+lensing}) \,.
\end{equation}
This constraint slightly improves on the corresponding \Planck\ 2015 95\,\% CL bound, 
i.e., $r_{0.002} < 0.11$ \citepalias{planck2014-a24}, and is unchanged when 
high-$\ell$ polarization data are also combined. Note that by using \texttt{CAMspec} 
instead of \texttt{Plik} as the high-$\ell$ joint temperature-polarization likelihood, we 
obtain a slightly looser bound, i.e., $r_{0.002} < 0.14$ at 95\,\% CL. By including the 
\Planck\ $B$-mode information at $2 < \ell < 30$ in the low-$\ell$ polarization 
likelihood, the 95\,\% CL constraint is essentially unchanged.

Since
inflationary gravitational waves contribute to CMB temperature anisotropies mostly 
at $\ell \lesssim 100$,  the low-$\ell$ temperature deficit contributes in a 
nontrivial way to the \Planck\ bound on $r$. By excising the $2 \le \ell \le 29$ 
temperature data, the constraint on $r$ with \Planck\ TT,TE,EE+lensing+lowEB relaxes 
to
\begin{equation}
r_{0.002} < 0.16 \quad (95\,\%\ \text{CL}) \,. \label{Planck_bound_no_lowell}
\end{equation}
This result improves on the 2015 95\,\% CL result, i.e., $r \lesssim 
0.24$ \citepalias{planck2014-a24}, because of the inclusion of high-$\ell$ polarization and 
of the improved determination of $\tau$.

Since this \Planck\ constraint on $r$ relies on temperature and $E$-mode 
polarization, the \Planck-only limit depends somewhat on the underlying 
cosmological model. Table~\ref{tab:r_extended} shows the constraints on $n_{\mathrm 
s}$ and $r$ for a few important extensions of $\Lambda$CDM plus tensors, which 
include a non-zero running, a non-zero spatial curvature, and a non-minimal neutrino 
sector. We observe that the bound on $r$ is relaxed by at most 30\,\% when the scale 
dependence of the scalar tilt is allowed to vary. In all the other extensions the 
\Planck\ $r$ bound is modified at most by 10\,\%, demonstrating the constraining power 
of the \Planck\ 2018 release in reducing the degeneracy of the tensor-to-scalar ratio 
with other cosmological parameters. As far as the scalar tilt is concerned, we find 
the largest shift (by roughly $1\sigma$ higher) when the assumption of spatial flatness 
is relaxed.

A $B$-mode polarization measurement 
can further tighten the constraint on $r$ and help in reducing its degeneracies with 
other cosmological parameters that may appear when using only temperature and 
$E$-mode polarization data. After the release of the first BICEP-Keck Array-\Planck\ 
\citepalias{pb2015} joint cross-correlation, constraints on $r$ from $B$-mode 
polarization data alone have become tighter than those based on \Planck\ data alone, 
thanks to the inclusion of the 95-GHz channel 
\citepalias{Array:2015xqh} and of the 220-GHz channel \citepalias{bk15}.
By combining the \Planck\ 2018 and BK15 data we obtain
\begin{align}
r_{0.002} < 0.056 \quad
&(95\,\%\ \text{CL, \Planck\ TT,TE,EE} \nonumber \\
&\textrm{+lowE+lensing+BK15}) \,. \label{Planck_BK15}
\end{align}
This bound 
improves on the corresponding one obtained in combination with \citetalias{Array:2015xqh}, 
i.e., $r_{0.002} < 0.064$ at 95 \% CL.
Note that by using \texttt{CAMspec} instead of \texttt{Plik} as high-$\ell$ TT,TE,EE 
likelihood, we obtain a slightly looser bound, i.e., $r_{0.002} < 0.069$ at 95\,\% CL. 
The effectiveness of the combination with the \citetalias{bk15} likelihood in 
constraining $r$ is also remarkable in the extensions of $\Lambda$CDM plus tensors, 
as can be seen from Table~\ref{tab:r_extended}. 
By further combining with BAO the limits for $r$ are only slightly modified.
  
The \Planck\ 2018 baseline plus BK15 constraint on $r$ is equivalent to an upper bound 
on the energy scale of inflation
when the pivot scale exits the Hubble radius of
\begin{equation}
V_* = \frac{3 \pi^2 A_{\mathrm{s}}}{2} \, r \, M_{\mathrm {Pl}}^4 < (1.6 \times 10^{16}~{\mathrm{GeV}} )^4 \quad  (95\,\%\ \text{CL}) \,.
\end{equation}
Equivalently, this last result implies an upper bound
on the Hubble parameter during inflation of
\begin{equation}
\frac{H_*}{M_{\mathrm{Pl}}} < 2.5 \times 10^{-5} \quad (95\,\%\ \text{CL}) \,.
\end{equation}

\subsection{Beyond the tensor-to-scalar ratio consistency condition}
\label{sec:beyondConsistency}

The increasing constraining power of $B$-mode polarization data allows us to set 
upper bounds on $r$ without imposing the consistency condition for the tensor tilt, 
$n_{\mathrm t} = -r/8$, which is motivated by standard slow-roll single-scalar-field 
inflation. Deviations can occur in  multifield inflation
\citep{Bartolo:2001rt,Wands:2002bn,Byrnes:2006fr},
in the models with generalized Lagrangians
\citep{1999PhLB..458..219G,Kobayashi:2010cm}, in gauge inflation
\citep{Maleknejad:2012fw}, or in a more radical way in alternative models to
inflation \citep{Gasperini:1992em,Boyle:2003km,Brandenberger:2006xi}.

As the current data do not lead to a detection of a non-zero 
tensor amplitude, virtually any value of $n_{\mathrm t}$ would give a good fit as 
long as $r$ is close enough to zero. Therefore, as in \citetalias{planck2014-a24}, we 
characterize the tensor perturbations by two well-constrained parameters that we 
choose to be $r$ at two different scales, ($r_{k_1},r_{k_2}$), with $k_1 = 
0.002$\,Mpc$^{-1}$ and $k_2 = 0.02$\,Mpc$^{-1}$, and assume a power-law power 
spectrum. We call this two-parameter extension of the \LCDM\ model the 
``\LCDM$+r_{0.002}$+$r_{0.02}$'' model. We also quote our results in terms of 
$(r_{\tilde{k}}, n_{\mathrm t})$, calculated from the primary parameters as 
$n_{\mathrm t} = [\ln(r_{k_2}/r_{k_1})\,/\,\ln(k_2/k_1)] + n_\mathrm{s}-1$ and 
$r_{\tilde{k}} = r_{k_1} (\tilde{k} / k_1)^{n_\mathrm{t}-n_\mathrm{s}+1}$.
For $\tilde{k}$ we choose $0.01$\,Mpc$^{-1}$, which 
corresponds roughly to the decorrelation scale of $r$ and $n_{\mathrm t}$ when using 
the \Planck\ and BK15 data.

\begin{figure}
\includegraphics[width=\columnwidth]{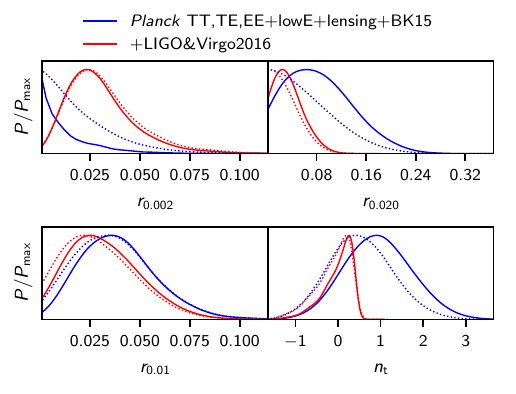}
\caption{Posterior probability density of the tensor-to-scalar ratio at two different 
scales in the \LCDM$+r_{0.002}$+$r_{0.02}$ model, i.e., when the inflationary 
consistency relation is relaxed (top panels). The solid contours show the results 
when $r_{0.002}$ and $r_{0.02}$ are used as sampling parameters with uniform priors, 
which leads to non-uniform priors for the derived parameters $r_{0.01}$ and 
$n_\mathrm{t}$ (bottom panels). The dotted contours indicate the results after 
weighting the posterior by the Jacobian 
$J=r_{0.01}/[r_{0.002}r_{0.02}\ln(0.02/0.002)]$ of the transformation 
$(r_{0.002},\,r_{0.02}) \rightarrow (r_{0.01},\,n_\mathrm{t})$, giving the result we 
would have obtained had we assigned uniform priors on $r_{0.01}$ and $n_\mathrm{t}$. 
\label{fig:freent_1d} }
\end{figure}
\begin{figure*}
\includegraphics[width=\columnwidth]{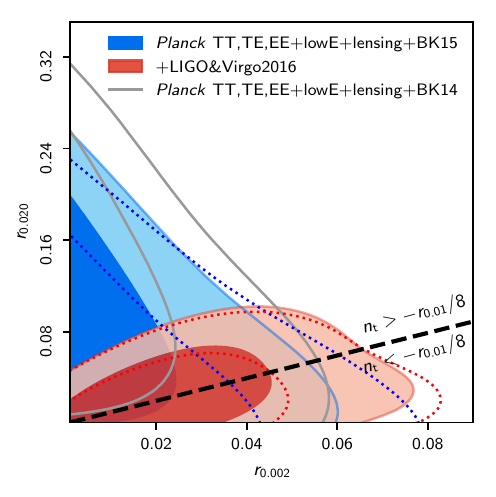}
\includegraphics[width=\columnwidth]{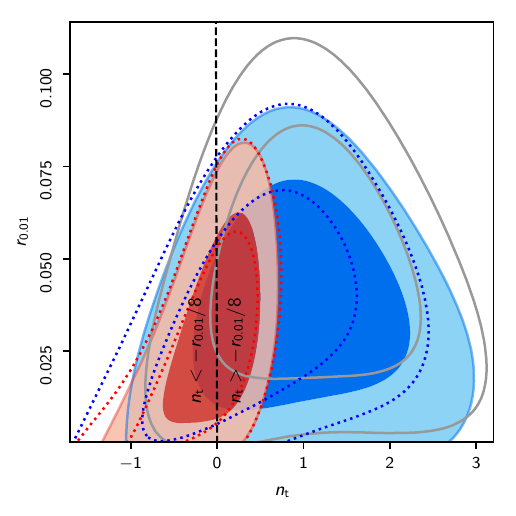}
\caption{68\,\% and 95\,\% CL constraints on tensor perturbations in the 
\LCDM$+r_{0.002}$+$r_{0.02}$ model, i.e., when the inflationary
consistency relation is relaxed.  Filled contours in the left panel show the results 
for our independent primary parameters $r_{0.002}$ and $r_{0.02}$, which have uniform 
priors,
and in the right panel for the derived parameters $n_\mathrm{t}$ and $r_{0.01}$, 
which have non-uniform priors. The dotted lines assume uniform priors on $r_{0.01}$ 
and $n_\mathrm{t}$, calculated as in Fig.~\ref{fig:freent_1d}. The scale 
$k=0.01\,$Mpc$^{-1}$ is near the decorrelation scale of \mbox{($n_\mathrm{t}$, $r$)} for 
the \Planck+BK15 data. In both panels the dashed black line indicates the 
inflationary consistency condition, $n_{\mathrm t} =-r_{0.01}/8$. (The
grey contours follow if we use the older BK14 data instead of the BK15 data.)
\label{fig:freent_2d} }
\end{figure*}

The one-dimensional posteriors are displayed in Fig.~\ref{fig:freent_1d} (which also 
shows an additional data set, ``LIGO\&Virgo2016,'' discussed at the end of this 
subsection).  We obtain for the \LCDM$+r_{0.002}$+$r_{0.02}$ model:
\begin{equation}
\left.\begin{aligned}
   r_{0.002} &< 0.044\\
   r_{0.02}  &< 0.184
\end{aligned}\ \right\}\ \ 
\mbox{\text{\parbox{4.2cm}{\begin{flushleft}
   (95\,\% CL, \Planck\ TT,TE,EE\\+lowE+lensing+BK15).
\end{flushleft}}}}
\label{eq:LCDMr1r2BK15}
\end{equation}
The constraints on the derived tensor parameters are
\mbox{$r_{0.01} < 0.076$} and $-0.55 < n_{\mathrm t} < 2.54$ at 95\,\% CL. 

The left and right panels of Fig.~\ref{fig:freent_2d} show the two-dimensional 
contours for the primary parameters $(r_{0.002},r_{0.02})$ and the derived ones 
$(r_{0.01},n_{\mathrm t})$, respectively. 
The consistency condition, $n_{\mathrm t} = -r/8$, denoted by the dashed lines, is 
fully compatible with \Planck+BK15 data. However, a very blue tensor tilt with 
$n_{\mathrm t} \simeq 2$ and $r_{0.01} \simeq 0.05$ is still within the 68\,\% CL 
region. Indeed, despite the larger amplitude of the primordial tensor power spectrum 
at small scales for blue $n_{\mathrm t}$,
the tensor modes are suppressed when re-entering the Hubble radius, which leads to 
damping of the observational signal at high $k$. This explains why the 68\,\% CL CMB 
constraint on $r_{0.02}$ is about
a factor of four
weaker than the one on 
$r_{0.002}$.  Fig.~\ref{fig:freent_2d} also shows a slight improvement
of constraints by BK15 compared to the older BK14 data.

A stochastic background of gravitational waves (GWs) with a blue tensor tilt could be 
further constrained at much shorter wavelengths, such as those probed by ground-based 
interferometers dedicated to the direct detection of GWs. For example, assuming a 
scale-invariant tensor spectrum and using the frequency range (20--85.8)\,Hz, which 
corresponds to the wavenumbers $k=2\pi f =(1.3$--$5.5)\times10^{16}\,$Mpc$^{-1}$, 
LIGO and Virgo set an upper bound on the GW density parameter of 
$\Omega_{\mathrm{GW}}(f) \le 1.7\times10^{-7}$ at 95\,\% CL 
\citep{TheLIGOScientific:2016dpb}.  While these scales are likely to be dominated by 
astrophysical sources, such as GWs from binary mergers, we next examine what 
constraints the LIGO\&Virgo upper bound sets on primordial tensor perturbations, if 
we assume that they had a power-law spectrum all the way from CMB scales to 
ultra-short scales. We refer the interested reader to \citet{Meerburg:2015zua} and 
\citet{Cabass:2015jwe} for the use of alternative data on short scales or of 
additional constraints on the effective energy-momentum tensor of the stochastic background of 
GWs averaged over wavelengths.  

We obtain a conservative upper limit on the \emph{primordial} contribution by 
demanding that the GW density from our scale-dependent primordial tensor 
perturbations \citep{Meerburg:2015zua,TheLIGOScientific:2016dpb,Cabass:2015jwe},
\begin{equation}
 \Omega_{\mathrm{GW}}(k) = 
\frac{k}{\rho_{\mathrm{critical}}}\frac{d\rho_{\mathrm{GW}}}{dk} = 
\frac{A_{\mathrm{t}}(k)}{24z_{\mathrm{eq}}}
 = \frac{A_{\mathrm{t}1}(k/k_1)^{n_{\mathrm{t}}}}{24z_{\mathrm{eq}}},\,
\label{eq:OmegaGW}
\end{equation}
stays below the above-quoted limit at least at $k\!=\!1.3\times10^{16}\,$Mpc$^{-1}$ 
($f\!=\!20\,$Hz). The posterior probability densities when this constraint is 
included in the analysis as a half-Gaussian prior are compared with those obtained by 
\Planck+BK15 alone in Figs.~\ref{fig:freent_1d} and \ref{fig:freent_2d}. 
LIGO\&Virgo sets a very high upper bound\footnote{%
Using  Eq.~(\ref{eq:OmegaGW}), the upper bound  
$\Omega_{\mathrm{GW}}(f=20\,\text{Hz}) \le 1.7\times10^{-7}$\linebreak corresponds to 
a tensor perturbation amplitude 
$A_\mathrm{t}(k=1.3\times10^{16}\,\text{Mpc}^{-1})\linebreak \le 24z_\mathrm{eq} 
\times 1.7\times 10^{-7} = 1.4\times 10^{-2}$, where we used $z_\mathrm{eq} \simeq 
3400$. Assuming further for the scalar perturbations that $n_\mathrm{s} = 0.9659$ and 
$\ln(10^{10}A_\mathrm{s}) = 3.044$ at $k=0.05\,\text{Mpc}^{-1}$, this can be 
converted into an upper bound $r \le 2.6 \times 10^{7}$ at 
$k=1.3\times10^{16}\,\text{Mpc}^{-1}$.
} on $r$ at ultra-high $k$, separated from CMB scales by a factor of $10^{18}$ in 
$k$. Due to the long arm length, this effectively provides a cutoff for 
$n_\mathrm{t}$ and excludes the bluest spectra that were allowed by the CMB alone, 
leading to
\begin{equation}
\left.\begin{aligned}\!\!\!\!\!
   r_{0.002} &< 0.064\\
   r_{0.02}  &< 0.081
\end{aligned}\ \right\}\ \ 
\mbox{\text{\parbox{5cm}{\begin{flushleft}
   (95\,\% CL, \Planck\ TT,TE,EE+lowE\\+lensing+BK15+LIGO\&Virgo2016),
\end{flushleft}}}}
\label{eq:LCDMr1r2BK15LIGO}
\end{equation}
or $r_{0.01} < 0.066$ and $-0.76< n_{\mathrm t} < 0.52$. 
The consistency condition $n_{\mathrm t} = -r/8$ is also compatible with these 
tighter constraints, as can be seen by comparing the red contours and dashed black 
lines in Fig.~\ref{fig:freent_2d}. As LIGO\&Virgo pushes $r_{0.02}$ down (and we 
assume a power-law tensor spectrum), the upper bound on $r_{0.002}$ becomes weaker 
than with the CMB alone. This is not surprising, since the system is analogous to a 
see-saw with a pivot point at $k\sim0.01\,$Mpc$^{-1}$, where the data are the most 
sensitive to the tensor perturbations (taking into account also the transfer function 
from primordial tensor perturbations to the observable $B$-mode signal). Once one end 
of the see-saw is pushed down the other end can go up without disturbing the spectrum 
too much at the middle scales. We will observe analogous behaviour with isocurvature 
perturbations, for which we also assume a power-law spectrum and have only an upper 
bound (not a detection); see Sect.~\ref{sec:isocVsAL}.

\FloatBarrier

\section{Implications for single-field slow-roll inflationary models \label{sec:single}}

\def\reff@jnl#1{{\rm#1\/}}
\def\apj{\reff@jnl{ApJ}}       
\def\apjs{\reff@jnl{ApJS}}     
\def\aaps{\reff@jnl{A\&AS}}    
\def\mnras{\reff@jnl{MNRAS}}   
\def\prd{\reff@jnl{Phys.\ Rev.\ D}}    

\newcommand{\Nside}{\ensuremath{N_{\mathrm{side}}}} 
\newcommand{\Npix}{\ensuremath{N_{\mathrm{pix}}}}   
\newcommand{\Ntau}{\ensuremath{N_{\tau}}}   
\newcommand{\vA}{\mathbf{A}}
\newcommand{\va}{\mathbf{a}}
\newcommand{\vB}{\mathbf{B}}
\newcommand{\vM}{\mathbf{M}}
\newcommand{\vN}{\mathbf{N}}
\newcommand{\vP}{\mathbf{P}}
\newcommand{\vS}{\mathbf{S}}
\newcommand{\vX}{\mathbf{X}}
\newcommand{\vY}{\mathbf{Y}}
\newcommand{\vd}{\mathbf{d}}
\newcommand{\vn}{\mathbf{n}}
\newcommand{\vs}{\mathbf{s}}
\newcommand{\vC}{\mathbf{C}}
\newcommand{\vI}{\mathbf{I}}
\newcommand{\vt}{\mathbf{t}}
\newcommand{\vE}{\mathbf{E}}
\newcommand{\vx}{\mathbf{x}}
\newcommand{\vphi}{\mathbf{\phi}}
\newcommand{\veta}{\mathbf{\eta}}
\newcommand{\vshat}{\vec{\hat{s}}}
\newcommand{\vxhat}{\vec{\hat{x}}}
\newcommand{\vnu}{\vn_{u}}
\newcommand{\vNu}{\vN_{u}}
\newcommand{\tr}{^{\mathrm{T}}} 
\newcommand{\beq}{\begin{equation}}
\newcommand{\eeq}{\end{equation}}
\newcommand{\rsht}{{MASTER}}
\newcommand{\tC}{\widetilde{C}}
\newcommand{\tN}{\widetilde{N}}
\renewcommand{\r}{{\bf{r}}}
\newcommand{\n}{{\bf{n}}}
\renewcommand{\k}{{\bf{k}}}
\newcommand{\fsky}{{f_{\rm sky}}}
\newcommand{\fzero}{{F^{(0)}}}
\newcommand{\fzerol}{{F^{(0)}_{\ell}}}
\newcommand{\npix}{{N_{\rm pix}}}
\newcommand{\nbins}{{n_{\rm bins}}}
\newcommand{\lmax}{{\ell_{\rm max}}}
\newcommand{\nmc}{{N_{\rm MC}}}
\newcommand{\nmcs}{{N_{\rm MC}^{\rm (s)}}}
\newcommand{\nmcn}{{N_{\rm MC}^{\rm (n)}}}
\newcommand{\nmcsn}{{N_{\rm MC}^{\rm (s+n)}}}
\newcommand{\ntau}{{N_{\tau}}}
\newcommand{\nfft}{{N_{\rm FFT}}}
\newcommand{\Cltheory}{{C_\ell^{\rm th}}}
\newcommand{\Ctheory}{{C^{\rm th}}}
\newcommand{\VEV}[1]{\langle#1\rangle}
\newcommand{\boom}{{\sc{BOOMERanG}}}
\newcommand{\bldb}{{Boom-LDB}}
\newcommand{\wjjj}[6]
{{
\left(
\begin{array}{lcr} #1 & #2 & #3 \\#4 & #5 & #6 \end{array}
\right)
}}
\newcommand{\niter}{{N_{\rm iter}}}
\newcommand{\col}[2]{\left[\begin{array}{c}{#1}\\{#2}\end{array}\right]}
\newcommand{\mat}[4]{\left[\begin{array}{cc}{#1}&{#2}\\{#3}&{#4}\end{array}
\right]}
\newcommand{\vsh}{\hat{\vs}}
\newcommand{\vxh}{\hat{\vx}}
\newcommand{\Nmap}{\vN_{\mathrm{map}}}
\newcommand{\Noff}{\vN_{\mathrm{off}}}
\newcommand{\Nmapz}{\Nmap^{(0)}}
\newcommand{\thalf}{\tfrac{1}{2}}
\newcommand{\be}{\begin{equation}}
\newcommand{\ee}{\end{equation}}
\newcommand{\bea}{\begin{eqnarray}}
\newcommand{\eea}{\end{eqnarray}}
\def\nn{\nonumber}
\def\L{\mathcal{L}}

\renewcommand{\dbltopfraction}{1.0}
\renewcommand{\textfraction}{0}

\newenvironment{myitem}%
{\begin{enumerate}\setlength{\itemsep}{0mm}}%
{\end{enumerate}}
\newenvironment{myenum}%
{\begin{enumerate}\setlength{\itemsep}{0mm}}%
{\end{enumerate}}

\def\nd#1#2{{d #1 \over d #2}}
\def\pd#1#2{{\upartial #1 \over \upartial #2}}
\def\spd#1#2#3{{\upartial ^2 #1 \over \upartial #2 \upartial #3}}
\def\sspd#1#2{{\upartial ^2 #1 \over \upartial #2^2}}
\def\tfrac#1#2{{\textstyle\frac{#1}{#2}}}
\def\vect#1{{\mathbf{#1}}}

\newcommand{\bc}{\begin{center}}
\newcommand{\ec}{\end{center}}
\newcommand{\bi}{\begin{itemize}}
\newcommand{\ei}{\end{itemize}}
\newcommand{\ben}{\begin{enumerate}}
\newcommand{\een}{\end{enumerate}}
\newcommand{\R}{\Re\textrm{e}}
\newcommand{\I}{\Im\textrm{m}}
\newcommand{\Ab}{\boldsymbol{A}}
\newcommand{\Mb}{\boldsymbol{M}}
\newcommand{\Tb}{\boldsymbol{T}}

 \newcommand{\mtc}[1]{\mathcal{#1}}
 \newcommand{\Pow}{{\mathcal P}}
 \newcommand{\nad}{n_\mathrm{ad}}
 \newcommand{\nadI}{n_\mathrm{ar}}
 \newcommand{\nadII}{n_\mathrm{as}}
 \newcommand{\niso}{n_\mathrm{iso}}
 \newcommand{\ncor}{n_\mathrm{cor}}
 \newcommand{\etzz}{\eta_{\sigma\sigma}}
 \newcommand{\etzs}{\eta_{\sigma s}}
 \newcommand{\etss}{\eta_{ss}}
 \newcommand{\veps}{\varepsilon}
\newcommand{\ff}[1]{{{\textcolor{blu}{#1}}}}

\newfont{\gwpfont}{cmssq8 scaled 1000}
\newcommand{\rexcess}{{\gwpfont REXCESS}}

\def\xmm{{\it XMM-Newton}}
\def\Mv {M_\mathrm{500}}
\def\msol {\mathrm{M}_{\odot}}
\def\YX {Y_\mathrm{X}} 
\def \Rv {R_{500}} 
\def\keV {\mathrm{keV}} 

\newcommand{\chisq}{\Delta \chi^2_\mathrm{eff}}
\newcommand{\kpiv}{k_{\mathrm{pivot}}}
\newcommand{\Mpl}{M_{\mathrm{pl}}}
\newcommand{\aend}{a_{\mathrm{end}}}
\newcommand{\aeq}{a_{\mathrm{eq}}}
\newcommand{\arh}{a_{\mathrm{th}}}
\newcommand{\wprim}{w_{\mathrm{prim}}}
\newcommand{\wint}{w_{\mathrm{int}}}
\newcommand{\rhorh}{\rho_{\mathrm{th}}}
\newcommand{\rhoeq}{\rho_{\mathrm{eq}}}
\newcommand{\rhoend}{\rho_{\mathrm{end}}}
\newcommand{\vend}{V_{\mathrm{end}}}
\newcommand{\ModeCode}{{\tt ModeCode}}
\newcommand{\MultiNest}{{\tt MultiNest}}
\newcommand{\CAMB}{{\tt CAMB}}
\newcommand{\CosmoMC}{{\tt CosmoMC}}
\newcommand{\codename}{{\ModeCode}}

\def\gtorder{\mathrel{\raise.3ex\hbox{$>$}\mkern-14mu
             \lower0.6ex\hbox{$\sim$}}}
\def\ltorder{\mathrel{\raise.3ex\hbox{$<$}\mkern-14mu
             \lower0.6ex\hbox{$\sim$}}}

\def\ba{\begin{eqnarray}}
\def\ea{\end{eqnarray}}

In this section we discuss the implications of the \Planck\ 2018 likelihood for standard 
single-field slow-roll inflation.  We first update the results for the Hubble flow 
functions (HFFs) $\epsilon_i$ 
and the potential slow-roll parameters obtained by the analytic perturbative expansion 
in terms of the HFFs for the primordial spectra of fluctuations. For 
definitions of the HFF 
hierarchy and the potential slow-roll parameters see Table~\ref{table:CPDefinitions}. 
We then present a Bayesian comparison for a representative selection of standard 
slow-roll inflationary 
models.

\begin{figure}[!t]
\centering
\includegraphics[width=6cm]{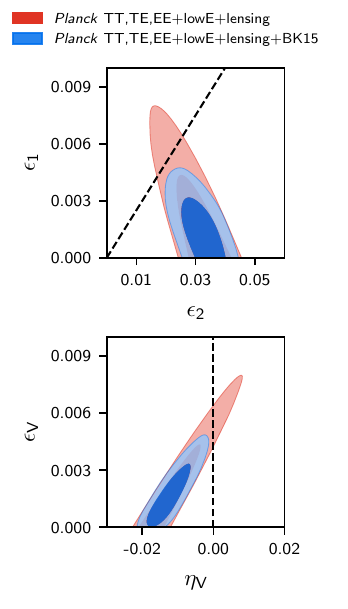}
\caption{Marginalized joint two-dimensional
68\,\% and 95\,\%~CL regions for $(\epsilon_1 \,, \epsilon_2)$ (top panel) and $(\epsilon_V \,, \eta_V)$ (bottom panel)
for \Planck\ TT,TE,EE+lowE+lensing (red contours), compared with \Planck\ TT,TE,EE+lowE+lensing+BK15 (blue contours).
The dashed lines divide between convex and concave potentials.} \label{fig:epsilon_no3}
\end{figure}

\subsection{Constraints on slow-roll parameters}
\label{sec:implns_slow_roll}

Exploiting the approximate analytic expressions for the primordial power spectrum 
of scalar and tensor fluctuations obtained by the Green's function 
method \citep{stewart:1993,Gong:2001he,Leach:2002ar}, 
we can construct constraints on the slow-roll parameters.

When restricting to parameters first order in the HFFs, we obtain 
with \Planck\ TT,TE,EE+lowE+lensing(+BK15)
\begin{align}
\phantom{0000}\epsilon_1 &< 0.0063                  &     (&0.0039)\phantom{00}           & &\text{(95\,\% CL)} \,,\\
\phantom{0000}\epsilon_2 &= 0.030^{+0.007}_{-0.005} &     (&0.031 \pm 0.005)\phantom{00}  & &\text{(68\,\% CL)} \,.
\end{align}
The \Planck\ TT,TE,EE+lowE+lensing(+BK15) 
constraints on the slow-roll potential parameters 
$\epsilon_V$ and $\eta_V$ can be obtained by an exact remapping of the constraints on 
the HFF parameters \citep{Leach:2002ar,Finelli:2009bs} given above:
\begin{align}
\phantom{00}\epsilon_V &< 0.0063                   & &(0.0039)\phantom{00}                             & &\text{(95\,\% CL)} \,, \\
\phantom{00}\eta_V     &= -0.010^{+0.004}_{-0.008} & &\Big({-0.012}^{+0.004}_{-0.005}\Big)\phantom{00} & &\text{(68\,\% CL)} \,.
\end{align}
As can be seen from Fig.~\ref{fig:epsilon_no3}, the 95\,\%~CL allowed contours are  
in the region of concave potentials 
when BK15 is combined with \Planck\ 2018 data.

\begin{figure*}
\begin{center}
\includegraphics[width=14cm]{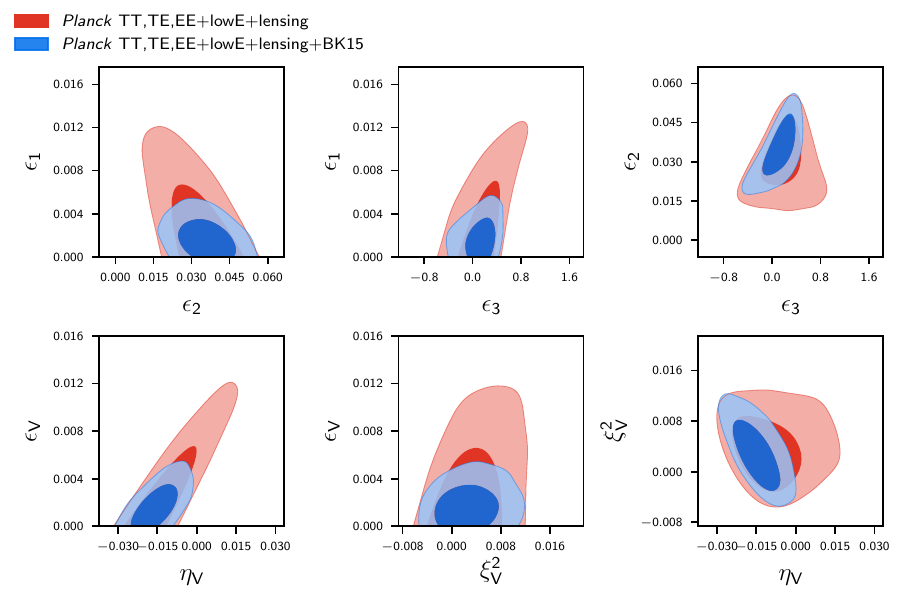}
\end{center}
\caption{Marginalized joint two-dimensional 68\,\% and 95\,\%~CL regions for combinations 
of $(\epsilon_1 \,, \epsilon_2 \,, \epsilon_3)$ 
(upper panels) and $(\epsilon_V \,, \eta_V, \xi^2_V)$ (lower panels) for \Planck\ TT,TE,EE+lowE+lensing 
(red contours), compared with \Planck\ TT,TE,EE+lowE+lensing+BK15
(blue contours). }
\label{eps2_bk15}
\end{figure*}

When contributions to the primordial power spectra that are second-order in the HFFs are included,
for \Planck\ TT,TE,EE+lowE+lensing(+BK15) we obtain the following constraints on the slow-roll HFFs:
\begin{align}
\phantom{00}\epsilon_1 &< 0.0097                  & &(0.0044)\phantom{00}                       & &\text{(95\,\% CL)} \,,\\
\phantom{00}\epsilon_2 &= 0.032^{+0.009}_{-0.008} & &(0.035 \pm 0.008)\phantom{00}              & &\text{(68\,\% CL)} \,,\\
\phantom{00}\epsilon_3 &= 0.19^{+0.55}_{-0.53}    & &\Big(0.12^{+0.36}_{-0.42}\Big)\phantom{00} & &\text{(95\,\% CL)} \,,
\end{align}
and on the slow-roll potential parameters we obtain:
\begin{align}
\phantom{0}\epsilon_V &< 0.0097                       & &(0.0044)\phantom{0}                             & &\text{(95\,\% CL)} \,,\\
\phantom{0}\eta_V     &= -0.010^{+0.007}_{-0.011}     & &({-0.015} \pm 0.006)\phantom{0}                 & &\text{(68\,\% CL)} \,,\\
\phantom{0}\xi^2_V    &= 0.0035^{+0.0078}_{-0.0072}   & &\Big(0.0029^{+0.0073}_{-0.0069}\Big)\phantom{0} & &\text{(95\,\% CL)} \,.
\end{align}

\noindent The marginalized 68\,\% and 95\,\%~CLs for the slow-roll HFF and 
potential parameters, allowing $\epsilon_3 \ne 0$, with \Planck\ data alone or in 
combination with BK15, are displayed in Fig.~\ref{eps2_bk15}.

\begin{figure*}[!t]
\includegraphics[width=18cm]{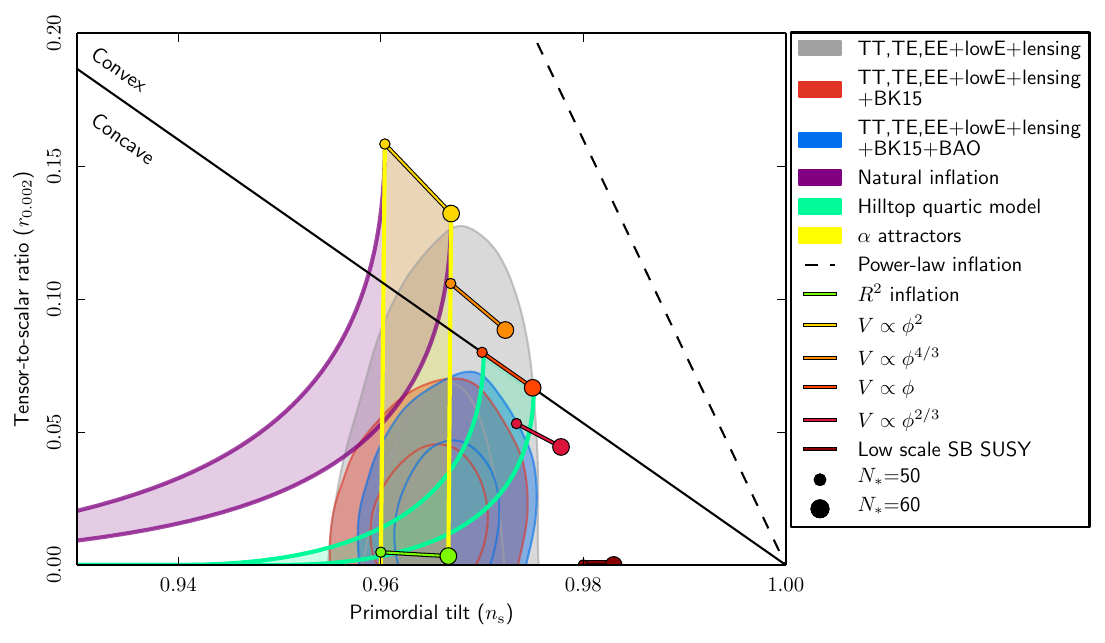}
\caption{Marginalized joint 68\,\% and 95\,\%~CL regions for $n_\mathrm{s}$ and $r$ at 
$k = 0.002$\,Mpc$^{-1}$ from \Planck\ alone and in combination 
with BK15 or BK15+BAO data, compared to the theoretical predictions of selected 
inflationary models. Note that the marginalized joint 68\,\% and 95\,\%~CL 
regions assume $d n_\mathrm{s}/d \ln k = 0$. } \label{fig:nsvsr}
\end{figure*}

\subsection{Implications for selected slow-roll inflationary models}

\label{sec:implns_sel_slow_roll}

The predictions for $(n_\mathrm{s},r)$ to first order in the slow-roll 
approximation for a few inflationary models are shown in 
Fig.~\ref{fig:nsvsr}, which updates figure~12 of \citetalias{planck2014-a24} 
and figure~1 of \citetalias{planck2013-p17} with the same notation.  
These predictions are calculated for scale $k = 0.002$\,Mpc$^{-1}$ and include 
an uncertainty in the number of $e$-folds of $50 < N_* < 60$.

\begin{table}[tb]                 
\begin{center}
  \begingroup
  \newdimen\tblskip \tblskip=5pt
  \nointerlineskip
  \vskip -3mm
  \footnotesize
  \setbox\tablebox=\vbox{
    \newdimen\digitwidth
    \setbox0=\hbox{\rm 0}
    \digitwidth=\wd0
    \catcode`*=\active
    \def*{\kern\digitwidth}
    \newdimen\signwidth
    \setbox0=\hbox{+}
    \signwidth=\wd0
    \catcode`!=\active
    \def!{\kern\signwidth}
    \halign{%
      \hfil#\hfil&
      \hfil#\hfil\cr
      \noalign{\doubleline}
      Parameter range &
      Prior type\cr
      \noalign{\vskip 3pt\hrule\vskip 5pt}
      $0.019< \Omega_\mathrm{b} h^2 <0.025$ &
      uniform
      \cr
      $0.095< \Omega_\mathrm{c} h^2 <0.145$ &
      uniform
      \cr
      $1.03< 100\theta_\mathrm{MC} <1.05$ &
      uniform
      \cr
      $0.01< \tau< 0.4$ &
      uniform
      \cr
      \noalign{\vskip 3pt\hrule\vskip 5pt}}}
    \endPlancktable                    
  \endgroup
\end{center}
\caption{%
Priors for cosmological parameters used in the Bayesian comparison of inflationary models.}
\label{tab:Bayesian_comparison_priors_six}                            
\end{table}                        

In the following we discuss the implications of the \Planck\ 2018 
data release by taking into account the uncertainties in the entropy generation stage 
for a selection of representative standard single-field slow-roll inflationary models, 
updating the analysis reported in \citetalias{planck2013-p17} and \citetalias{planck2014-a24}.
As in \citetalias{planck2014-a24}, we use the primordial power spectra of cosmological 
fluctuations generated during slow-roll inflation parameterized by the HFFs, $\epsilon_i$, to second order, 
which can be expressed in terms of the parameters of the inflationary model and 
the number of $e$-folds to the end of inflation, $N_*$ \citep{Liddle:2003as,Martin:2010kz}, 
given by \citepalias{planck2013-p17}
\begin{align}
N_* \simeq\; &67 - \ln \left(\frac{k_*}{a_0 H_0}\right)
+  \frac{1}{4}\ln{\left( \frac{V_*^2}{\Mpl^4\,\rhoend}\right) } \nonumber\\
&+ \frac{1-3w_\mathrm{int}}{12(1+w_\mathrm{int})} \ln{\left(\frac{\rhorh}{\rhoend} \right)} - \frac{1}{12} \ln (g_\mathrm{th} ) \; , \label{eq:nefolds}
\end{align}
where $\rhoend$ is the energy density at the end of inflation, $a_0 H_0$ is the 
present Hubble scale, 
$V_*$ is the potential energy when $k_*$ left the Hubble radius during inflation, 
$w_\mathrm{int}$ characterizes the effective equation of 
state between the end of inflation and the thermalization energy scale $\rhorh$, and $g_\mathrm{th}$ 
is the number of effective bosonic degrees of freedom at the energy scale $\rhorh$. 
We fix $g_\mathrm{th}=10^3$ and $\epsilon_\mathrm{end}=1$, and we use modified routines of
the public code {\tt ASPIC}\footnote{\url{http://cp3.irmp.ucl.ac.be/~ringeval/aspic.html}} \citep{Martin:2014vha}.
In order to make contact with Fig.~\ref{fig:nsvsr}, we consider the pivot scale 
$k_*=0.002$\,Mpc$^{-1}$ in this subsection. 
We assume the uniform priors for the cosmological
parameters listed in Table~\ref{tab:Bayesian_comparison_priors_six}, and 
logarithmic priors on $10^{10} A_\mathrm{s}$
(over the interval $[e^{2.5}, e^{3.7}]$) and $\rhorh$ (over the interval 
$[(1\,\mathrm{TeV})^4, \rhoend]$). Prior ranges for 
additional parameters in the inflationary models considered are listed in 
Table~\ref{table:model_compar}.  
In this paper we consider the implications of the \Planck\ 2018 data for the selection of representative models 
studied in \citetalias{planck2014-a24} by restricting ourselves to $w_\mathrm{int} = (p-2)/(p+2)$, when 
the potential can be approximated as $V(\phi) \propto \phi^p$ during the coherent 
oscillation regime after inflation, or simply $w_\mathrm{int} = 0$ when the 
potential considered describes only the inflationary stage.\footnote{Note 
that some inflationary potentials in this selection are
a valid model for all stages, from the slow-roll phase all the way to
coherent oscillations around the minimum during reheating, while
others are ``incomplete'' in the sense that they only describe the
slow-roll regime. The hilltop, D-brane, potential with exponential
tails, and spontaneously broken SUSY models fall into the latter category and rely on
additional terms, denoted by the ellipses, to complete the potential
at the end of inflation.  With the increasing precision of CMB data and
accompanying accuracy requirements for theoretical predictions, the
precise form of the additional terms may affect the scientific
interpretation of some incomplete models, as pointed out for the case
of quadratic hilltop and double-well inflationary models in \citetalias{planck2014-a24}.}  
For data we use the 
full constraining power of \Planck, i.e., \Planck\ TT,TE,EE+lowE+lensing, 
in combination with BK15.

\begin{figure}[!h]
\includegraphics[width=8.2cm]{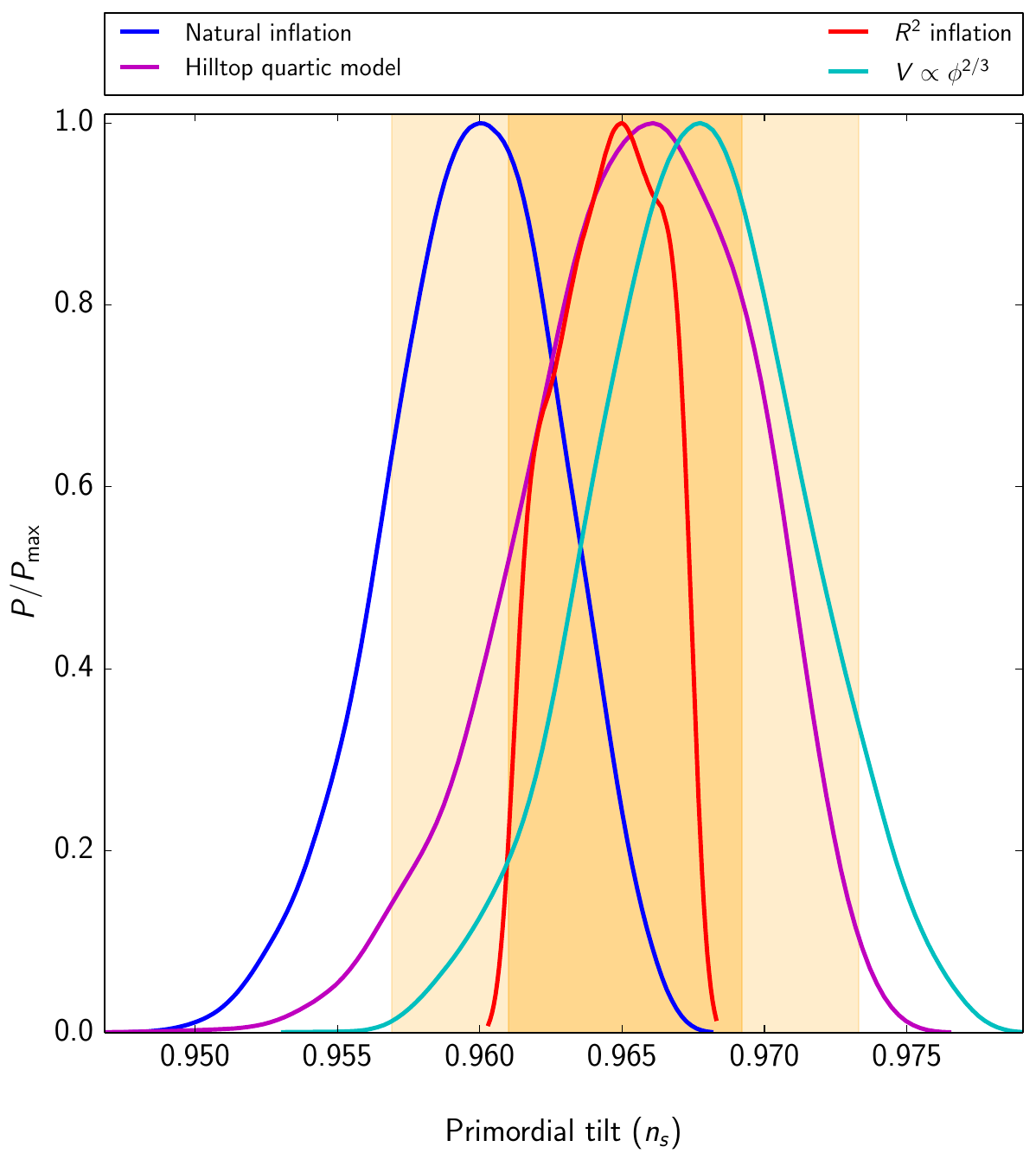} \\
\includegraphics[width=8.2cm]{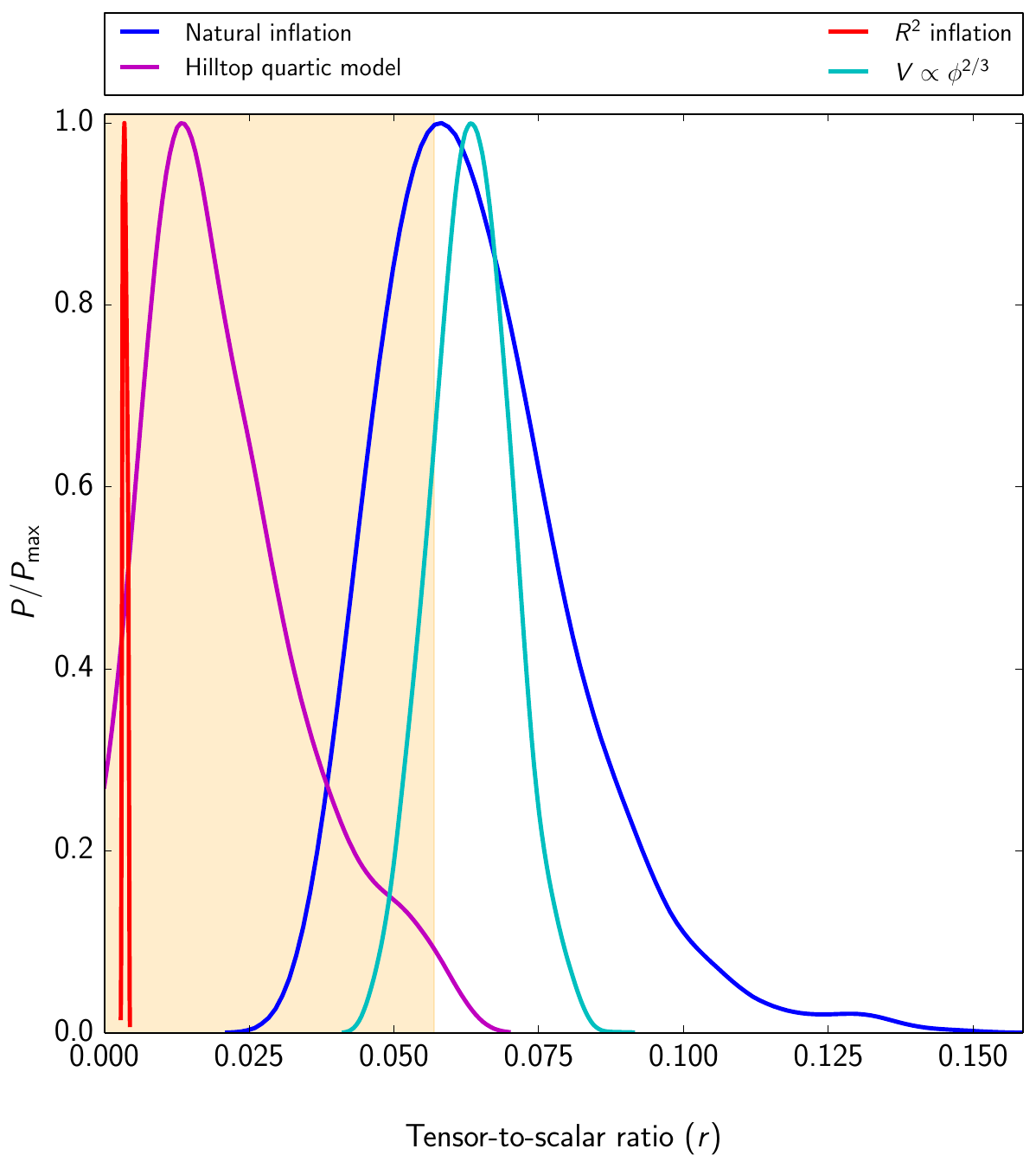}
\caption{Marginalized probability densities of the scalar tilt $n_\mathrm{s}$ (top panel) and $r$ (bottom panel) 
at $k=0.002$\,Mpc$^{-1}$ for natural, $R^2$, hilltop quartic, and $V(\phi) \propto \phi^{2/3}$ inflation, 
obtained by marginalizing over the uncertainties 
in the entropy generation stage, compared to 
the corresponding 68\,\% and 95\,\% CL limits obtained from a $\Lambda$CDM-plus-tensor 
fit.} \label{fig:ns_reh}
\end{figure}


\begin{table*}
\begin{center}
\begin{tabular}{lccrr}
\noalign{\hrule\vskip 2pt} \noalign{\hrule\vskip 3pt}
Inflationary model & Potential $V(\phi)$ & Parameter range & $\Delta\chi^2$ & $\ln B$ \\
\noalign{\hrule\vskip 3pt}
$R + R^2/(6 M^2)$ & $\Lambda^4 \left( 1 - e^{- \sqrt{2/3} \phi/M_\mathrm{Pl} } \right)^2$ & \dots & \dots & $ \dots $ \\
\noalign{\vskip 2pt}
Power-law potential & $\lambda M_\mathrm{Pl}^{10/3} \phi^{2/3}$ & \dots & $4.0$ & $-4.6$ \\
\noalign{\vskip 2pt}
Power-law potential & $\lambda M_\mathrm{Pl}^3 \phi$ & \dots & $6.8$ & $-3.9$ \\
\noalign{\vskip 2pt}
Power-law potential & $\lambda M_\mathrm{Pl}^{8/3} \phi^{4/3}$ & \dots & $12.0$ & $-6.4$ \\
\noalign{\vskip 2pt}
Power-law potential & $\lambda M_\mathrm{Pl}^2 \phi^2$ & \dots & $21.6$ & $-11.5$ \\
\noalign{\vskip 2pt}
Power-law potential & $\lambda M_\mathrm{Pl} \phi^3$ & \dots & $44.7$ & $-13.2$ \\
\noalign{\vskip 2pt}
Power-law potential & $\lambda \phi^4$ & \dots & $75.3$ & $-56.0$ \\
\noalign{\vskip 2pt}
Non-minimal coupling & $\lambda^4 \phi^4 + \xi \phi^2 R/2 $ & $-4 < \log_{10} \xi < 4$ & $0.4$ & $-2.4$ \\
\noalign{\vskip 2pt}
Natural inflation & $\Lambda^4 \left[ 1+\cos \left( \phi / f \right) \right]$ & $0.3 < \log_{10} (f/M_\mathrm{Pl}) < 2.5$ & $9.9$ & $-6.6$ \\
\noalign{\vskip 2pt}
Hilltop quadratic model & $\Lambda^4 \left( 1 - \phi^2 / \mu_2^2 + \dots \right)$ & $0.3 < \log_{10} (\mu_2 /M_\mathrm{Pl}) < 4.85$ & $1.3$ & $-2.0$ \\
\noalign{\vskip 2pt}
Hilltop quartic model & $\Lambda^4 \left( 1 - \phi^4 / \mu_4^4 + \dots \right)$ & $-2 < \log_{10} (\mu_4 /M_\mathrm{Pl}) < 2$ & $-0.3$ & $-1.4$ \\
\noalign{\vskip 2pt}
D-brane inflation ($p=2$) & $\Lambda^4 \left( 1 - \mu^2_{{\rm D} \, 2} / \phi^p + \dots \right)$ & $- 6  < \log_{10} (\mu_{{\rm D} \, 2} /M_\mathrm{Pl}) < 0.3$ & $-2.0$ & $0.6$ \\
\noalign{\vskip 2pt}
D-brane inflation ($p=4$) & $\Lambda^4 \left( 1 - \mu^4_{{\rm D} \, 4} / \phi^p + \dots \right)$ & $- 6  < \log_{10} (\mu_{{\rm D} \, 4} /M_\mathrm{Pl}) < 0.3$ & $-3.5$ & $-0.4$ \\
\noalign{\vskip 2pt}
Potential with exponential tails & $\Lambda^4 \left[ 1 - \exp{\left(-q \phi / M_\mathrm{Pl}\right)} + \dots \right]$ & $- 3  < \log_{10} q < 3$ & $-0.4$ & $-1.0$ \\
\noalign{\vskip 2pt}
Spontaneously broken SUSY & $\Lambda^4 \left[ 1 + \alpha_h \log \left(\phi / M_\mathrm{Pl}\right) + \dots  \right]$ & $- 2.5  < \log_{10} \alpha_h < 1$ & $6.7$ & $-6.8$ \\
\noalign{\vskip 2pt}
E-model $(n = 1)$ & $\Lambda^4 \left\{ 1 - \exp{\left[- \sqrt{2} \phi \left(\sqrt{3 \alpha_1^{\rm E}} M_\mathrm{Pl}\right)^{-1}\right] }\right\}^{2 n} $ & $-2 < \log_{10} \alpha_1^{\rm E} < 4$ & $0.8$ & $-0.3$ \\
\noalign{\vskip 2pt}
E-model $(n = 2)$ & $\Lambda^4 \left\{ 1 - \exp{\left[- \sqrt{2} \phi \left(\sqrt{3 \alpha_2^{\rm E}} M_\mathrm{Pl}\right)^{-1}\right] }\right\}^{2 n} $ & $-2 < \log_{10} \alpha_2^{\rm E} < 4$ & $0.8$ & $-1.6$ \\
\noalign{\vskip 2pt}
T-model $(m = 1)$ & $\Lambda^4 \tanh^{2 m} \left[ \phi \left(\sqrt{6 \alpha_1^{\rm T}} M_\mathrm{Pl}\right)^{-1}\right] $ & $-2 < \log_{10} \alpha_1^{\rm T} < 4$ & $-0.1$ & $-1.2$ \\
\noalign{\vskip 2pt}
T-model $(m = 2)$ & $\Lambda^4 \tanh^{2 m} \left[ \phi \left(\sqrt{6 \alpha_2^{\rm T}} M_\mathrm{Pl}\right)^{-1}\right] $ & $-2 < \log_{10} \alpha_2^{\rm T} < 4$ & $0.8$ & $-0.6$ \\
\noalign{\vskip 2pt\hrule}
\end{tabular}
\end{center}
\caption{\label{table:model_compar} 
Bayesian comparison for a selection of slow-roll inflationary models with $w_\mathrm{int}$ fixed (see text for more details). 
We quote 0.3 as the error on the Bayes factor. Models are strongly disfavoured when $\ln B < -5$. }
\end{table*}

The $\Delta \chi^2$ and the Bayesian evidence values for a selection of 
inflationary models with respect to the 
$R^2$ model \citep{Starobinsky:1980te,Mukhanov:1981xt,Starobinsky:1983zz}
are shown in Table~\ref{table:model_compar}. Figure~\ref{fig:ns_reh} shows the 
resulting marginalized probability densities 
of $n_\mathrm{s}$ and $r$ at $k=0.002$\,Mpc$^{-1}$ for a few inflationary models with the above specified priors, compared to 
the corresponding 68\,\% and 95\,\% CL limits obtained from a $\Lambda$CDM-plus-tensor fit.
We refer the interested 
reader to \citetalias{planck2014-a24} for a concise 
description of the inflationary models studied here 
and we limit ourselves here to a summary of the main results of this analysis.

\begin{itemize}

\item{The inflationary predictions \citep{Mukhanov:1981xt,Starobinsky:1983zz}
originally computed for the $R^2$ model \citep{Starobinsky:1980te}
to lowest order,
\be
n_\mathrm{s} - 1 \simeq - \frac{2}{N} \,, \quad r \simeq \frac{12}{N^2},
\ee
are in good agreement with \Planck\ 2018 data, confirming the previous 2013 and 2015 results. 
The 95\,\% CL allowed range $49<N_*<59$ is compatible 
with the $R^2$ basic predictions $N_* = 54$, corresponding to $T_\mathrm{reh} 
\sim 10^9$\,GeV \citep{Bezrukov:2011gp}. 
A higher reheating temperature $T_\mathrm{reh} \sim 10^{13}$\,GeV, as predicted 
in Higgs inflation \citep{Bezrukov:2007ep}, 
is also compatible with the \Planck\ data. }

\item{
Monomial potentials \citep{Linde:1983gd} $V(\phi) = \lambda M_\mathrm{Pl}^4 
\left( \phi / M_\mathrm{Pl} \right)^p$ with $p \ge 2$ are strongly disfavoured 
with respect to the $R^2$ model. For these values the Bayesian evidence 
is worse than in 2015 because of the smaller level of tensor modes allowed by \citetalias{bk15}.
Models with $p=1$ or $p=2/3$ \citep{Silverstein:2008sg, McAllister:2008hb,
McAllister:2014mpa} are more compatible with the data.
}

\item{There are several mechanisms which could lower the predictions 
for the tensor-to-scalar ratio for a given potential $V(\phi)$ in single-field inflationary models.
Important examples are a subluminal inflaton speed of sound due to a 
non-standard kinetic term \citep{1999PhLB..458..219G},
a non-minimal coupling to gravity \citep{Spokoiny:1984bd,Lucchin:1985ip,Salopek:1988qh,Fakir:1990eg}, 
or an additional damping term for
the inflaton due to dissipation in other degrees of freedom, as in warm inflation \citep{Berera:1995ie,Bastero-Gil:2016qru}.
In the following we report on the constraints for a non-minimal coupling to 
gravity of the type $F(\phi) R$, with $F(\phi) = M^2_\mathrm{Pl}
+ \xi \phi^2$, and a quartic potential.
For this model we compute the theoretical predictions in terms of HFFs and
number of $e$-folds to the end of inflation in the Einstein frame as for the 
$R^2$ model above, but we omit the technical details for the sake of 
brevity.\footnote{In this model the potential in the Einstein
frame is known only in implicit form (see, for instance, \citet{GarciaBellido:2008ab}) and the algebra is therefore more complicated.}  
Our results show that a quartic potential, which would be 
excluded at high statistical significance
for a minimally-coupled scalar inflaton as seen from Table~\ref{table:model_compar}, 
can be reconciled with the \Planck\ and BK15
data for $\xi > 0$: we obtain a 95\,\% CL lower limit $\log_{10} \xi > - 1.5$ 
with $\ln B = - 2.4$.}

\item{Natural inflation \citep{1990PhRvL..65.3233F,Adams:1992bn} is strongly disfavoured by 
the \Planck\ 2018 plus BK15 data with a Bayes factor $\ln B = -6.6$. 
}

\item{Within the class of hilltop inflationary models \citep{Boubekeur:2005zm} 
we find that a quartic potential provides a better fit than 
a quadratic one. In the quartic case we find the 95\,\% CL lower limit 
$\log_{10} (\mu_2 / M_\mathrm{Pl}) > 1.0$.}

\item{D-brane inflationary models \citep{Kachru:2003sx,Dvali:2001fw,GarciaBellido:2001ky}
provide a good fit to \Planck\ and BK15 data for a large portion of their parameter space. 
}

\item{For the simple class of inflationary potentials with exponential tails 
\citep{Goncharov:1985yu,Stewart:1994ts,Dvali:1998pa,Burgess:2001vr,Cicoli:2008gp} 
we find $\ln B = -1.0$.}

\item{
\Planck\ 2018 and BK15 data strongly disfavour the hybrid model driven by logarithmic quantum 
corrections in spontaneously broken supersymmetric (SB SUSY) theories \citep{Dvali:1994ms}, 
with $\ln B = -6.8$.}

\item{\Planck\ and BK15 data set tight constraints on $\alpha$ 
attractors \citep{Kallosh:2013yoa,Ferrara:2013rsa}. 
We obtain $\log_{10} \alpha_1^{\rm E} < 1.3$ and $\log_{10} \alpha_2^{\rm E} < 1.1$ 
at 95\,\% CL for the E-model. 
We obtain slightly tighter 95\,\% CL bounds for the T-model, i.e., 
$\log_{10} \alpha_1^{\rm T} < 1.0$ and $\log_{10} \alpha_2^{\rm T} < 1.0$.
Given the relation $|R_K| = 2/(3 \alpha)$ between the curvature 
of the K\"{a}hler geometry $R_K$ and $\alpha$ in some of the T-models 
motivated by supergravity, \Planck\ and BK15 data imply a lower 
bound on $|R_K|$, which is still in the low-curvature regime. 
The discrete set of values $\alpha = i/3$ with an integer $i$ 
in the range $[1,7]$ motivated by maximal supersymmetry 
\citep{Ferrara:2016fwe,Kallosh:2017ced} is compatible with the current data.}

\end{itemize}



\section{Reconstruction of the inflaton potential \label{sec:taylor}}

\subsection{Taylor expansion of $V($\texorpdfstring{$\phi$}{ϕ}$)$ in the observable
region \label{Sec:TaylorV}}

In this section, as in section~6 of \citetalias{planck2013-p17} and section~7.1 of
\citetalias{planck2014-a24}, we try to reconstruct the inflaton potential only in its
observable window, making no assumptions about the end of inflation. The motivation
for being so conservative is that what happens after the inflaton rolls down beyond
this range might not be captured by the simplest descriptions.  More elaborate
treatments would be required, for instance, in the case of a non-trivial potential
shape before the end of inflation, a waterfall transition involving extra scalar
fields, or several short inflationary stages between the time at which CMB scales
exit the Hubble radius and the nucleosynthesis epoch. The analysis of this section
relies, however, on the assumption that the potential is smooth enough inside the
observable window to be described by a Taylor expansion up to order four. Note that
this assumption is much weaker than assuming that a Taylor expansion is valid up to
the end of inflation. However, it excludes from the analysis potentials with sharp
features in the observable window, such as those studied in the next sections.

We perform the Taylor expansion around the value $\phi_*$ of the inflaton field
evaluated precisely at the time $t_*$ when the pivot scale
$k_*=0.05\,\mathrm{Mpc}^{-1}$ fulfills the relation $k_*=a(t_*)H(t_*)$. We
separately study the cases where the expansion is performed at order $n=2$, $n=3$, or
$n=4$. We compute the primordial spectrum with a full integration of the Fourier mode
evolution, using the inflationary module of the {\tt CLASS} code. Although this
method assumes no slow-roll approximation at any point, we speed up the convergence
of the Markov Chain by taking flat priors not directly on the five Taylor
coefficients $\{ V, V_{\phi}, \dots, V_{\phi\phi\phi\phi}\}$, but on combinations of
them matching the definitions of the potential slow-roll parameters $\{ \epsilon_V,
\eta_V, \xi_V^2, \varpi_V^3\}$ presented in table~2 of \citetalias{planck2013-p17}.
Even beyond the slow-roll approximation, these combinations provide nearly linear
contributions to the tilt, running, running of the running, etc., of the scalar and
tensor spectrum. Hence, they are directly related to observable quantities and well
constrained by the data. Instead, if we ran with flat priors on $\{ V, V_{\phi},
\dots, V_{\phi\phi\phi\phi}\}$, the convergence would be plagued by complicated
parameter degeneracies.

\begin{figure}[!ht]
\includegraphics[width=\columnwidth]{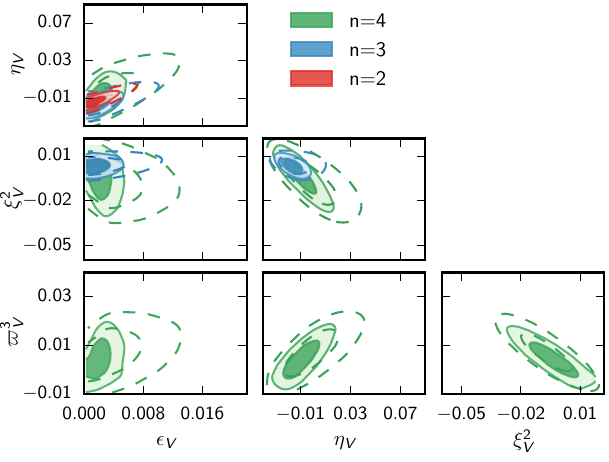}
\caption{Taylor expansion of $V(\phi)$ at order $n=2$, $3$, and $4$ in the observable
region, making no assumption about the end of inflation. The parameters are
combinations of Taylor coefficients with flat priors. Dashed contours are \Planck\
TT,TE,EE+lowE, while solid contours are \Planck\ TT,TE,EE+lowE+lensing+BK15.  The scales
are the same as in \citetalias{planck2014-a24}.} \label{fig:psr}
\end{figure}

\begin{figure}[!ht]
\includegraphics[width=\columnwidth]{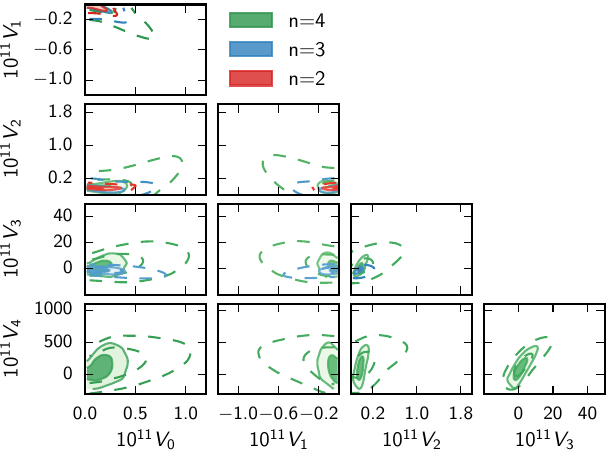}
\caption{Taylor expansion of $V(\phi)$ at order $n=2$, $3$, and $4$ in the observable
region, making no assumption about  the end of inflation.  In natural units (where
$\sqrt{8 \pi} M_\mathrm{Pl}=1$).  The parameters are the Taylor coefficients,
obtained here as derived parameters with non-flat priors.  Dashed contours are
\Planck\ TT,TE,EE+lowE, while solid contours are \Planck\ TT,TE,EE+lowE+lensing+BK15.
The scales are the same as in \citetalias{planck2014-a24}.} \label{fig:psr_V}
\end{figure}

The results of this analysis are shown in the panels of Fig.~\ref{fig:psr} and
Table~\ref{tab:V_phi} for $n=2$, $3$, and $4$, using two data sets for each: \Planck\
TT,TE,EE+lowE alone; or \Planck\ TT,TE,EE+lowE+lensing+BK15. The plot in
Fig.~\ref{fig:psr} deliberately has a lot of white space because, for the sake of
comparison, we plotted it over the same parameter ranges as the same plot in
\citetalias{planck2014-a24}. We notice some significant improvement. Comparing
\Planck\ TT,TE,EE+lowE results from 2015 and 2018, we find that error bars on
individual parameters have typically been reduced by 30\,\% thanks to improved
polarization data. Including BK data provides further constraining power. Comparing
\Planck\ 2015 TT+lowP+BAO and \Planck\ 2018 TT,TE,EE+lowE+lensing+BK15, we find that
the error bars on $\{ \epsilon_V,
\eta_V, \xi_V^2, \varpi_V^3\}$ shrink by factors of 2 to 4. The new
data tend to resolve degeneracies which previously appeared in the $n=4$ case and
could be understood as a compensation mechanism between potentials with large running
of the tilt, running of the running, tensor contribution, etc. The parameters
$\xi_V^2$ and $\varpi_V^3$ are perfectly compatible with zero (see Fig.~\ref{fig:psr}
and Table~\ref{tab:V_phi}), and so are $V_{\phi\phi\phi}$ and  $V_{\phi\phi\phi\phi}$
(see the contours on the parameters $\{ V, V_{\phi}, \dots, V_{\phi\phi\phi\phi}\}$
in Fig.~\ref{fig:psr_V}). This is consistent with the fact that the new data set
brings no evidence for running or running of the running. It also explains why the
results of this section are close to those of Sect.~\ref{sec:implns_slow_roll},
obtained under the slow-roll approximation. Similar to 2015, the best-fit value of
running for $n=3$ is negative, but has moved down from $-0.013$ to $-0.007$, and
remains compatible with zero at the 1.0$\sigma$ level. For $n=4$, the trend observed
in 2015 to fit the data slightly better with a non-zero tensor contribution has
disappeared.
The decrease of the minimum effective $\chi^2$ when moving from $n=2$ to $n=3$ is
insignificant and even smaller than in 2015, showing that the data do not require
anything more complicated than an approximately parabolic shape for the inflaton
potential within the observable window.

\begin{figure}[!ht]
\includegraphics[width=8cm]{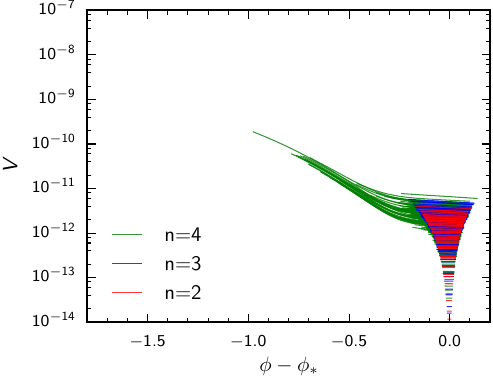}
\caption{Representative sample of the observable region of inflaton potentials
allowed at the 95\,\% CL, when the potential is Taylor-expanded at order $n=2$, $3$,
and $4$ in the observable region, making no assumption about the end of inflation,
and using \Planck\ TT,TE,EE+lowE+lensing+BK15.  In natural units (where $\sqrt{8 \pi}
M_\mathrm{Pl}=1$).  We use the same scales as in \citetalias{planck2014-a24}. Note
that there is another branch of solutions that is symmetric under $(\phi-\phi_*)
\rightarrow -(\phi-\phi_*)$. } \label{fig:psr_Vphi}
\end{figure}

This can be checked by considering the random sample of well-fitting potentials
presented in Fig.~\ref{fig:psr_Vphi}. Actually, for $n=4$, a few of the plotted
potentials have a non-parabolic ``spoon-like'' shape (with a kink and a plateau),
because non-negligible values of $|V_{\phi\phi\phi\phi}|$ are still allowed.
However, this sub-class of models is by no means preferred over simpler
parabolic-like potentials with a negligible $|V_{\phi\phi\phi\phi}|$; otherwise, we
would have obtained a better $\chi^2_{\rm eff}$ for $n=4$. Hence one should {\it not}
take from Fig.~\ref{fig:psr_Vphi} the message that special potentials with a kink and
a plateau are favoured by the \Planck\ data. Comparing this plot to figure~15 of
\citetalias{planck2014-a24}, we see that the models with the largest $V(\phi)$
amplitude are excluded by stronger bounds on the tensor modes.

Finally, it is interesting to notice that the predictions for the parameters of the
minimal $\Lambda$CDM model, such as $n_{\rm s}$ or $\tau$, remain extremely stable
when increasing the freedom in the inflaton potential.

\begin{table}
\begin{center}
\begin{tabular}{c c c c}
\noalign{\hrule\vskip 2pt}
\noalign{\hrule\vskip 3pt}
$n$ & 2 & 3 & 4  \\
\hline
&&&\\[-1.5ex]
$\epsilon_V$  & $<0.0042$ & $<0.0045$ &$<0.0048$ \\[1.1ex]
$\eta_V$      & $-0.0124_{-0.0052}^{+0.0033}$ & $-0.0163_{-0.0063}^{+0.0057}$ & $-0.0082_{-0.0120}^{+0.0096}$ \\[1.1ex]
$\xi_V^2$      & \dots & $\phantom{-}0.0036_{-0.0037}^{+0.0035}$ & $-0.004_{-0.009}^{+0.011}$  \\[1.1ex]
$\varpi_V^3$    & \dots & \dots & $0.0048_{-0.0067}^{+0.0052}$ \\[0.5ex]
\hline
&&&\\[-1.5ex]
$\tau$         & $0.0546_{-0.0075}^{+0.0072}$ & $0.0559_{-0.0080}^{+0.0074}$ & $0.0571_{-0.0085}^{+0.0077}$\\[1.1ex]
$n_\mathrm{s}$ & $0.9650_{-0.0042}^{+0.0042}$ & $0.9639_{-0.0048}^{+0.0043}$ & $0.9623_{-0.0050}^{+0.0047}$\\[1.1ex]
$10^3 \frac{dn_\mathrm{s}}{d\ln k}$  & $-0.37_{-0.19}^{+0.29}$ & $-7.3_{-6.7}^{+7.0}$ & $-1.9_{-9.2}^{+9.0}$ \\[1.1ex]
$r_{0.002}$ & $<0.060$ & $<0.063$ & $<0.069$ \\[0.5ex]
\hline
&&&\\[-1.5ex]
$\Delta \chi_\mathrm{eff}^2$ & \dots  & $\Delta \chi^2_{3/2}= -0.22$ &
$\Delta \chi^2_{4/3}=-0.82$
\\[1.1ex]
\end{tabular}
\end{center}
\caption{Numerical reconstruction of the potential slow-roll parameters {\it beyond} any slow-roll approximation, when the potential is Taylor-expanded to $n$th order, using \Planck\ TT,TE,EE+lowE+lensing+BK15. We also show the corresponding bounds on some related parameters (here $n_\mathrm{s}$, $dn_\mathrm{s}/d\ln k$, and $r_{0.002}$ are derived from the numerically computed primordial spectra). All error bars are at the 68\,\% CL and all upper bounds at the 95\,\% CL. The effective $\chi^2$ value of model $n$ is given relative to model $n-1$.
\label{tab:V_phi}}
\end{table}

\subsection{Taylor expansion of $H($\texorpdfstring{$\phi$}{ϕ}$)$ in the observable
region \label{Sec:TaylorH}}

To assess the robustness of our method, in this section we repeat the analysis with a
Taylor expansion of the Hubble function $H(\phi)$ in the observable window, as we did
in 2015.  We refer the reader to section~7.2 of \citetalias{planck2014-a24} for a
precise description of this analysis, and we recall that the difference with respect
to the $V(\phi)$ reconstruction is more than a mere change of priors. For each value
of $n$, the new parameterization covers a slightly different range of potentials,
and, more importantly, it naturally includes a marginalization over the uncertainty
in the initial value of the derivative $\dot{\phi}$ when the inflaton enters the
observable window. Instead, in the previous analysis, $\dot{\phi}$ was assumed to
have reached the inflaton attractor solution, i.e., there was an implicit assumption
that inflation started well before that time. In the analysis based on $H(\phi)$,
inflation models with a minimal duration are not excluded by the priors.

\begin{figure}[!ht]
\includegraphics[width=\columnwidth]{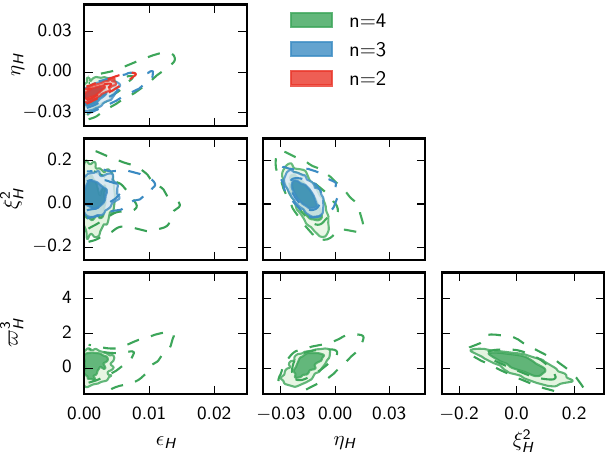}
\caption{Taylor expansion of $H(\phi)$ at order $n=2$, $3$, and $4$ in the observable
region, making no assumption about  the end of inflation. The parameters are
combinations of Taylor coefficients with flat priors.  Dashed contours are \Planck\
TT,TE,EE+lowE, while solid contours are \Planck\ TT,TE,EE+lowE+lensing+BK15.  The scales
are the same as in \citetalias{planck2014-a24}.} \label{fig:hsr}
\end{figure}

The improvement with respect to the 2015 results is even more impressive in this
case. Bounds on the $n=4$ parameters are typically 3 to 4 times stronger compared
with 2015, as can be checked from  Table~\ref{tab:H_phi} and Fig.~\ref{fig:hsr}. We
found that a factor of 2 improvement comes from switching to the new set of
low-$\ell$ likelihoods, and another factor of 2 from adding the BK likelihood. On the
other hand, the use of more recent high-$\ell$ and lensing likelihoods has a modest
impact.

\begin{figure}[!ht]
\includegraphics[width=8cm]{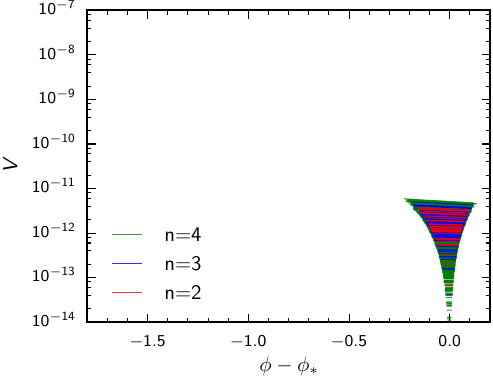}
\caption{Representative sample of the observable region of inflaton potentials
allowed at the 95\,\% CL, inferred from $H(\phi)$ when that function is
Taylor-expanded at order $n=2$, $3$, and $4$ in the observable region, making no
assumption about  the end of inflation, and using \Planck\ TT,TE,EE+lowE+lensing+BK15.
In natural units (where $\sqrt{8 \pi} M_\mathrm{Pl}=1$).   The scales are the same as
in \citetalias{planck2014-a24}. Note that there is another branch of solutions
symmetric under $(\phi-\phi_*) \rightarrow -(\phi-\phi_*)$.} \label{fig:hsr_Vphi}
\end{figure}

A consequence of these improved constraints can be seen in Fig.~\ref{fig:hsr_Vphi},
when we compare it to its counterpart from 2015 (figure~20 in
\citetalias{planck2014-a24}). Again, for a better comparison Fig.~\ref{fig:hsr_Vphi}
uses the same scale as figure~20 of \citetalias{planck2014-a24}. For $n=4$, the
previously best-fitting models included many scenarios starting with a fast-roll
stage, producing a tail with large $V(\phi)$ before pivot-scale crossing. These
models are now excluded by better polarization data and tensor constraints.

Going beyond the parabolic approximation for $H(\phi)$ does not improve the
goodness-of-fit: as in the potential-based analysis of Sect.~\ref{Sec:TaylorV}, the
$\Delta \chi^2$s between $n=2$, $n=3$, and $n=4$ are negligible, and the parameters
$\xi_H^2$ and $\varpi_H^3$ related to $H_{\phi\phi\phi}$ and $H_{\phi\phi\phi\phi}$
are compatible with zero.

\begin{table}
\begin{center}
\begin{tabular}{c c c c}
\noalign{\hrule\vskip 2pt}
\noalign{\hrule\vskip 3pt}
$n$ & 2 & 3 & 4  \\
\hline
&&&\\[-1.5ex]
$\epsilon_H$  & $<0.0041 $& $<0.0046$ & $<0.0041$ \\[1.1ex]
$\eta_H$  & $-0.0139_{-0.0038}^{+0.0026}$ & $-0.0170_{-0.0048}^{+0.0044}$ & $-0.0158_{-0.0056}^{+0.0057}$ \\[1.1ex]
$\xi_H^2$  & \dots & $0.046_{-0.045}^{+0.043}$ & $0.021_{-0.076}^{+0.071}$  \\[1.1ex]
$\varpi_H^3$  & \dots & \dots & $0.16_{-0.37}^{+0.64}$ \\[0.5ex]
\hline
&&&\\[-1.5ex]
$\tau$             & $0.0548_{-0.0074}^{+0.0075}$ & $0.0556_{-0.0078}^{+0.0076}$ & $0.0563_{-0.0078}^{+0.0073}$ \\[1.1ex]
$n_\mathrm{s}$     & $0.9651_{-0.0044}^{+0.0040}$ & $0.9637_{-0.0046}^{+0.0042}$ & $0.9637_{-0.0048}^{+0.0042}$  \\[1.1ex]
$10^3 \frac{dn_\mathrm{s}}{d\ln k}$ & $-0.25_{-0.12}^{+0.20}$ & $-7.5_{-6.7}^{+7.0}$ & $-5.1_{-8.1}^{+7.8}$ \\[1.1ex]
$r_{0.002}$         & $<0.059$ & $<0.065$ & $<0.057$ \\[0.5ex]
\hline
&&&\\[-1.5ex]
$\Delta \chi_\mathrm{eff}^2$ & \dots  & $\Delta \chi^2_{3/2}= -1.60$ & $\Delta \chi^2_{4/3}= -2.32$
\\[0.5ex]
\end{tabular}
\end{center}
\caption{Numerical reconstruction of the Hubble slow-roll parameters {\it beyond} any slow-roll approximation, using \Planck\ TT,TE,EE+lowE+lensing+BK15. We also show the corresponding bounds on some related parameters (here $n_\mathrm{s}$, $dn_\mathrm{s}/d\ln k$, and $r_{0.002}$ are derived from the numerically computed primordial spectra).  All error bars are at the 68\,\% CL and all upper bounds at the 95\,\% CL. The effective $\chi^2$ value of model $n$ is given relative to model
$n-1$.
\label{tab:H_phi}}
\end{table}

\subsection{Taylor expansion of full $V($\texorpdfstring{$\phi$}{ϕ}$)$
\label{Sec:TaylorVend}}

We now present a new analysis with less conservative assumptions than in the previous
subsections. We switch to the hypothesis that the inflaton potential is very smooth
not only within its observable window, but also until the end of inflation, such that
its whole shape can be captured by a Taylor expansion. We further assume that
inflation ends when the first slow-roll condition is violated ($\epsilon_V=1$),
without invoking any other field. Finally, we fix the number of $e$-folds between
Hubble crossing of the pivot scale and the end of inflation to $N_*=55$, which
implicitly relies on the hypothesis that no further inflationary stage took place at
a later epoch.

Technically, the analysis pipeline for this case is similar to that of
Sect.~\ref{Sec:TaylorV}, except for an extra step in which the {\tt CLASS}
inflationary module integrates the background equations until the end of inflation,
goes backwards in time by 55 $e$-folds, and imposes that the Hubble crossing for the
pivot scale $k_*=0.05\,\mathrm{Mpc}^{-1}$ matches that time.

These models are much more constrained than those of Sects.~\ref{Sec:TaylorV}
and~\ref{Sec:TaylorH}, since the $e$-fold condition is imposed in addition to having
a potential with a good shape within the observable window. The constraining power is
then sufficient for running the MCMC chains directly with
flat priors on $\{ V, V_{\phi}, \dots, V_{\phi\phi\phi\phi}\}$.

\begin{figure}[!ht]
\includegraphics[width=\columnwidth]{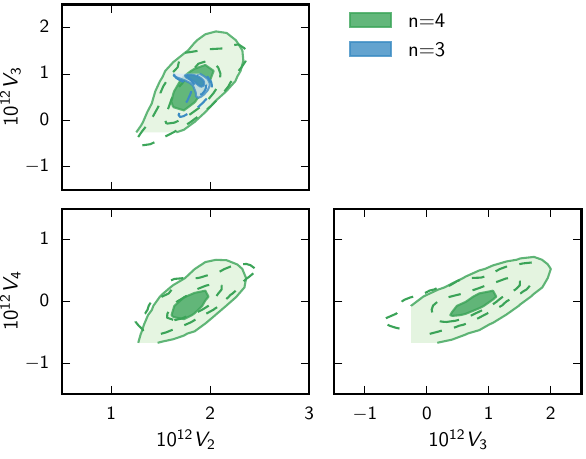}
\caption{Taylor expansion of the full $V(\phi)$ at order $n=3$ and $4$, trusted until
the end of inflation, in natural units (where $\sqrt{8 \pi} M_\mathrm{Pl}=1$). The
parameters are the Taylor coefficients with flat priors. Dashed contours are \Planck\
TT,TE,EE+lowE, while solid contours are \Planck\ TT,TE,EE+lowE+lensing+BK15.}
\label{fig:VE}
\end{figure}

\begin{figure*}[!ht]
\hspace*{1cm}\includegraphics[height=6.3cm]{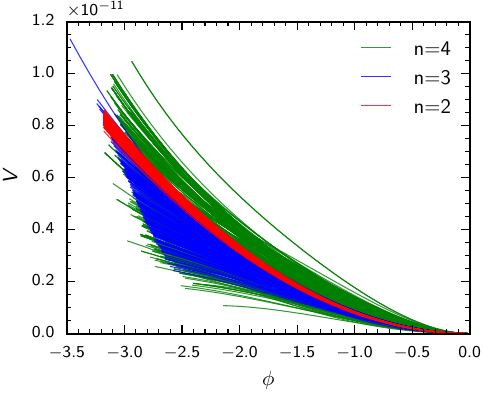}
\hspace*{2mm}\includegraphics[height=6.3cm]{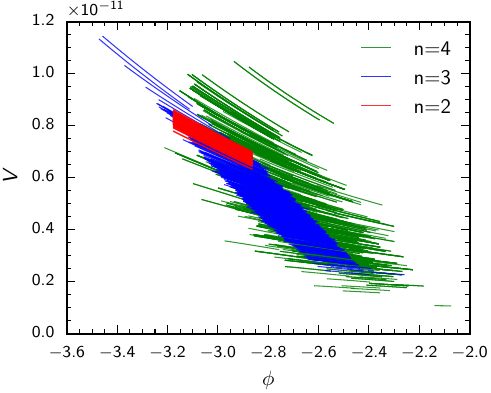}\\
\hspace*{1cm}\includegraphics[height=6.3cm]{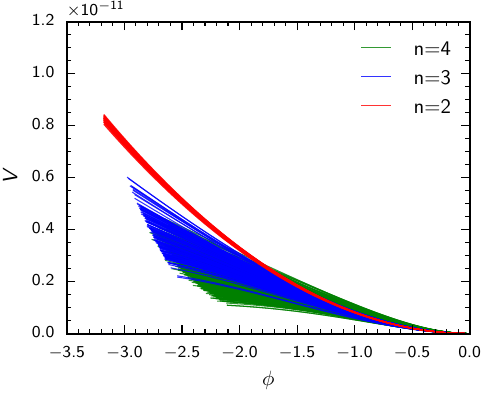}
\hspace*{2mm}\includegraphics[height=6.3cm]{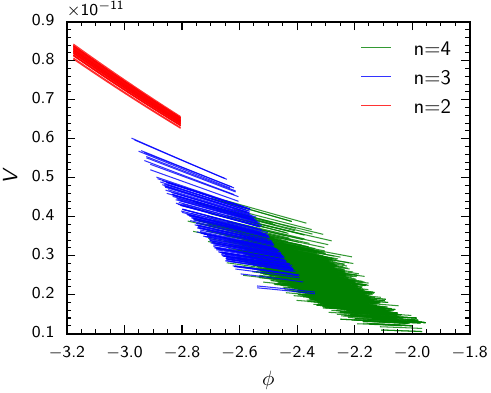}
\caption{Representative sample of the inflaton potentials allowed at the 95\,\% CL,
when the potential is Taylor-expanded at order $n=2$, $3$, and $4$ and trusted until
the end of inflation, and under the assumption of $N_*=55$ $e$-folds of inflation
between Hubble-radius crossing for the pivot scale and the end of inflation.  In
natural units (where $\sqrt{8 \pi} M_\mathrm{Pl}=1$). {\it Left panels:} Full
potential from the beginning of the observable window till the end of inflation.
{\it Right:} Zoom on the observable window directly constrained by inflation. {\it
Top:} \Planck\ TT,TE,EE+lowE.  {\it Bottom:} \Planck\ TT,TE,EE+lowE+lensing+BK15. Note
that there is another branch of solutions that is symmetric under $\phi \rightarrow
-\phi$.} \label{fig:VEphi}
\end{figure*}

Our results are presented in Figs.~\ref{fig:VE} and \ref{fig:VEphi} and in
Table~\ref{tab:VE_phi}. For models with a purely quadratic potential, the numerically computed tilt and tensor-to-scalar ratio depend
almost exclusively on $N_*$, thus they remain fixed to $n_s=0.963$ and $r_{0.002}=0.136$. Such a large $r$
is in tension with the \Planck\ data, and
even more so with the \Planck+BK data. Thus the effective $\chi^2$ is poor in the
$n=2$ case and improves considerably when adding some freedom in going to $n=3$.
Indeed, the presence of an additional cubic term allows us to reach smaller values of
the tensor-to-scalar ratio for roughly the same scalar tilt, and lowers $\chi^2_{\rm
eff}$ by more than 13. Instead, when also adding a quartic term, we find no
significant improvement in the goodness of fit, and the coefficient of the $\phi^4$
term is consistent with zero.

These findings are consistent with the global picture that \Planck\ data prefer
potentials which are concave in the observable window. The blue and green curves in
the lower left panel of Fig.~\ref{fig:VEphi}  illustrate the preference of the
\Planck+lensing+BK15 data for potentials with an inflection point, appearing
qualitatively similar to scalar field potentials associated with spontaneous symmetry
breaking models, hilltop models, new inflation, natural inflation, etc.

In these runs, the value of the scalar tilt running is always very precisely
constrained around a value of $dn_{\rm s}/d\ln k \simeq -6\times 10^{-4}$. This does
not come as a surprise if we keep in mind that these bounds are not imposed directly
by the data, but rather by the class of inflationary potentials considered here, with
potential parameters fixed by observational bounds on the {\it amplitude} and {\it
tilt} of the scalar and tensor spectra. In other words, the running is not directly
measured, but rather predicted as a function of the scalar/tensor amplitudes and
scalar tilt. Interestingly, if combinations of future CMB and large-scale structure
data with a wide lever arm in wavenumber space could become directly sensitive to
such tiny values (which would require a factor of around 10 improvement in
sensitivity compared to current CMB+BAO data), a very large class of currently
successful inflationary models could be either confirmed or ruled out.

\begin{table}
\begin{center}
\begin{tabular}{c c c c}
\noalign{\hrule\vskip 2pt}
\noalign{\hrule\vskip 3pt}
$n$ & 2 & 3 & 4  \\
\hline
&&&\\[-1.5ex]
$10^{12}V_2$  &$1.631_{-0.022}^{+0.022}$  &$1.81_{-0.06}^{+0.12}$ & $1.86_{-0.23}^{+0.25}$ \\[1.1ex]
$10^{12}V_3$      & \dots & $0.89_{-0.03}^{+0.10}$ & $0.85_{-0.64}^{+0.46}$ \\[1.1ex]
$10^{12}V_4$    & \dots & \dots & $0.044_{-0.35}^{+0.26}$\\[0.5ex]
\hline
&&&\\[-1.5ex]
$\tau$                  & $0.0518_{-0.0066}^{+0.0066}$ & $0.0501_{-0.0069}^{+0.0078}$ &$0.05628_{-0.0087}^{+0.0075}$  \\[1.1ex]
$n_\mathrm{s}$           & $0.963$ & $0.9599_{-0.0018}^{+0.0034}$&  $0.9656_{-0.0043}^{+0.0035}$\\[1.1ex]
$10^3 \frac{dn_\mathrm{s}}{d\ln k}$  &$-0.6731_{-0.0005}^{+0.0005}$  & $-0.534_{-0.096}^{+0.079}$&  $-0.74_{-0.13}^{+0.16}$\\[1.1ex]
$r_{0.002}$ &  $0.136$ & $0.066_{-0.016}^{+0.010}$ & $0.042_{-0.014}^{+0.009}$  \\[0.5ex]
\hline
&&&\\[-1.5ex]
$\Delta \chi_\mathrm{eff}^2$ & \dots  & $\Delta \chi^2_{3/2}= -13.18$ & $\Delta \chi^2_{4/3}=-3.50$
\\[1.1ex]
\end{tabular}
\end{center}
\caption{Numerical reconstruction of the potential parameters {\it beyond} any slow-roll approximation, when the potential is Taylor-expanded to $n$th order, trusted until the end of inflation, and using \Planck\ high-$\ell$ TT,TE,EE+lowE+lensing+BK15. We also show the corresponding bounds on some related parameters (here $n_\mathrm{s}$, $dn_\mathrm{s}/d\ln k$, and $r_{0.002}$ are derived from the numerically computed primordial spectra). All error bars are at the 68\,\% CL and all upper bounds at the 95\,\% CL. The effective $\chi^2$ value of model $n$ is given relative to model $n-1$.
\label{tab:VE_phi}}
\end{table}

\renewcommand{\d}[2][]{d^{#1}{#2}}

\subsection{Free-form potential reconstruction}
\label{sec:vphi_bayesian_reconstruction}

As a complementary analysis to the previous three subsections, we next perform
a free-form reconstruction of the inflationary potential with cubic splines, in
a manner akin to the reconstructions of \citetalias{planck2014-a24} and
Sect.~\ref{sec:ppsr_bayesian_reconstruction}. Further plots and theoretical detail can be found in~\citet{Handley2018b}. 

A free-form reconstruction usually proceeds by parameterizing the 
function of interest via a spline and taking the locations of the interpolation knots 
as free parameters in a posterior distribution. These are then varied along with any 
other model parameters, and then marginalized out to yield a model-independent 
reconstruction of the function of interest. The analysis is run for differing numbers 
of knots, $N$, and the Bayesian evidence is computed to allow for model comparison to 
determine how many knots are appropriate from the perspective of the data.

To reconstruct the inflationary potential $V(\phi)$, one cannot take a linear 
interpolating spline (as in Sect.~\ref{sec:ppsr_bayesian_reconstruction}), since the 
equations of motion in general depend on first (and sometimes second) derivatives of 
$V$.  We therefore choose to parameterize the {\em second derivative\/} of the 
log-potential as a linear spline. The log-potential is computed by integrating this 
function twice, yielding a function with two additional free parameters---a global 
offset and a gradient. Our reconstruction function is therefore
\begin{align}
    \ln V &= \ln V_* + (\phi-\phi_*)\frac{\d{\ln V_*}}{\d{\phi}}
    \nonumber\\
    &+ \int_{\phi_*}^\phi\d{{\phi^\prime}}\int_{\phi_*}^{\phi^\prime} 
\d{\phi^{\prime\prime}} \:\mathrm{LS}(\phi^{\prime\prime};\theta),\\
    \theta &= \left(\phi_1,\ldots,\phi_N,\frac{\d[2]{\ln 
V_1}}{\d{\phi}^2},\ldots,\frac{\d[2]{\ln V_N}}{\d{\phi}^2}\right),\\
    \mathrm{LS}(\phi;\theta) &=
    \left\{
        \begin{array}{r}
            \frac{\d[2]{\ln V_{i}}}{\d{\phi}^2} 
\frac{\phi-\phi_{i+1}}{\phi_{i}-\phi_{i+1}}
            +\frac{\d[2]{\ln V_{i+1}}}{\d{\phi}^2} 
\frac{\phi-\phi_i}{\phi_{i+1}-\phi_{i}}
            \\
            : \phi_i < \phi < \phi_{i+1}
        \end{array}
    \right..
\end{align}
Here $\mathrm{LS}(\phi;\theta)$ is a standard linear spline dependent on $N$ knots, 
$\ln V_*$ is the potential at the pivot scale, and $\d{\ln V_*}/\d{\phi}$ is the  
gradient of the log-potential at the pivot scale.

In general, any reconstruction of the potential will be sensitive only to the 
observable window of inflation in $\phi\in[\phi_{\min{}},\phi_{\max{}}]$, where  
$\phi_{\min{}}$ and $\phi_{\max{}}$ are defined as the field values when the 
largest and smallest observable scales $k_{\min{}}$ and $k_{\max{}}$ exit the 
Hubble radius during inflation. 
Regions of the potential outside these $\phi$ values are unconstrained by current CMB 
data. In our analysis, we take $k_{\min{}}=10^{-4}\,\mathrm{Mpc}^{-1}$ and 
$k_{\max{}}=10^{-0.3}\,\mathrm{Mpc}^{-1}$, which encompasses the multipole range 
constrained by \Planck\ (see Sect.~\ref{sec:ppsr_bayesian_reconstruction}). The 
locations $\phi_1,\ldots,\phi_N$ of the reconstruction knots should be distributed 
throughout this observable window. Whilst the locations $\phi_1,\ldots\phi_N$ and 
heights ${d^2{\ln V_1}}/{d{\phi}^2},\ldots,{d^2{\ln V_N}}/{d{\phi}^2}$ 
themselves influence the size of the observable window, a reasonable approach is to 
first estimate it using the unperturbed potential (i.e., setting $N=0$). This gives 
an alternative range $\phi\in[\tilde{\phi}_{\min{}},\tilde{\phi}_{\max{}}]$. The 
priors on all our variables are indicated in Table~\ref{tab:prior}.

\begin{table}
\begin{center}
    \begin{tabular}{lll}
        \noalign{\hrule\vskip 2pt}
        \noalign{\hrule\vskip 3pt}
        Parameters& Prior type& Prior range \\
        \hline
        &&\\[-1.5ex]
        $N$ & Discrete uniform & $[0, 8]$ \\
        $\ln V_*$ & Uniform & $[-25, -15]$ \\
        ${d\ln V_*}/{d{\phi}}$ & Log-uniform & $[10^{-3}, 10^{-0.3}]$ \\
        ${d^2{\ln V_1}}/{d{\phi}^2},\ldots,{d^2{\ln V_N}}/{d{\phi}^2}$ & Uniform & $[-0.5,0.5]$ \\
        $\phi_1,\ldots,\phi_N$ & Sorted uniform & $[\tilde{\phi}_{\min{}},\tilde{\phi}_{\max{}}]$ \\
        $\ln 10^{10}\mathcal{P}_\mathcal{R}(k)$ & Indirect constraint & $[2,4]$ \\
        \hline
    \end{tabular}
\end{center}
    \caption{Parameters of the free-form potential reconstruction analysis and 
details of the priors. There is a further prior constraint in that we require that 
the inflaton should evolve in an inflating phase throughout the observable window, 
that the inflaton should be rolling downhill from negative to positive $\phi$ 
throughout, and that any primordial power spectra generated sit in the range ${2 < 
\ln 10^{10} \mathcal{P}_\mathcal{R}(k) < 4}$.\label{tab:prior}}
\end{table}

\begin{figure*}[!h]
\begin{center}
    \includegraphics[scale=1]{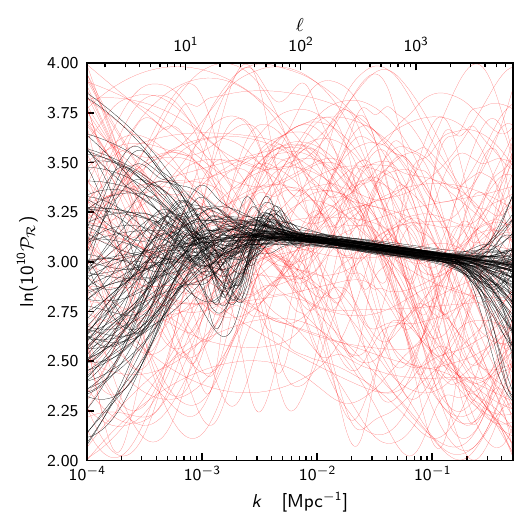}
    \hspace{0mm}\includegraphics[scale=1]{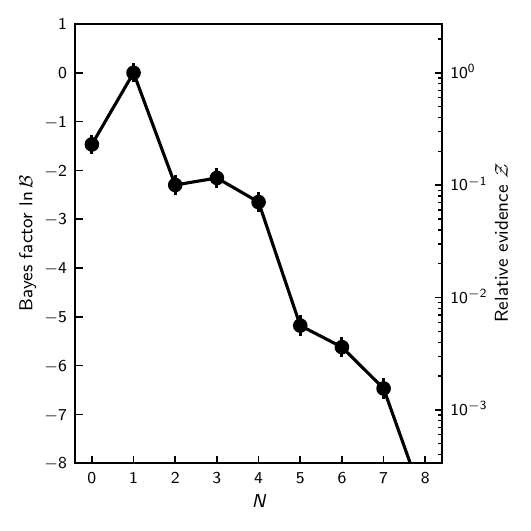}
    \includegraphics[scale=1]{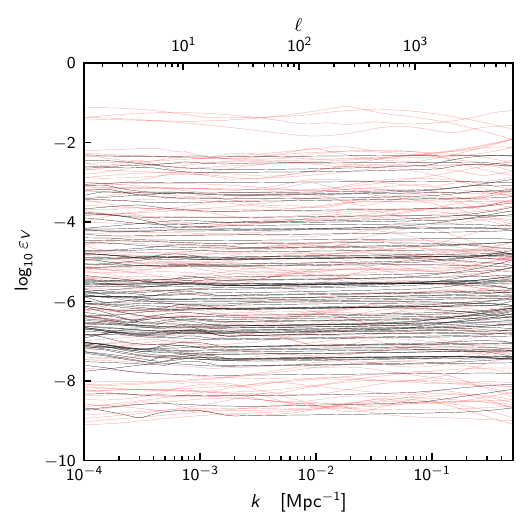}
    \includegraphics[scale=1]{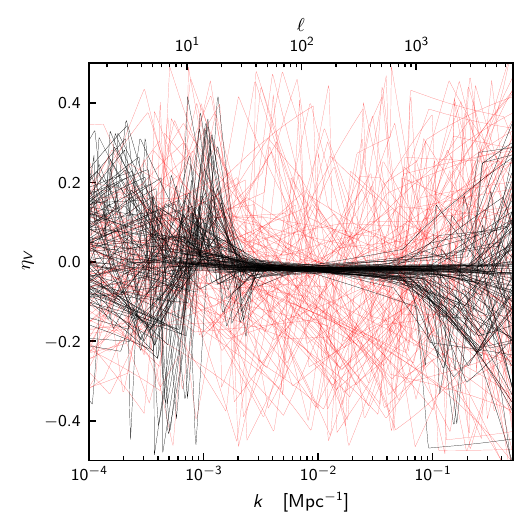}
\end{center}
    \caption{ Free-form potential reconstructions using \Planck\
TT+TE+EE+lowl+lowE+lensing (Sect.~\protect\ref{sec:vphi_bayesian_reconstruction}).
The top-right panel shows Bayes factors for the free-form potential reconstruction. The preferred
reconstruction has $N=1$, corresponding to a constant non-zero
${d^2{\ln V_1}}/{d{\phi}^2}$. The remaining panels show reconstructions for the $N=8$
knot case, focusing on the scalar primordial power spectrum, and the inflationary
slow-roll parameters $\varepsilon_V$ and $\eta_V$.  Red lines indicate sample
trajectories from the prior, whilst black lines are from the posterior. Technically
the slow-roll parameters are defined as functions of $\phi$, but we instead
substitute this for the Hubble-radius-exit value to make for clearer comparison between
posterior samples.  In all plots, the approximate link between $\ell$ and $k$ is via
the Limber approximation, $\ell \simeq k/D_\mathrm{A}$, where $D_A=r_*/\theta_*$ is
the comoving angular distance to recombination, which is at comoving distance
$r_*$.\label{fig:vphi_pps}}
\end{figure*}

Our results are detailed in Fig.~\ref{fig:vphi_pps}. The Bayesian evidence shows 
that the reconstruction preferred by the data is that using $N=1$, corresponding 
to a constant 
non-zero ${d^2{\ln V}}/{d{\phi}^2}$. This indicates that the \Planck\ data do not 
significantly constrain the inflationary potential within the window any further than 
up to a quadratic term in a Taylor expansion. It is illuminating, however, to 
consider adding further structure to the potential, and Fig.~\ref{fig:vphi_pps} 
shows reconstructions for $N=8$.

Considering the predictive posterior of the primordial power spectrum, we see that 
our parameterization is sufficient to exhibit the deficit at $\ell\simeq30$, cosmic 
variance at low $\ell$, and the loss of resolution at high $\ell$, as seen in 
Sect.~\ref{sec:ppsr_bayesian_reconstruction}. Consistent with the rest of the 
analyses, $\varepsilon_V$ is unconstrained, whilst the \Planck\ data provide 
relatively powerful constraints on $\eta_V$ within the observable window of 
inflation.

\section{Primordial power spectrum reconstruction \label{sec:reconstruction}}

\label{sec:section_five}

This section reports results for the non-parametric reconstruction 
of the primordial scalar power spectrum using the new \Planck\ 2018 
likelihoods, as well as comparisons with the previously reported results 
for the \Planck\ 2013 and 2015 releases.  
The objective here is to search for deviations from a simple power-law 
primordial power spectrum (i.e., ${\cal P}_{\cal R}(k) = 
A_{\rm s}(k/k_*)^{n_{\rm s} - 1}$) in a manner that does not presuppose
any particular theoretical model giving rise to such deviations.
This work is complementary to the searches
considered in Sect.~\ref{sec:parametrized}, where particular functional forms 
for such deviations motivated by theory are investigated.

Here we apply three distinct nonparametric methods. 
In 
2013 only the first method was used to reconstruct the 
primordial power spectrum, the so-called 
``penalized likelihood'' method, for which the 
2018 results are presented in Sect.~\ref{penalizedLikelihood}. In 2015 
two additional methods were also used: a linear spline method (discussed 
in Sect.~\ref{bayesRecon}) for which both the number of knots and 
their positions were allowed to vary, and ideas from Bayesian model 
selection were applied to determine the appropriate number of knots; and a 
method using cubic splines
(discussed in Sect.~\ref{cubicSplineRecon}).  
Although the discussion below includes some 
description of each method in order to make the paper self-contained, for 
details the reader is referred to the 2013 and 2015 papers. Here we specify 
only those details specific to the 2018 analysis or different from the 
choices in the 2013 and 2015 analyses. See references in \citetalias{planck2013-p17} 
and \citet{Hunt:2013bha}, \citet{Hazra:2014jwa}, and \citet{Hunt:2015iua} for other approaches
to non-parametric reconstruction of the primordial power spectrum.

\subsection{Penalized likelihood \label{penalizedLikelihood}}

\noindent 
The underlying idea of the penalized likelihood approach is to add a term to the log-likelihood that
penalizes deviations from a perfect power-law spectrum. We parameterize the power spectrum as
\begin{equation}
{\cal P}_{\cal R}(k)= {\cal P}_0(k) \exp \bigl[ f(k) \bigr],
\end{equation}
where ${\cal P}_0(k) = A_{\rm s}(k/k_*)^{n_{\rm s} - 1}$, and add the following term to $-2\ln {\cal L}$:
\begin{align}
\mathbf{f}^{\rm T}~\mathbf{R}(\lambda , \alpha )~\mathbf{f}
&\equiv
\lambda  \int _{\kappa _{\rm min}}^{\kappa _{\rm max}}d\kappa ~ 
\left(
\frac{\partial ^2f(\kappa )}{\partial \kappa ^2}
\right) ^2\nonumber\\
&+\alpha \int _{-\infty }^{\kappa _{\rm min}}~f^2(\kappa )
+\alpha \int ^{+\infty }_{\kappa _{\rm max}}~f^2(\kappa ),
\label{penLike}
\end{align}
where $\kappa =\ln k.$ The interval $[\kappa _{\rm min}, \kappa _{\rm max}]$ is chosen to cover the range over 
which the likelihood is able to constrain the data. The two $\alpha $ terms serve to pin the reconstruction 
to the simple power law where the data have almost no constraining power. One may imagine that $\alpha >0$ should be 
infinite, but for numerical reasons a large but finite value is used to simplify the numerics.  
Numerically, for each $\lambda $ the dimension of $\mathbf{f}$ is chosen to be so large that 
the continuum version of the penalty given in Eq.~(\ref{penLike}) has been accurately approximated. 
For more details see 
\cite{Gauthier} and the extensive references therein to prior literature, 
as well as \citetalias{planck2013-p17} and \citetalias{planck2014-a24}.

\begin{figure*}[tp]
\centering
\includegraphics[scale=0.95]{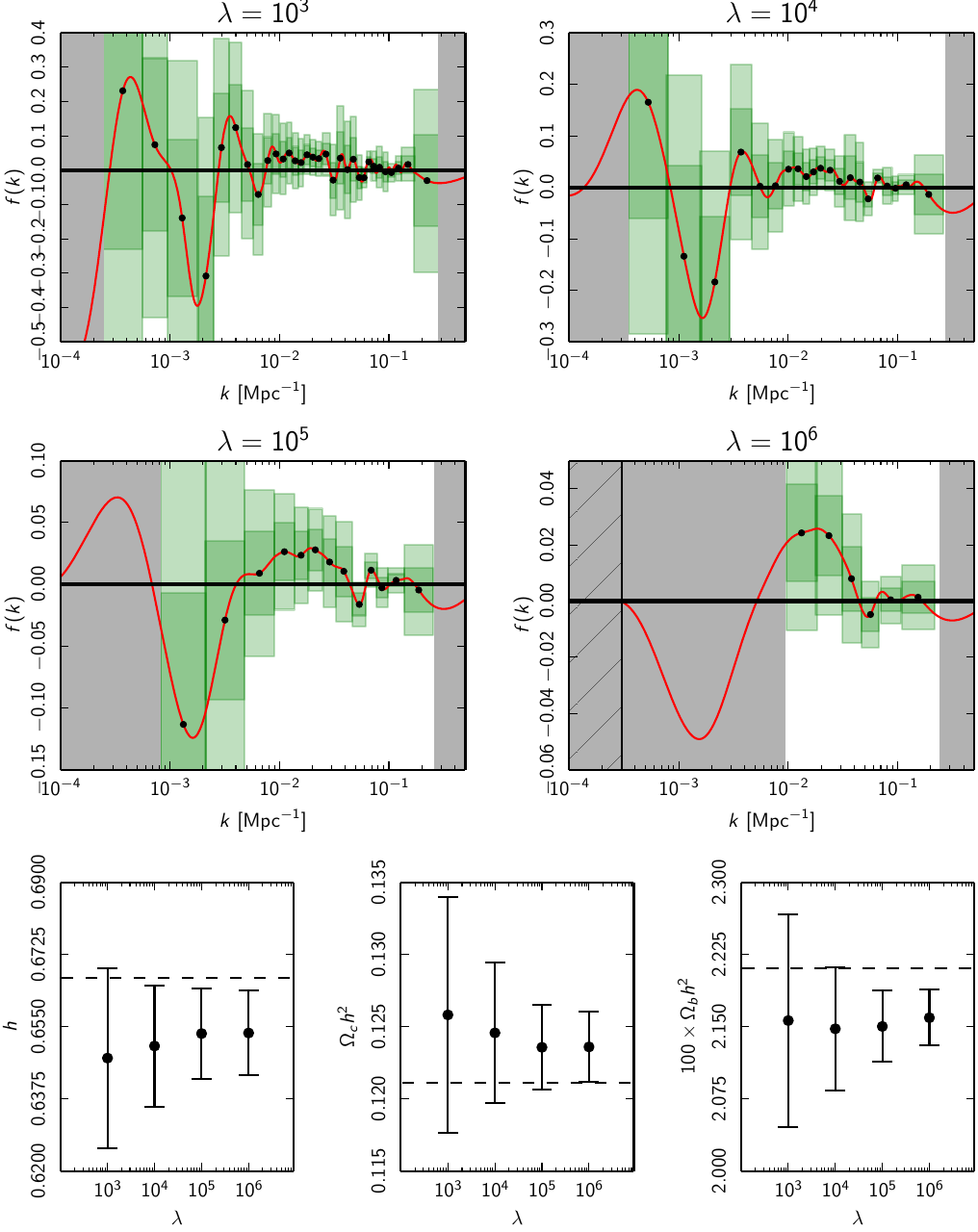}
\caption{%
\Planck\ TT+lowE penalized likelihood primordial power spectrum reconstruction.  
{\it Top four panels}: The deviation $f(k)$ for four different roughness penalties.  
The red curves indicate the best-fit deviation, while the vertical extents of the 
dark and light green error bars indicate the $\pm1\sigma$ and $\pm2\sigma$ 
errors, respectively.  The width of the error bars indicates the minimum 
reconstructible width (the minimum width for a Gaussian feature such that the mean 
square deviation of the reconstruction is less than 10\,\%).  The grey regions 
display where the minimum reconstructible width is undefined, meaning that the 
reconstruction in these regions is untrustworthy.  The hatched region in the 
$\lambda = 10^{6}$ plot indicates where the fixing penalty has been applied.  
{\it Lower three panels:} $\pm1\sigma$ error bars for the three 
non-primordial-specctrum cosmological parameters included in the reconstruction.  The 
respective best-fit fiducial model values are indicated by the dashed lines.
}
\label{fig:PPSpenLike}
\vspace{2cm}
\end{figure*}

\begin{figure*}[tp]
\centering
\includegraphics[scale=0.95]{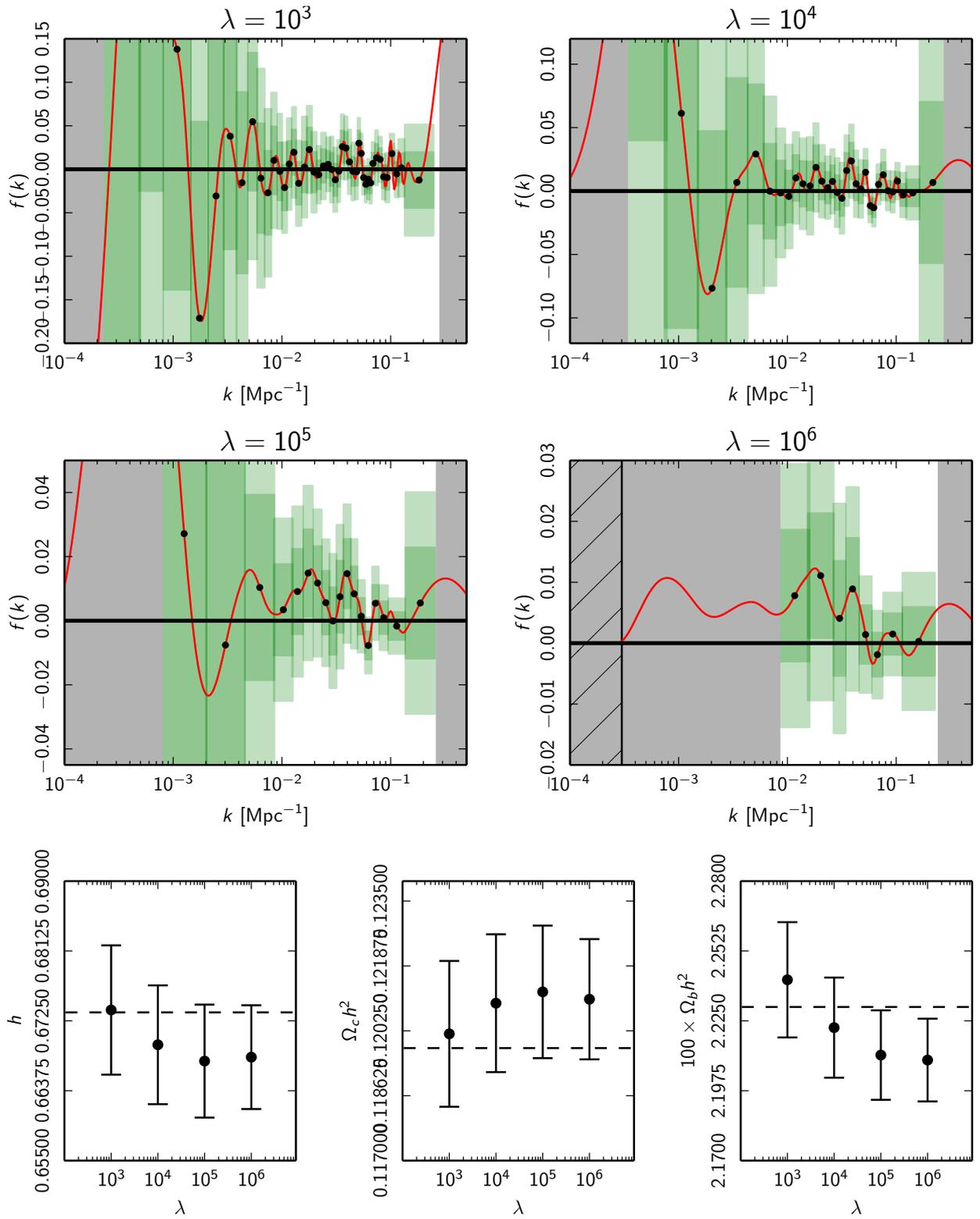}
\caption{%
Penalized likelihood reconstruction, as Fig.~\ref{fig:PPSpenLike} but for \Planck\ TT,TE,EE+lowE.
}
\label{fig:PPSpenLikePol}
\vspace{2cm}
\end{figure*}

In Fig.~\ref{fig:PPSpenLike}
we show the results using \Planck\ TT+lowE and in Fig.~\ref{fig:PPSpenLikePol}
we show the 
results for \Planck\ TT,TE,EE+lowE.  
In both cases we have assumed the usual base-$\Lambda$CDM model specified in 
\citetalias{planck2016-l06}, except that the power spectrum is 
now parameterized by a set of spline points. In addition to these 
spline points, we also maximize the likelihood with respect to the dimensionless Hubble 
parameter, $h$, and the baryon, $\Omega_{\rm b} h^{2}$, and CDM, $\Omega_{\rm c} h^{2}$, densities. 
All other cosmological and nuisance parameters are the same as those quoted in 
\citetalias{planck2016-l06}.

For the TT-only case, the maximum deviations are 
$1.55\sigma$, $2.10\sigma$, $1.80\sigma$, and $1.65\sigma$ for $\lambda = 10^{3}$, $10^{4}$, $10^{5}$, and $10^{6}$,
respectively, for which the probabilities to exceed are 
13\,\%, 28\,\%, 31\,\%, and 23\,\% (where we have taken into account
the look-elsewhere effect). Similarly, for the TT,TE,EE 
case, the maximum deviations are 
$2.07\sigma$, $1.77\sigma$, $1.77\sigma$, and $1.08\sigma$ for $\lambda = 10^{3}$, $10^{4}$, $10^{5}$, and $10^{6}$,
respectively, for which the probabilities to exceed are 
29\,\%, 23\,\%, 32\,\%, and 25\,\%. We consequently find no
statistically significant evidence for a deviation
from the simple power-law hypothesis. This result
is consistent with the results previously reported
for the \Planck\ 2013 and 2015 releases using
essentially the same method. It is likewise
consistent with the results below in Sects.~\ref{bayesRecon} and~\ref{cubicSplineRecon}, 
which use different methods.

\subsection{Bayesian reconstruction \label{bayesRecon}}

\newcommand{\PPS}{\mathcal{P}}
\newcommand{\Nknots}{N}
\newcommand{\thetamax}{\theta_{\max{}}}
\newcommand{\kmax}{k_{\max{}}}

To reconstruct the primordial power spectrum of curvature perturbations, we 
follow the methodology of section~8.2 of \citetalias{planck2014-a24}, using an 
\(\Nknots\)-point interpolating logarithmic spline  with the positions of the knots 
considered as free parameters in the full posterior distribution. The positions of 
the points in the \((k,\PPS)\) plane are treated as likelihood parameters with 
log-uniform priors. Further, the \(k\)-positions are sorted a priori such 
that \({k_1<k_2<\cdots<k_{\Nknots}}\), with \(k_1\) and \(k_\Nknots\) fixed. We 
compute posteriors and evidence values (conditioned on \(\Nknots\)) using 
{\tt PolyChord}~\citep{Handley2015a,Handley2015b}, also varying all cosmological and 
nuisance parameters. We then use evidence values for each model to correctly 
marginalize out the number of knots \(\Nknots\).

To plot our reconstructions of \(\PPS(k)\), we compute the marginalized 
posterior distribution of \(\ln\PPS\) conditioned on \(k\). The iso-probability 
confidence intervals are then plotted in the \((k,\PPS)\) plane (see, e.g., 
Fig.~\ref{fig:PPSR_comparison}), using code recorded in~\cite{Handley2018a}. To 
quantify the constraining power of a given experiment, we use the conditional 
Kullback-Leibler (KL) divergence as exemplified by~\cite{Hee2016}. For two 
distributions $P(\theta)$ and $Q(\theta)$, the KL divergence is defined as
\begin{equation}
    D_{KL}(P|Q) = \int \ln \left[ \frac{P(\theta)}{Q(\theta)} \right] P(\theta) d\theta,
\end{equation}
and may be interpreted as the information gain in moving from a prior \(Q\) to a 
posterior \(P\)~\citep{Raveri2016}. For our reconstructions, we compute the KL 
divergence of each distribution conditioned on \(k\) and \(\Nknots\), and then 
marginalize over \(\Nknots\) using evidence values to produce a \(k\)-dependent number 
which quantifies the compression or information that each data set provides at 
each value of \(k\). Further plots and theoretical detail can be found
in~\citet{Handley2018b}.

\subsubsection{Update on \Planck\ 2015}
\label{sec:ppsr_bayesian_reconstruction}

\begin{figure*}
    \centering
    \includegraphics{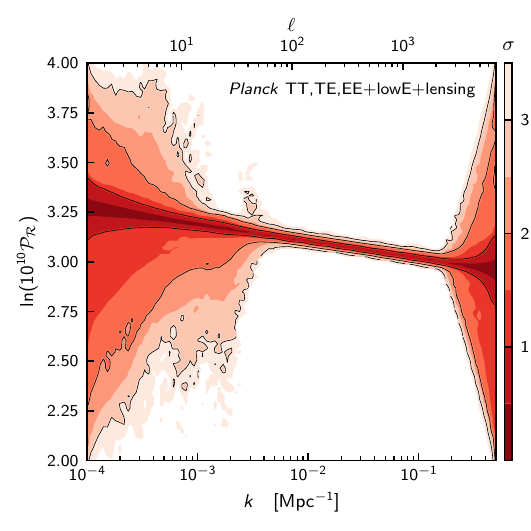}
    \includegraphics{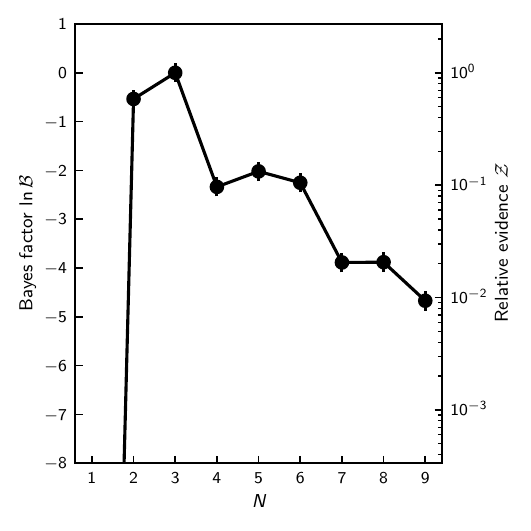}
    \includegraphics{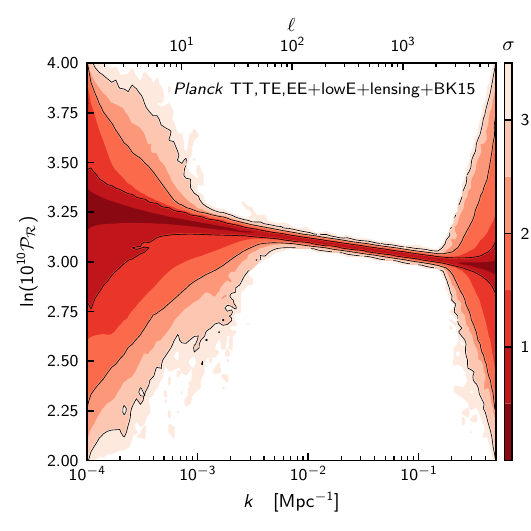}
    \includegraphics{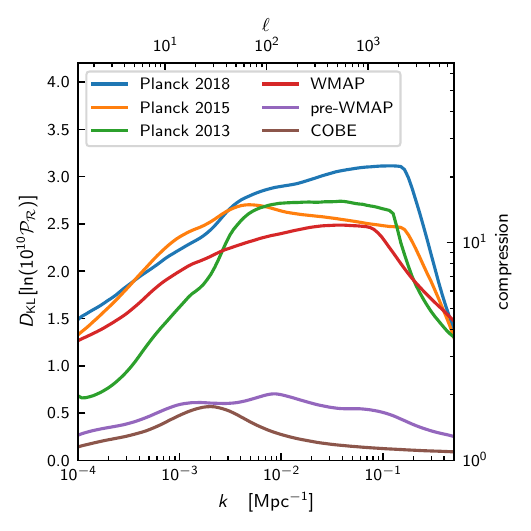}
    \caption{Free-form Bayesian reconstruction of the primordial power spectrum 
(Sect.~\protect\ref{sec:ppsr_bayesian_reconstruction}) using 
\Planck\ TT,TE,EE+lowE+lensing.
        {\it Top-right:} Evidence values for each $N$-knot reconstruction. The evidence 
is maximal for the $N=2$ and $N=3$ knot cases, and semi-competitive for the remaining 
higher knots. Marginalizing over the number of knots produces a predictive posterior 
plot, shown in the top-left panel. Here we see generic features, with the limit of 
resolution of \Planck\ at $\ell\simeq2400$ and cosmic variance at low~$\ell$. 
        {\it Bottom-left:} Same as top-left, but using the additional BK15 data and 
allowing $r$ to vary. {\it Bottom-right:} Kullback-Leibler 
divergence conditional on $k$, marginalized over the number of 
knots, showing the increase in compression of the primordial power 
spectrum over several past CMB missions. The difference in 
constraining power between \Planck\ 2013 and 2015 is driven entirely 
by the shift in the $\tau$ constraint. }\label{fig:PPSR_comparison}
\vspace{4cm}
\end{figure*}

\begin{figure*}
    \centering
    \includegraphics{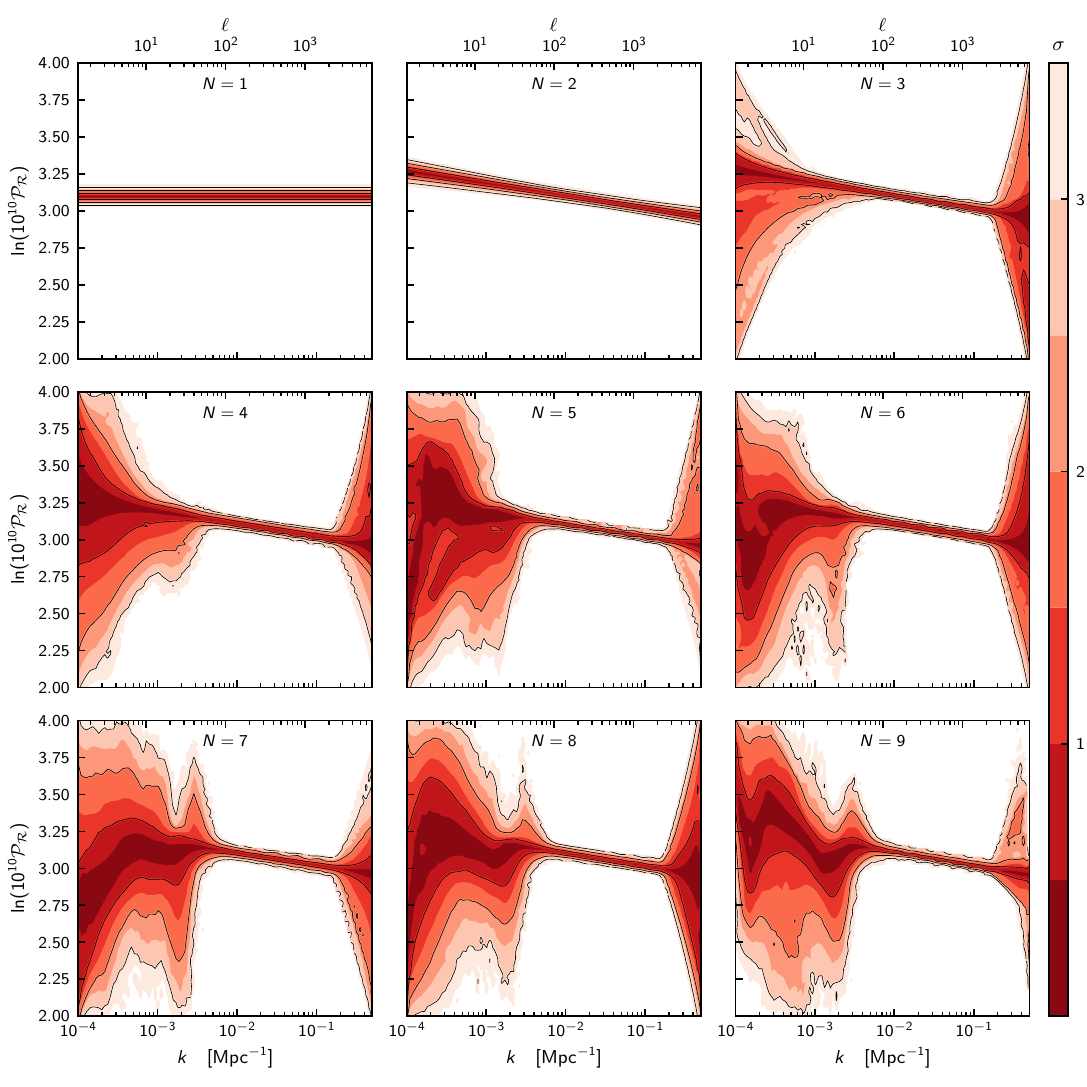}
    \caption{Free-form Bayesian reconstruction of the primordial power spectrum for 
varying numbers of knots (Sect.~\protect\ref{sec:ppsr_bayesian_reconstruction}) using 
TT,TE,EE+lowE+lensing. The amplitude and tilt are consistent with the rest of 
the results with the same combination of likelihoods. As more knots are added, the 
$\ell\simeq30$ feature in the $C_\ell$ temperature spectrum is visible as a dip to 
lower power.}\label{fig:PPSR_grid}
\vspace{4cm}
\end{figure*}

In \citetalias{planck2014-a24}, our analysis focussed predominantly on the TT+lowTEB 
data set. Here we present results for TT,TE,EE+lowE+lensing.  First, in updating to 
the lowE likelihood, we find that there is a marked tightening in the constraint on the 
amplitude of the reconstructed spectrum at all values of $k$. The improvement in the 
constraint can be seen 
directly in the predictive posterior plots (Fig.~\protect\ref{fig:PPSR_comparison}, 
top-left panel, and Fig.~\ref{fig:PPSR_grid}), and is quantified in 
Fig.~\ref{fig:PPSR_comparison} (bottom-right) via the KL divergence. The reason for 
the high-$\ell$ constraint provided by a low-$\ell$ likelihood change is due to the 
reduced uncertainty on $\tau$ that {\tt SimAll} EE provides. This can be seen by 
examining the shifts in the underlying cosmological parameters in Fig.~\ref{fig:TvsP}. 

Upon adding TE and EE data, we find that the hint of a feature at $\ell\simeq30$ is 
still present, in spite of the additional constraining power provided by 
polarization. Using polarization data, the $N=3$ case is now the most strongly 
favoured model by the evidence criterion. This indicates that there is some scope for 
models which account for low-$\ell$ cosmic variance to be preferred in a Bayesian sense. 
The other underlying cosmological parameters are unaffected by the additional degrees 
of freedom in the primordial power spectrum provided by the reconstruction.

In order to combine \Planck\ polarization data with BK15, we also allow the 
tensor-to-scalar ratio $r$ to vary, and fix the tensor tilt $n_{\rm t}$ via the 
inflationary consistency condition. As can be seen in the bottom-left panel of 
Fig.~\ref{fig:PPSR_comparison}, upon adding BK15, the effect of the low-$\ell$ 
deficit is softened, but with otherwise little change to the reconstruction. We
repeated our analysis with \texttt{CamSpec} in place of \texttt{Plik} and found
our results to be qualitatively and quantitatively unchanged.

\subsubsection{Free-form search for features}
\label{sec:free_form_features}

Next we examine the effect that sharp features in the primordial power spectrum 
can have on cosmological parameters. We model sharp features in the spectrum 
as a variable number of top-hat functions with varying widths, heights, and 
locations. On top of the traditional \(A_\mathrm{s}, n_\mathrm{s}\) 
parameterization of the power spectrum, we place \(N\) sharp top-hat features 
into the spectrum at locations \(k_i\) with widths \(d_i\) and heights 
\(h_i\) (\(i=1,\ldots, N\)).  That is, we set
\begin{align}
    \ln \mathcal{P}_\mathcal{R}(k) &= \ln A_\mathrm{s} + (n_\mathrm{s} -1) \ln\left( \frac{k}{k_*} \right)\nonumber\\
    &+ \sum\limits_{i=1}^N h_i \left[ |k-k_i|< \frac{d_i}{2} \right],
    \label{eqn:parameterisation}
\end{align}
where the square brackets in the summation denote a logical truth function as 
introduced by~\citet{ConcreteMathematics}. For values of \(N=0,\ldots,8\), we treat 
the variables in parameterization~\eqref{eqn:parameterisation} as parameters in a 
posterior distribution along with the traditional cosmological and \Planck\ nuisance 
parameters, with priors as detailed in Table~\ref{tab:priors}. We run with both 
linear and logarithmic priors on the $k$-locations of the features, as this alters 
the sensitivity to the type of features uncovered. We sample the posteriors using 
{\tt PolyChord}~\citep{Handley2015a,Handley2015b}.

\begin{table}
    \begin{center}
    \begin{tabular}{lll}
        \noalign{\hrule\vskip 2pt}
        \noalign{\hrule\vskip 3pt}
        Parameter & Prior & Range\\
        \hline
        &&\\[-1.5ex]
        \(k_i,\kmax\) & Sorted uniform & \( 0<k_1<\cdots<k_N<0.2\,\mathrm{Mpc}^{-1} \)\\
        \(d_i\) & Uniform        & \( 0<d_1,\ldots,d_N<0.01\,\mathrm{Mpc}^{-1}\)\\
        \hline
        \(k_i,\kmax\) & Sorted log-uniform & \( -4\!<\!\log_{10}k_1\cdots\log_{10}k_N\!<\!-0.3 \)\\
        \(d_i\) & Log-uniform    & \( 0<\log_{10}d_1,\cdots,\log_{10}d_N<1 \)\\
        \hline
        \(h_i\) & Uniform    & \(-1<h_1,\ldots,h_N<1 \)\\
        \({\ln}(10^{10} A_{\rm s})    \)& Uniform &\( 2   < {\ln}(10^{10} A_{\rm s}) <4      \) \\
        \(n_{\rm s}                   \)& Uniform &\( 0.8 < n_{\rm s}                <1.2    \) \\
        \hline
    \end{tabular}
    \end{center}
    \caption{Priors for the search for sharp features in the primordial power 
spectrum.  Units for $k$ and $d$ are $\mathrm{Mpc}^{-1}$.}\label{tab:priors}
\end{table}

\begin{figure*}
    \centering
    \includegraphics{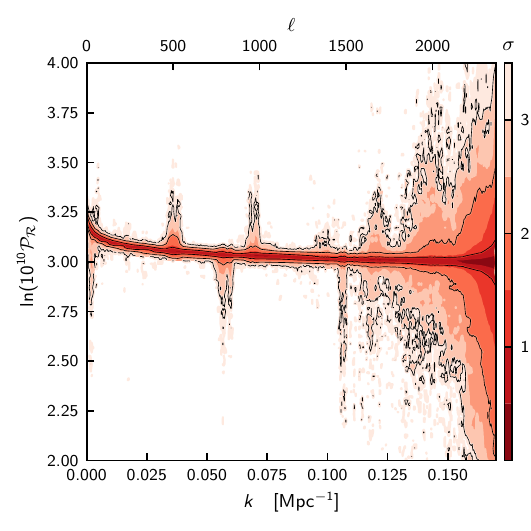}
    \includegraphics{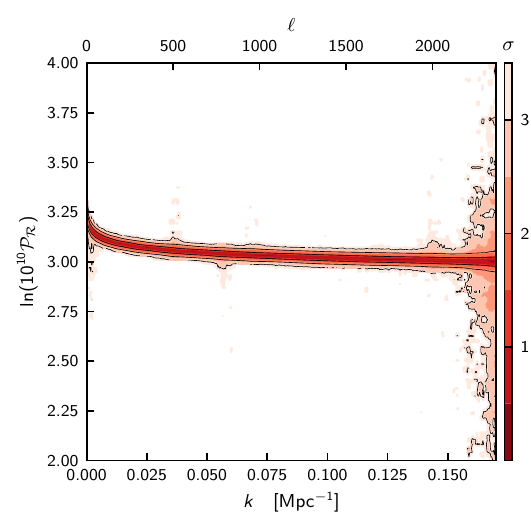}
    \includegraphics{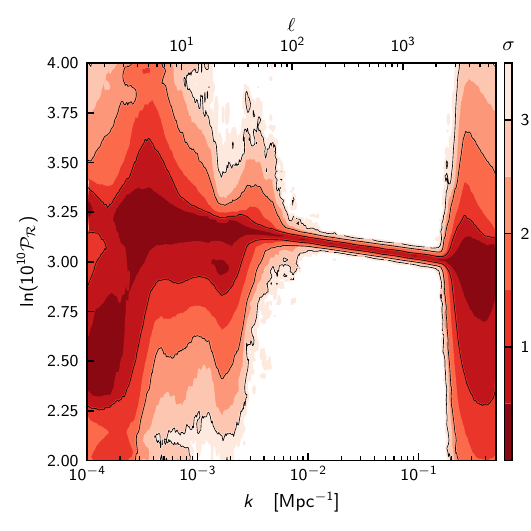}
    \includegraphics{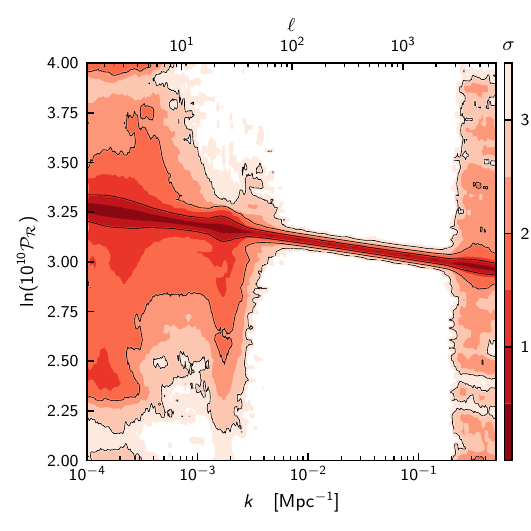}
    \caption{Free-form Bayesian search for features 
(Sect.~\protect\ref{sec:free_form_features}) with \Planck\ TT,TE,EE+lowE+lensing. 
The upper panels show runs with linear priors on the $k$-locations. The lower panels 
use logarithmic priors on the $k$-features. Left panels show the reconstruction 
for $N=8$ features, while the right panels show the reconstruction marginalized over 
$N=0,\ldots,8$ features. }\label{fig:features_grid}
\vspace{4cm}
\end{figure*}

\begin{figure*}
\centering
    \includegraphics[width=\textwidth]{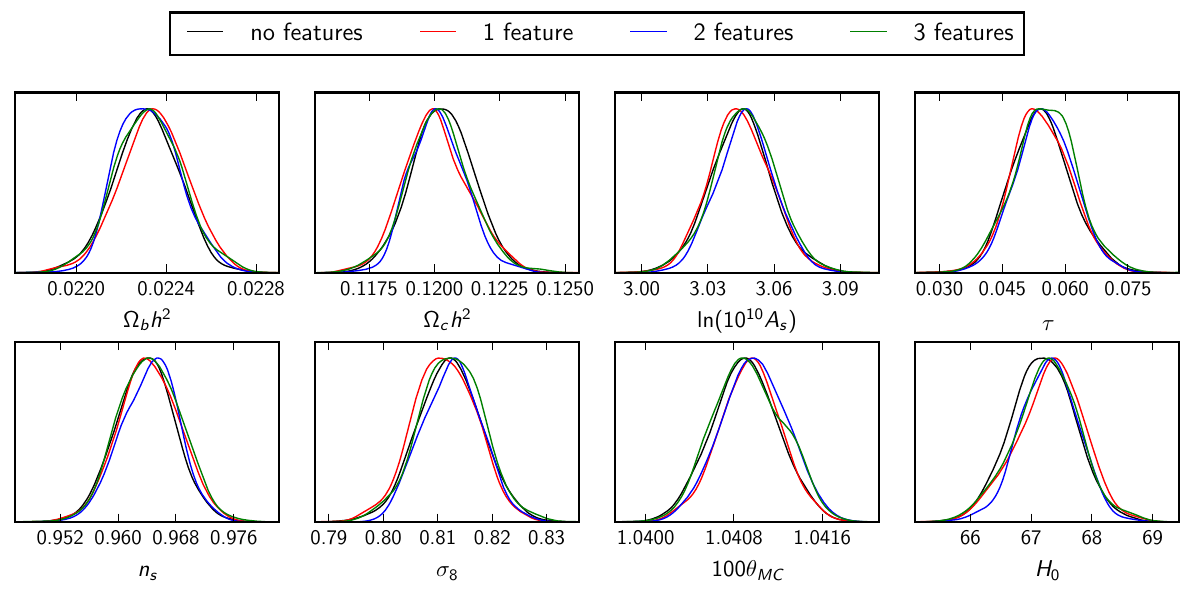}
    \caption{The effect on the underlying cosmological parameters of the free-form 
Bayesian search for features (Sect.~\protect\ref{sec:free_form_features}), for 
$N=0,\ldots,3$ features with linear $k$-priors.  The parameters remain stable up to 
$N=8$ features, and when changing to logarithmic $k$-priors.}\label{fig:features_parameters}
\end{figure*}

Figures~\ref{fig:features_grid} and~\ref{fig:features_parameters} show our results. 
With the linear priors case, there are statistically insignificant features 
corresponding to the peaks of the $TT$ spectrum, which arise due to the enhanced cosmic 
variance at these locations. With the logarithmic priors case, a stronger but still 
statistically insignificant feature is detected at $\ell\simeq30$, with a small 
deficit and surrounding enhancement of power. This case reproduces the results found 
in Sect.~\ref{sec:ppsr_bayesian_reconstruction}. In both cases, the Bayesian evidence 
shows preference for a no-features spectrum, and steadily declines as more features 
are added. The cosmological parameters remain unperturbed despite the introduction of 
features.



\subsection{Cubic spline reconstruction \label{cubicSplineRecon}}

\def\gtorder{\mathrel{\raise.3ex\hbox{$>$}\mkern-14mu \lower0.6ex\hbox{$\sim$}}}
\def\ltorder{\mathrel{\raise.3ex\hbox{$<$}\mkern-14mu \lower0.6ex\hbox{$\sim$}}}

In this subsection we update the third method of reconstruction used in \citetalias{planck2014-a24}, in 
which $\ln \mathcal{P}_{\cal R}(\ln (k))$ was expanded in cubic splines localized in $\ln (k)$ about 
uniformly spaced ``knots,'' $\{\ln (k_b), b =1,\dots,N\}$, whose range was chosen to cover all relevant 
cosmological scales, from $10^{-4}$\,Mpc$^{-1}$ to $O(1)$\,Mpc$^{-1}.$ We single out the standard scalar 
power spectrum pivot scale as a ``pivot knot'' $b\!=\!p$, with $k_p=k_*=0.05\,\mathrm{Mpc}^{-1}.$ Its 
associated power $\ln A_\mathrm{s} = \ln \mathcal{P}_{\cal R} (k_*)$ is assigned a uniformly 
distributed prior.  A tilted primordial power spectrum $\mathcal{P}_{{\cal R},\mathrm{fid}} \equiv 
A_\mathrm{s}(k/k_*)^{n_{\mathrm{s}, \mathrm{fid}}-1}$, with fixed spectral index $n_{\mathrm{s}, \mathrm{fid}}$ is used 
as the fiducial baseline from which deviations are measured, expressed in terms of $N-1$ relative 
spectral shape parameters: $q_b = \ln \left( \mathcal{P}_{\cal R} (k_b)/ \mathcal{P}_{\mathrm{{\cal 
R},fid}} (k_b)\right)$ for $\ b\ne p$.  For the results presented here, $n_{\mathrm{s}, \mathrm{fid}} = 
0.967$ was chosen.
As in \citetalias{planck2014-a24} we continue to use cubic splines for 
the $k$-space modes we expand in, with natural boundary conditions (i.e., vanishing second 
derivatives at the first and last knots). The treatment here is therefore quite analogous to that in 
Sect.~\ref{sec:vphi_bayesian_reconstruction}, where the inflaton potential rather than the curvature power 
spectrum is expanded in cubic splines.  Knot numbers up to 18 were reported in 
\citetalias{planck2014-a24}, and it was shown that 12 were sufficient to capture the variations desired 
by the \Planck\ CMB data. The mode functions were also varied. For example,
linear interpolation leads to 
similar reconstructions as long as enough knots are used. A weak uniform prior ($-1\le q_b\le 1$) was 
imposed on $q_b$. Outside of the spline coverage region $[k_1,k_N]$ we set $\ln \left( 
\mathcal{P}_{\cal R} (k)/ {\mathcal{P}_{\mathrm{{\cal R},fid}}}(k)\right)$ to be $q_1$ for $k < k_1$ and 
$q_N$ for $k>k_N$. The prior on $q_b$ and boundary condition choices
have little impact on the reconstructions over most of the $k$-range.

The current \Planck\ TT,TE,EE+lowE+lensing+BK15 data give only upper limits to the allowed values of the 
tensor amplitude, $r < 0.06$. Consequently, adding shape degrees of freedom to the tensor power spectrum would 
yield a completely prior-driven result. Instead we adopt the standard model power-law parameterization 
for tensors, $\mathcal{P}_\mathrm{t}(k) = r A_\mathrm{s}(k/k_*)^{n_\mathrm{t}}$, with the tensor spectral 
index constrained by the consistency relation $n_\mathrm{t} = -r/8$. Without $B$-mode constraints and with 
enough knots one could deform the primordial scalar spectrum to mimic a tensor contribution to the CMB 
power. However, this near-degeneracy is broken with direct $B$-mode observations, effectively so even if 
there are only upper limits as for the BK15 data. Our reconstructions here focus on letting $r$ float 
over a prior range $0\le r\le 1$, but the posterior is strongly constrained by the BK15 data.

\begin{figure}[p!]
\begin{center}
 \includegraphics[width= 0.45\textwidth]{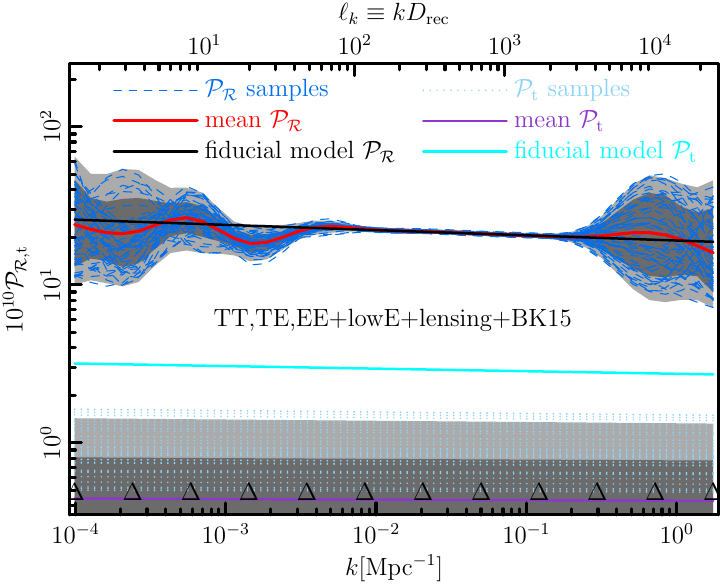}
  \includegraphics[width= 0.45\textwidth]{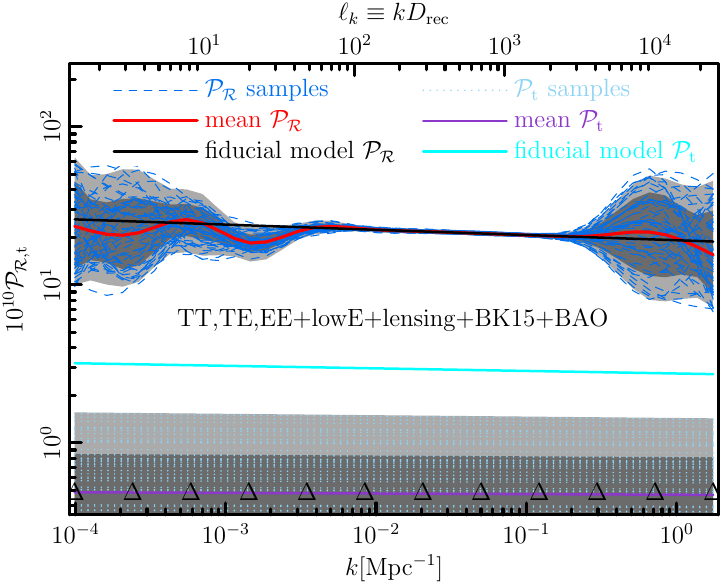}
  \includegraphics[width= 0.45\textwidth]{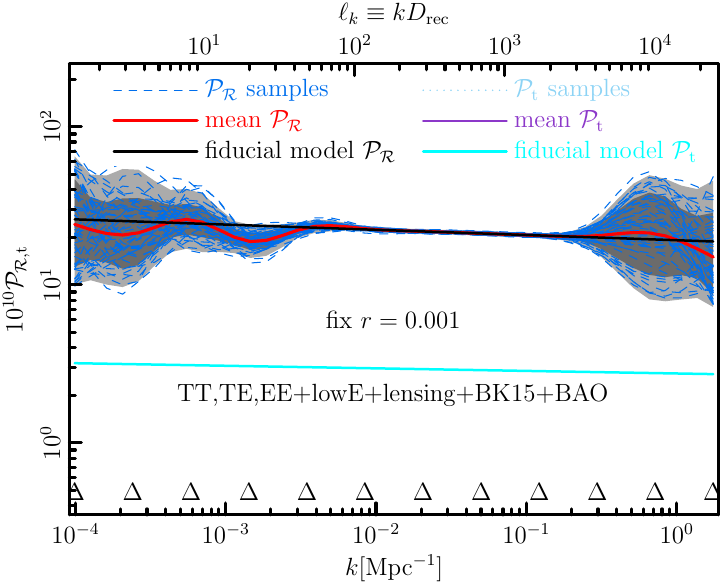}
\end{center}
  \caption{Reconstructed primordial scalar power spectrum derived using \Planck\ 
TT,TE,EE+lowE+lensing+BK15 data and 12 knots for the cubic spline interpolation (with positions marked as 
$\Delta$ at the bottom of each panel). Mean (ensemble-averaged) spectra are heavy lines, allowed 
$\pm1\sigma$ and $\pm2\sigma$ regions for trajectories are the shaded regions, and the dashed lines 
denote selected trajectories with parameters sampled within the $\pm1\sigma$ posterior. Below the 
scalar power is the tensor power reconstruction. The addition of the BAO likelihood shown in the middle 
panel makes almost no visual difference to the reconstructions.  In the bottom panel, fixing the 
tensor-to-scalar ratio to $r=0.001$ also produces only small differences in reconstruction.  Knot 
positions in $k$ roughly translate to multipoles through $k D_{\mathrm{rec}}$, where $D_{\mathrm{rec}}$ 
is the comoving distance to recombination.
\label{fig:traj_power_spline}}
\end{figure}

The joint probability distributions of $\{q_b , b\ne p\}$, $\ln A_\mathrm{s}$, and the other cosmic and 
nuisance parameters are determined by {\tt CosmoMC} modified to incorporate the $N$-knot 
parameterization for fixed knot number $N$. Figure~\ref{fig:traj_power_spline} shows the reconstruction. 
Apart from the mean and $1\sigma $ and $2\sigma $ limits on the ensemble of trajectories allowed by 
the posterior probability, we also show a set of individual trajectories with parameters taken from 
$1\sigma$ samples to illustrate the knot-to-knot coherence (dashed curves). The tensor trajectories are 
straight lines, as required by the adopted tensor power model.

\begin{figure}[t!]
\begin{center}
 \includegraphics[width= 0.45\textwidth]{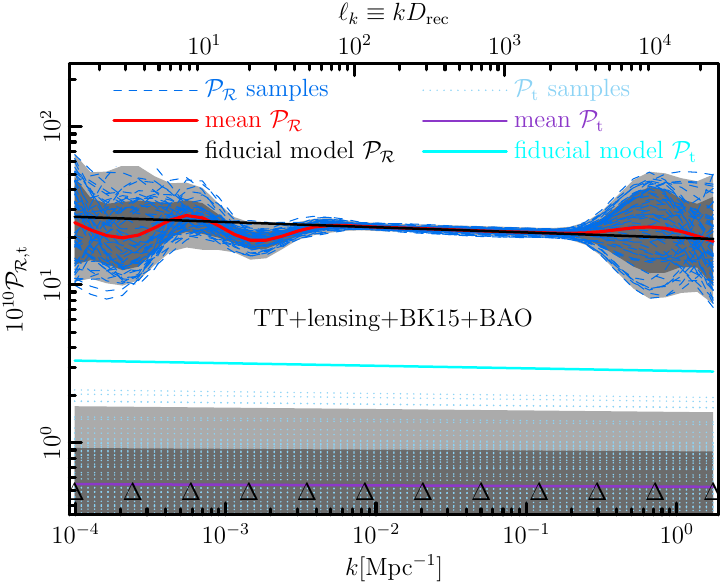}
 \includegraphics[width= 0.45\textwidth]{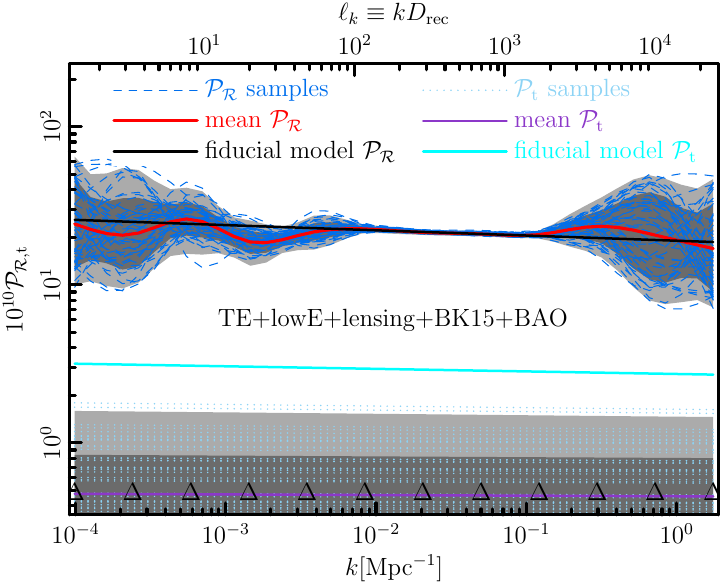}
 \includegraphics[width= 0.45\textwidth]{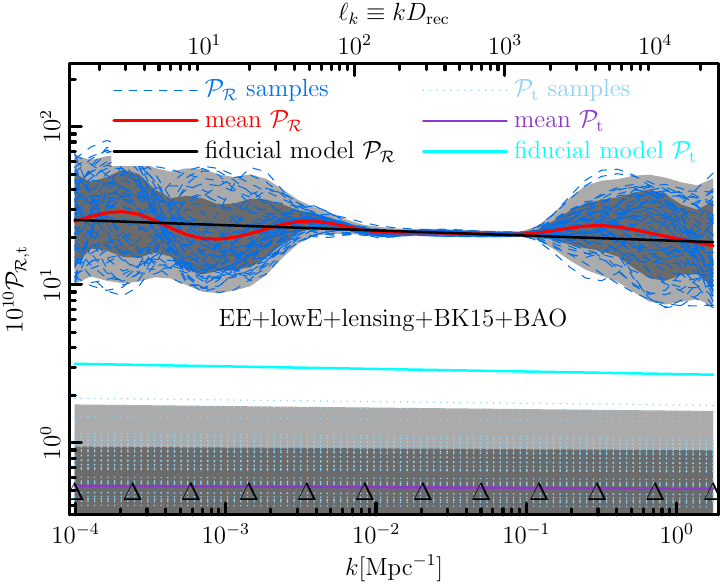}
\end{center}
  \caption{ Reconstructed 12-knot power spectra.  The robustness of the reconstruction 
is apparent when sub-selections of the \Planck\ 
data are used: \Planck\ TT+lowE+lensing+BK15 (top); \Planck\ TE+lowE+lensing+BK15 (middle);
and \Planck\ EE+lowE+lensing+BK15 (bottom). 
\label{fig:traj_compare_datasets}}
\vspace{1cm}
\end{figure}

In spite of the extra scalar shape freedom in the $k$-space region over which the tensor modes affect the 
CMB, the 12 knot reconstruction still leads to a strong constraint of $r < 0.069$, rather close to the $r 
< 0.06$ limit obtained if the only shape parameter is $n_{\mathrm{s}}$. In fact we find that the current 
limits on $r$ are such that the scalar-power reconstructions are not sensitive to the details of the 
$r$ distribution. To illustrate this, the lower panel of Fig.~\ref{fig:traj_power_spline} shows the 
spectrum when $r$ is fixed at the tiny value of $r=0.001$. One could regard this as a 
theoretical prior for low-energy inflation models or a forecast for a future in which $r$ is measured or 
tightly constrained by $B$-mode experiments.

In \citetalias{planck2014-a24}, the main cubic spline reconstruction included non-CMB data to help pin 
down cosmological parameters such as $H_0$, $\tau$, and the late-time expansion history. The improvements 
in the data from 2015 to 2018, especially the decreased errors on $\tau$, result in no non-CMB data being 
needed. Although $\tau$ and $\ln A_{\rm s}$ are about 90\,\% correlated, as they are in the standard power-law 
model, neither are very correlated with the $q_b.$ The strongest is about 40\,\% for $q_3$ at $k\simeq 
0.0006\,\mathrm{Mpc}^{-1}$, corresponding to $\ell \sim 10$ where reionization is kicking in. The second 
strongest is about $30\,\%$ for $k\simeq 0.02\,\mathrm{Mpc}^{-1}$, similar to the correlation of $\tau$ and 
$n_{\rm s}$ in the standard power-law model. (The correlations among the $q_b$ are also relatively small, except for the 
high $k$ bands $b=10$ and $11$, where the data are not constraining.)

The middle panel in Fig.~\ref{fig:traj_power_spline} shows the effect of adding the BAO constraint.
Although apparently visually identical, there are slight differences. For example,
the 1$\sigma $ error on $q_b$ at 
$k\simeq 0.02\,\mathrm{Mpc}^{-1}$ decreases by about $7\,\%$, from $ 0.0090$ to $ 0.0084$, while at 
$k\simeq 0.1\,\mathrm{Mpc}^{-1}$ the decrease is about $5\,\%$. The restriction to $r=0.001$ does not change the error 
bars over the floating $r$ case.  At intermediate $k$ for modes $5$ to $8$ the errors on $q_b$ are so 
close to zero that the reconstruction is quite compatible with a simple power law, corresponding to a 
straight line in Fig.~\ref{fig:traj_power_spline}. This was also a main result of the 2015 \Planck\ 
reconstructions.

The errors on the $q_b$ grow above $\pm 0.1$ for $b=4$ as a consequence of increased cosmic variance, 
giving more freedom in the allowed trajectories. Unfortunately this is also the region of relevance to the $TT$ 
power spectrum deficit in the $\ell \simeq 20$--$30$ range. The most significant deviation from zero occurs for $q_4$ at $k\simeq 
0.0014\,\mathrm{Mpc}^{-1}$: $ -0.254 \pm 0.127 $, $ -0.255 \pm 0.125$, and $-0.235\pm 0.128$ for the three 
cases. Thus the anomaly in terms of deviation from the power law of the standard model hovers at around 
the $2\sigma$ level.  More precisely, the 95\,\% upper confidence limits on $q_4$ are $-0.011$, $-0.018$, and $+0.017$, 
for the respective cases. This $2\sigma $ level of the anomaly was also the conclusion of the 2015 
\Planck\ reconstructions. Therefore, even though the low-$k$ deficit is robust against the various choices 
for the reconstruction, we conclude that it is not statistically significant. The associated $TT$, $TE$, 
$EE$, and $BB$ power spectra responses to the allowed primordial power variations are derived from the mode 
expansion, and match the $\mathcal{D}_\ell^{XY}$ data well, in particular following the dip in $TT$ at 
$\ell \simeq 20$--$30$.

In Fig.~\ref{fig:traj_compare_datasets} we show the reconstructed power spectra using only the TT, TE, and 
EE data in conjunction with BK15. The fixed $r=0.001$ cases look very similar. Except at high~$k$, the 
polarization data using either EE or TE also enforce a nearly uniform $n_\mathrm{s} (k)$ over a broad 
range in $k$, with values in excellent agreement with those obtained from TT alone, from TE and EE 
used in combination, and from the combined TT,TE,EE results. For example, the $ \pm 0.0087$ and $ \pm 
0.0060$ $1\sigma $ errors at $k\simeq 0.02\,\mathrm{Mpc}^{-1}$ and $k\simeq 0.1\,\mathrm{Mpc}^{-1}$, 
respectively, increase only slightly for TT only, to $ \pm 0.012$ and $ \pm 0.0068$, but, more significantly, to 
$ \pm 0.017$ and $ \pm 0.069$ with EE alone. The deficit region remains about the same, with the TT,TE,EE 
result for $q_4$ of $ -0.255 \pm 0.125 $ quoted above changing slightly for TT alone, to $ -0.252 \pm 
0.130$, but with no hint of an anomaly for EE alone, at $ -0.126 \pm 0.460$. If just the TE cross data are used, 
the values are closer to the TT case, namely, $ -0.232 \pm 0.163$, now under a $2\sigma $ excursion from the tilted 
fiducial model.

\begin{figure}
\begin{center}
\includegraphics[width= 0.42\textwidth]{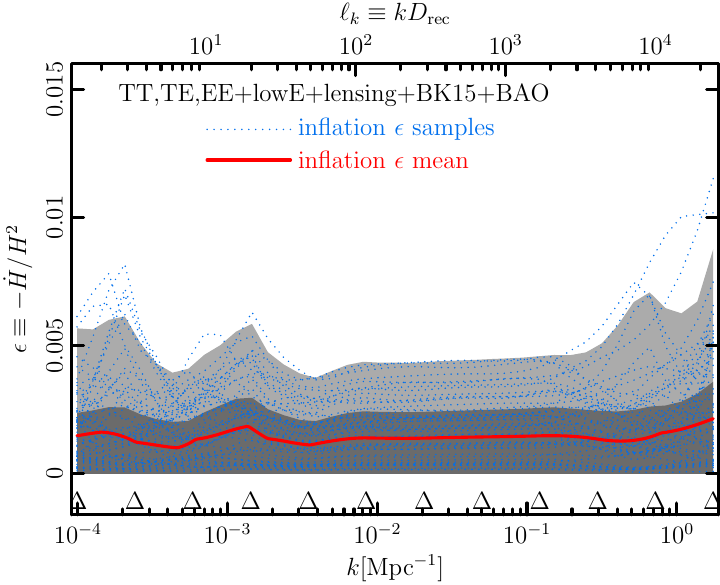}
\includegraphics[width= 0.42\textwidth]{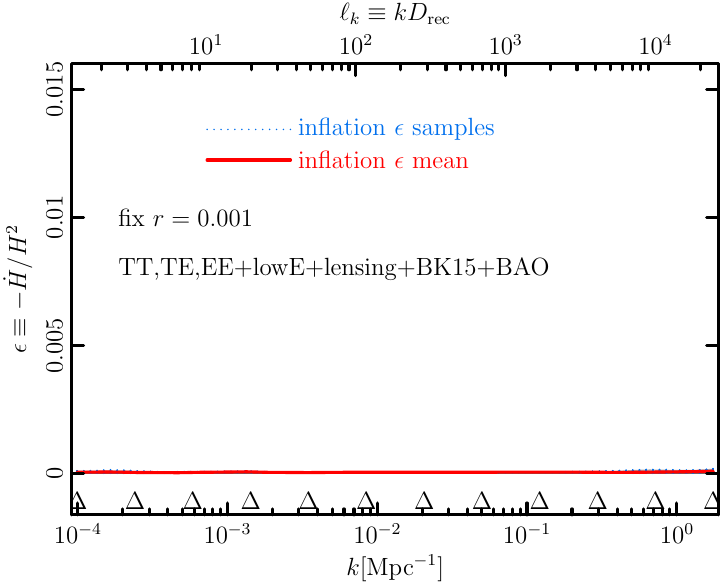}
\end{center}
\caption{Acceleration history $\epsilon (k)$ for reconstructed trajectories using 12 knots (marked as 
$\Delta$ at the bottom of the figure), with cubic-spline interpolation and the \Planck\ 
TT,TE,EE+lowE+lensing+BK15+BAO data for the two cases of floating $r$ and $r$ fixed at 0.001. Sample 
$1\sigma$ trajectories for the floating $r$ case allow wide variability, which is naturally greatly 
diminished if $r$ is fixed to $r=0.001$. \label{fig:traj_eps_spline}}
\label{SlowRollBond} 
\end{figure}

\begin{figure}
\begin{center}
  \includegraphics[width= 0.42\textwidth]{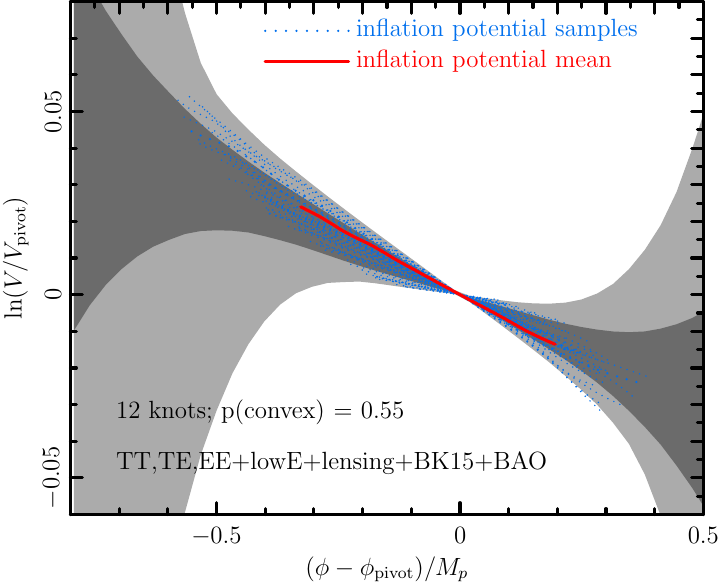}
  \includegraphics[width= 0.42\textwidth]{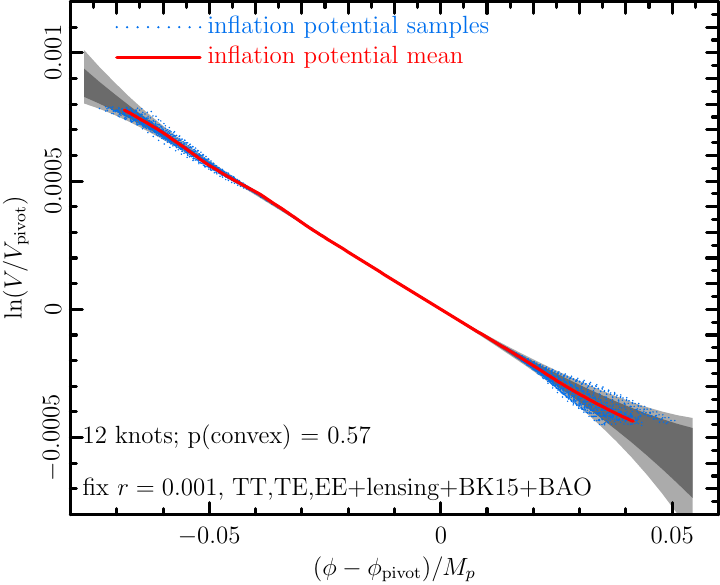}
\end{center}
  \caption{ {\it Top:} Reconstructed shape of the single-field inflaton potential from the 
cubic-spline power spectra mode-expansion using 12 knots and the \Planck\ TT,TE,EE+lowE+lensing+BK15+BAO 
data. {\it Bottom:} Result when $r$ is fixed at 0.001. Instead of plotting as a function of 
wavenumber $k$ we plot $\ln V(\phi )/V_{\rm pivot}$ about a pivot field value $\phi_{\rm pivot}$. Note that the 
range on the $ \phi$ axis is quite different for the small $r$ case than the floating case.  The probability of local convexity evaluated at $\phi_{\rm pivot}$ is denoted as $p(\mathrm{convex})$.}
\label{fig:traj_pot_spline}
\end{figure}

As in \citetalias{planck2014-a24}, we can use the ${\cal P}_{\cal R}(k) 
\propto H^2/\epsilon$ and ${\cal P}_{t}(k) /{\cal P}_{\cal R}(k)  \simeq 16 \epsilon$ 
reconstructions to get an idea of the history of the acceleration of 
the Universe as a function of time over the significant number of $e$-folds 
of the cosmic expansion that the CMB data probe, codified by the dynamical 
slow-roll parameter $\epsilon (k)= -\dot H/H^2\vert_{k=aH}$, considered as 
a function of $aH$, the value of the wavenumber at Hubble crossing. Results 
with floating $r$ and $r$ fixed to $0.001$ are shown in 
Fig.~\ref{fig:traj_eps_spline}. For the dynamical time variable we use $k=aH$ 
for the horizontal axis for ease in comparing with the ${\cal P}_{\cal R}(k)$ 
curves of Fig.~\ref{fig:traj_power_spline}. The wide spread in the $\epsilon $ 
trajectories for the floating $r$ case is a consequence of being able to 
fit the $n_{\rm s} (k)$ shape by a combination of $\epsilon (k)$ and 
$d \ln \epsilon (k)/d \ln k$. When $r$ is small, $n_{\rm s} (k)$ is almost entirely 
determined by $d \ln \epsilon /d \ln k$, and the $\epsilon (k)$ values cluster 
near $r/16$. 

The Hamilton-Jacobi energy constraint equation relates the potential to $\epsilon$ and $H$ via 
$V= 3 M_P^2 H^2 (1-\epsilon /3)$. Figure~\ref{fig:traj_pot_spline} shows the reconstructed inflationary potential shapes 
in the region over which the allowed inflationary potentials are constrained by the data for the floating 
$r$ and fixed $r$ cases. Instead of using $k$ for the horizontal axis, we translate into inflaton-field 
$\phi$-space using the relation between $\phi$ and $ \sqrt{\epsilon}$, referenced to the pivot 
position $\phi_{\rm pivot}$. For the vertical axis we plot $\ln V/V_{\rm pivot}$, with the overall normalization 
$V_{\rm pivot}$ removed. Its value is set by $r$, hence there is a distribution of constant $V_{\rm pivot}$ 
amplitudes to superimpose if we want the total $V$.  The radically different visual appearance for the 
floating $r$ and fixed $r$ cases is due to the observable $k$ range being compressed through the smallness 
of $\epsilon$ into a small precisely determined field range, whereas this range has a distribution in the 
floating $r$ case. One can monitor whether the shapes of the individual realizations of the potential 
trajectories bend upwards or downwards or do both, an indication of convexity. The sample trajectories shown 
are not exclusively convex or concave, and a measure of the probability that they are convex can be made from the 
ensemble. As indicated in Fig.~\ref{fig:traj_pot_spline} for the 12 knot case, the ensemble-averaged 
potentials are roughly exponential, with individual trajectories bending away from the mean, but with no 
strong tendency for convexity or concavity. (The roughly $50\,\%$ probability changes somewhat depending upon 
the combination of data used, whether TT,TE,EE or the individual data sets.)

The standard cosmological parameter determinations are highly robust to the addition of these spline shape 
degrees of freedom. The mean values change little and the error bars grow slightly, by around $10\,\%$ for $\ln 
A_{\rm s}$, $\tau$, and $H_0$. The largest error increase is for $\sigma_8,$ with $ \sigma_8 = 0.812 \pm 0.0058$ 
becoming $0.814 \pm 0.0096$. The main conclusions of this section on $\epsilon$ and $V$, and ${\cal 
P}_{\cal R}(k)$, remain as in \citetalias{planck2014-a24}, but the results have been 
noticeably sharpened by the improvements in the \Planck\ 2018 data sets.



\section{Search for primordial features in the \Planck\ power spectrum \label{sec:parametrized}}

The ``bottom-up" power spectrum reconstruction methods of the previous section are an 
excellent way to search for coarse features in the spectrum, but lack the resolution 
to detect the higher-frequency features generically predicted by various physical 
mechanisms [see, e.g., \cite{Chluba:2015bqa} for a review]. It is therefore useful to 
complement power spectrum reconstruction with a ``top-down" approach by fitting 
specific feature models to the data.  In this section we will analyse 
a representative range of power spectrum templates which parameterize features in 
terms of a handful of new parameters.

With \Planck's temperature and polarization data, we have two essentially independent 
probes of features at our disposal and will pay particular attention to examining the 
consistency between the two~\citep{Miranda:2014fwa}.

\subsection{Power spectrum templates with features \label{Sec:PStemplates}}

\begin{table*}[!ht]
\centering~
\begin{tabular}{lccccccc}
\noalign{\hrule\vskip 2pt} \noalign{\hrule\vskip 3pt}
 & Log osc & Running log osc & Lin osc & Step & Kin cutoff & Rad cutoff & Kink cutoff \\
\noalign{\hrule\vskip 3pt}
$\mathcal{A}_X$ & [0,0.5] & [0,0.5] & [0,0.5] & \dots & \dots & \dots & \dots \\
$\log_{10} \omega_X$ & [0,2.1] & [0,2.1] & [0,2] & \dots & \dots & \dots & \dots \\
$\varphi_X/(2 \pi)$ & [0,1] & [0,1] & [0,1] & \dots & \dots & \dots & \dots \\
$\alpha_\mathrm{rf}$ & \dots & [$-0.1$,0.1] & \dots & \dots & \dots & \dots & \dots \\
\noalign{\hrule\vskip 2pt}
$\mathcal{A}_\mathrm{s}$ & \dots & \dots & \dots & [0,1] & \dots & \dots & \dots \\
$\log_{10} \left(k_\mathrm{s}/\mathrm{Mpc}^{-1} \right)$  & \dots & \dots & \dots & [$-5$,$-1$] & \dots & \dots & \dots \\
$\ln x_\mathrm{s}$ & \dots & \dots & \dots & [$-1$,5] & \dots & \dots & \dots \\
\noalign{\hrule\vskip 2pt}
$\log_{10} \left( k_Y^\mathrm{c}/\mathrm{Mpc}^{-1} \right)$ & \dots & \dots & \dots & \dots & [$-5$,$-3$] & [$-5$,$-3$] & [$-5$,$-3$] \\
$R_\mathrm{c}$ & \dots & \dots & \dots & \dots & \dots & \dots & [$-1$,0.7] \\
\noalign{\vskip 2pt\hrule}
\end{tabular}
\vspace{.4cm} \caption{\label{tab:features_priors} Prior ranges for the parameters of
the feature model templates of Sect.~\ref{Sec:PStemplates}.}
\end{table*}

\subsubsection{Global oscillation models} 
Periodic or quasi-periodic modulations of the power spectrum which extend over the 
entire observable range of wavenumbers can occur in a variety of models [cf., e.g., \citet{Danielsson:2002kx,Martin:2003kp,Bozza:2003pr,Chen:2011zf,Jackson:2013vka}]. 
A general parameterization of models with a sinusoidal modulation of the primordial power 
spectrum reads
\begin{equation} 
\mathcal{P}_\mathcal{R}^{\, X}(k) = \mathcal{P}_\mathcal{R}^{\, 0}(k) \left[ 1 + 
\mathcal{A}_X \cos \left( \omega_X \Xi_X(k) + \varphi_X \right) \right], 
\end{equation} 
where $X \in \{\mathrm{log},\mathrm{lin},\mathrm{rf}\}$.  Defining $\kappa \equiv 
k/k_*$,\footnote{Throughout this section, we take $k_* = 0.05\,\mathrm{Mpc}^{-1}$.} we 
consider the logarithmic oscillation model, given by $\Xi_\mathrm{log} \equiv \ln 
\kappa$, and the linear oscillation model, $\Xi_\mathrm{lin} \equiv \kappa$. In 
addition, we investigate a logarithmic model with running frequency, $\Xi_\mathrm{rf} 
\equiv \ln \kappa \left( 1 + \alpha_\mathrm{rf} \ln \kappa \right)$.  For $0 \leq 
\alpha_\mathrm{rf} \lsim 0.01$, this is a good approximation for the scalar power 
spectrum in the axion monodromy model~\citep{Flauger:2014ana}, which will be analysed 
in more detail below, but here we allow for a wider range of the running parameter 
$\alpha_\mathrm{rf}$, including negative values (i.e., decreasing frequency with 
increasing $k$).

\subsubsection{Localized oscillatory features: inflation with a step}
A sudden transient event in the evolution of the inflation field, triggered by a 
sharp feature in the inflaton potential, or a sharp turn in field space, generically 
leads to a localized oscillatory feature in the power spectrum 
\citep{Adams:2001vc,Chen:2006xjb,Achucarro:2010da,Miranda:2012rm,Bartolo:2013exa}.  
As an example of this class of feature models, we consider here the case of 
a $\tanh$-step in an otherwise smooth inflaton potential~\citep{Adams:2001vc}, 
whose power spectrum can be parameterized as \citep{Miranda:2013wxa}
\begin{equation}
\label{eq:step1} \ln \mathcal{P}_\mathcal{R}^\mathrm{s}(k) = \ln 
\mathcal{P}_\mathcal{R}^0(k) +  \mathcal{I}_0(k) + \ln \left(1 + 
\mathcal{I}_1^2(k)\right),
\end{equation}
where the first- and second-order terms are given by
\begin{align}
\mathcal{I}_0 &= \mathcal{A}_\mathrm{s} \, \mathcal{W}_0(k/k_\mathrm{s}) \, 
\mathcal{D}\left( \frac{k/k_\mathrm{s}}{x_\mathrm{s}} \right),\\
\mathcal{I}_1 &= \frac{1}{\sqrt{2}} \left[ \frac{\pi}{2} (1 - n_\mathrm{s}) +
 \mathcal{A}_\mathrm{s} \, \mathcal{W}_1(k/k_\mathrm{s}) \, \mathcal{D}\left( 
\frac{k/k_\mathrm{s}}{x_\mathrm{s}} \right) \right],
\end{align}
with window functions
\begin{align}
\mathcal{W}_0(x) &= \frac{1}{2 x^4} \left[ \left( 18 x - 6 x^3 \right) \cos 2x + 
\left( 15 x^2 - 9 \right) \sin 2x \right],\\
\mathcal{W}_1(x) &= -\frac{3}{x^4} (x \cos x - \sin x) \left[3 x \cos x + \left(2 x^2 
-3\right) \sin x\right],
\end{align}
and damping function
\begin{equation}
\label{eq:step2} \mathcal{D}(x) = \frac{x}{\sinh x}.
\end{equation}
In this model, the parameter $\mathcal{A}_\mathrm{s}$ determines the amplitude of the 
oscillatory feature, the step scale $k_\mathrm{s}$ sets the position of the step in 
$k$-space, and the damping parameter $x_\mathrm{s}$ determines the width of the 
envelope function.

\subsubsection{Models with suppressed power at large scales}
The apparent lack of power at the largest scales in the temperature power spectrum 
with respect to the expectation of base $\Lambda$CDM serves as a motivation for 
models with a suppression of primordial perturbations below a cutoff scale 
$k^\mathrm{c}$. Physically, this effect may be due to fluctuations at the largest 
observable scales being generated at the onset of the inflationary phase after a 
prior era of, e.g., kinetic or radiation 
domination~\citep{Vilenkin:1982wt,Contaldi:2003zv}, or due to an isolated event such 
as a kink in the inflaton potential~\citep{Starobinsky:1992ts}.

In these scenarios, the primordial spectrum can generally be analytically 
approximated by an expression of the form
\begin{equation}
\ln \mathcal{P}_\mathcal{R}^{\, Y}(k) = \ln \mathcal{P}_\mathcal{R}^{\, 0}(k)  + \ln 
\Upsilon_Y(k/k_Y^\mathrm{c}),
\end{equation}
with $Y \in \{\mathrm{kin},\mathrm{rad},\mathrm{kink}\}$, where $\Upsilon_Y$ is a 
function with $\ln \Upsilon_Y \rightarrow 0$ in the limit $k \gg k_Y^\mathrm{c}$ that 
describes the shape of the cutoff and the transition to a power-law spectrum at 
smaller scales.

\subsubsection*{Initial kinetic domination}
If inflation is preceded by an era dominated by the kinetic energy of the inflaton 
field (i.e., fast roll), we have
\begin{equation}
\Upsilon_\mathrm{kin}(y) = \frac{\pi}{16} \, y \left| C_\mathrm{c}(y) - 
D_\mathrm{c}(y) \right|^2,
\end{equation}
with
\begin{align}
C_\mathrm{c}(y) &= e^{-i\hspace{.08em} y} \, 
\left[H_0^{(2)}\left(\frac{y}{2}\right) - \left( \frac{1}{y} + i \right) 
H_1^{(2)}\left(\frac{y}{2}\right) \right],\\
D_\mathrm{c}(y) &= e^{i\hspace{.08em} y} \, 
\left[H_0^{(2)}\left(\frac{y}{2}\right) - \left( \frac{1}{y} - i \right) 
H_1^{(2)}\left(\frac{y}{2}\right) \right],
\end{align}
where $H_n^{(2)}$ denotes the Hankel function of the second 
kind~\citep{Contaldi:2003zv}.

\subsubsection*{Initial radiation domination}
If inflation begins immediately after a radiation-dominated phase, the cutoff 
function reads~\citep{Vilenkin:1982wt}
\begin{equation}
\Upsilon_\mathrm{rad}(y) =  \frac{1}{4 y^4} \left| e^{-2i\hspace{.08em} y} 
\left( 1 + 2i \hspace{0.08em} y \right) - 1 - 2 y^2 \right|^2.
\end{equation}

\subsubsection*{Kink in the inflaton potential (Starobinsky model)}
A kink in the inflaton potential, first discussed by~\cite{Starobinsky:1992ts}, leads 
to a spectrum approximately given by
\begin{align}
&\Upsilon_\mathrm{kink}(y) = 1 - 3(R_\mathrm{c} - 1) \frac{1}{y} \left[  \left( 1 - 
\frac{1}{y^2}  \right) \sin 2 y  + \frac{2}{y} \cos 2 y   \right]\\
&+  \frac{9}{2} (R_\mathrm{c} - 1)^2 \frac{1}{y^2} \left( 1 + \frac{1}{y^2}  \right) 
\left[ 1 + \frac{1}{y^2} +  \left( 1 - \frac{1}{y^2}  \right) \cos 2 y - \frac{2}{y} 
\sin 2 y  \right],\nonumber
\end{align}
with the parameter $R_\mathrm{c}$ expressing the ratio of the slopes of the inflaton 
potential before and after the kink~\citep{Sinha:2005mn}.\\

\subsection{Data analysis} 

We employ a modified version of {\tt CAMB} with suitably increased numerical 
precision settings to calculate the CMB angular power spectra for the feature models.  
Since variations of the primordial spectrum may be degenerate with late-time 
cosmology parameters~\citep{Obied:2017tpd}, we explore a parameter space consisting 
of the base-$\Lambda$CDM parameters and the respective additional free parameters of 
the feature models (see Table~\ref{tab:features_priors} for the prior ranges). Note 
that we take primordial tensor perturbations to be absent in our analysis. In the 
results presented in Sect.~\ref{sec:features_results}, nuisance parameters are assumed 
to be uncorrelated with the feature parameters and kept fixed to their 
base-$\Lambda$CDM best-fit values.

In order to maximize sensitivity to narrow features, we use only the \emph{unbinned} 
versions of the \Planck\ high-$\ell$ likelihoods in the following combinations: 
(i)~temperature data, \Planck~TT(unbinned)+lowE; (ii)~$E$-polarization data only, 
\Planck~EE(unbinned)+lowE; and (iii)~temperature plus polarization data, 
\Planck~TT,TE,EE(unbinned)+lowE.

For all combinations of feature models and data, the parameter space is sampled with 
the nested sampling algorithm as implemented in {\tt MultiNest}.  The improvement in 
the fit due to the introduction of a feature is quantified by the effective $\Delta 
\chi^2 \equiv -2 (\ln \mathcal{L}_{\Lambda\mathrm{CDM}}^\mathrm{best\,fit} - \ln 
\mathcal{L}_\mathrm{feature}^\mathrm{best\,fit})$.  Being more complex than a power-law 
spectrum, feature models will in general have a negative $\Delta \chi^2$.  However, 
determining whether the improvement in fit is due to overfitting scatter in the data 
or due to an actual feature is not straightforward and requires model-dependent 
simulations~\citepalias{planck2014-a24} or analytic estimates~\citep{Fergusson:2014hya} 
to determine the expected $\Delta \chi^2$ under the null-hypothesis of an underlying 
power-law spectrum. In the Bayesian approach, a feature model's general performance 
relative to base $\Lambda$CDM can be expressed in terms of the Bayesian evidence 
$\mathcal{E}$~\citep{Trotta:2005ar}, which is also evaluated by {\tt MultiNest}.

\subsection{Feature candidates and their evidence \label{sec:features_results}}

\begin{table*}[thb]
\centering
\begin{tabular}{lccccccccccccccccccccc}
\noalign{\hrule\vskip 2pt} \noalign{\hrule\vskip 3pt}
\multirow{2}{*}{}  & \multicolumn{3}{c}{Step} &  \multicolumn{3}{c}{Kin cutoff} & \multicolumn{3}{c}{Rad cutoff} & \multicolumn{3}{c}{Kink cutoff} \\
\omit&\multispan3\hspace{0.1cm}\hrulefill\hspace{0.1cm}&\multispan3\hspace{0.1cm}\hrulefill\hspace{0.1cm}&\multispan3\hspace{0.1cm}\hrulefill\hspace{0.1cm}&\multispan3\hspace{0.1cm}\hrulefill\hspace{0.1cm}\cr
\noalign{\vskip 2pt}
& TT & EE & TT,TE,EE & TT & EE & TT,TE,EE & TT & EE & TT,TE,EE & TT & EE & TT,TE,EE  \\
\noalign{\hrule\vskip 3pt}
$\Delta \chi^2_\mathrm{eff}$ & $-7.0$ & $-5.2$ & $-5.4$ & $-1.2$ & 0.0 & $-0.9$ & $-0.2$ & $-4.7$ & $-0.0$ & $-2.1$ & $-7.4$ & $-1.1$ \\
$\ln B$ & 0.0 & $-0.2$ & 0.1 & 0.0 & $-0.2$  & 0.0 & $-0.8$ & 0.0 & $-0.7$ & $-0.4$ & 0.1 & $-0.4$ \\
\noalign{\hrule\vskip 3pt}
$\mathcal{A}_\mathrm{s}$                  & 0.29 & 0.19 & 0.38 & \dots & \dots & \dots & \dots & \dots & \dots & \dots & \dots & \dots \\
$\log_{10} \left(k_\mathrm{s}\right)$  &$-3.11$ &$-3.47$ &$-3.09$ & \dots & \dots & \dots & \dots & \dots & \dots & \dots & \dots & \dots \\
$\ln x_\mathrm{s}$                    & 0.57 & 2.17 & 0.15 & \dots & \dots & \dots & \dots & \dots & \dots & \dots & \dots & \dots \\
\noalign{\hrule\vskip 2pt}
$\log_{10} \left( k_Y^\mathrm{c}\right)$ & \dots & \dots & \dots & $-3.70$ & $-4.98$ & $-3.72$ & $-4.87$ & $-3.48$ & $-4.86$ & $-3.05$ & $-3.48$ & $-3.91$ \\
$R_\mathrm{c}$                          & \dots & \dots & \dots & \dots & \dots & \dots & \dots & \dots & \dots & $-0.02$ & 0.33  & $-0.22$ \\
\noalign{\vskip 2pt\hrule}
\end{tabular}
\vspace{.4cm} \caption{\label{tab:features_parameters1} Best-fit effective $\Delta
\chi^2$ and logarithm of the Bayes factors with respect to a featureless power spectrum, as well as
best-fit feature parameters, for the step and cutoff models. Negative values of $\ln B$ indicate a preference for a power-law spectrum, while positive ones
prefer the feature model.  Wavenumbers are in units of $\mathrm{Mpc}^{-1}$.
}
\end{table*}

\begin{table*}[!h]
\centering
\begin{tabular}{lccccccccc}
\noalign{\hrule\vskip 2pt} \noalign{\hrule\vskip 3pt}
\multirow{2}{*}{}  & \multicolumn{3}{c}{Log~osc} & \multicolumn{3}{c}{Running log osc} & \multicolumn{3}{c}{Lin osc} \\
\omit&\multispan3\hspace{0.1cm}\hrulefill\hspace{0.1cm}&\multispan3\hspace{0.1cm}\hrulefill\hspace{0.1cm}&\multispan3\hspace{0.1cm}\hrulefill\hspace{0.1cm}\cr
\noalign{\vskip 2pt}
& TT & EE & TT,TE,EE & TT & EE & TT,TE,EE & TT & EE & TT,TE,EE \\
\noalign{\hrule\vskip 3pt}
$\Delta \chi^2_\mathrm{eff}$ & $-8.5$ & $-13.5$ & $-11.0$ & $-9.3$ & $-16.5$  & $-11.4$ & $-4.2$ & $-9.0$ & $-10.8$ \\
$\ln B$ & $-1.5$ & $-0.2$ & $-0.9$ & $-1.3$ & 0.2 & $-0.5$ & $-1.8$ & $-1.3$  & $-0.8$  \\
\noalign{\hrule\vskip 3pt}
$\mathcal{A}_X$      & 0.024 & 0.073 & 0.014 & 0.028 & 0.082 & 0.016 & 0.024 & 0.046 & 0.015 \\
$\log_{10} \omega_X$     & 1.51  & 1.72  & 1.26  & 1.50  & 1.71  & 1.26  & 1.74  & 1.84  & 1.05  \\
$\varphi_X/(2 \pi)$    & 0.60  & 0.07  & 0.07  & 0.68  & 0.62  & 0.11  & 0.34  & 0.81  & 0.56  \\
$\alpha_\mathrm{rf}$      & \dots & \dots & \dots &$-0.028$ & 0.022 &$-0.021$ & \dots & \dots  & \dots  \\
\noalign{\hrule\vskip 2pt}
\end{tabular}
\vspace{.4cm} \caption{\label{tab:features_parameters2} Same as
Table~\ref{tab:features_parameters1}, but for the oscillatory feature models.}
\end{table*}

We list the best-fit effective $\Delta \chi^2$ and Bayes factors with respect to a 
power-law spectrum in Tables~\ref{tab:features_parameters1} 
and~\ref{tab:features_parameters2}. Examining the effective $\Delta \chi^2$ for the 
feature models previously considered in \citetalias{planck2014-a24} reveals only minor 
differences, with a general trend towards smaller improvements due to features.  The 
$\Delta \chi^2$ of the oscillation and step models fall well within the expected 
range of $\Delta \chi^2 \sim 10$ found in~\citetalias{planck2014-a24}. Of note are the 
relatively high values of the radiation and kink cutoff models for polarization-only 
data, partially driven by the high quadrupole of the $EE$ data.  However, the best-fit 
parameters and spectra (see Fig.~\ref{fig:cutoffandstep}) do not match their 
counterparts in the temperature data at all, which strongly suggests that this is not 
a physical effect. The same observation can also be made for the step model: the 
best fit to the $EE$ data is clearly out of phase with the temperature best fit.

We find a similar conclusion for the three oscillation models.  As can be seen from 
the profile likelihood of the frequency parameters in Fig.~\ref{fig:frequencylike}, 
the likelihood peaks in the modulation frequencies do not match up between the TT and 
EE data sets.  Furthermore, the preferred modulation amplitude for the EE data is in 
all cases much larger than that for the TT or TT,TE,EE data---given that the 
polarization data are noisier, this behaviour would be expected for a procedure that 
is overfitting the data.

Consequently, the Bayesian evidence for all combinations of models and data lies 
between barely worth mentioning and substantial evidence against the feature model on 
the Jeffreys scale. This implies that, currently, the \Planck\ data do not 
show a preference for the feature models considered here.

\begin{figure}[!ht]
\centering \hspace*{-18mm}
\begin{center}
\includegraphics[height=33mm]{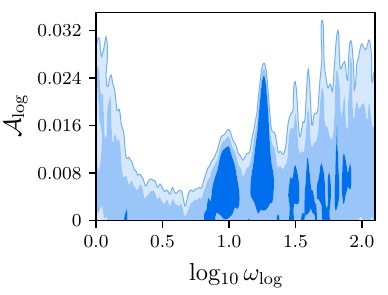}
\includegraphics[height=33mm]{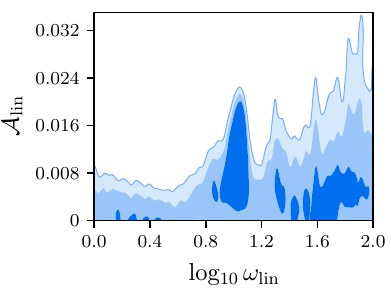}
\end{center}
\caption{\label{fig:ampvsfreq}
Marginalized joint 68\,\%, 95\,\%, and 99\,\%~CL regions of the modulation amplitude versus frequency parameter 
using the TT,TE,EE data set for the logarithmic ({\it left}) and linear ({\it right}) oscillation models.
}
\vspace{0cm}
\end{figure}

Conversely, within the frequency ranges given by our priors, the relative modulation of the power 
spectrum is constrained to not exceed roughly $3\,\%$, as shown in Fig.~\ref{fig:ampvsfreq} for the logarithmic 
and linear oscillation models.

\begin{figure}[!ht]
\centering \hspace*{-18mm}
\begin{center}
\includegraphics[height=33mm]{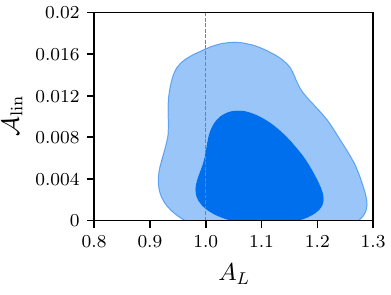}
\includegraphics[height=33mm]{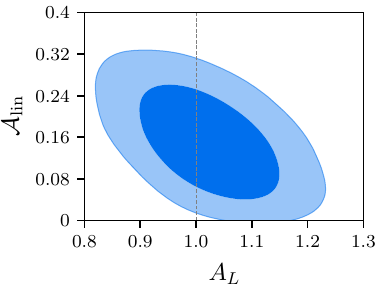}
\end{center}
\caption{\label{fig:lensingwiggles}
Marginalized joint 68\,\% and 95\,\%~CL regions for the lensing parameter $A_\mathrm{L}$ and the modulation amplitude parameter $\mathcal{A}_\mathrm{lin}$ using the TT data set.
{\it Left:} Linear oscillation model with $\log_{10} \omega_\mathrm{lin} = 1.158$ and $\varphi_\mathrm{lin} = \pi$.
{\it Right:}  Modified linear oscillation model with a Gaussian envelope function (see text) and \mbox{$\log_{10} \omega_\mathrm{lin} = 1.158$}, $\varphi_\mathrm{lin} = \pi$,
$\mu_\mathrm{env} = 0.2\,\mathrm{Mpc}^{-1}$, and \mbox{$\sigma_\mathrm{env} = 0.057\,\mathrm{Mpc}^{-1}.$}
}
\vspace{0cm}
\end{figure}

It may also be worth pointing out that in models with oscillations linear in $k$, the wavelength 
of the corresponding modulation of the angular power spectra matches that of the CMB's acoustic 
oscillations, $\Delta \ell \simeq 300$, if $\log_{10} \omega_\mathrm{lin} \simeq 1.158$.  One 
might therefore suspect that features with frequencies around this value and carefully tuned 
amplitudes and phases could in principle mimic the (unphysical) effect of a lensing parameter, 
$A_\mathrm{L} \neq 1$.  However, for a model with a modulation at the BAO frequency
and a $k$-independent modulation amplitude 
$\mathcal{A}_\mathrm{lin}$, it can be seen in the left panel of Fig.~\ref{fig:lensingwiggles} that 
we find no correlation between $A_\mathrm{L}$ and $\mathcal{A}_\mathrm{lin}$.  This is due to a 
different $\ell$-dependence of the respective $\Delta \mathcal{D}_\ell$'s.  Explaining the lensing 
discrepancy would thus require a model with a carefully arranged scale-dependent linear modulation of the 
primordial spectrum.  We demonstrate this possibility for a shaped modulation with a Gaussian envelope 
of the form 
$\mathcal{P}_\mathcal{R}(k) = \mathcal{P}_\mathcal{R}^{\, 0}(k) \left[ 1 +
\mathcal{A}_\mathrm{lin} \, \exp (- (k - \mu_\mathrm{env})^2/2\sigma_\mathrm{env}^2 ) \, \cos \left( \omega_\mathrm{lin} k/k_* + \varphi_\mathrm{lin} \right) \right]$
in the right panel of Fig.~\ref{fig:lensingwiggles}, but it should be noted that this particular
example is of course highly tuned to produce the desired effect.

Additionally, while the phenomenology of $A_\mathrm{L}$ and linear modulation models is similar 
for temperature and polarization spectra individually, the two scenarios are in principle 
distinguishable by a combination of temperature and polarization data.  This is due to the phase 
difference of the acoustic peaks in $TT$, $TE$, and $EE$, which leads to similar phase 
differences for the residuals when varying $A_\mathrm{L}$---unlike modifications of the 
primordial spectrum which do not shift phase in the same way.  However, for features with an amplitude chosen to resemble the apparent 
lensing excess in the \Planck~TT data, the \Planck~TE and EE data are not sensitive enough to 
make this distinction.

\begin{figure}[!ht]
\centering \hspace*{-12mm}
\includegraphics[height=88mm,angle=270]{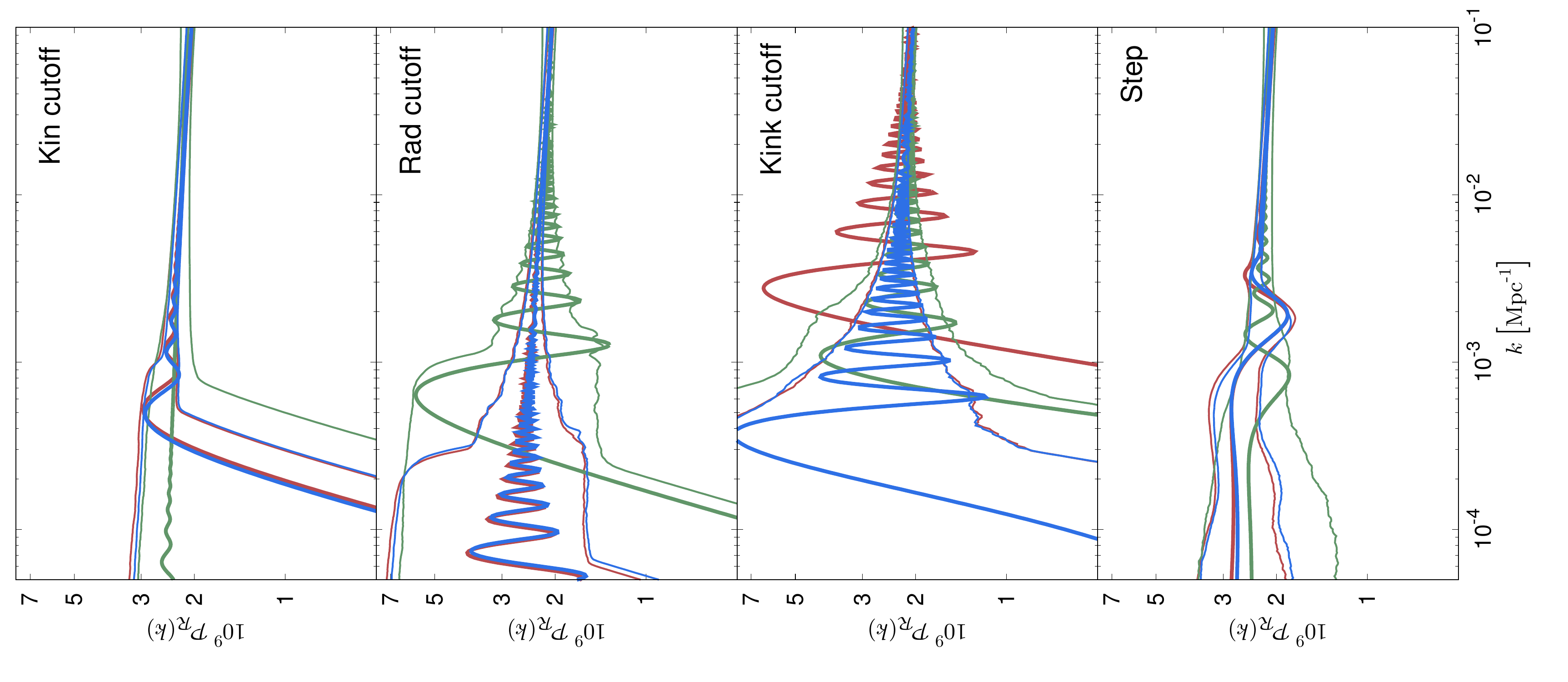}
\caption{\label{fig:cutoffandstep} Best-fit and central 95\,\% CL regions for the 
primordial power spectrum in the three cutoff and the step models for TT data 
({red} curves), EE data ({green}), and TT,TE,EE data ({blue}).  Note 
that for the combination of kink cutoff model and TT data, the best-fit value for the 
cutoff scale $k^{c}_\mathrm{kink}$ lies close to the prior boundary, and therefore 
the best-fit spectrum does not fall within the \textit{central} 95\,\%-credible band.}
\vspace{1cm}
\end{figure}

\begin{figure}[!ht]
\centering \hspace*{-18mm}
\includegraphics[height=88mm,angle=270]{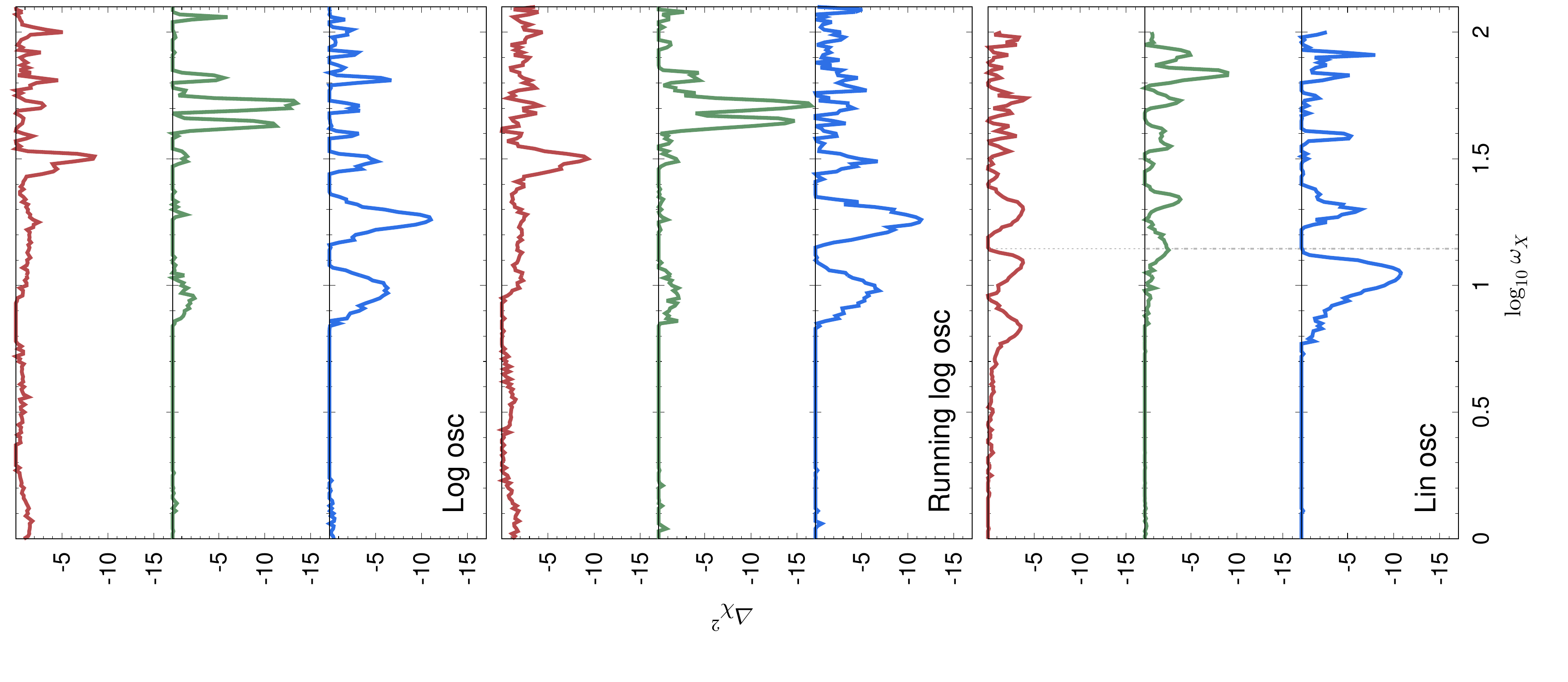}
\caption{\label{fig:frequencylike} Profile likelihood of the frequency parameter in 
the three oscillatory feature models for TT~({red} curves), EE~({green}), and 
TT,TE,EE~data~({blue}).  The dotted grey line in the bottom panels marks the 
frequency for which the linear oscillation model leads to a modulation of the angular 
power spectra whose wavelength roughly matches that of the CMB's acoustic 
oscillations. Note the lack of alignment between the temperature and polarization 
likelihood peaks in the vicinity of this frequency.}
\vspace{0cm}
\end{figure}

\subsection{Axion monodromy}

As in section~10.3 of \citetalias{planck2014-a24}, we next derive constraints on the 
underlying parameters in axion monodromy inflation \citep{Silverstein:2008sg, 
McAllister:2008hb, Kaloper:2011jz, Flauger:2014ana}, which within string theory 
motivates a broad class of inflationary potentials of the form
\begin{equation}
\label{eq:am} V(\phi) = \mu^{4-p}\phi^p+\Lambda_0^4 
e^{-C_0\left(\phi/\phi_0\right)^{p_{\Lambda}}}\cos{\left(\gamma_0+\frac{\phi_0}{f}\left(\frac{\phi}{\phi_0}\right)^{p_f+1}\right)},
\end{equation}
where $\mu, \ \Lambda_0, \ f,$ and $\phi_0$ are constants which have dimensions of 
mass, while $C_0, \ p, \ p_{\Lambda}, \ p_f$, and $\gamma_0$ are dimensionless. In 
the literature, one can find theoretically motivated models with $p \ = \ 3, \ 2, \ 
4/3, \ 1,$ and $ 2/3$ \citep{Silverstein:2008sg, McAllister:2008hb, 
McAllister:2014mpa}. In the following, we neglect a possible amplitude drift in the 
modulation amplitude by fixing $C_0=p_{\Lambda} = 0$, focussing instead on a possible 
frequency drift $p_f$, as was done in previous analyses \citep{Peiris:2013opa, 
Easther:2013kla, Jackson:2013mka, Meerburg:2012id, Meerburg:2013dla, 
Meerburg:2014kna, Meerburg:2013cla}.

Due to its oscillating nature, a numerical study of this model is restrictive 
\citep{Peiris:2013opa}. As such, we employ the semi-analytic template 
\citep{Flauger:2014ana} used in previous analyses, namely
\begin{equation}
\label{eq:sat} 
\mathcal{P_R}(k)=\mathcal{P_R}(k_{*})\left(\frac{k}{k_{*}}\right)^{n_{\rm 
s}-1}\left\lbrace 1+\delta n_{\rm s}\cos{\left[ \frac{\phi_0}{f}\left( 
\frac{\phi_k}{\phi_0}\right)^{p_f+1}+\Delta\phi \right]} \right\rbrace.
\end{equation}
We neglect the effect of small oscillations in the tensor primordial spectrum, and 
approximate it as a power law with a very small spectral index $n_\mathrm{t}$ (fixed 
by the single-field slow-roll self-consistency condition). The most well studied case 
to date is for $p=4/3$, but given the high tensor-to-scalar ratio predicted by this 
model and the current upper bounds on $r$ given in Sect.~\ref{sec:rconstraints}, we 
extend our study to the cases of $p=1$ and $p=2/3$. Furthermore, to completely 
specify this template, we assume instantaneous reheating, which, for a pivot scale of 
$k_{*} = 0.05\,\text{Mpc}^{-1}$, corresponds to $N_{*} \approx 57.5$, and $\phi_0 = 
12.38M_{\rm Pl}$ with $\phi_{\rm end} = 0.59M_{\rm Pl}$. This leads to definite 
predictions for $(r, n_{\rm s})$; namely, $(0.0922,0.971)$ for $p=4/3$, 
$(0.0692,0.974)$ for $p=1$, and $(0.0462,0.977)$ for $p=2/3$.

To constrain this model, we carry out a Bayesian analysis using a modified version of 
{\tt CLASS} \citep{Lesgourgues:2011re, Blas:2011rf}, which has been adapted to allow 
for a full parameter exploration, using the aforementioned template. As part of these 
modifications, special care needs to be taken to ensure that a correct sampling 
$\Delta k$ in wavenumber space is chosen, at two different levels in the Boltzmann 
code: when computing an interpolation table for the primordial spectrum of scalars 
and tensors; and when performing the integral over the squared photon transfer 
functions multiplied by the primordial spectra to get the multipoles $C_{\ell}$. This 
sampling needs to be fine enough to guarantee that no features are smoothed out or 
lost in this convolution, and we checked carefully that this is the case in our runs. 
The grid of $\ell$ values at which the $C_\ell$'s are actually computed and not just 
interpolated also needs to be refined.

We fit to the data the five cosmological parameters $\{\omega_\mathrm{b}, 
\omega_\mathrm{c}, \theta, A_\mathrm{s}, \tau\}$ plus the frequency $f$ of the 
underlying axion decay constant, the frequency drift $p_f$, and the oscillation 
amplitude $\delta n_{\rm s}$. We adopt the same priors used in previous analyses: 
$-4\leq \log_{10}{\left(f/M_{\rm Pl}\right)}\leq -1$ for the frequency; $-0.75< 
p_f<1$ for the frequency drift; and an upper bound on the amplitude of  $\delta 
n_{\rm s}< 0.5$. Furthermore, for the phase parameter $\Delta \phi$ we take a uniform 
prior of $-\pi < \Delta \phi < \pi$.

\begin{figure}[!t]
\centering \hspace*{0mm}\includegraphics[width=\columnwidth]{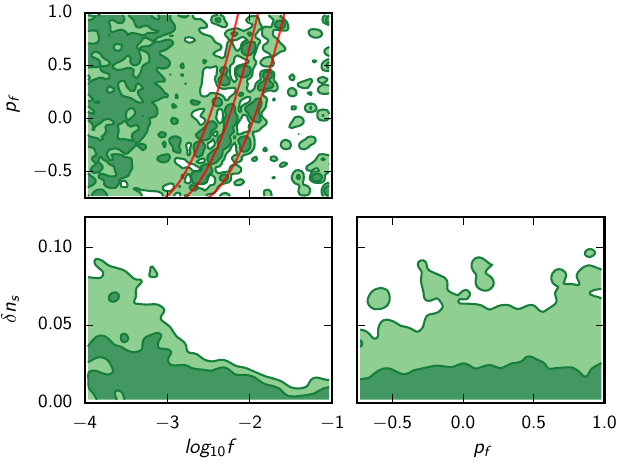} 
\caption{\label{fig:high} Joint $68\,\%$ and $95\,\%$ CL constraints on the axion 
monodromy parameters using \Planck\ (unbinned) TT,TE,EE+lowE+BK14, for the case of 
$p=4/3$. All smoothing has been turned down in the $p_f-\log_{10}{\left(f/M_{\rm 
Pl}\right)}$ posterior to avoid smoothing the features highlighted in red.}
\end{figure}

\begin{figure}[!t]
\centering \hspace*{0mm}\includegraphics[width=\columnwidth]{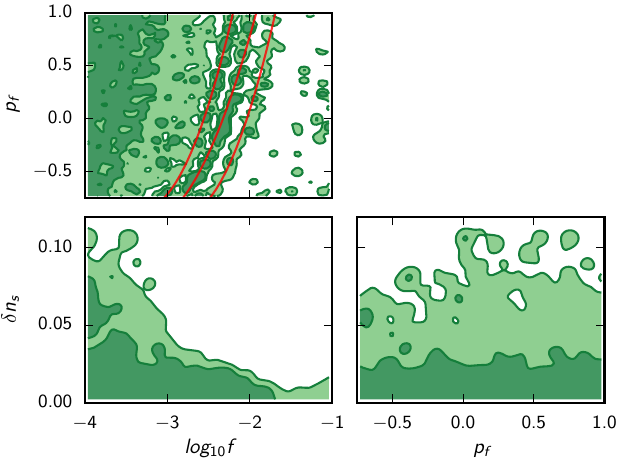}
\caption{\label{fig:med} Same as Fig.~\ref{fig:high}, but for the case of $p=1$.}
\end{figure}

\begin{figure}[!h]
\centering \hspace*{0mm}\includegraphics[width=\columnwidth]{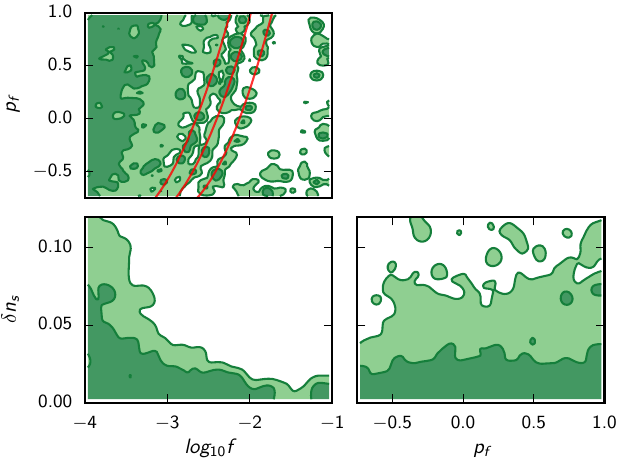}
\caption{\label{fig:low} Same as Fig.~\ref{fig:high}, but for the case of $p=2/3$.}
\end{figure}

In Figs.~\ref{fig:high}, \ref{fig:med}, and \ref{fig:low} we show the joint posterior 
constraints on pairs of primordial parameters for the semi-analytic template, for 
$p=4/3$, $p=1$, and $p=2/3$, respectively.

In all three cases, we find two expected asymptotic behaviours. First, when the 
frequency is very high (which means that $f$ is small in our parameterization), the 
oscillations in the primordial spectrum are smoothed out in the angular power 
spectrum, and the oscillation amplitude parameter $\delta n_{\rm s}$ becomes 
irrelevant and unconstrained. Second, in the limit of a very small amplitude 
parameter $\delta n_{\rm s}$, the oscillations become undetectable and the parameter 
$f$ is also unconstrained. In all cases, no preferred frequency drift is found, which 
is compatible with previous analyses.

We recover the complex structures (highlighted in red in Figs.~\ref{fig:high}, 
\ref{fig:med}, and \ref{fig:low}) found in previous analyses in the 
frequency-frequency drift parameter space, which, as was discussed in 
\citetalias{planck2014-a24}, arise due to underlying modulations in the data and the 
model \citep{Easther:2004vq}. These structures become more apparent as we reduce the 
index $p$.

We perform a $\chi^2$ comparison with the minimal 6-parameter $\Lambda$CDM model, and 
find $\Delta \chi^2_{(p=4/3)/\Lambda{\rm CDM}} = 0.4$, $\Delta \chi^2_{(p=1)/\Lambda 
{\rm CDM}} = 0.6$ and $\Delta \chi^2_{(p=2/3)/\Lambda{\rm CDM}} = 1.2$. The reason 
for higher $\chi^2$ in the axion monodromy models, despite the addition of extra 
parameters, is that the predicted $r$ values are in tension with the CMB data. This 
shows that, overall, axion monodromy models are disfavoured due to their high 
tensor-mode amplitude.

In order to check specifically whether the data give any hint of oscillatory patterns 
in the primordial spectrum matching the axion monodromy template, as well as to 
compare with the results discussed in the previous subsection, we fitted the data 
with $\Lambda$CDM+$r$ models in which $r$ and $n_{\rm s}$ were fixed to the same 
values as in the axion monodromy model with $p=2/3$, $1$, and $4/3$. In each case, 
the comparison between axion monodromy and $\Lambda$CDM+$r$ with the same ($r$, 
$n_{\rm s}$) gives $\Delta \chi^2_{(p=4/3)/\Lambda{\rm CDM}+r} = -7.8$, $\Delta 
\chi^2_{(p=1)/\Lambda{\rm CDM}+r} = -7.6$ and $\Delta \chi^2_{(p=2/3)/\Lambda{\rm 
CDM}+r} = -8$. That is, in all cases we find $\Delta \chi^2 \sim 10$, which is 
compatible with the general results  shown in Table~\ref{tab:features_parameters2}. 
With three more free parameters, these improvements are statistically insignificant, 
and we conclude that the data show no preference for axion monodromy models.




\section{Combined power spectrum and bispectrum analysis for oscillatory features \label{sec:combined}}

\subsection{Approach}\label{sec:combined_intro}

This section establishes constraints on oscillatory models using the 
power spectrum and the bispectrum simultaneously.
Oscillatory features can appear in multiple correlation functions 
\citep{Chen:2008wn,Meerburg:2009ys,monodromyIII,Flauger:2010ja,Achucarro:2010da,Adshead:2011jq,Achucarro:2013cva,Flauger:2014ana,Flauger:2016idt} 
(see, e.g., \cite{Chluba:2015bqa} for a recent review).
More powerful constraints result when spectra of various orders are combined 
\citep{Palma:2014hra,Mooij:2016dsi,Gong:2017yih}. 
Past work has suggested that the statistical weight of the oscillations 
in the bispectrum (or higher-order correlation functions) is less than that 
in the power spectrum \citep{Behbahani:2011it}; however, counterexamples exist as well (see, 
e.g., \citealt{Behbahani:2012be}). The analysis in Sect.~\ref{sec:parametrized} used 
the \Planck\ data to establish 
stringent constraints on the presence of features in the power spectrum. 
The 2015 
\Planck\ data were analyzed to constrain non-Gaussianties containing features 
\citep{planck2014-a19}, where, as in the power spectrum analysis, several candidate features were 
identified at low statistical significance. 
The analysis here focuses on the {\it location,} or frequency, of the feature.
Joint analyses of the power spectrum and bispectrum were discussed in 
several studies \citep{Fergusson:2014hya,2015PhRvD..91l3506F,Meerburg:2015owa}.
We apply some of the tools developed there to the \Planck\ temperature and polarization data.

\begin{figure}[h]
\begin{center}
\includegraphics[height=72.0mm]{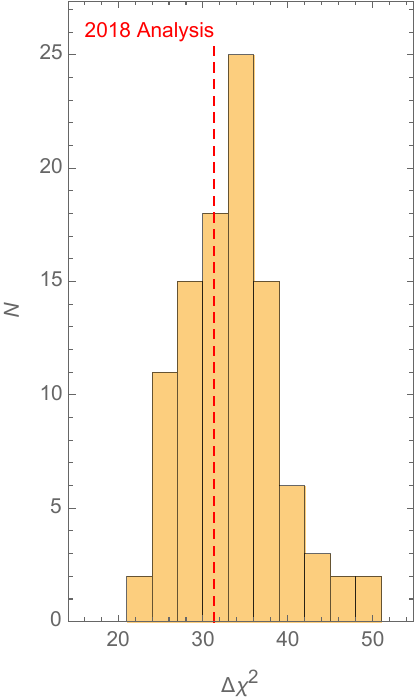}
\includegraphics[height=72.0mm]{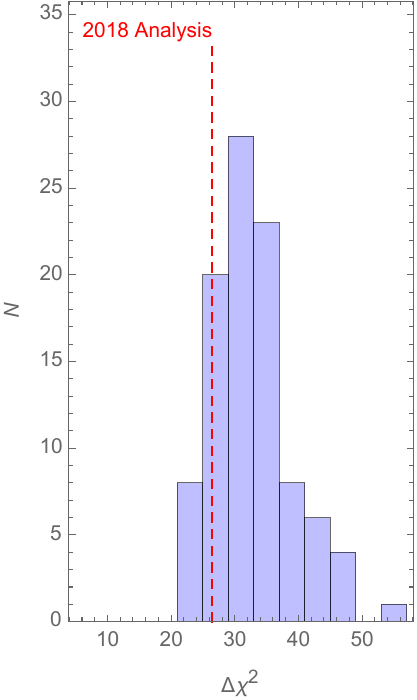}
\end{center}
\caption{Typical best-fit improvement in units of $\Delta \chi^2$ in 100 simulations compared to the real data (red dashed lines) for the log feature (left) and the linear feature (right) models. } \label{fig:BestFitPSBS}
\end{figure}

The analysis here is incomplete and limited in several respects.
First, we analyse the bispectrum keeping all cosmological parameters fixed. Second, the parameters 
varied in the bispectrum are not varied in the Bayesian sense. The bispectrum is 
analyzed using a best-fit analysis based on 
how well a template shape fits the data. Third, the data suggest the primary 
bispectrum is close to zero and its covariance dominated by the scalar contributions in the power 
spectrum. 
We find that an ideal Bayesian analysis is not computationally feasible (see, e.g., 
\citealt{Verde:2013gv}).

The output from the bispectrum analysis for features provides us with a map that specifies the 
significance of a feature in units of $\sigma$, given the location (frequency) and the phase of the 
feature. We can turn this map into a likelihood, which we can simply add to that of the 
power spectrum; in other words, we take
\begin{align} 
\ln \mathcal{L}_{\rm tot} &= \ln \mathcal{L}_{\rm PS}( c,f,\omega_{\rm P},A_{\rm P},\phi_{\rm P}| {\rm dat })\nonumber\\
   &+ \ln \mathcal{L}_{\rm BS}(A_{\rm B},\omega_{\rm B},\phi_{\rm B} | {\rm dat }),
\end{align} 
where $c$ represents the 
standard cosmological parameters, $f$ the foregrounds, $\omega_{\rm P,B}$ the frequency, $A_{\rm P,B}$ the 
amplitude, and $\phi_{\rm P,B}$ the phase of the modulation in, respectively, the power spectrum and the 
bispectrum.  We assume vanishing covariance between the power spectrum and the bispectrum, which 
has been shown to be a good approximation \citep{Fergusson:2014hya,2015PhRvD..91l3506F,Meerburg:2015owa}. 
Furthermore, the likelihood $\ln \mathcal{L}_{\rm BS}$ is not normalized (more precisely $\ln 
\mathcal{L}_{\rm tot}$ is not normalized in a universe with a non-zero bispectrum).

Strictly speaking, we do not have a likelihood that measures $A_{\rm B}$ with a certain probability. Furthermore, 
several studies have shown that the frequency parameter, in combination with the amplitude and the phase, 
does not obey a $\chi^2$ fitting to the data 
\citep{Hamann:2009bz,Meerburg:2013cla,Meerburg:2013dla,Meerburg:2014kna,Meerburg:2014bpa,Easther:2013kla}. 
Removing the frequency from the search results in a $\chi^2$ distribution with two degrees of 
freedom. Since $\ln \mathcal{L}_{\rm tot} $ is rather large (of order $10^4$ when combining all data), we 
can change the equation above by limiting ourselves only to improvements that are driven by $\omega_{\rm B}\sim 
\omega_{\rm P} \equiv \omega$, that is,
\begin{align} 
\ln \mathcal{L}_{\rm tot} &= \ln \mathcal{L}_{\rm PS}( 
c,f,\omega,A_{\rm P},\phi_{\rm P} | {\rm dat })\nonumber\\
&+ \Delta \ln \mathcal{L}_{\rm BS}( A_{\rm B},\omega,\phi_{\rm B}| {\rm dat }).
\end{align}

Assuming that $\phi_{\rm B}$ and $A_{\rm B}$ are well described by a two-parameter $\chi^2$ distribution, we can now 
convert our $\sigma$ map into a $\chi^2$ improvement via
\be 
\chi^2 = -2 \log\left[{\rm Erf}\left(\sigma/\sqrt{2}\right)+1\right], 
\ee 
or, in terms of the likelihoods,
\begin{align}
 -2 \ln \mathcal{L}_{\rm tot} &= -2\ln \mathcal{L}_{\rm PS}
\left( c,f,\omega,A_{\rm P},\phi_{\rm P} | {\rm dat }\right)\nonumber\\
&+ 2 \log \left\{{\rm Erf}\left[\sigma(\omega,A_{\rm B},\phi_{\rm B})/\sqrt{2}\right]+1\right\}.
\label{eq:TotLike}
\end{align}
We will use the above expression to derive the posterior of the joint fit. 

\subsection{Models}

We will focus on two models: the local or linear feature model and the log feature model.  For the log 
model we set $\{A_{\rm P},\omega,\phi_{\rm P}, A_{\rm B}, \phi_{\rm B}\} = \{A_{\rm log},\omega_{\rm log}, \phi_{\rm log}, B_{\rm 
log}, \tilde{\phi}_{\rm log}\}$ and use for the power spectrum
\be
\mathcal{P_{\rm log}}(k)=\mathcal{P}_{0}(k)
\left[ 1+ A_{\rm log}\cos{\left(\omega_{\rm log} 
\log \frac{k}{\tilde{k}_0}+\phi_{\rm log} \right)} \right],
\ee
with $\mathcal{P}_{0}(k) = A_{\rm s}(k/k_*)^{n_{\rm s}-1}$.  For the bispectrum we use \citep{2010AdAst2010E..72C}
\be
B_{\rm log}(k_1,k_2,k_3)= \frac{B_{\rm log} A_{\rm s} }{k_1^2k_2^2k_3^2} 
\cos{\left( \omega_{\rm log} \log \sum \frac{k_i}{\tilde{k}_0} 
+ \tilde{\phi}_{\rm log}\right)}. 
\label{eq:BisLog}
\ee
The above parameterized spectra are examples that could be generated in axion monodromy 
inflation \citep{2010JCAP...06..009F,Flauger:2010ja}, but generally are expected to appear in models where there exists an oscillatory potential. 

For the linear model we follow \citep{2015PhRvD..91l3506F} with $\{A_{\rm P},\omega_{\rm P},\phi_{\rm P},A_{\rm B}, \phi_{\rm B}\} = 
\{A_{\rm lin},\omega_{\rm lin}, \phi_{\rm lin}, B_{\rm lin}, \tilde{\phi}_{\rm lin}\}$ and write 
\be
\mathcal{P_{\rm lin}}(k)=\mathcal{P}_{0}(k)\left[ 1+ A_{\rm lin} \sin{\left(2 \omega_{\rm lin} \frac{k}{\tilde{k}_0}+\phi_{\rm lin} \right)} \right]
\ee
and \citep{Chen:2006xjb}
\be
B_{\rm lin}(k_1,k_2,k_3)= \frac{B_{\rm lin} A_{\rm s}}{k_1^2k_2^2k_3^2} \cos{\left[ \omega_{\rm lin} \left(\sum \frac{k_i }{\tilde{k}_0}\right) + \tilde{\phi}_{\rm lin}\right]}. \label{eq:BisLin}
\ee
In both models we choose $\tilde{k}_0 = 1$\,Mpc$^{-1}$, {\it which is different from the choice in 
Sect.~\ref{sec:parametrized} for the linear model}. As a result the linear frequencies can be related 
using $\omega_{\rm lin, Sect8} = 10 \omega_{\rm lin, Sect7}$. The pivot scale is set to the usual value 
$k_* = 0.05$\,Mpc$^{-1}$. The above parameterization is a proxy for models that contain sharp oscillatory 
features \citep{2010AdAst2010E..72C,Hu:2011vr,Adshead:2014sga}, although typically such effects would generate decaying features, which will not be considered here. 

\begin{figure}
\begin{center}
\includegraphics[width=43mm]{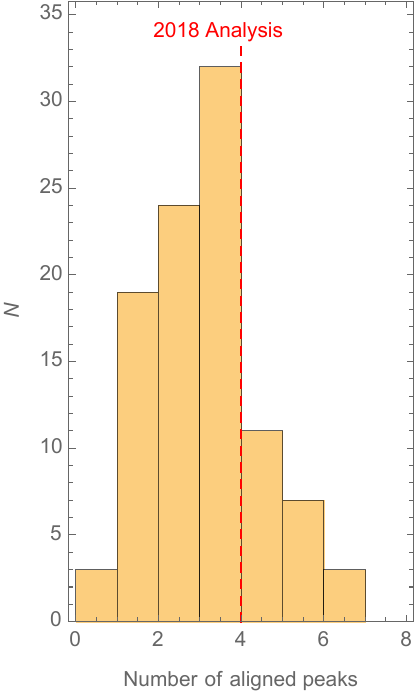}
\includegraphics[width=43mm]{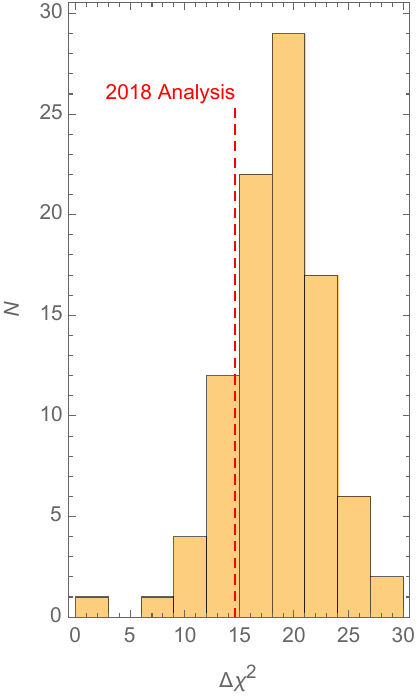}
\end{center}
\caption{{\it Left:} Number of aligned peaks in the power spectrum and the bispectrum for the log feature model. {\it Right:} Mean improvement of those same peaks in 100 simulated bispectra combined with the unbinned high-$\ell$ likelihood.  } \label{fig:PeakAlignmentLogFeatures}
\end{figure}

\subsection{Data analysis}

\subsubsection{Power spectrum}

Our analysis uses a modified version of {\tt CAMB} \citep{CAMB}, which is capable of adaptively 
changing the sampling in both $k$ and $\ell$ depending on the frequency of the feature,
allowing 
us to scan a wide range of frequencies. As in the previous section, we use the unbinned versions of 
the \Planck\ high-$\ell$ likelihoods for temperature plus polarization, in combination with lensing and 
large-scale temperature and polarization (i.e., lowE). We compared the power spectrum results for the limited 
frequency range considered in Sect.~\ref{sec:parametrized} for the log and linear model, and find 
excellent agreement (sampled with {\tt Multinest}). We developed a 
bispectrum 
likelihood module 
based on Eq.~\eqref{eq:TotLike} using the 2015 data analysis \citep{planck2014-a19} for both the 
log and linear feature models, obtained using optimal estimators following 
\citet{Munchmeyer:2014nqa,Munchmeyer:2014cca} and \citet{Meerburg:2015yka}. For the log model, 
the frequency range is set to $10\leq\omega_{\log} \leq 1000$, while for the linear model we consider the 
frequency range $10\leq\omega_{\rm lin} \leq 3000$.  This joint analysis excludes the very low frequencies 
known to (weakly) correlate with cosmological parameters. 
The cosmological parameters are held fixed in the bispectrum analysis.
We consider amplitudes $0\leq A_{\log,\rm lin} \leq 0.9$; the highest amplitudes will only 
be allowed for high frequencies where projection suppresses the power of the oscillating part in the 
power spectrum significantly. The phase is varied and marginalized over in the joint analysis. We use the 
{\tt PolyChord} sampler \citep{Handley2015a,Handley2015b}, which is powerful enough to include 
foregrounds (with $n_{\rm live} = 512$).

\subsubsection{Bispectrum}

The bispectrum likelihood is derived from the posterior distributions generated in \cite{planck2016-l09}. 
Although the linear bispectrum of Eq.~\eqref{eq:BisLin} can easily be factorized, the log 
bispectrum of Eq.~\eqref{eq:BisLog} is not of the factorized form. Using modal techniques developed by 
\citet{Fergusson:2008ra} and \citet{2010PhRvD..82b3502F,2012JCAP...12..032F}, any shape can be 
factorized, with a close-to-optimal estimator. The modal method converts the angular-average bispectrum 
into a set of factorizable orthogonal mode functions. These functions can be directly constrained using 
foreground-cleaned CMB maps. From these measurements, a large number of bispectra can be reconstructed 
and constrained by appropriately weighting the mode functions. The convergence of this method, in terms of 
how many mode functions are required to accurately reconstruct the shape of interest, depends on the 
choice of the mode functions. In the 2015 analysis, two different mode functions were used: a 
polynomial-based reconstruction; and a trigonometric-based reconstruction.  The latter was developed 
by \cite{Munchmeyer:2014nqa,Munchmeyer:2014cca} and relies on expanding around linear oscillations. The 
polynomial-based reconstruction is extremely powerful for most bispectra, but is non-optimal for 
oscillatory bispectra, which require a large number of modes (e.g., more than $2000$ for $\omega_{\log} = 
50$). Trigonometric modes allow for faster convergence and provide good reconstruction for much higher 
frequencies, both for linear- and log-type modulations. For low frequencies, both methods can be compared 
and results show excellent agreement \citep{planck2014-a19}. In addition, both methods were developed 
independently, which provides further confidence in the results. In the analysis presented here, we use 
the results obtained using the trigonometric mode functions. Further details can be found in 
\citet{planck2016-l09}.

\subsection{Estimating significance}

Next we will estimate the significance of the improvements driven by the joint analysis. For this 
purpose we generate 100 mock spectra as in \cite{Meerburg:2015owa} {\it without 
features} and perform an analysis jointly with true CMB power spectrum data, i.e., we use the same power 
spectrum likelihood (real data) in combination with the simulated bispectrum likelihoods (mock data).  We 
will do this for both the linear and the log models, with 100 simulations in total. Each analysis 
requires a similar amount of time as does the real data analysis, using about $12\,000$ CPU hours for the 
linear feature and about $40\,000$ CPU hours for the log feature {\it per simulation}. More details on 
the simulated spectra can be found in \citep{planck2016-l09}. 

These simulations help us assess the statistical significance of our results. 
Improvement in fit is given in units of $\chi^2$ compared to a no-feature model as defined the previous 
section [i.e., $\Delta \chi^2 \simeq -2 (\ln \mathcal{L}_{\Lambda\mathrm{CDM}}^\mathrm{best\,fit} - \ln 
\mathcal{L}_\mathrm{feature}^\mathrm{best\,fit})$].  The left panel of Fig.~\ref{fig:BestFitPSBS} shows 
the typical best-fit improvement from a set of simulations for the log feature model.  This first 
analysis shows that the best fit in the data is perfectly consistent with a standard $\Lambda$CDM 
universe, without features, with $P(\Delta \chi^2\geq \Delta \chi_{\rm data}^2) = 28\,\%$. This outcome 
is not unexpected, given earlier analyses for the power spectrum (see, e.g., \citealt{Meerburg:2013dla, 
Easther:2013kla,Benetti:2013cja,Miranda:2013wxa,Fergusson:2014hya,planck2014-a24,Hazra:2016fkm}) and the 
significance of features in the bispectrum alone \citep{planck2014-a15}. 
The look-elsewhere effect lowers the significance of features and by 
jointly constraining features in the power spectrum and the bispectrum it is possible to alleviate some 
of this suppression. To quantify this, we consider the following two questions: 1) considering the 
various frequencies with $\Delta \chi^2$ improvement over no features in the joint analysis, how many of 
those were present in the power spectrum analysis only; and 2) what is the mean improvement, in units of 
$\Delta \chi^2$, of these fits? We compare the results of the simulations, which do not contain any real 
features, to the data.

Before we answer these questions we need a criterion to decide if two frequencies will 
be considered the same or not, i.e., we need a frequency 
correlation measure. We will consider a simple ansatz, which will have an analytical solution and will 
serve to estimate the correlation between frequencies, by defining
\be
f_{\log}[\omega_1, \omega_2,\phi ] \simeq \int dx \cos[\omega_1 \log x + \phi] \cos[\omega_2 \log x + \phi].
\ee
Next we marginalize over phase, defining
\begin{align}
g_{\log}[\omega_1,\omega_2] &= \int d\phi f_{\log}[\omega_1,\omega_2, \phi] \cr 
  &= \frac{\pi}{1+\Delta \omega_{12}^2} \left\{ x_{\max } \cos \left[\Delta \omega_{12} \log \left(x_{\max
   }\right)\right] \right.  \nonumber \\ 
   &~~~- \cos \left[\Delta \omega_{12} \log \left(x_{\min}\right)\right]x_{\min}  \nonumber \\ 
   &~~~+ \Delta \omega_{12}\left[ \sin \left[\Delta \omega_{12} \log \left(x_{\max}\right)\right]x_{\max } \right. \nonumber \\
   & ~~~\left. \left.- \sin \left[\Delta \omega_{12} \log \left(x_{\min}\right)\right]x_{\min} \right] \right\},
\end{align}
where $\Delta \omega_{12} = \omega_1 -\omega_2$.  For linear oscillations we can derive a similar 
measure, with
\begin{align}
g_{\rm lin}[\omega_1,\omega_2] &= \frac{\pi}{2 \Delta \omega_{12}}
\left[\sin\left(2 \Delta \omega_{12} x_{\max}\right)\right.\nonumber\\
   &- \left.\sin \left(2 \Delta \omega_{12} x_{\min}\right)\right].
\end{align}
The correlation is given by
\begin{align}
{\rm Cor} (\omega_1,\omega_2) 
\quad = \frac{g[\omega_1,\omega_2]}{\sqrt{g[\omega_1,\omega_1] g[\omega_2,\omega_2]}}.
\end{align}
The parameters $x_{\min}$ and $x_{\max}$ play a role in determining the correlation length. Although 
strictly speaking they correspond to the minimum and maximum scales observable in the CMB, they can be 
used to model the correlator to allow for shifts in the frequency coming from a non-optimal analysis. We 
argue that this is reasonable given the low number of peaks in the analysis. We tested the above on 
various nearby peaks in the data and found that demanding ${\rm Cor}(\omega_1,\omega_2) \leq 0.1$ is 
generally sufficient to effectively identify independent peaks. We explored the sensitivity of the 
results to the correlation criterion of 0.1. First we increased it to 0.3 and found that in this case 
many peaks were missed when counting the number of aligned peaks (with little effect on determining the 
peaks). When we lowered the criterion to 0.01, we obtained many aligned peaks that should not be aligned. 
Small changes in the correlation criterion have minimal effect on the results presented here.  
Ideally, more simulations should be generated, which would help to establish the best choice for the correlation criterion. 
We found that the choice of\ $x_{\min}$ does not affect the correlator as long as $x_{\min}\ll 1$. We set 
$x_{\max} = 0.05$ for linear oscillations, which roughly correlates peaks with $\Delta \omega_{\rm lin} 
\sim 10$, which is within the tails of the observed widths of the peaks in the power and bispectrum 
analysis. For log oscillations we choose $x_{\max} = 1$, which has the advantage that the correlator has 
no zero-crossings near the peak (but hardly effects the correlation length). We find $\Delta 
\omega_{\log} \sim 1$ at $\omega_{\log} = 100$, which seems reasonable in light of the power and 
bispectrum peak widths (see Fig.~\ref{fig:frequencylike}).

In Fig.~\ref{fig:PeakAlignmentLogFeatures} we show the number of peaks in the joint analysis that have 
improved (left panel) as well as their mean improvement (right) over a no-feature analysis. We find 
$P(\#{\rm peaks} \geq \#{\rm peaks}_{\rm data}) = 16\,\%$ and those peaks do not lead to significant 
improvements in the joint $\chi^2$, with $P(\Delta \chi_{\rm peaks}^2\geq \Delta \chi_{\rm peaks, 
data}^2) = 83\,\%$. Assuming that these 100 simulations provide a fair sample of the noisy 
data, we conclude that there are no significant features present in both the power spectrum and the 
bispectrum for the models considered within the chosen range of feature parameters.

\begin{figure}
\begin{center}
\includegraphics[width=43mm]{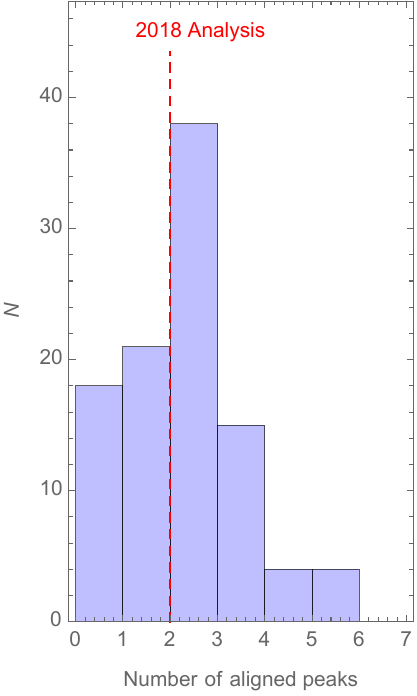}
\includegraphics[width=43mm]{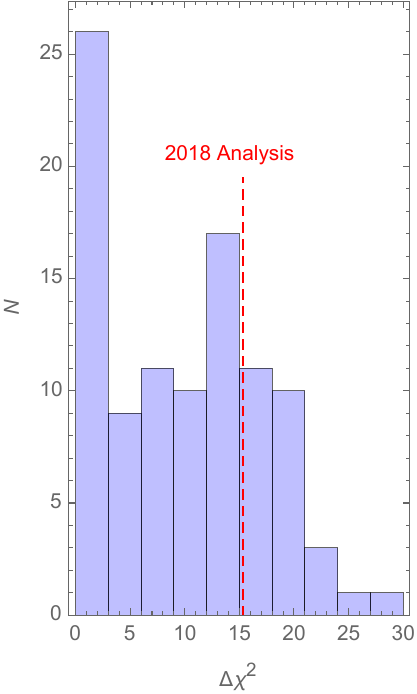}
\end{center}
\caption{{\it Left:} Number of aligned peaks in the power spectrum and 
the bispectrum for the linear feature model. {\it Right:} Mean improvement 
of those same peaks in 100 simulated bispectra combined with the unbinned 
high-$\ell$ likelihood.  } \label{fig:PeakAlignmentLinFeatures}
\end{figure}

We carry out the same analysis for linear features and show the results in 
Fig.~\ref{fig:BestFitPSBS} (right panel), deriving a typical best 
fit from 100 simulated noisy spectra. The true best 
fit, as derived from the joint analysis of the 2018 bispectrum and the 2018 power spectrum, shows a 
relatively small improvement 
with $P(\Delta \chi^2\geq \Delta \chi_{\rm data}^2) = 85\,\%$. 
Further correlated-features searches find  2 features within the frequency window which
may be considered aligned, with a mean $\Delta \chi^2$ of 12.3, as illustrated in 
Fig.~\ref{fig:PeakAlignmentLinFeatures}. Compared to 100 simulated noisy spectra, we obtain $P(\#{\rm 
peaks} \geq \#{\rm peaks}_{\rm data}) = 42\,\%$ and $P(\Delta \chi_{\rm peaks}^2\geq \Delta \chi_{\rm 
peaks, data}^2) = 26\,\%$. Since the overall improvement from 
fitting these aligned peaks does not exceed the $3\sigma$ threshold, we conclude that there is no 
statistically significant evidence for any of these features.

We conclude that the simple parameterization considered in this analysis does not provide any evidence 
for features. The two models analyzed 
are representive of a broad class and have well-studied phenomenological 
spectra; however, other classes of models exist. 
Features, for example, can have
scale dependence [i.e., an 
``envelope'' \citep{2010AdAst2010E..72C,Achucarro:2014msa,Torrado:2016sls}]. Likewise, more realistic 
modelling of axion models shows that the frequency could depend on scale [i.e., ``running'' 
\citep{Flauger:2014ana,Flauger:2016idt}]. Both these possibilities could substantially change the spectra 
and likely the joint analysis and significance. 

Here we have not imposed 
an explicit relation between the amplitude of the 
bispectrum and the frequency of the power spectrum. In the simplest form of axion monodromy, one 
has 
$f_{\rm NL} = A_{\log} \omega_{\log}^2/8$. On account of the quadratic scaling with frequency 
and the fact that the constraint on the amplitude tends to become poorer as the frequency increases due to 
projection (see, e.g., Fig.~\ref{fig:high}), it was already pointed out in \cite{planck2014-a24} that 
there is no evidence for this relation in the data, and with  
the current data set this situation remains unchanged. 




\section{Constraints on isocurvature fluctuations \label{sec:isocurvature}}

\newcommand{\dms}{\Delta^2_\mathrm{rms}}
\newcommand{\jvc}[1]{\textcolor{red}{{JV: ``#1''}}}
\newcommand{\pzJ}{\phantom{0}}
\newcommand{\pzzJ}{\phantom{00}}
\newcommand{\pdJ}{\phantom{-}}

\begin{table*}
\begin{center}
\footnotesize
\setlength{\tabcolsep}{1.88mm}
\begin{tabular}{lrrrrrrrrr}
\hline
\hline
\noalign{\vskip 1pt}
& & \multicolumn{3}{c}{$100\beta_\mathrm{iso}$ at} \cr
\omit&\omit&\multispan3\hrulefill\cr
\omit\hfil Model and data\hfil & \omit\hfil$\Delta n$\hfil & {$k_{\mathrm{low}}$} & $k_{\mathrm{mid}}$ & {$k_{\mathrm{high}}$} & \omit\hfil$100\cos\Delta$\hfil & \omit\hfil$100\alpha _{\mathrm{non-adi}}$\hfil & $\Delta\chi^2$& \omit\hfil$\ln B$\hfil \cr
\noalign{\vskip 1pt}
\noalign{\hrule\vskip 3pt}
\multicolumn{8}{l}{General models ({\it three} isocurvature parameters):}\cr
\noalign{\vskip 3pt}
$\quad $ CDI \Planck\ 2015 TT+lowP & $3$ & $4.1$ & $37$ & $57$ & $[-30 : 20]$  & $[-1.48 : 1.91]$  & $-2.1$ &  \cr  
$\quad $ CDI \Planck\ TT+lowE & $3$& $ 3.6$& $38$& $61$& $[-23 :  27]$& $[-0.76 :  2.05]$& $     -0.7$& $    -12.6$ \cr
$\quad $ CDI \texttt{CamSpec} TT+lowE & $3$& $3.8$& $35$& $56$& $[-22 :  23]$& $[-0.62 :  2.12]$& $     -0.7$& $    -13.4$ \cr
$\quad $ CDI \Planck\ TT+lowP & $3$& $ 4.2$& $35$& $56$& $[-25 :  23]$& $[-1.03 :  1.98]$& $     -0.5$& $    -12.6$ \cr
$\quad$ CDI \Planck\ TT + $\tau$ prior & $3$& $ 8.4$& $27$& $40$& $[-21 :  29]$& $[-0.83 :  5.35]$&& \cr

\noalign{\vskip 3pt}
$\quad $ CDI \Planck\ 2015 TT+lowP+lensing  & $3$  &       $4.5$ & $[1 : 40]$ & $[ 1 : 62]$ & $ [-28 : 17] $ & $[-1.05 : 1.86]$ & $-1.2$ &  \cr 
$\quad $ CDI \Planck\ TT+lowE+lensing & $3$& $ 4.0$& $35$& $57$& $[-28 :  23]$& $[-1.20 :  2.04]$& $     -0.6$& $    -12.3$ \cr
$\quad $ CDI \texttt{CamSpec} TT+lowE+lensing & $3$& $ 3.7$& $34$& $55$& $[-24 :  24]$& $[-0.96 :  2.10]$& $     -0.5$& $    -12.8$ \cr

\noalign{\vskip 3pt}
$\quad $ CDI \Planck\ 2015 TT,TE,EE+lowP  & $3$  &      $2.0$ & $[ 3 : 28]$ & $[5 : 52]$ & $[-6 : 20]$ & $[0.09 : 1.51]$  & $-5.3$  & \cr 
$\quad $ CDI \Planck\ TT,TE,EE+lowE & $3$& $ 2.1$& $[ 1 : 31]$& $58$& $[-11 :  15]$& $[-0.18 :  1.24]$& $     -3.0$& $    -12.8$ \cr
$\quad $ CDI \texttt{CamSpec} TT,TE,EE+lowE & $3$& $ 2.8$& $21$& $38$& $[-12 :  20]$& $[-0.20 :  1.67]$& $     -1.2$& $    -14.0$ \cr
$\quad $ CDI \Planck\ TT,TE,EE+lowP & $3$& $ 2.4$& $27$& $50$& $[-11 :  17]$& $[-0.16 :  1.45]$& $     -2.3$& $    -13.4$ \cr
$\quad$ CDI \Planck\ TT,TE,EE + $\tau$ prior & $3$& $ 6.2$& $17$& $30$& $[-13 :  14]$& $[-0.48 :  3.94]$&& \cr

\noalign{\vskip 3pt}
$\quad $ {\bf CDI \Planck\ TT,TE,EE+lowE+lensing} & $3$& $ 2.5$& $[ 1 : 26]$& $47$& $[-12 :  15]$& $[-0.25 :  1.31]$& $     -2.8$& $    -12.8$ \cr
$\quad $ CDI \texttt{CamSpec} TT,TE,EE+lowE+lensing & $3$& $ 3.0$& $19$& $33$& $[-16 :  18]$& $[-0.38 :  1.54]$& $     -0.9$& $    -14.1$ \cr
$\quad $ CDI \Planck\ TT,TE,EE+lowP+lensing & $3$& $ 2.2$& $[ 1: 27]$& $50$& $[-11 :  16]$& $[-0.16 :  1.36]$&& \cr

\noalign{\vskip 3pt}
$\quad$ CDI WMAP-9 & $3$& $20.1$& $[ 2 : 50]$& $66$& $[-38 :  34]$& $[-1.79 :  6.46]$& $     -0.2$& $     -9.6$ \cr

\noalign{\hrule\vskip 5pt}
$\quad $ NDI \Planck\ 2015 TT+lowP+lensing   & $3$ & $15.8$ & $[2 : 24]$ & $[2 : 29]$ & $[-32 : 0]$ & $[-4.04 : 1.37]$ &$-2.8$ & \cr 
$\quad $ NDI \Planck\ TT+lowE+lensing & $3$& $15.3$& $17$& $21$& $[-36 :   4]$& $[-4.20 :  1.53]$& $     -1.9$& $    -10.8$ \cr
$\quad $ {\bf NDI \Planck\ TT,TE,EE+lowE+lensing} & $3$& $7.4$& $[ 3 : 17]$& $[ 2 : 23]$& $[-13 :   8]$& $[-0.76 :  1.74]$& $     -5.3$& $    -10.9$ \cr

\noalign{\hrule\vskip 5pt}
$\quad $ NVI \Planck\ 2015 TT+lowP+lensing   & $3$ & $9.8$ & $[1 : 12]$ & $14$ & $[-23 :  7]$ & $[ -2.03 : 2.95]$ &  $-2.5$ & \cr 
$\quad $ NVI \Planck\ TT+lowE+lensing & $3$& $ 7.1$& $10$& $12$& $[-36 :   3]$& $[-3.34 :  1.71]$& $     -2.5$& $    -12.6$ \cr
$\quad $ {\bf NVI \Planck\ TT,TE,EE+lowE+lensing} & $3$& $6.8$& $[ 1 :  8]$& $10$& $[-20 :   0]$& $[-1.66 :  1.29]$& $     -5.2$& $    -12.0$ \cr

\noalign{\hrule\vskip 5pt}
$\quad $ CDI+$A_{\rm L}$ \Planck\ TT+lowE & $4$& $ 9.4$& $28$& $41$& $[-41 :  10]$& $[-2.32 :  2.29]$& $     -9.2$& $    -10.1$ \cr
$\quad $ CDI+$A_{\rm L}$ \Planck\ TT+lowE+lensing & $4$& $ 6.0$& $36$& $57$& $[-27 :  18]$& $[-1.16 :  2.19]$& $     -4.1$& $    -13.1$ \cr
$\quad $ CDI+$A_{\rm L}$ \Planck\ TT,TE,EE+lowE & $4$& $ 3.3$& $20$& $36$& $[-12 :  19]$& $[-0.24 :  1.89]$& $    -10.6$& $    -11.2$ \cr
$\quad $ {\bf CDI+$A_{\rm L}$ \Planck\ TT,TE,EE+lowE+lensing} & $4$& $ 2.7$& $[ 1 : 27]$& $49$& $[-10 :  16]$& $[-0.12 :  1.53]$& $     -8.1$& $    -13.5$ \cr

\noalign{\hrule\vskip 2pt}
\noalign{\hrule\vskip 5pt}
\multicolumn{8}{l}{Special CDI cases ({\it one} isocurvature parameter):}\cr
$\quad $ Uncorrelated, $n_\mathcal{II}=1$ & & & & & & & \cr
 $\quad\quad$ ``axion I'' \Planck\ 2015 TT+lowP+lensing  & $1$   & $3.9$ & $4.3$ & $4.4$ &  $0$ & $[0 : 1.70]$  &  $0$  & \cr 
$\quad\quad$ ``axion I'' \Planck\ TT+lowE+lensing & $1$& $ 3.5$& $ 3.9$& $ 3.9$& $0$& $[0 : 1.58]$& $      0$& $     -5.7$ \cr
$\quad\quad$ {\bf ``axion I'' \Planck\ TT,TE,EE+lowE+lensing} & $1$& $ 3.5$& $ 3.8$& $ 3.9$& $0$& $[0 : 1.55]$& $      0$& $     -5.5$ \cr
\noalign{\vskip 2pt}
$\quad $ Fully correlated, $n_\mathcal{II}=n_\mathcal{RR}$  & & & & & & & \cr
 $\quad\quad$ ``curvaton I'' \Planck\ 2015 TT+lowP+lensing  & $1$   & $0.2$ & $0.2$ & $0.2$ & $+100$ & $[ 0.30 : 2.70 ]$ &  $0$ & \cr 
$\quad\quad$ ``curvaton I'' \Planck\ TT+lowE+lensing & $1$& $ 0.2$& $ 0.2$& $ 0.2$& $+100$& $[ 0.09 :  2.91]$& $      0$& $     -8.9$ \cr
$\quad\quad$ {\bf ``curvaton I'' \Planck\ TT,TE,EE+lowE+lensing} & $1$& $ 0.1$& $ 0.1$& $ 0.1$& $+100$& $[ 0.07 :  1.81]$& $      0$& $     -9.7$ \cr
\noalign{\vskip 2pt}
 $\quad $ Fully anti-correlated, $n_\mathcal{II}=n_\mathcal{RR}$ & & &  & & & & \cr
 $\quad\quad$  ``curvaton II'' \Planck\ 2015 TT+lowP+lensing   & $1$  & $0.5$ & $0.5$ & $0.5$  & $-100$ & $[ -4.40 : -0.40 ]$ & $-0.6$ & \cr 
$\quad\quad$ ``curvaton II'' \Planck\ TT+lowE+lensing & $1$& $ 0.5$& $ 0.5$& $ 0.5$& $-100$& $[-4.40 : -0.35]$& $     -0.4$& $     -7.2$ \cr
$\quad\quad$ {\bf ``curvaton II'' \Planck\ TT,TE,EE+lowE+lensing} & $1$& $ 0.1$& $ 0.1$& $ 0.1$& $-100$& $[-2.04 : -0.13]$& $      0$& $     -9.2$ \cr
\noalign{\hrule\vskip 2pt}
\noalign{\hrule\vskip 5pt}
\multicolumn{8}{l}{Special CDI cases ({\it two} isocurvature parameters):}\cr 
 $\quad $ Uncorrelated, $n_\mathcal{II}$ free & & & & & & & \cr 
$\quad\quad$ ``axion II'' \Planck\ TT+lowE+lensing & $2$& $ 2.3$& $[ 3 : 43]$& $[ 6 : 75]$& $0$& $[ 0.03 :  1.24]$& $     -0.5$& $     -6.7$ \cr
$\quad\quad$ {\bf ``axion II'' \Planck\ TT,TE,EE+lowE+lensing} & $2$& $ 1.1$& $[ 5 : 38]$& $[ 10 : 77]$& $0$& $[ 0.07 :  0.66]$& $     -2.8$& $     -6.3$ \cr
\noalign{\vskip 2pt}
$\quad $ Arbitrarily correlated, $n_\mathcal{II}=n_\mathcal{RR}$  & & & & & & & \cr 
$\quad\quad$ ``curvaton III'' \Planck\ TT+lowE+lensing & $2$& $ 4.7$& $ 4.7$& $ 4.7$& $[-75 :  28]$& $[-3.38 :  1.99]$& $     -0.4$& $    -10.0$ \cr
$\quad\quad$ {\bf ``curvaton III'' \Planck\ TT,TE,EE+lowE+lensing} & $2$& $ 3.9$& $ 3.9$& $ 3.9$& $[-41 :  31]$& $[-1.30 :  2.10]$& $     0$& $    -10.5$ \cr
\noalign{\vskip 2pt}
 $\quad $ Fully correlated, $n_\mathcal{II}$ free & & & & & & & \cr
$\quad\quad$ \Planck\ TT+lowE+lensing & $2$& $ 0.1$& $ 4.6$& $16.0$& $+100$& $[ 0.28 :  2.15]$& $     -0.5$& $    -13.2$ \cr
$\quad\quad$ {\bf \Planck\ TT,TE,EE+lowE+lensing} & $2$& $ 0.02\!\!\!$& $ 1.5$& $ 6.2$& $+100$& $[ 0.14 :  0.99]$& $     -0.3$& $    -16.0$ \cr
\noalign{\vskip 2pt}
$\quad $ Fully anti-correlated, $n_\mathcal{II}$ free & & & & & & & \cr
$\quad\quad$ \Planck\ TT+lowE+lensing & $2$& $ 0.6$& $ 0.9$& $ 1.3$& $-100$& $[-5.56 : -0.53]$& $     -1.3$& $    -13.8$ \cr
$\quad\quad$ {\bf \Planck\ TT,TE,EE+lowE+lensing} & $2$& $ 0.3$& $ 0.2$& $ 0.2$& $-100$& $[-4.56 : -0.16]$& $     -0.6$& $    -16.7$ \cr
\noalign{\hrule\vskip 2pt}
\vspace{-7mm}
\end{tabular}
\end{center}
\caption{\label{tab:how_much_ic} Constraints on mixed adiabatic
  and isocurvature models. We report 95\,\% CL intervals or upper bounds on the isocurvature fraction
$\beta_\mathrm{iso}$ at three scales ($k_{\mathrm{low}}= 0.002\,$Mpc$^{-1}$,
$k_{\mathrm{mid}}= 0.050\,$Mpc$^{-1} ,$ and $k_{\mathrm{high}}=
0.100\,$Mpc$^{-1}$), the scale-independent 
correlation fraction, $\cos\Delta$, and the non-adiabatic contribution to the CMB temperature variance, 
$\alpha_{\text{non-adi}}$.
Here $\Delta \chi^2$ is the difference between the $\chi^2$ of the
best-fit mixed and pure adiabatic models.
In the last
column we give the difference between the log of Bayesian
evidences. (A negative $\ln B$ 
means that Bayesian model comparison disfavours the mixed model.)
The number of extra parameters compared with 
\LCDM\ is denoted by $\Delta n$ in the first column.
Note
that the uniform priors on the primordial powers at two scales lead to non-uniform priors on the parameters reported in this table.
This is particularly significant for  $\beta_\mathrm{iso}(k_{\mathrm{mid}})$, where the prior peaks at a non-zero value.
The baseline \Planck\ 2018 TT,TE,EE+lowE+lensing results are highlighted in bold.}
\end{table*}

\begin{figure*}
\begin{center}
(a)\hspace{140mm}$\phantom{1}$\\
\vspace{-5mm}
\includegraphics[width=12.4cm]{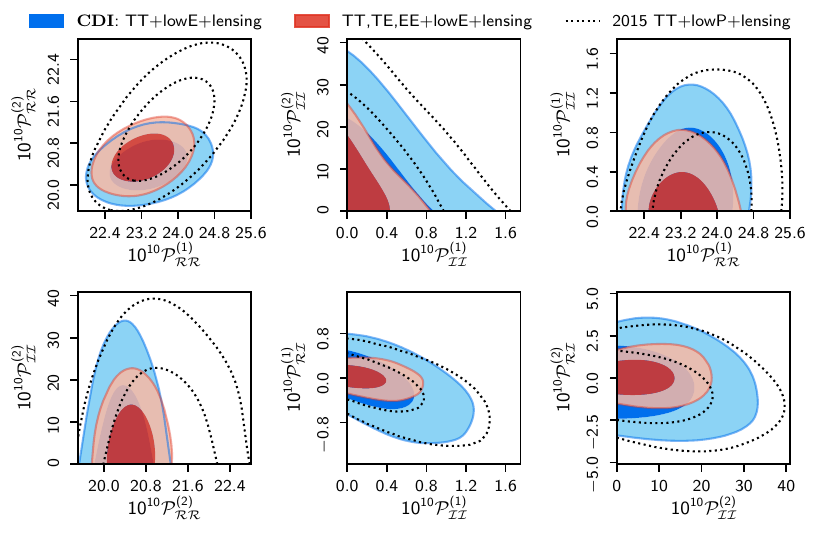}\\
(b)\hspace{140mm}$\phantom{1}$\\
\vspace{-5mm}
\includegraphics[width=12.4cm]{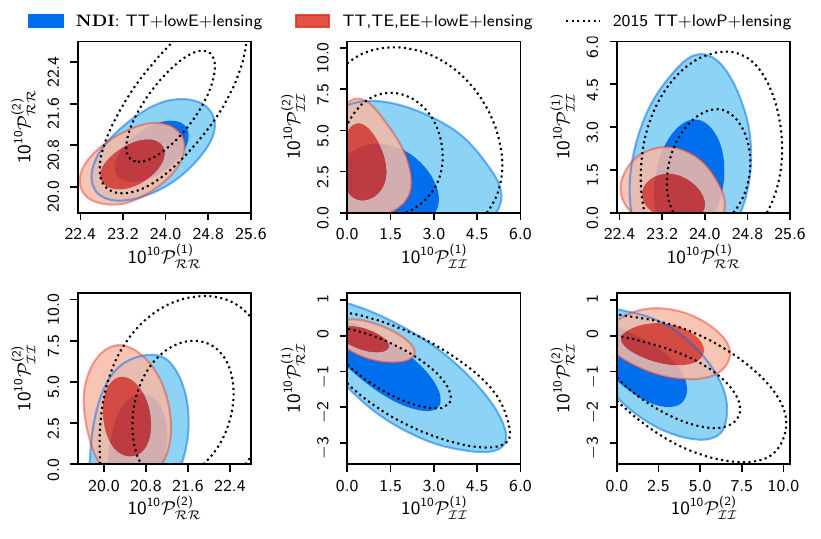}\\
(c)\hspace{140mm}$\phantom{1}$\\
\vspace{-5mm}
\includegraphics[width=12.4cm]{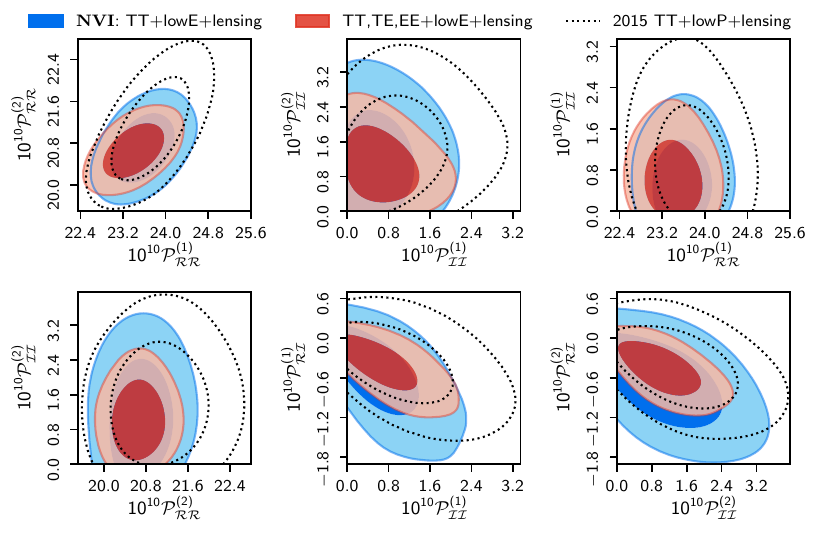}\\
\vspace{-8mm}
\end{center}
\caption{Constraints on the primordial perturbation power in 
generally correlated ADI+CDI (a), ADI+NDI (b), and ADI+NVI (c) models at two 
scales, $k_1\!=\!0.002\,\mathrm{Mpc}^{-1}$ (1) and $k_2\!=\!0.100\,\mathrm{Mpc}^{-1}$ (2).
Note that in our modelling $\mathcal{P}_{\mathcal{RI}}^{(2)}$ is not an independent 
parameter. 
\label{fig:JVprimordialPowers}}
\end{figure*}
\begin{figure*}
\begin{center}
(a)\hspace{178.5mm}$\phantom{1}$\\
\vspace{-5.5mm}
\includegraphics[width=17.65cm]{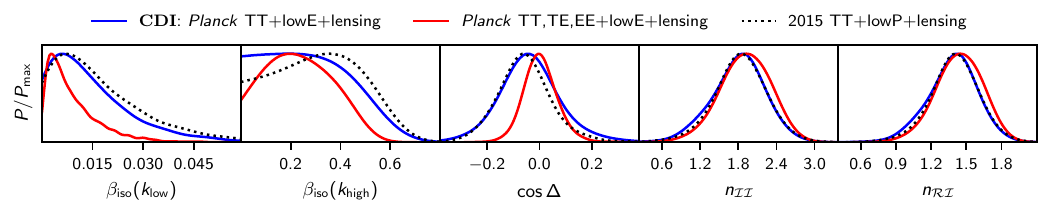}\\
(b)\hspace{178.5mm}$\phantom{1}$\\
\vspace{-5.5mm}
\includegraphics[width=17.65cm]{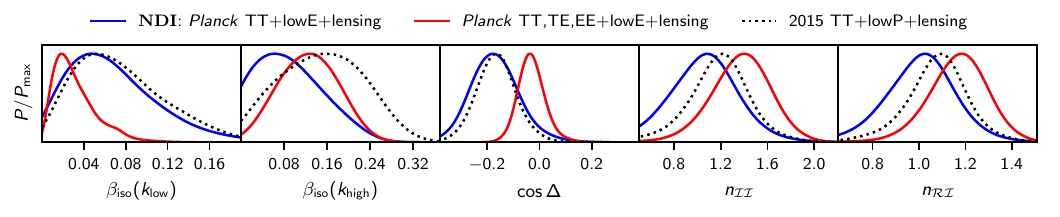}\\
(c)\hspace{178.5mm}$\phantom{1}$\\
\vspace{-5.5mm}
\includegraphics[width=17.65cm]{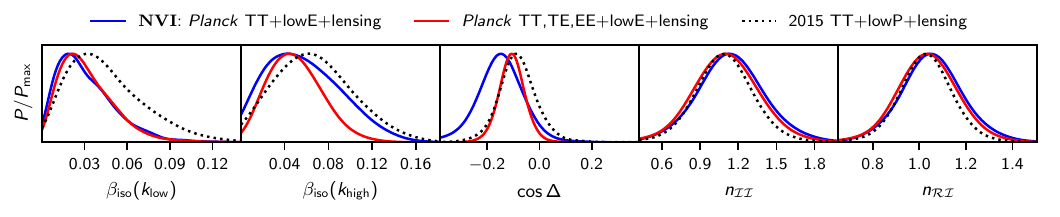}\\
\vspace{-7mm}
\end{center}
\caption{Constraints on the primordial isocurvature fraction, $\beta_{\mathrm{iso}}$, 
at $k_{\mathrm{low}}=0.002\,\mathrm{Mpc}^{-1}$ and 
$k_{\mathrm{high}}=0.100\,\mathrm{Mpc}^{-1}$; the primordial correlation fraction, $\cos\Delta$; 
the isocurvature spectral index, $n_{\cal I \cal I}$; and the correlation spectral 
index, $n_{\cal R \cal I} = (n_{\cal R \cal R} + n_{\cal I \cal I}) / 2$, for the 
generally correlated mixed ADI+CDI model (a), for the ADI+NDI model (b), and for 
the ADI+NVI model (c). All these parameters are derived, and the distributions 
shown here result from a uniform prior on the primary parameters shown in 
Fig.~\ref{fig:JVprimordialPowers}. However, the effect of the non-flat 
derived-parameter priors is negligible for all parameters except for 
$n_{\cal I \cal I}$ (and $n_{\cal R \cal I}$) where the prior biases the 
distribution toward unity. Note that these spectral indices are not well constrained, 
since we do not have a detection of non-zero isocurvature or correlation amplitude. 
With a sufficiently small isocurvature or correlation amplitude, an arbitrarily small or 
large spectral index leads to a very good fit to the data, since the model is then 
practically adiabatic over the range covered by the \Planck\ data. \label{fig:JVprimordialFractions} }
\end{figure*}

\subsection{Background and modeling}

Single-field inflation with a canonical kinetic term gives rise to primordial 
super-Hubble comoving curvature perturbations, $\mathcal{R}$. In this case the relative 
number densities of the various particle species are spatially constant, i.e., the 
perturbations are adiabatic. Typically photons are chosen as a reference species. Then 
adiabaticity implies that for every particle species with number density $n_i$ the 
quantity $\delta(n_i/n_\gamma)$ vanishes. However, in addition to $\mathcal{R}$, 
multi-field inflation can stimulate isocurvature modes, 
$\mathcal{I}_i$, where at primordial times $n_i/n_\gamma$ varies spatially 
\citep{Linde:1985yf,Polarski:1994rz,Linde:1996gt,GarciaBellido:1995qq}. In this section we consider all possible 
non-decaying modes of this type \citep{Bucher:1999re}: cold dark matter density isocurvature (CDI); baryon 
density isocurvature (BDI); and neutrino density isocurvature (NDI) modes. For 
completeness, we also constrain the fourth non-decaying mode, 
neutrino velocity isocurvature (NVI), although there are no known mechanisms to excite 
it. Finally, we consider compensated isocurvature perturbations (CIP) between 
baryons and CDM \citep{Grin:2011tf,Grin:2011nk}. In this case, opposite BDI and CDI perturbations cancel in 
such a way that the total matter isocurvature perturbation vanishes and there is no first-order 
isocurvature signal in the CMB. However, we utilize a higher-order lensing-like effect 
from this mode to obtain constraints on CIP from \Planck\ temperature and 
polarization power spectra. We find the most powerful power-spectra-based constraints on 
this mode by exploiting the cosmological information in the low-$L$ lensing potential
reconstruction in Sect.~\ref{sec:CIP}, but leave the 
use of \Planck\ trispectra in constraining CIP for future work.

As the positions of the peaks and dips of the CMB angular power spectra in the density 
isocurvature models are roughly in opposite phase compared to the pure adiabatic (ADI) 
spectrum, the primordial CDI, BDI, and NDI modes leave a very distinctive observational 
imprint on the CMB, whereas the imprint of the NVI mode more closely resembles the pure ADI mode; 
see, e.g., figure~43 in \citetalias{planck2014-a24}. Prior to the detection of CMB 
anisotropies, studies such as \citet{Peebles:1970ag} and \citet{Efstathiou:1986,Efstathiou:1987} 
discussed the possibility that isocurvature perturbations were the sole source of 
cosmological fluctuations. However, at least after the detection of the first acoustic 
peak in $TT$, it became clear that the density isocurvature mode(s) had to be subdominant 
\citep{Enqvist:2000hp,Enqvist:2001fu}, while the adiabatic mode led to a 
good agreement with observations. Several pre-\Planck\ isocurvature 
constraints were obtained \citep{Stompor:1995py,Pierpaoli:1999zj,Langlois:2000ar,Amendola:2001ni,Peiris:2003ff, Valiviita:2003ty,Bucher:2004an,Moodley:2004nz,Beltran:2004uv,KurkiSuonio:2004mn,Dunkley:2005va,Bean:2006qz,Trotta:2006ww,Keskitalo:2006qv,Komatsu:2008hk,Valiviita:2009bp}.

The mixture of curvature and isocurvature perturbations can be uncorrelated, but 
typically an arbitrary amount of correlation arises between them if the trajectory in 
field space is curved between Hubble radius exit and the end of multi-field inflation 
\citep{Gordon:2000hv}. In extreme cases, such as the simplest curvaton models, there is 
full correlation or full anticorrelation between $\mathcal{R}$ and $\mathcal{I}$. In the 
following subsections, we start with the generic case of generally correlated adiabatic 
and CDI, NDI, or NVI perturbations. Then we deal with various special CDI (or BDI) cases 
with no correlation or full (anti)correlation.

We parameterize the primordial perturbations as in 
\citetalias{planck2014-a24}, following the notation described there. The primary 
perturbation parameters scanned by \texttt{MultiNest} (in addition to the four standard 
\LCDM\ background cosmological parameters and the \Planck\ nuisance parameters) are the 
primordial abiabatic perturbation power and isocurvature perturbation power at two 
scales, corresponding to $k_1=k_\mathrm{low}=0.002\,$Mpc$^{-1}$ and 
$k_2=k_\mathrm{high}=0.1\,$Mpc$^{-1}$, namely, $\mathcal{P}_{\mathcal{RR}}^{(1)}$, 
$\mathcal{P}_{\mathcal{RR}}^{(2)}$, $\mathcal{P}_{\mathcal{II}}^{(1)}$, 
$\mathcal{P}_{\mathcal{II}}^{(2)}$, and the correlation power between $\mathcal{R}$ and 
$\mathcal{I}$ at $k_1$, i.e., $\mathcal{P}_{\mathcal{RI}}^{(1)}$. We assume a power-law 
form for the adiabatic and isocurvature power spectra and denote the spectral indices 
that can be calculated from the primary parameters by $n_\mathcal{RR}$ and 
$n_\mathcal{II}$. The correlation spectrum is also assumed to obey a power law, with 
spectral index $n_\mathcal{RI} = (n_\mathcal{RR} + n_\mathcal{II})/2$. Thus 
$\mathcal{P}_{\mathcal{RI}}^{(2)}$ is not an independent parameter. This ensures that 
the correlation fraction $\cos\Delta = {\mathcal{P}_{\mathcal{RI}}}/{\left( 
\mathcal{P}_{\mathcal{RR}} \mathcal{P}_{\mathcal{II}} \right)^{1/2}}$ stays inside
the interval
$(-1,\,1)$ at every $k$, as long as we reject any $\mathcal{P}_{\mathcal{RI}}^{(1)}$ 
which does not obey this requirement. While the correlation fraction is $k$-independent 
in our modelling, the primordial isocurvature fraction $\beta_\mathrm{iso}(k) = 
\mathcal{P}_{\mathcal{II}}(k) / \left[ \mathcal{P}_{\mathcal{RR}}(k) + 
\mathcal{P}_{\mathcal{II}}(k) \right]$ depends on $k$, unless $n_\mathcal{II} = 
n_\mathcal{RR}$.  We also report $\beta_\mathrm{iso}$ at an intermediate scale, 
$k_{\mathrm{mid}}= 0.05\,$Mpc$^{-1}$.

We do not separately quote constraints on BDI or the total matter density isocurvature 
(MDI), since these modes are observationally indistinguishable from 
the CDI case.\footnote{%
If we assume no NVI or NDI perturbations, then the MDI perturbation 
(i.e., the spatial perturbation in the relative number densities of matter particles and 
photons) is
\begin{equation}
   \mathcal{I}_\mathrm{MDI} = \frac{\Omega_\mathrm{c}}{\Omega_\mathrm{m}} \mathcal{I}_\mathrm{CDI} +  \frac{\Omega_\mathrm{b}}{\Omega_\mathrm{m}} \mathcal{I}_\mathrm{BDI}\,.
\label{eq:MDIperturbation}
\end{equation}
As we will see, 
the posteriors for $\Omega_\mathrm{c} h^2$ and $\Omega_\mathrm{b} h^2$ are 
insensitive to the assumed initial conditions. 
Thus it is a good approximation to use
the mean values obtained in the generally correlated mixed adiabatic and CDI 
model with TT,TE,EE+lowE+lensing data, 
namely $\Omega_\mathrm{c} / \Omega_\mathrm{m} \simeq 0.842$, 
$\Omega_\mathrm{b} / \Omega_\mathrm{m} \simeq 0.158$, 
$\Omega_\mathrm{c} / \Omega_\mathrm{b} \simeq 5.33$, and  
$(\Omega_\mathrm{c} / \Omega_\mathrm{b})^2 \simeq 28.4$. 
For example, to convert our CDI upper bound on $\mathcal{P_{II}}$ 
to a BDI bound, we should multiply the constraint 
by $(\Omega_\mathrm{c} / \Omega_\mathrm{b})^2 = 28.4$, and  
to convert the CDI $\mathcal{P_{RI}}$ to BDI, 
we should multiply the constraint by 
$\Omega_\mathrm{c} / \Omega_\mathrm{b} \simeq 5.33$. 
If $\beta_{\mathrm{iso}} \ll 1$, then this also can be converted to a 
BDI constraint by multiplying the CDI constraint by $28.4$. 
The constraint on $\cos\Delta$ will be the same for the CDI and 
BDI cases, since the conversion factor cancels out.
}

Numerical results for various isocurvature models and selected derived parameters are 
reported in Table~\ref{tab:how_much_ic},
utilizing various data combinations. The table 
is divided into three main sections: generally correlated models (discussed in 
Sect.~\ref{sec:genIsoc}); one-isocurvature-parameter CDI models (discussed in 
Sects.~\ref{sec:axionI} and \ref{sec:curvatonIandII}); and, finally, two-isocurvature-parameter 
CDI models (discussed in Sects.~\ref{sec:axionII}, \ref{sec:curvatonIII}, and 
\ref{sec:FCandACarbCor}). For generally correlated CDI we study the 
stability of constraints (see Sect.~\ref{sec:isocVsAL}) by using several different 
subsets of the \Planck\ data: (1) only 
high-$\ell$ TT; 
(2) high-$\ell$ TT+lensing; 
(3) TT,TE,EE; 
and 
(4) TT,TE,EE+lensing. 
For comparison, some \Planck\ 2015 and WMAP results are also cited.
Table~\ref{tab:how_much_ic} also includes comparisons to the pure adiabatic 
model in terms of the difference in the best-fit $\chi^2$ and the natural 
logarithm of the Bayesian evidence (``model probability'') ratios
$\ln B$, negative $\ln B$ being evidence against the mixed
models.\footnote{The values of $\ln B$ depend on the priors. We adopt
uniform priors in the range $(15,\,40)\times10^{-10}$ for the adiabatic,
$(0,\,100)\times10^{-10}$ for the isocurvature, and
$(-100,\,100)\times10^{-10}$ for the primordial correlation power parameter.}

\begin{figure}
\includegraphics{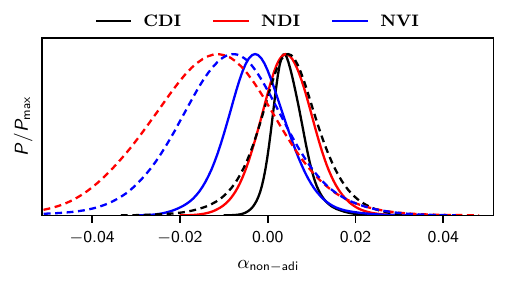}
\caption{Posterior probability density of the observable non-adiabatic fraction of 
the CMB temperature variance, assuming a generally correlated mixed adiabatic 
and isocurvature model.  These results used \Planck\ TT+lowE+lensing data (dashed lines) 
and TT,TE,EE+lowE+lensing data (solid lines). \label{fig:JVobsFraction} }
\end{figure}

\begin{figure*}
\includegraphics{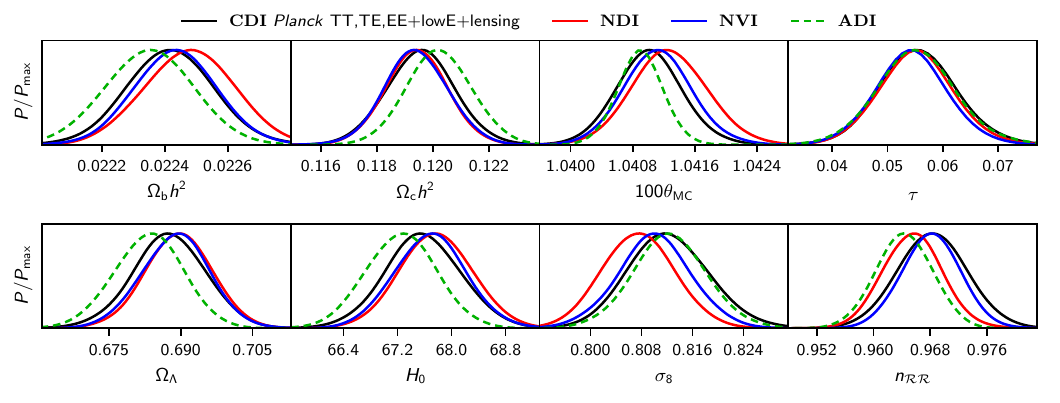}
\caption{Comparison of posterior probability density for the standard cosmological 
parameters in mixed adiabatic and isocurvature models (solid lines) to those in 
pure adiabatic \LCDM\ model (ADI, dashed green lines), using \Planck\ 
TT,TE,EE+lowE+lensing data. \label{fig:JVadiParams} }
\end{figure*}

\begin{figure*}
\includegraphics{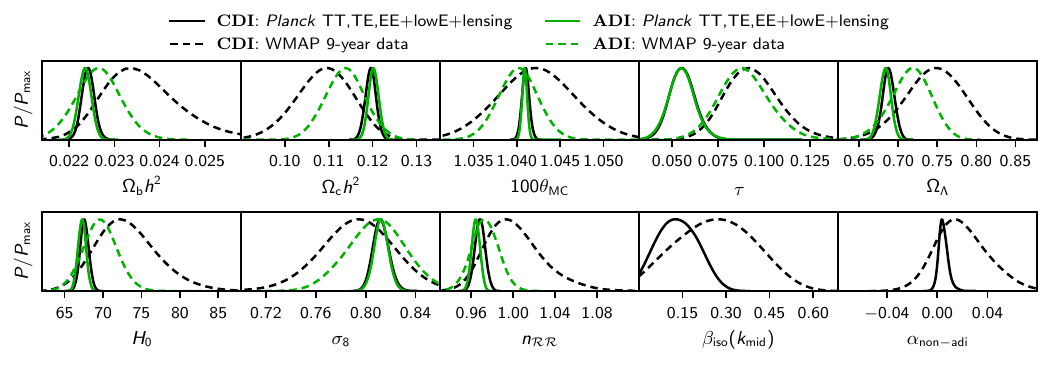}
\caption{Comparison of posterior probability density for selected cosmological 
parameters with \Planck\ data and pre-\Planck\ (i.e., WMAP 9-year) 
data. Black lines indicate the results obtained for the generally correlated 
mixed CDI+ADI model, and green lines for the pure adiabatic model.\label{fig:JVWMAPparams}}
\end{figure*}

\subsection{Results for generally correlated adiabatic and isocurvature modes}
\label{sec:genIsoc}

This subsection explores mixed adiabatic and isocurvature models where only one 
isocurvature mode at a time is considered. We consider
the CDI, NDI, and NVI modes using the \Planck\ 
2018 TT(,TE,EE)+lowE(+lensing) data. All five primordial perturbation power
amplitudes (of 
which three describe the isocurvature perturbations) are free parameters. It 
follows that $n_\mathcal{II}$ and $n_\mathcal{RR}$ are independent and $\cos\Delta$ 
varies between $-1$ and $+1$. The constraints for the primary perturbation parameters 
and the derived parameter $\mathcal{P}_{\mathcal{RI}}^{(2)}$ are shown in 
Fig.~\ref{fig:JVprimordialPowers}.

In all three cases the \Planck\ TT+lowE+lensing results are very similar to the 
previous results 
from the \Planck\ 2015 TT+lowP+lensing likelihood. 
As expected, the lower value of $\tau$ 
preferred by the 2018 (lowE) data is reflected in the adiabatic amplitudes 
$\mathcal{P}_{\mathcal{RR}}^{(1)}$ and $\mathcal{P}_{\mathcal{RR}}^{(2)}$. For CDI 
and NDI there is no significant shift in the constraints on isocurvature 
parameters, but we find slightly tighter constraints than in 2015. For NVI, a minor 
shift towards more negative correlations is observed (see the last two panels of 
Fig.~\ref{fig:JVprimordialPowers}c). As in 2015, adding the high-$\ell$ 
TE,EE data significantly tightens the constraints in all three cases.

When fitting the generally correlated three-isocurvature-parameter models,
the \Planck\ data are consistent with null detection, i.e., with the pure adiabatic model, 
$\left(\mathcal{P}_{\mathcal{II}}^{(1)}, \mathcal{P}_{\mathcal{II}}^{(2)}\right)=(0,0)$ and 
$\mathcal{P}_{\mathcal{RI}}^{(1)}=0$ (and $\mathcal{P}_{\mathcal{RI}}^{(2)}=0$). The 
natural logarithm of the ratio of model probabilities [i.e., the Bayes factor $\ln B = 
\ln(P_{\mathrm{ISO}}/P_{\mathrm{ADI}})$] is below $-10.9$, corresponding to odds of less 
than 1\,:\,$54\,000$ for all three (CDI, NDI, NVI) models. If there were an 
undetected subdominant isocurvature contribution to the primordial perturbations, a 
negative correlation between $\mathcal{R}$ and $\mathcal{I}$ would be favoured,
in particular for NDI and NVI (see the last two panels of 
Figs.~\ref{fig:JVprimordialPowers}a,b,c).  With our sign convention, this leads to a 
negative contribution to the Sachs-Wolfe effect and hence reduces the amplitude of the 
temperature angular power spectrum at low multipoles.

Figure~\ref{fig:JVprimordialFractions} updates the 2015 \Planck\ constraints on the derived 
primordial fractions and spectral indices. At large scales we find with \Planck\ 
TT,TE,EE+lowE+lensing that $\beta_\mathrm{iso}(k_{\mathrm{low}}) < 2.5\,\%$ for the CDI, 
$7.4\,\%$ for the NDI, and $6.8\,\%$ for the NVI model, all at 95\,\% CL. 
Figure~\ref{fig:JVobsFraction} shows the non-adiabatic fraction in the observed CMB 
temperature variance, defined as 
\begin{equation}
\alpha_\mathrm{non-adi} =1- \frac{(\Delta T)^2_\mathcal{RR}(\ell=2,2500)} {(\Delta T)^2_\mathrm{tot}(\ell=2,2500)}, \label{eq:FracDef}
\end{equation}
where
\begin{equation}
{(\Delta T)^2_\mathrm{X}(\ell=2,2500)}= \sum _{\ell =2}^{2500} (2\ell +1)C_{\mathrm{X}, \ell }^{TT}.
\end{equation}
The non-adiabatic fraction $|\alpha_\mathrm{non-adi}|$ is below 1.7\,\% with \Planck\ TT,TE,EE+lowE+lensing data for 
all three cases at 95\,\% CL.

Since the \Planck\ data do not allow a significant isocurvature contribution, the determination 
of standard cosmological parameters 
depends only very weakly on 
the assumed initial 
conditions, as seen in Fig.~\ref{fig:JVadiParams}. We place this result in 
historical perspective in Fig.~\ref{fig:JVWMAPparams} (and Table~\ref{tab:how_much_ic}) where the 
parameter
determinations
of the mixed CDI model and the 
pure adiabatic model 
are compared to the pre-\Planck\ constraints set by the WMAP 9-year 
data.\footnote{%
The pivot scales $k_\mathrm{low}$ and 
$k_\mathrm{high}$ used here
to parameterize the primary perturbation amplitudes 
are not optimal for WMAP, since the WMAP data extend only to $k\simeq k_\mathrm{mid}$. 
Nevertheless, an analysis tailored to WMAP
[see \citet{Savelainen:2013iwa}, who used $k_\mathrm{mid}$ as an upper pivot $k$]
gives a similar posterior range for $\alpha_\mathrm{non-adi}.$ 
}
\Planck\ has dramatically tightened the constraint on the adiabatic
spectral index.
Its value is now $8.4\sigma$ below unity (scale-invariance) 
in the pure ADI case. Allowing for generally correlated CDI reduces the significance of this 
detection only slightly, to $7\sigma$, whereas the
WMAP 9-year data were consistent with a blue tilt as large as 
$n_\mathcal{RR} = 1.06$ at 95\,\% CL. The non-adiabatic 
contribution to the CMB temperature variance 
is constrained (about zero) 5 times more tightly than by WMAP. 
Finally, the allowed range 
for the sound horizon angle has shrunk by a factor of $10$ in the CDI case, 
thanks to the \Planck\ data covering more acoustic peaks beyond 
the first three peaks detected by WMAP.

\subsection{Role of lensing parameter $A_\mathrm{L}$ and likelihood choices}
\label{sec:isocVsAL}

\begin{figure*}
\includegraphics{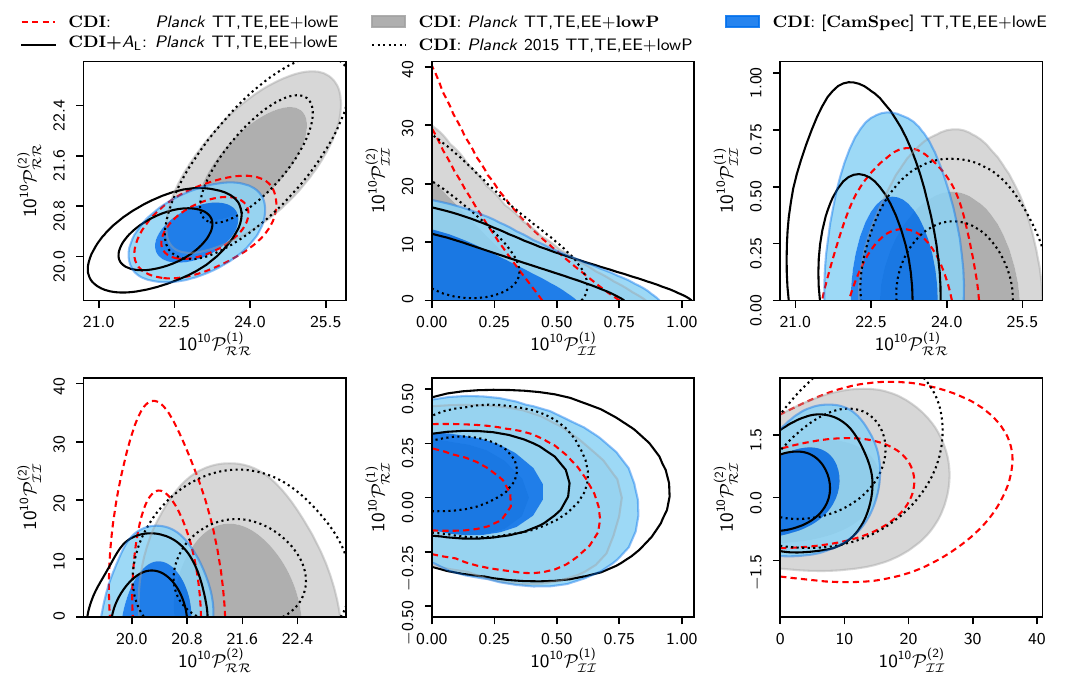}
\caption{
Comparison of the effect of \Planck\ 2018 likelihood choices and phenomenological 
lensing amplitude, $A_\mathrm{L}$, on the constraints on the generally correlated mixed 
adiabatic and CDI model. The reference case, indicated by the red dashed curves, is for 
the 2018 baseline \texttt{Plik} high-$\ell$ likelihood supplemented by the low-$\ell$ 
\texttt{Commander} TT likelihood and the low-$\ell$ \texttt{SimAll} EE likelihood. This 
combination is the same as in Fig.~\ref{fig:JVprimordialFractions}a (red curves), 
except now without the lensing likelihood, which to some extent hides the 
differences between other likelihoods and the effect of $A_\mathrm{L}$. Black solid 
contours show the results using the same likelihood as in the reference case, but now 
for the mixed adiabatic and CDI model when simultaneously allowing $A_\mathrm{L}$ to 
vary. The remaining two 
curves are for the mixed adiabatic and CDI model (with $A_\mathrm{L}=1$), but now 
changing the low-$\ell$ likelihood from \texttt{Commander} TT+\texttt{SimAll} EE 
to LFI 70\,GHz T,E,B (grey), or high-$\ell$ likelihood from \texttt{Plik} 
to \texttt{CamSpec} (blue). \label{fig:CDIandAlens}}
\end{figure*}

The small-scale primordial CDI amplitude is extremely sensitive to the
details of the high-$\ell$ temperature and polarization power spectra 
and to choices made in constructing the likelihoods. 
Therefore the general CDI model serves as a robustness test 
of the \Planck\ data and likelihoods. We now 
discuss a few curious aspects related to CMB lensing and likelihoods.

Lensing smooths the peaks of the CMB power spectra. 
This effect is taken 
into account in our theoretical predictions for the mixed adiabatic and isocurvature 
models by first calculating the total unlensed CMB spectra as a sum of adiabatic, 
isocurvature, and correlation $C_\ell$'s, and performing a similar summation for the lensing 
potential power spectrum \citep{Seljak:1995ve,Lewis:2006fu}.
The total lensing potential is then used to lens 
the total CMB spectra. 
Starting with the WMAP data, accounting for CMB lensing 
became necessary for calculating constraints
on isocurvature models, as
\cite{Valiviita:2012ub} showed that there 
is a strong degeneracy between the lensing effect and the CDI contribution in the 
generally correlated mixed models.
Fixing $n_\mathcal{II}=1$ or $n_\mathcal{II} = 
n_\mathcal{RR}$ (as is done in the next subsection) makes 
this degeneracy disappear. This is because in 
these models the CDI contribution modifies only the low-$\ell$ part of the angular power 
spectra. The transfer function mapping the primordial CDI mode to the $TT$ (and 
$EE$) angular power is suppressed by a factor $(k/k_\mathrm{eq})^{-2} \sim 
(\ell/\ell_\mathrm{eq})^{-2} $ relative to the adiabatic mode. Therefore, to be 
observable at high $\ell$, the CDI mode must be blue tilted ($n_\mathcal{II} > 1$). 
A blue-tilted CDI mode affects the total angular power spectra in a manner 
somewhat similar to lensing. Since the acoustic peaks of the CDI mode have the 
opposite phase compared to the adiabatic mode, a CDI admixture can ``smooth'' the peaks and dips of 
adiabatic acoustic oscillations.
The NDI mode does not have precisely the 
opposite phase and is not damped relative to the adiabatic mode (see figure~43 in 
\citetalias{planck2014-a24}). Thus we expect a weaker impact of
lensing on the \emph{primordial} NDI amplitude than in the CDI case. 
Therefore in this subsection we explore the general CDI model as an example.

\begin{figure}
\includegraphics[width=88mm]{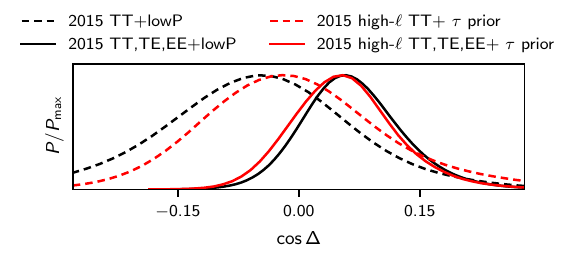}
\includegraphics[width=88mm]{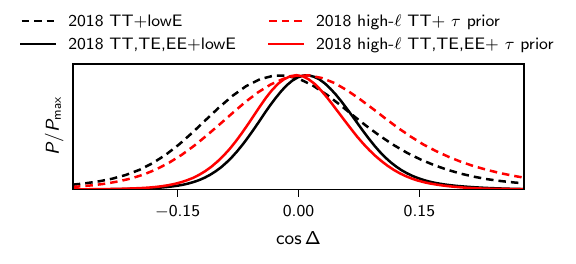}
\caption{Scale-independent primordial correlation fraction in the mixed 
adiabatic and CDI model. The black curves are with \Planck\ high-$\ell$ and 
low-$\ell$ data, while the red ones result from using only the high-$\ell$ 
\texttt{Plik} likelihood supplemented with a Gaussian prior on the optical depth. 
For the 2015 data (top panel) this prior was $\tau=0.078\pm0.019$, whereas 
for the 2018 case (bottom panel) we have adopted $\tau=0.055\pm0.007$ from 
the \Planck\ 2018 TT,TE,EE+lowE+lensing CDI chain. 
\label{fig:cosDelta4CDI}}
\end{figure}

Starting with the \Planck\ 2013 release,
the consistency of the smoothing effect with the adiabatic \LCDM\ model has 
been routinely tested by multiplying the lensing power spectrum by a phenomenological 
lensing consistency parameter, $A_\mathrm{L}$, prior to lensing the unlensed CMB spectra 
\citep{Calabrese:2008rt}. The expectation is that $A_\mathrm{L}=1$. 
However, the \Planck\ temperature and polarization data prefer a higher level of 
lensing-like smoothing 
($A_\mathrm{L}>1$) than expected in the adiabatic \LCDM\ model. 
In the 2018 release \citepalias{planck2016-l06} we have
\begin{eqnarray}
A_\mathrm{L} & = & 1.243 \pm 0.096\ \ \text{(68\,\% CL, \Planck\ TT+lowE)},\\
A_\mathrm{L} & = & 1.180 \pm 0.065\ \ \text{(68\,\% CL, \Planck\ TT,TE,EE+lowE).}
\end{eqnarray}
Adding the \Planck\ CMB lensing likelihood pulls these constraints towards $A_\mathrm{L}=1$
(see also Table~\ref{tab:CIP}). The measurement of $A_\mathrm{L}$ when TT,TE,EE data 
are included depends on the calibration of the polarization channels. This 
procedure and the details of the sky masks
differ between the \Planck\ baseline \texttt{Plik} TT,TE,EE and the 
alternative \Planck\ \texttt{CamSpec} TT,TE,EE likelihood, 
as discussed in \citetalias{planck2016-l05} and \citetalias{planck2016-l06}.  
The \Planck\ \texttt{CamSpec} TT,TE,EE likelihood prefers a smaller 
value of $A_\mathrm{L}$ than \texttt{Plik}, but still lying 
about $2\sigma$ above unity. 

Given the above motivation, we check the response
of the generally correlated 
CDI model to the various possible choices of likelihoods available in the \Planck\ 2018 release, and, on 
the other hand, we gauge how the baseline \texttt{Plik} likelihood reacts when allowing 
$A_\mathrm{L}$ to vary. For clarity, in Fig.~\ref{fig:CDIandAlens} we restrict the 
analysis to high-$\ell$ TT,TE,EE and low-$\ell$ TT,EE(,BB) data without the lensing 
reconstruction data, but in Table~\ref{tab:how_much_ic} we also report TT+lowE and 
TT+lowP results, and include \Planck\ lensing in some cases.

We notice a 
considerable variation in the constraints on the isocurvature power at high $k$, 
$\mathcal{P}_{\mathcal{II}}^{(2)}$, which corresponds to the high-$\ell$ region in the 
observed power spectra. 
In Fig.~\ref{fig:CDIandAlens} we can compare the red dashed reference contours (obtained with 
the \Planck\ baseline \texttt{Plik} TT,TE,EE+lowE likelihood) for the CDI model (where 
$A_\mathrm{L}=1$) with the solid black contours (obtained with the same data) for the 
CDI+$A_\mathrm{L}$ model, where $A_\mathrm{L}$ is allowed to vary. 
In many cases, adding an extra free parameter is expected to 
weaken the constraints on the other parameters, but in this case adding $A_\mathrm{L}$ 
tightens the 95\,\% CL constraint on $\mathcal{P}_{\mathcal{II}}^{(2)}$ by a factor of $2.5$ 
from $28.6\times10^{-10}$ to $11.4\times10^{-10}$. This is reflected in the derived 
primordial isocurvature fraction $\beta_\mathrm{iso}(k_{\mathrm{high}})$, whose upper 
bound changes from $0.58$ to $0.36$, according to Table~\ref{tab:how_much_ic}. 
Therefore, we conclude that, when $A_\mathrm{L}=1$, the CDI mode partially accounts for 
the extra lensing-like smoothing effect required by the \Planck\ TT(,TE,EE) data. Once 
we allow the lensing amplitude to vary, there is not much need for the CDI 
contribution at high $\ell$, which should be kept in mind when interpreting the results. 
In Table~\ref{tab:how_much_ic} we report four cases where $A_\mathrm{L}$ 
is allowed to vary. In all the other cases we have fixed $A_\mathrm{L}=1$. In these cases 
the constraints at high $k$ are ``conservative'', i.e., weaker than the \Planck\ data 
were expected to be capable of \citep{Finelli:2016cyd}, due to CDI (or NDI) partially 
fitting the lensing anomaly. Furthermore, again comparing the red dashed and black solid 
contours, we observe a slight weakening of the constraint for 
$\mathcal{P}_{\mathcal{II}}^{(1)}$ in the CDI+$A_\mathrm{L}$ model. This is due to the 
rigidity of the assumed power-law spectrum. When the high-$\ell$ data allow much 
less CDI, the low-$\ell$ (low-$k$) CDI amplitude can be larger without much affecting 
the middle-$\ell$ range of the CMB power spectra between the first and third 
acoustic peaks, which is the most sensitive region to departures from 
adiabaticity. (This is the same see-saw effect discussed in the end of 
Sect.~\ref{sec:beyondConsistency} in the case of tensor perturbations.) Finally, in the 
$\left(\mathcal{P}_{\mathcal{RR}}^{(1)},\,\mathcal{P}_{\mathcal{RR}}^{(2)}\right)$ panel we see a 
minor shift toward smaller amplitudes, which is an indication of the well-known degeneracy 
between $A_\mathrm{L}$ and the overall primordial perturbation amplitude.

From Table~\ref{tab:how_much_ic} it is obvious that adding the lensing data 
reduces the differences discussed above. This is again as expected, 
since the lensing data favour values of $A_\mathrm{L}$ only mildly above unity. For 
example, with \Planck\ TT,TE,EE+lowE+lensing we obtain 
$\beta_\mathrm{iso}(k_{\mathrm{high}}) < 0.49$ for CDI+$A_\mathrm{L}$ and $0.47$ for 
the CDI model.

We now proceed to a comparison of the likelihoods. The grey shaded contours in 
Fig.~\ref{fig:CDIandAlens} indicate the results for the same CDI model as the red dashed 
contours, but changing the low-$\ell$ likelihood from the combination \texttt{Commander} 
TT+\texttt{SimAll} EE to the LFI 70\,GHz pixel-based low T,E,B, which 
is by its methodology and construction very similar to the 2015 baseline low-$\ell$ 
likelihood (dotted contours). Indeed, this can be seen in the results: most of the 
isocurvature parameters follow more closely the 2015 results with this likelihood 
combination than with the 2018 baseline. This implies that when it comes to 
isocurvature, not much has changed in high-$\ell$ TT. 2018 lowP favours slightly smaller 
values of the optical depth $\tau$ than the 2015 version, hence the small shift towards 
smaller values of the adiabatic amplitude in the 
$\left(\mathcal{P}_{\mathcal{RR}}^{(1)},\,\mathcal{P}_{\mathcal{RR}}^{(2)}\right)$ panel. With 
respect to the red dashed contours, the grey contours prefer higher adiabatic amplitudes 
and have a long degeneracy line in the 
$\left(\mathcal{P}_{\mathcal{RR}}^{(1)},\,\mathcal{P}_{\mathcal{RR}}^{(2)}\right)$ plane. This is 
due to lowP having a higher central value and larger uncertainty on $\tau$.

Finally, the blue shaded contours in Fig.~\ref{fig:CDIandAlens} represent the results 
when using the \texttt{CamSpec} likelihood, to be compared to the red dashed contours 
obtained by the baseline \texttt{Plik} likelihood. All the other parameters shown are 
relatively stable against the high-$\ell$ likelihood, but 
$\mathcal{P}_{\mathcal{II}}^{(2)}$ stands out. \texttt{CamSpec} leads to an upper bound of 
$12.6\times10^{-10}$, whereas the baseline \texttt{Plik} result was 
$28.6\times10^{-10}$, or for $\beta_\mathrm{iso}(k_{\mathrm{high}})$ $0.38$ versus $0.58$ 
at 95\,\% CL, according to Table~\ref{tab:how_much_ic}. This difference is not surprising, 
given the different responses of these likelihoods to $A_\mathrm{L}$ in the adiabatic 
\LCDM+$A_\mathrm{L}$ model, and keeping in mind the $A_\mathrm{L}$--CDI degeneracy in the 
CDI model. However, 
this difference is not as 
concerning as it might appear at first sight: all the cases shown in 
Fig.~\ref{fig:CDIandAlens} are fully consistent with zero isocurvature. It is only the 
upper bound that varies, with the baseline \texttt{Plik} likelihood and CDI model with 
$A_\mathrm{L}=1$ leading to the most conservative (i.e., weakest or safest) upper 
bounds.

For CDI the 2015 release \Planck\ high-$\ell$ TT data favoured a negative correlation 
fraction but the preliminary high-$\ell$ TT,TE,EE data favoured a slightly positive correlation.
This was confirmed using only the high-$\ell$ \texttt{Plik} likelihood (and a prior 
on $\tau$), as shown by the red curves in the top panel of Fig.~\ref{fig:cosDelta4CDI}. 
Including the low-$\ell$ data (black curves) did not significantly alter this 
tension between TT and TT,TE,EE results. In the present 2018 \Planck\ release this tension 
has disappeared. Both high-$\ell$ TT and TT,TE,EE data lead to a 
correlation fraction posterior peaking at zero, as demonstrated in the bottom panel of 
Fig.~\ref{fig:cosDelta4CDI}. Including the low-$\ell$ TT data (black dashed 
curve) still shifts the posterior slightly towards negative values, due to the low $TT$ 
power at low multipoles in the data.


\subsection{Specific CDI models}

In this subsection we constrain CDI models with only one or two isocurvature parameters. 
The two-parameter cases were not studied in the 2013 and 2015 \Planck\ releases.

First we fix $n_{\cal I \cal I}$ to unity and assume no correlation between the CDI and 
adiabatic modes (``axion''), or we fix $n_{\cal I \cal I} = n_{\cal R \cal R}$ and assume full 
(anti)correlation between the CDI and adiabatic modes (``curvaton I/II''). These models are 
less sensitive to any residual systematic effects in the high-$\ell$ data (such as the 
determination of polarization efficiencies or foreground modeling) than the generally 
correlated models, since CDI now modifies the angular power spectra insignificantly at $\ell 
\gtrsim 200$ (see figure~43 in \citetalias{planck2014-a24}). As seen in the middle section of 
Table~\ref{tab:how_much_ic}, the Bayesian evidence values for the one-parameter extensions of the 
adiabatic \LCDM\ model are higher than for the three-parameter extensions, but all Bayes 
factors fall below $-5$.  None of the one-parameter extensions
improve $\chi^2$ over the adiabatic \LCDM\ model.
The two-parameter extensions in the bottom section of 
Table~\ref{tab:how_much_ic} are even more strongly disfavoured, except for the uncorrelated case 
with free $n_{\cal I \cal I}$ (``axion II''), which is actually the
only model that improves the best-fit $\chi^2$ by slightly more than the number
of extra parameters.

\subsubsection{Uncorrelated ADI+CDI (``axion I'')}
\label{sec:axionI}

Particularly insensitive to any $\ell \gtrsim 30$ data is the ``axion I'' case, since the CDI 
transfer function has a $(k/k_\mathrm{eq})^{-2}$ suppression and there is no correlation 
component whose amplitude would be higher than that of the isocurvature alone and hence would modify 
the adiabatic spectrum. The ``axion I'' case is achieved in our parameterization by setting 
$\mathcal{P}_{\cal R\cal I}=0$ and $\mathcal{P}_{\cal I\cal I}^{(2)} = \mathcal{P}_{\cal 
I\cal I}^{(1)}$. Thus the only varied isocurvature parameter is $\mathcal{P}_{\cal I\cal 
I}^{(1)}$. This uncorrelated case with $n_{\cal I \cal I}=1$ is a good approximation for many 
multi-field inflationary models where the slow-roll parameter (in the isocurvature field 
perturbation direction) $\eta_{ss}$ is negligible and the background trajectory in field 
space is straight between Hubble radius exit and the end of inflation.  The predictions for 
the spectral indices (to first order in the slow-roll parameters) are $ n_\mathcal{RR} = 
1-6\epsilon + 2\eta_{\sigma\sigma}$ and $n_\mathcal{II} = 1-2\epsilon + 2\eta_{ss}$, where 
$\epsilon \ge 0$ and $\eta_{\sigma\sigma}$ is the second slow-roll parameter in the 
``adiabatic'' direction (i.e., along the trajectory) in the field space. (An exact match with 
our model would require $\eta_{ss} = \epsilon$.) The axion model
[see, e.g., a recent review by \citet{Marsh:2015xka} and references therein],
which was originally proposed to solve the strong CP 
problem and provides a dark matter candidate, can produce this type of isocurvature modes 
with $n_{\cal I \cal I} \simeq 1$ under the following assumptions 
(\citetalias{planck2013-p17,planck2014-a24}): the Peccei-Quinn symmetry should be broken 
before inflation; it should not be restored by quantum fluctuations of the inflaton nor by 
thermal fluctuations when the Universe reheats; and axions produced through the misalignment 
angle should form a significant fraction of the dark matter.

Table~\ref{tab:how_much_ic} indicates a slight tightening of the ``axion I'' constraints using 
TT+lowE+lensing with respect to 2015 TT+lowP+lensing. This is due to the change of the 
baseline low-$\ell$ data from the 2015 LFI 70-GHz pixel-based T,E,B to the 2018 
combination of \texttt{Commander} TT and \texttt{SimAll} EE, which in the generally 
correlated cases also gave tighter constraints at low $k$. As expected, the addition of 
high-$\ell$ TE,EE data only marginally improves the constraints, since the standard 
(non-isocurvature) parameters are better constrained now. 
For $n_\mathcal{II}=1$ uncorrelated CDI, we obtain
\begin{equation}
\left.\begin{aligned}
\beta_\mathrm{iso}(k_{\mathrm{mid}}) & < 0.038\\
0 \le \alpha_{\mathrm{non-adi}} & < 1.55\,\%
\end{aligned}\ \right\}\ \ 
\mbox{\text{\parbox{4.2cm}{\begin{flushleft}
   (95\,\% CL, \Planck\ TT,TE,EE\\
\ +lowE+lensing).
\end{flushleft}}}}
\end{equation}
Using equation~(73) of \citetalias{planck2013-p17}, 
we convert the constraint on the primordial isocurvature 
fraction to a bound on the inflationary energy scale.
If all the dark matter 
is in axions, the above $\beta_\mathrm{iso}(k_{\mathrm{mid}})$ constraint 
corresponds to the same limit we quoted in 2015, that is,
\begin{equation}
H_\mathrm{inf} < 0.86 \times 10^7\,\mathrm{GeV} \left( \frac{f_a}{10^{11}\,\mathrm{GeV}} 
\right)^{0.408} \quad \mbox{ (95\,\% CL)}\,,
\end{equation}
where $H_\mathrm{inf}$ is the expansion rate at Hubble radius exit of the scale corresponding 
to $k_\mathrm{mid}$ and $f_a$ is the Peccei-Quinn symmetry-breaking energy scale.

\subsubsection{Fully (anti)correlated ADI+CDI (``curvatonI/II'')}
\label{sec:curvatonIandII}

If
$n_{\cal I \cal I} = n_{\cal R \cal R}$, the low-$\ell$ data are maximally 
sensitive to the fully correlated isocurvature perturbations.
In this case the 
correlation component is a geometric average of the adiabatic and isocurvature components,
and 
hence much larger than the isocurvature component alone. We achieve this case in our 
parameterization by setting $\mathcal{P}_{\cal I\cal I}^{(2)} = \left(\mathcal{P}_{\cal R\cal 
R}^{(2)} / \mathcal{P}_{\cal R \cal R}^{(1)}\right) \mathcal{P}_{\cal I\cal I}^{(1)}$ and 
$\mathcal{P}_{\cal R\cal I}^{(1)} = \pm \left( \mathcal{P}_{\cal R\cal R}^{(1)} 
\mathcal{P}_{\cal I\cal I}^{(1)} \right)^{1/2}$, i.e., $\cos\Delta=\pm 1$. The only 
isocurvature parameter to be varied is again $\mathcal{P}_{\cal I\cal I}^{(1)}$. Since 
$n_{\cal I \cal I} = n_{\cal R \cal R}$, the derived isocurvature fraction 
$\beta_\mathrm{iso}$ is independent of $k$. A physically motivated example of this type of 
model is the simplest curvaton model, where a light scalar field $\chi$ that is subdominant 
(and hence irrelevant for the inflationary dynamics) starts to oscillate at the bottom of its 
potential after the end of inflation, causing its average energy density to evolve like 
non-relativistic matter. Once fully (or almost fully) dominating the energy density of the 
Universe, this curvaton field decays either to CDM or to other species 
\citep{Mollerach:1989hu,Linde:1996gt,Enqvist:2001zp,Moroi:2001ct,Lyth:2001nq,Bartolo:2002vf,Lyth:2002my}. 
The amount of isocurvature and non-Gaussianity present after curvaton decay depends on the 
``curvaton decay fraction,'' $r_D = 3\bar\rho_\chi / (3\bar\rho_\chi + 
4\bar\rho_{\mathrm{radiation}})$, evaluated at curvaton decay time. Under a number of (very) 
restrictive assumptions discussed in \citetalias{planck2014-a24}, the curvaton model can lead 
to fully (anti)correlated CDI (or BDI) and adiabatic perturbations.

Not surprisingly, both in the fully correlated and anticorrelated cases, the constraint 
on $\beta_\mathrm{iso}$ is much (about 40 times) stronger than in the uncorrelated case. At 95\,\% CL, 
\Planck\ TT,TE,EE+lowE+lensing leads to
\begin{align}
\beta_\mathrm{iso} &< 0.00095\ \ \text{(Fully correlated, ``curvaton I'')},\\ 
\beta_\mathrm{iso} &< 0.00107\ \ \text{(Fully anticorrelated, ``curvaton II''),}
\end{align}
both rounded to $0.001$ in Table~\ref{tab:how_much_ic}. As in 2015, the TT data favour anticorrelation, 
due to the low power in the low-$\ell$ temperature compared to the expectation of the adiabatic \LCDM\ 
model. But when the TE,EE data (which do not particularly favour negative correlation) are added,
a very tight (one part per thousand) constraint on the primordial isocurvature 
fraction results.

Fully correlated perturbations are obtained, e.g., in case~4 described in 
\citet{Gordon:2002gv}. Many models giving anticorrelation produce too large 
an isocurvature fraction to be consistent with the above limit, but case~9 of 
\citet{Gordon:2002gv} survives. 
After the curvaton decay, 
the primordial isocurvature fraction in these models 
will be $\beta_{\mathrm{iso}} \simeq 9(1-\tilde{r})^2 / [r_D^2 + 
9(1-\tilde{r})^2]$, where $\tilde{r} = r_D$ for the fully correlated CDI case and $\tilde{r} 
= r_D/R_\mathrm{c} \ge 1$ for the fully anticorrelated CDI case, 
and $R_\mathrm{c} = \bar\rho_\mathrm{c} / (\bar\rho_\mathrm{c}+\bar\rho_\mathrm{b})$
is the CDM fraction of the total non-relativistic matter.

On the other hand, the nonlinearity parameter describing non-Gaussianity is 
\citep{Sasaki:2006kq}
\begin{equation}
f_{\mathrm{NL}}^\mathrm{local}=\left(1+\Delta_s^2\right)\frac{5}{4 r_D} - 
\frac{5}{3} - \frac{5 r_D}{6}, \label{eq:fnlcurvaton}
\end{equation}
where $\Delta_s^2 = \langle \delta\chi^2 \rangle_s / \bar\chi^2$ is the small-scale variance 
of the curvaton perturbations, or the ratio of the energy density carried by the curvaton 
particles to the energy density of the curvaton field (if there is significant production of 
curvaton particles). The parameter $f_{\mathrm{NL}}^\mathrm{local}$ cannot be smaller than $-5/4$, which is 
obtained when $r_D = 1$ and $\Delta_s^2 = 0$, as implicitly assumed, for example, in 
\citet{Bartolo:2004ty,Bartolo:2003jx}. The above $\beta_\mathrm{iso}$ limits correspond to 
the following $r_D$ and $f_{\mathrm{NL}}^\mathrm{local}$ constraints (assuming $\Delta_s^2 = 
0$):\footnote{%
The quoted precision of these constraints does not reflect the precision of the 
sampling of the likelihood surface, but we report several digits here since the constraints are so 
tight that rounding to, e.g., two significant digits would give empty or almost empty 
ranges.
}
\begin{align}
&0.98982 < r_D  \le 1  &\text{(``curvaton I'')\phantom{I}}\nonumber\\
&\quad \Rightarrow -1.2500  \le  f_{\mathrm{NL}}^\mathrm{local}  < -1.2287,\\
&0.84347 \le  r_D  <  0.85129 &\text{(``curvaton II'')}\label{eq:rDcurvatonII}\nonumber\\
&\quad \Rightarrow -0.9077  <  f_{\mathrm{NL}}^\mathrm{local}  \le  -0.8876.
\end{align}
Even with the maximal allowed isocurvature fraction,
the local non-Gaussianity in the curvaton model 
is well within the observational \Planck\ limits presented in 
\citet{planck2016-l09}. The residual isocurvature peturbations in the two studied curvaton 
models 
set much tighter constraints on the curvaton 
decay fraction than do constraints on the observed (consistent with zero) non-Gaussianity.

\subsubsection{Uncorrelated ADI+CDI with free $n_{\cal I \cal I}$ (``axion II'')}
\label{sec:axionII}

Axion models do not necessarily produce nearly scale-invariant isocurvature 
perturbations. In particular, even highly blue-tilted spectra (in the observable CMB range) 
are possible. For example, \citet{Kasuya:2009up} construct a model with $n_\mathcal{II} = 2$--$4$. 
This motivates studying a two-isocurvature-parameter model, where adiabatic and isocurvature 
modes are uncorrelated, but the isocurvature fraction and spectral index are free to vary. In our 
parameterization this is achieved by setting $\mathcal{P}_{\cal R\cal I}^{(1)} = 0$, and 
varying $\mathcal{P}_{\cal I\cal I}^{(1)} $ and $\mathcal{P}_{\cal I\cal I}^{(2)}$ 
independently. The results for this model are presented in the first two rows of the third section of 
Table~\ref{tab:how_much_ic}. The low-$\ell$ temperature data do not favour any extra 
contribution beyond the (already too high) abiabatic contribution,
whereas the fit to the 
high-$\ell$ temperature and polarization data can be 
improved 
slightly 
by the ``smoothing'' 
caused by the CDI mode. This leads to a very blue isocurvature spectrum. \Planck\ 
TT,TE,EE+lowE+lensing gives at 95\,\% CL $1.55 < n_\mathcal{II} < 3.67$, consistent with the 
recent findings of \citet{Chung:2017uzc}. Even the very 
large upper bound $\beta_\mathrm{iso}(k_\mathrm{high}) < 77\,\%$ corresponds to 
a contribution of less than order $1\%$ to the observable CMB $TT$ (or $EE$) power spectra at $\ell 
\simeq 1400$. The uncertainty in the \Planck\ $TT$ spectrum at these high multipoles is 
$\Delta\mathcal{D}_\ell^{TT} \sim 10\,\muK^2$ and the actual spectrum is 
$\mathcal{D}_\ell^{TT} \sim 1000\,\muK^2$. 
Thus the allowed CDI contribution is only of the same 
$1\,\%$ order as the observable uncertainty.
Consequently the  non-adiabatic contribution to the observed CMB temperature variance,
$\alpha_{\rm non-adi}$, is also vanishingly small,
between $7\times10^{-4}$ and $7\times10^{-3}$.

\subsubsection{Arbitrarily correlated ADI+CDI with $n_{\cal I \cal I}=n_{\cal R \cal R}$ 
(``curvaton III'')}
\label{sec:curvatonIII}

Apart from the extremes of $\pm100\,\%$ correlation, some curvaton models predict
an arbitrary degree of correlation. The generic feature of most curvaton models is
that the isocurvature and adiabatic spectral indices are equal.  This is because 
both perturbations typically arise from the same source. In the next-to-simplest 
models, the correlation fraction can be written as $\cos\Delta = \sqrt{\lambda / 
(1+\lambda)}$, where $\lambda = (8/9)r_D^2 \epsilon_\ast (M_\mathrm{Pl} / \bar\chi_\ast)^2$. 
Therefore, the model is fully correlated only if $\lambda \gg 1$, in which case the results of 
``curvaton I'' apply. If the slow-roll parameter $\epsilon_\ast$ is very close to zero or the 
curvaton field value $\bar\chi_\ast$ is large compared to the Planck mass, this model leads 
to almost uncorrelated perturbations and the constraints are well approximated by ``axion 
I.'' Any other case leads to an arbitrary degree of positive correlation between the CDI and 
adiabatic modes.

Modulated reheating with thermal or non-thermal production of gravitinos can lead to positive 
or negative correlation, respectively \citep{Takahashi:2009cx}. While the correlation could 
in principle be arbitrarily large, the observational constrains on $\beta_\mathrm{iso}$ 
favour only small correlations.

Arbitrarily correlated ADI+CDI with $n_{\cal I \cal I}=n_{\cal R \cal R}$ is also a good 
approximation for those two-field (or multi-field) slow-roll models [e.g., double quadratic 
inflation \citep{Langlois:1999dw,Beltran:2005gr}] where the trajectory in field space is 
curved between the Hubble radius exit of perturbations during inflation and the end of 
inflation. The fraction of isocurvature perturbations converted to adiabatic depends on 
how the trajectory is curved and this part of the adiabatic perturbations will be fully 
(anti)correlated with the isocurvature modes, whereas the adiabatic perturbations 
already present at Hubble radius exit are uncorrelated with isocurvature modes to 
first order in the slow-roll parameters, and only slightly correlated to second order.
[See, e.g., \citet{Gordon:2000hv}, \citet{Amendola:2001ni}, \citet{vanTent:2003mn}, 
and \citet{Byrnes:2006fr}.] 
The result is a non-zero correlation between isocurvature and total adiabatic perturbations.
The spectral indices of both components are typically $1 - \mathcal{O}(\text{slow-roll 
parameters})$, which is well approximated by $n_{\cal I \cal I}=n_{\cal R \cal R}$ since the 
data indicate $n_{\cal R \cal R} \simeq 0.965$.

As expected, the \Planck\ data favour negative correlations, since these $n_{\cal I \cal 
I}=n_{\cal R \cal R}$ models modify only the low-$\ell$ part of the CMB spectra, where $TT$ 
power is lower than predicted by the adiabatic \LCDM\ model. With TT,TE,EE+lowE+lensing we 
find, at 95\,\% CL, $\beta_\mathrm{iso} < 0.039$ and $-0.41 < \cos\Delta < 0.31$.

\subsubsection{Fully (anti)correlated ADI+CDI with free $n_{\cal I \cal I}$}
\label{sec:FCandACarbCor}

The remaining two-parameter CDI extensions of the adiabatic \LCDM\ model are those where the 
perturbations are fully (anti)correlated, as in the simplest curvaton models, but the 
isocurvature spectral index is not fixed to the adiabatic one. In this case the free 
isocurvature parameters are $\mathcal{P}_{\cal I\cal I}^{(1)} $ and $\mathcal{P}_{\cal I\cal I}^{(2)}$, 
while $\mathcal{P}_{\cal R\cal I}^{(1)} = \pm \left( \mathcal{P}_{\cal R\cal R}^{(1)} 
\mathcal{P}_{\cal I\cal I}^{(1)} \right)^{1/2}$. These models are somewhat difficult to 
motivate, since full (anti)correlation typically implies that the curvature and 
isocurvature perturbations have their origin in (the decay products of) the same field. Then 
one would expect equal spectral indices, as in the curvaton model. The conversion 
of isocurvature perturbations to adiabatic ones (e.g., between Hubble radius exit and the 
end of inflation, or by curvaton-type decay, or by reheating/thermalization) should be 
scale dependent in order to obtain $n_{\cal I \cal I} \neq n_{\cal R \cal R}$. Slow-roll 
two-field inflation leads to an exact match, $n_{\cal I \cal I} = n_{\cal R \cal R}$, in the 
case where $\cos^2\Delta = 1$. [See, e.g., 
\citet{Byrnes:2006fr}]. Nevertheless, for completeness 
we report constraints on these phenomenological models in the last four rows of 
Table~\ref{tab:how_much_ic}.  Since the low-$\ell$ TT data favour negative correlation, a 
larger isocurvature fraction is allowed in the fully anticorrelated case at low $k$. This 
leads to scale-invariant isocurvature perturbations being in the favoured region of parameter 
space, namely $-0.28 < n_\mathcal{II} < 1.86$ with TT,TE,EE+lowE+lensing at 95\,\% CL. In 
contrast, in the fully correlated case the low-$\ell$ TT data disfavour any isocurvature 
contribution, and hence prefer a blue spectrum, with $1.37 < n_\mathcal{II} < 3.65$.


\subsection{Compensated BDI--CDI  mode}
\label{sec:CIP}

\begin{figure}
\includegraphics[width=88mm]{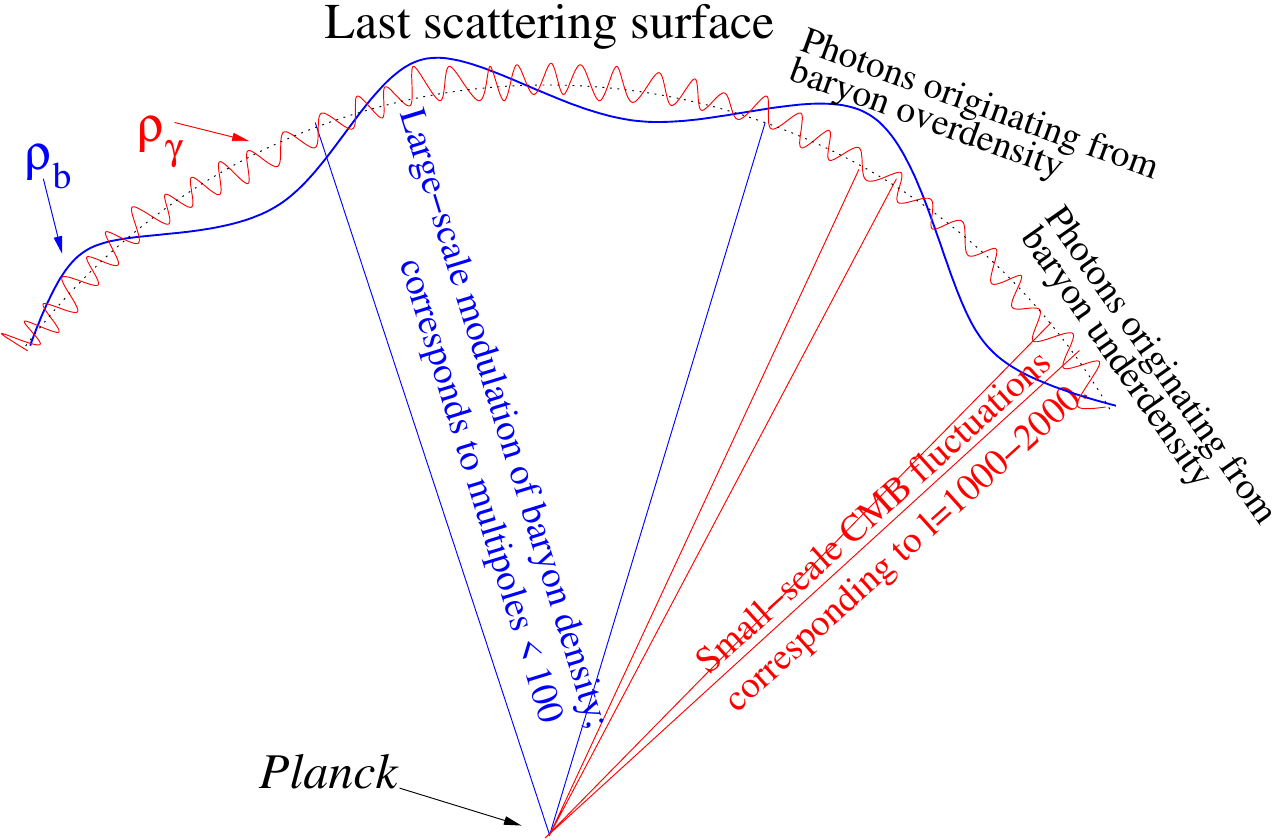}
\caption{Illustration of how the large-scale modulation of the baryon 
density by CIP gets converted into a small-scale ``smoothing'' effect of 
the temperature and polarization anisotropies.\label{fig:CIPschematic}}
\end{figure}

This subsection presents constraints on uncorrelated adiabatic and scale-invariant CIP 
modes and discusses the strong degeneracy between the phenomenological lensing parameter 
$A_\mathrm{L}$ and the CIP amplitude \citep{Valiviita:2017fbx}.  Assuming that there are 
no NVI or NDI perturbations, the total matter density isocurvature perturbation 
$\mathcal{I}_\mathrm{MDI}$, given by Eq.~(\ref{eq:MDIperturbation}), vanishes if
\begin{equation}
   \mathcal{I}_\mathrm{CDI} = -\frac{\Omega_\mathrm{b}}{\Omega_\mathrm{c}} \mathcal{I}_\mathrm{BDI}.
\label{eq:CIP1}
\end{equation}
This mode, where the anticorrelated CDI and BDI perturbations cancel even though their individual 
amplitudes can be large, is called a compensated baryon and cold dark matter 
isocurvature mode. The CIP mode does not leave a linear-order isocurvature signal in the 
CMB or matter power spectra \citep{Gordon:2002gv}, although it modifies the trispectrum 
\citep{Grin:2011tf,Grin:2013uya}. However, at the next order there is
a smoothing effect on the high-$\ell$ $TT$, $TE$, and $EE$ spectra. 
A formal derivation can be found, for example, in \citet{Smith:2017ndr}.
Here we summarize the heuristic arguments of \citet{Munoz:2015fdv}.

\begin{figure}
\centering 
\includegraphics[width=88mm]{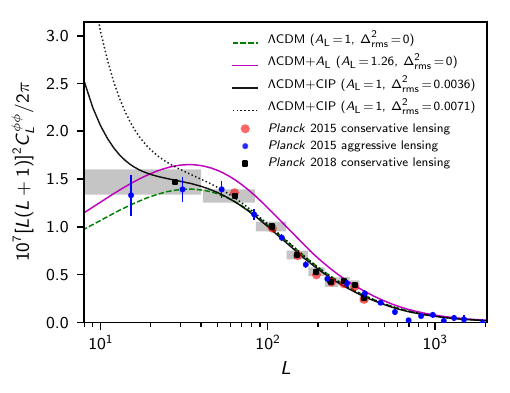}
\caption{Conservative \Planck\ 2015 lensing data (red points), 
aggressive \Planck\ 2015 lensing data (blue points with error bars), 
and conservative \Planck\ 2018 lensing data (black squares in grey boxes), 
along with the best-fit models to the \Planck\ data: 
the best-fit adiabatic \LCDM\ model to 2018 TT+lowE (green dashed line); 
the best-fit \LCDM+$A_\mathrm{L}$ model to 2018 TT+lowE (magenta solid line,
$A_\mathrm{L}=1.26$); and the best-fit \LCDM+CIP model to 2018 TT,TE,EE+lowE 
and conservative lensing data (black solid line, $\dms=0.0036$) and
to 2015 TT,TE,EE+lowP and conservative lensing data  (black dotted line, $\dms=0.0071$).
As CIP modifies only the very low-$L$ 
part of the lensing power spectrum, the conservative 2015 lensing data 
($40 \le L \le 400$) are insensitive to CIP even when $\dms=0.0071$.
On the other hand, the first two 
data points of the 2015 aggressive lensing data disfavour the large 
CIP amplitude \citep{Smith:2017ndr}, 
which gives a very good fit to all the other data. \Planck\ 2018 
conservative lensing data cover the range $8 \le L \le 400$ and 
consequently disfavour CIP variances  
$\dms \gtrsim 0.004$.\label{fig:conservativeVSaggressive}}
\end{figure}
\begin{figure}
\flushright{\includegraphics[width=88mm]{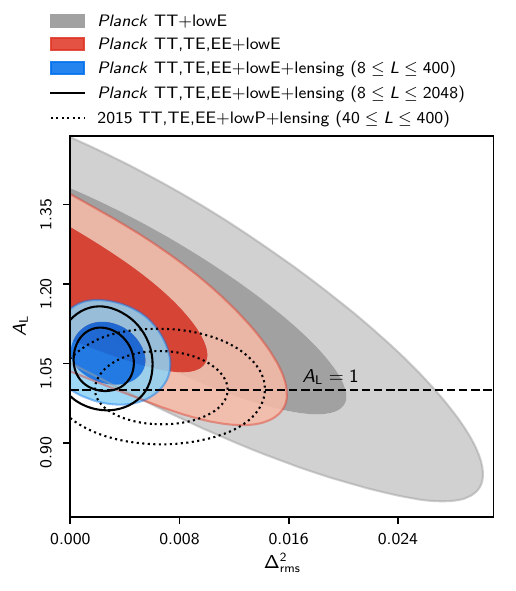}\\
\vspace{-1.2cm}
\includegraphics[width=83mm]{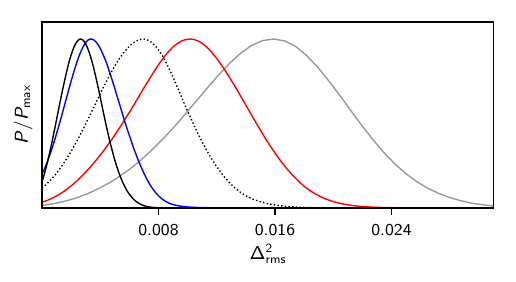}}
\caption{Degeneracy between $A_\mathrm{L}$ and $\dms$ in the \LCDM+$A_\mathrm{L}$+CIP 
  model (top panel) and constraints on $\dms$ in 
the \LCDM+CIP model, where  $A_\mathrm{L}=1$ (bottom panel).\label{fig:CIP2d}}
\end{figure}

On scales larger than the sound horizon, condition (\ref{eq:CIP1}) is preserved until 
last scattering and can be written as
\begin{equation}
\delta\rho_\mathrm{c}(\vec{x}) \simeq -\delta\rho_\mathrm{b}(\vec{x})\,. \label{eq:CIP2}
\end{equation}
Consequently, CIP can be described as a large scale modulation of the 
baryon and CDM density \citep{Munoz:2015fdv,Heinrich:2016gqe,Valiviita:2017fbx}, with
\begin{equation}
\label{eq:CIP3} \Omega_\mathrm{b}(\vec{\hat n})  =  
[1+\Delta(\vec{\hat n})]\bar\Omega_\mathrm{b}\,,\quad\quad\quad
 \Omega_\mathrm{c}(\vec{\hat n})  =  \bar\Omega_\mathrm{c} -\Delta(\vec{\hat n})\bar\Omega_\mathrm{b}\,.
\end{equation}
Here the overbar denotes an average over the whole sky and $\Delta(\vec{\hat n})$ a small 
perturbation about this average in the direction $\vec{\hat n}$, as illustrated in 
Fig.~\ref{fig:CIPschematic}.  In patches of sky where the CMB photons originate from 
baryon-overdense regions, the odd acoustic peaks at high-$\ell$ are more 
pronounced relative to the even peaks compared to the patches where the photons originate from 
baryon-underdense regions. Averaging over the sky leads to a lensing-like smoothing 
of the high-$\ell$ peaks.

A convenient measure of CIP is the variance 
$\dms \equiv \langle |\Delta(\hat n)|^2 \rangle 
\simeq \mathcal{P}_{\mathcal{I}_\mathrm{BDI}\mathcal{I}_\mathrm{BDI}}$. 
If $\Delta$ is a Gaussian random variable, the observed angular power 
of $TT$, $TE$, or $EE$ will be
\begin{align}
&C_\ell^\mathrm{obs}(\bar\Omega_\mathrm{b},\bar\Omega_\mathrm{c},
\tau,H_0,n_\mathrm{s},A_\mathrm{s})\\
  &= \frac{1}{\sqrt{2\pi\Delta^2_\mathrm{rms}}} \int  
\!\!C_\ell\big(\Omega_\mathrm{b}(\Delta),
\Omega_\mathrm{c}(\Delta),\tau,H_0,n_\mathrm{s},A_\mathrm{s}\big) \, 
e^{-\Delta^2/(2\Delta^2_\mathrm{rms})} d\Delta\,,\nonumber
\end{align}
where $\Omega_\mathrm{b}(\Delta) =  (1+\Delta)\bar\Omega_\mathrm{b}$ and 
$\Omega_\mathrm{c}(\Delta) = \bar\Omega_\mathrm{c} -\bar\Omega_\mathrm{b}\Delta$.  
For brevity, we will denote the power spectrum in the integrand by 
$C_\ell|_{\Delta=\delta}$. For each $\delta$ it can be calculated by 
assuming adiabatic initial conditions. Approximating the integrand by the first 
three terms of its Taylor series about $\Delta=0$, we end up with
\begin{equation}
C_\ell^\mathrm{obs} \simeq C_\ell|_{\Delta=0} + \frac{1}{2} 
\left.\frac{d^2 C_\ell}{d\Delta^2}\right|_{\Delta=0}\Delta^2_\mathrm{rms}\,. \label{eq:CIPobs}
\end{equation}

In the following we describe parameter scans where we vary the six standard (adiabatic) 
\LCDM\ parameters, the \Planck\ nuisance parameters, and the CIP variance $\dms$, 
calling this one-parameter extension of the \LCDM\ model the ``\LCDM+CIP'' model. We evaluate 
the right-hand side of  Eq.~(\ref{eq:CIPobs}) at each point in parameter space using a finite-difference 
approximation for the second derivative:
\begin{equation}
\left.\frac{1}{2} \frac{d^2 C_\ell}{d\Delta^2}\right|_{\Delta=0}\Delta^2_\mathrm{rms}
   \simeq  \frac{\textstyle{\frac{1}{2}} C_\ell|_{\Delta = \delta} -  C_\ell|_{\Delta = 0} + \textstyle{\frac{1}{2}} C_\ell|_{\Delta = -\delta}}{\delta^2}\,, \label{eq:finintedif}
\end{equation}
where $\delta$ should be ``sufficiently small.'' 
In practice, good numerical accuracy is achieved if $\delta$ is of order 
$\sqrt{\dms}$. So at each point in our \texttt{MultiNest} 
scan we set $\delta=\sqrt{\dms}$ for the point currently under evaluation, 
and thus the result of Eq.~(\ref{eq:CIPobs}) simplifies to
\begin{equation}
C_\ell^\mathrm{obs} \simeq  \frac{C_\ell\big|_{\Delta = \sqrt{\dms}} + C_\ell\big|_{\Delta = -\sqrt{\dms}}}{2}\,. \label{eq:CIPobs2}
\end{equation}
With this method each angular power spectra evaluation takes twice as long as for
the pure adiabatic case since the spectra are now an average of two spectra, resulting 
from different values of $\Omega_\mathrm{b}$ and $\Omega_\mathrm{c}$.\footnote{Actually, 
we call \texttt{CAMB} three times (i.e., also with $\bar\Omega_\mathrm{b}$ and 
$\bar\Omega_\mathrm{c}$, or $\Delta=0$), in order to obtain some auxiliary parameters such 
as $\sigma_8$ correctly, although we do not report them here.}

Unlike the high-$\ell$ $TT$, $TE$, and $EE$ spectra, the high-$L$ lensing potential 
power spectrum is virtually unaffected by CIP. Instead, CIP modifies the low 
multipoles of $[L(L+1)]^2C_L^{\phi\phi}/(2\pi)$ by, approximately, adding a term 
$\dms\times(L/0.053)^{-2}$.
For details, see table~II in \citet{Smith:2017ndr}. As illustrated in 
Fig.~\ref{fig:conservativeVSaggressive}, when using the \Planck\ 2015 conservative 
lensing data ($40\le L \le 400$) this term does not affect the results. In contrast, the 
\Planck\ 2018 conservative lensing data also contain the range $8\le L < 40$ and thus 
CIP variances $\dms \gtrsim 0.004$ fit the first data point of the 2018 lensing 
power spectrum ($8 \le L \le 400$) worse than in \LCDM.  However, even in this case the joint fit of 
the \LCDM+CIP model to the TT, TE, EE, and lensing data is better than that of the 
\LCDM\ model, the improvement being of the same order as for the \LCDM+$A_\mathrm{L}$ 
model.

\begin{table}
\begin{center}
\footnotesize
\setlength{\tabcolsep}{0.9mm}
\begin{tabular}{lrrrrlr}
\hline 
\hline
\noalign{\vskip 2pt}
& \multicolumn{2}{c}{Constraints} & \multicolumn{3}{c}{Best fit} &  \cr
\omit&\multispan2\hspace{0.1cm}\hrulefill\hspace{0.1cm}&\multispan3\hspace{0.1cm}\hrulefill\hspace{0.1cm}\cr
\noalign{\vskip 2pt}
\multicolumn{1}{c}{Data and model} &  \multicolumn{1}{c}{$1000\dms$} &  \multicolumn{1}{c}{$A_\mathrm{L}$} & \multicolumn{1}{c}{$1000\dms$} & \multicolumn{1}{c}{$A_\mathrm{L}$} & \multicolumn{1}{c}{$\Delta\chi^2$} & $\ln B$ \cr
\noalign{\vskip 2pt}
\hline 
\noalign{\vskip 2pt}
\multicolumn{3}{l}{TT+lowE} \cr
\noalign{\vskip 2pt}
$\quad$ \LCDM+CIP & $     \mathbf{15.5^{+      5.3}_{-      5.4}}$ & &      15.5 &      1.00 & $     -6.9$ & $      1.7$  \cr
\noalign{\vskip 2pt}
 $\quad$ \LCDM+$A_\mathrm{L}$ & & $     1.24^{+     0.10}_{-     0.10}$ &       0.0 &      1.26 & $     -8.7$ & $      2.1$  \cr
\noalign{\vskip 2pt} 
$\quad$ \LCDM+$A_\mathrm{L}$+CIP & $<     24.4$ & $     1.12^{+     0.14}_{-     0.12}$ &       3.4 &      1.23 & $     -8.8$ & $      0.6$  \cr
\noalign{\vskip 2pt}
\multicolumn{3}{l}{TT,TE,EE+lowE}\cr
\noalign{\vskip 2pt}
 $\quad$ \LCDM+CIP & $      \mathbf{10.1^{+      3.9}_{-      3.9}}$ & &      10.0 &      1.00 & $     -5.7$ & $      0.9$  \cr
\noalign{\vskip 2pt} 
$\quad$ \LCDM+$A_\mathrm{L}$ & & $     1.18^{+     0.07}_{-     0.07}$ &       0.0 &      1.19 & $     -9.7$ & $      2.5$  \cr
\noalign{\vskip 2pt}
$\quad$ \LCDM+$A_\mathrm{L}$+CIP & $<     12.7$ & $     1.13^{+     0.09}_{-     0.08}$ &       0.3 &      1.19 & $     -9.7$ & $      0.2$  \cr
\noalign{\vskip 2pt}
\multicolumn{3}{l}{TT,TE,EE+lowE+lensing (conserv.)}\cr
\noalign{\vskip 2pt}
 $\quad$ \LCDM+CIP & $      \mathbf{ 3.7^{+      1.6}_{-      2.1}}$ & &       3.6 &      1.00 & $     -3.3$ & $     -1.4$  \cr
\noalign{\vskip 2pt} 
$\quad$ \LCDM+$A_\mathrm{L}$ & & $     1.07^{+     0.04}_{-     0.04}$ &       0.0 &      1.07 & $     -3.4$ & $     -1.2$  \cr
\noalign{\vskip 2pt} 
$\quad$ \LCDM+$A_\mathrm{L}$+CIP & $      3.1^{+      1.4}_{-      2.0}$ & $     1.07^{+     0.04}_{-     0.04}$ &       2.9 &      1.07 & $     -6.4$ & $     -2.8$  \cr
\noalign{\vskip 2pt}
\multicolumn{3}{l}{TT,TE,EE+lowE+lensing (aggr.)}\cr
\noalign{\vskip 2pt}
 $\quad$ \LCDM+CIP & $       \mathbf{ 2.8^{+      1.2}_{-      1.5}}$ & &       2.7 &      1.00 & $     -4.0$ & $     -1.2$  \cr
\noalign{\vskip 2pt} 
$\quad$ \LCDM+$A_\mathrm{L}$ & & $     1.06^{+     0.04}_{-     0.04}$ &       0.0 &      1.05 & $     -2.2$ & $     -1.8$  \cr
\noalign{\vskip 2pt}
$\quad$ \LCDM+$A_\mathrm{L}$+CIP & $      2.6^{+      1.2}_{-      1.6}$ & $     1.06^{+     0.04}_{-     0.04}$ &       2.4 &      1.06 & $     -6.2$ & $     -3.0$  \cr
\noalign{\vskip 2pt}
\hline
\end{tabular}
\end{center}
\caption{Comparison of  \LCDM+CIP, \LCDM+$A_\mathrm{L}$, and
  \LCDM+$A_\mathrm{L}$+CIP  models with various \Planck\ datasets,
  when using the baseline \texttt{Plik} likelihood at high $\ell$. The
  first two columns (``Constraints'') are the 68\% CL ranges or the
  95\% CL upper bounds on  $1000\dms$  (highlighted in bold for  \LCDM+CIP) and  $A_\mathrm{L}$. The remaining columns give the best-fit pameter values, and the difference of the best-fit $\chi^2$ and the difference of the log of the Bayesian
evidence with respect to the pure adiabatic \LCDM\ model. A negative $\Delta\chi^2$ means that the quoted model fits the data better than \LCDM, while a positive $\ln B$ means that the Bayesian model comparison favours the quoted model, when adopting the uniform priors $0 \le 1000\dms < 75$ and $0.3 \le A_\mathrm{L} \le 1.7$.
\label{tab:CIP}}
\end{table}

The top panel of Fig.~\ref{fig:CIP2d} shows the $A
_\mathrm{L}$--$\dms$ degeneracy in the \LCDM+$A_\mathrm{L}$+CIP model and 
how it can be broken by the lensing data. The value $A _\mathrm{L}=1$ provides a good 
fit to the TT+lowE data, if $\dms \simeq 0.016$, and to the TT,TE,EE+lowE data, 
if $\dms \simeq 0.010$. The \LCDM+CIP model (where $A _\mathrm{L}=1$) 
with $\dms \simeq 0.008$ provides a better simultaneous fit to the 
\Planck\ 2015 TT,TE,EE and conservative lensing data ($40 \le L \le 400$) 
than does the \LCDM+$A_\mathrm{L}$ model.
When using the \Planck\ 2018  conservative lensing data ($8 \le L \le 400$), 
the best-fit value of $\dms$ decreases to $0.0036$.
This is due to the extra term $\propto L^{-2}$ brought by CIP to the lensing 
power estimator, as discussed above and shown in Fig.~\ref{fig:conservativeVSaggressive}.

Since \Planck\ TT,TE,EE+lowE and lensing data can be fit well by $A _\mathrm{L}=1$ in 
the CIP model, we show in the bottom panel of Fig.~\ref{fig:CIP2d} the one-dimensional posterior of 
$\dms$ in the \LCDM+CIP model. A non-zero value of $\dms$ is preferred at the $2.9\sigma$ 
($2.5\sigma$) level by \Planck\ 2018 (2015) TT+lowE(lowP) data and at the $2.6\sigma$ ($1.8\sigma$) 
level by the TT,TE,EE+lowE(lowP) data. Without lensing the 2018 data thus more 
strongly favour the non-zero CIP amplitude than the 2015 data, which is as we would expect, 
since the favoured $A _\mathrm{L}$ value in the \LCDM+$A_\mathrm{L}$ model has also increased. 
When using 2018 TT,TE,EE+lowE and the 2018 conservative lensing data the significance 
decreases to $2.0\sigma$, while switching to the aggressive lensing data ($8 \le L \le 
2048$) leads to $2.1\sigma$. The 68\,\% CL ranges of $\dms$ in the \LCDM+CIP model, 
obtained with the baseline high-$\ell$ \texttt{Plik} likelihood in combination of other 
\Planck\ data, are highlighted in Table~\ref{tab:CIP}. Replacing \texttt{Plik} with 
\texttt{CamSpec} (in particular \texttt{CamSpec} TT,TE,EE) leads to somewhat lower values,
\begin{equation}
\hspace{-2mm}
1000\dms \!=\!
\left\{\begin{aligned}
14.1^{+5.2}_{-5.2} & \ \text{TT+lowE,}\\
6.5^{+3.0}_{-4.2}   & \ \text{TT,TE,EE+lowE,}\\
2.8^{+1.2}_{-2.2}   & \ \text{TT,TE,EE+lowE+lensing (conserv.),}\\
2.2^{+1.0}_{-1.5}   & \ \text{TT,TE,EE+lowE+lensing (aggr.),}
\end{aligned}\right.
\end{equation}
and reduced significance above zero:
$2.7\sigma$, $1.9\sigma$, $1.7\sigma$, and $1.9\sigma$, respectively.

In order to check that the preference for $\dms > 0$ or
$A_\mathrm{L}>1$ is not just a parameter-space volume effect
upon marginalization over other parameters, we also report in Table~\ref{tab:CIP} 
the difference of $\chi^2$ between the best fit in extended models
and the base adiabatic \LCDM\ model. With all data sets, all three extended models
lead to an improvement of $\chi^2$ which clearly exceeds the number of extra
parameters of the model (1 for \LCDM+CIP and \LCDM+$A _\mathrm{L}$, and 2 for
\LCDM+$A_\mathrm{L}$+CIP). Although the inclusion of lensing data reduces
this improvement of fit, the \LCDM+CIP model gives a rather impressive $\Delta\chi^2 = -4$
with \Planck\ TT,TE,EE+lowE and aggressive lensing data.

Since we observe a moderate preference for a non-zero CIP amplitude, it might be tempting 
to ``solve'' the \Planck\ lensing anomaly by using CIP. However, this explanation seems quite 
unlikely, since in our treatment the CIP and adiabatic perturbations should be 
uncorrelated with each other, whereas CDI and BDI should be fully anticorrelated (and 
have a few orders of magnitude larger amplitude than the adiabatic modes while
keeping the perturbations nearly Gaussian). It is 
difficult to imagine a physical model that could lead to this situation. For example, 
some variants of curvaton model would naturally lead to anticorrelated CDI and BDI, 
but in these models there would be a correlation with the adiabatic mode too 
\citep{Gordon:2002gv,He:2015msa}. The above-studied compensated BDI-CDI mode falls into 
a similar category to NVI: it is an interesting theoretical setup, but a compelling 
early-Universe model for stimulating this mode has still to be discovered.

Nevertheless, the baseline \Planck\ \texttt{Plik} TT,TE,EE+lowE plus conservative lensing result, 
$\dms = 0.0037^{+0.0016}_{-0.0021}$, 
is fully compatible with current complementary observations, in particular, the WMAP 95\,\% CL trispectrum constraint, 
$\dms\lesssim0.012$ \citep{Grin:2013uya}, and the upper bound, $\dms \lesssim 0.006$, 
following from the  direct measurements of the variation of the baryon fraction in galaxy clusters 
\citep{Holder:2009gd,Grin:2013uya}.
It will be interesting to learn what other future CMB anisotropy 
\citep{Abazajian:2016yjj, Valiviita:2017fbx,Finelli:2016cyd} and complementary 
measurements, such as observations of the distribution of neutral hydrogen
using 21\,cm absorption lines \citep{Gordon:2009wx}, BAO \citep{Soumagnac:2016bjk,Soumagnac:2018atx}, or
CMB spectral distortion anisotropies \citep{Haga:2018pdl}, will tell us about the possible 
contribution of CIP to the primordial perturbations.

\FloatBarrier


\section{Constraints on anisotropic models of inflation \label{sec:anisotropic}}

   In this section we will test specific physical models for statistical 
anisotropy in the primordial fluctuations.  More phenomenological 
multipole- or map-space tests are performed in the companion paper, 
\citet{planck2016-l07}.  Here we update the results of the 2015 release 
\citepalias{planck2014-a24} with polarization and new temperature analyses.  
Incorporating polarization 
into these tests is particularly important, due to the mild statistical 
significance of temperature anomalies such as the dipolar asymmetry.  
Polarization offers the potential to confirm or refute a physical origin 
for such anomalies via the measurement of independent fluctuation modes.  
We perform such a new test with $k$-space dipolar modulation models.  In 
cases such as quadrupolar asymmetry, where no detection has been claimed 
with temperature, polarization offers the prospect of tightening existing 
constraints.

   Some asymmetry models predict a modification to the {\em isotropic} power 
spectra, in addition to a dipolar or quadrupolar asymmetry.  In other 
words, for these models, as well as non-zero {\em off-diagonal} 
multipole covariance elements, we expect departures in the {\em diagonal} 
elements relative to the standard $\Lambda$CDM prediction.  Therefore the 
isotropic spectra can provide independent tests of such models even using 
temperature data alone~\citep{chmsz17}.  The curvaton dipole modulation model 
we examine in Sect.~\ref{sec:curvaton} exhibits this property, and can be 
constrained via its predictions for isotropic isocurvature power.  
Similarly, some versions of the quadrupolar modulation model we study in 
Sect.~\ref{sec:quadrupoleasym} modify the isotropic spectra via a monopole 
term.  In both cases these isotropic constraints will be important in 
narrowing the viable parameter space.

\subsection{Dipolar asymmetry}

   A dipolar temperature power asymmetry has long been observed at the 
largest scales in the CMB~\citep{ehbgl04}, although its statistical 
significance is not high and is subject to a posteriori (look-elsewhere) 
corrections~\citep{wmap7anomalies,planck2013-p09,planck2014-a18}.  
Nevertheless, its large-scale character suggests potential links with 
inflationary physics and various models have been proposed to explain it.  
In this subsection we examine several physical models for a dipolar 
modulation.  Some models where a generic CDM density isocurvature (CDI) or 
tensor component is dipole 
modulated have already been ruled out due to their isotropic 
predictions~\citep{chmsz17}, so we do not consider these further here.

\subsubsection{Curvaton model}
\label{sec:curvaton}

   First we update our 2015 study \citepalias{planck2014-a24} of a specific 
inflationary model for the dipolar asymmetry: namely, the modulated 
curvaton model of \citet{ehk09}.  In that study we 
showed that that model could not explain the observed asymmetry.  Here, we 
generalize the curvaton model to allow for a non-scale-invariant uncorrelated 
CDI component.  In addition, we treat the power spectrum (isotropic) 
constraints in a fully unified way with the asymmetry likelihood.  Finally, 
we incorporate polarization.

   The modulated curvaton model employs a gradient in a background curvaton 
field to explain the observed large-scale power asymmetry.  The curvaton, 
via coupling $\kappa$, produces nearly scale-invariant CDI fluctuations, as 
well as a fraction, $\xi$, of the adiabatic fluctuations.  Both of these 
components will be modulated.  Up to a sign, $\xi$ is equal to the 
correlation parameter, and is also a measure of the amplitude of dipolar 
modulation.  The isocurvature fraction can be written in terms of these two 
parameters as
\begin{equation}
\beta_{\mathrm{iso}} = \frac{9\kappa^2\xi}{1 + 9\kappa^2\xi}.
\label{betadef}
\end{equation}
Full details of this model and our treatment of it can be found in 
\citet{ehk09} and \citetalias{planck2014-a24}.

\begin{figure}
\begin{center}
\includegraphics[width=\columnwidth]{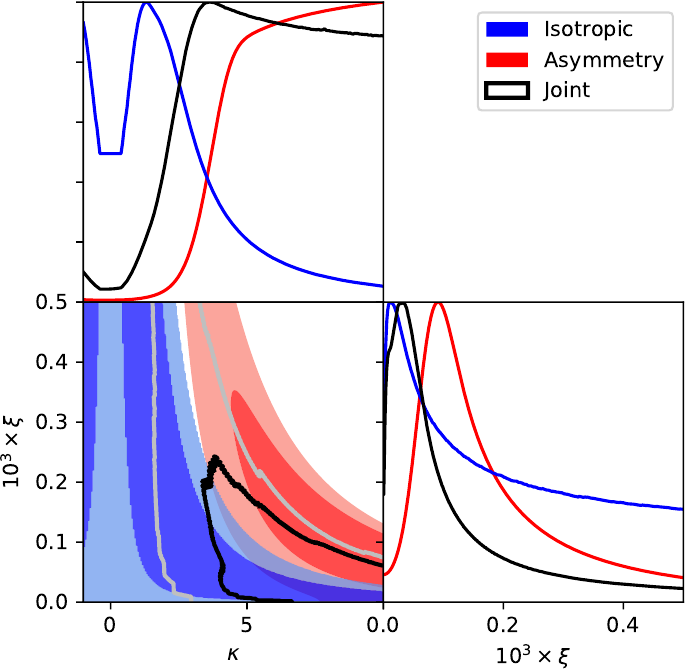}
\end{center}
\caption{Posteriors for the curvaton dipolar modulation model 
parameters $\kappa$ and $\xi$.  Contours enclose 68\,\% and 95\,\% 
of the posteriors.  The model can explain the well-known dipolar 
asymmetry: note the preference for $\xi > 0$ in the asymmetry constraint 
(red contours and curves).  However, the modulation preferred by the 
asymmetry constraint is 
reduced substantially when the isotropic constraint (blue) is added 
(black).  The asymmetry constraint here uses \texttt{SMICA}, while 
the isotropic constraint uses \Planck\ TT,TE,EE+lowE+lensing.  Resolution 
is reduced at very small $\kappa$ due to the sampling in $\beta_{\mathrm{iso}}$.}
\label{fig:xi_kappa_all}
\end{figure}

   Using the dipolar asymmetry estimator from \citetalias{planck2014-a24} we 
find the posteriors 
for the dipolar modulation parameters $\kappa$ and $\xi$; the results are 
presented in Fig.~\ref{fig:xi_kappa_all} (red contours).  We see that a 
substantial amount of asymmetry (as measured by amplitude $\xi$) can be 
captured by the model.  This preference for asymmetry simply means that the 
curvaton model can explain the well-known dipolar asymmetry in temperature.  
However, isocurvature constraints from the power spectra via 
Eq.~(\ref{betadef}), which we refer to as the {\em isotropic} 
constraints, can provide independent information~\citep{chmsz17}.  This is 
also shown in Fig.~\ref{fig:xi_kappa_all}, with the blue contours.  Here we 
see that the asymmetry and isotropic posteriors only weakly overlap, and 
the independent isotropic data do not support the presence 
of asymmetry for this model.  No evidence for asymmetry (i.e., no preference 
for $\xi > 0$) is present in the joint constraints, which treat the isotropic 
and asymmetry data as independent.  In other words, we have no reason to 
prefer this model over base $\Lambda$CDM.

\subsubsection{Adiabatic models}

   In the presence of a sufficiently large bispectrum it is possible that a 
long-wavelength mode can induce a dipolar asymmetry in the two-point function 
across our observable volume, although such scenarios appear to require fine 
tuning~\citep{brst16}.  Nevertheless, examples have been constructed which 
satisfy the \Planck\ $f_{\rm NL}$ constraints~\citep{brst16b}.  In this 
subsection we consider adiabatic models of this type, in which the isotropic 
power spectra agree with standard $\Lambda$CDM, while a scale-dependent 
dipolar asymmetry is present in the off-diagonal multipole 
covariance~\citep{czsbg17,chmsz17}.  As proposed in~\citet{czsbg17}, we 
fit the asymmetry model parameters to the temperature data and then use 
those parameters to predict the asymmetry in polarization.  We then compare 
those predictions with the \Planck\ polarization data as a test for a 
physical modulation.  Importantly, a position- (or $k$-) space model for 
the modulation is needed for reliable polarization predictions---it is not 
enough to restrict considerations to multipole space~\citep{czsbg17}.\footnote{E.g., if we 
observe a dipolar modulation in $T$ at, say, 5\,\% to $\ell = 65$, there is no 
reason to expect a modulation of the same amplitude and to the same scale in 
$E$, due to the different $T$ and $E$ transfer functions~\citep{czsbg17}.}

   As discussed in detail in~\citet{czsbg17}, we take a portion 
$\widetilde{\cal{R}}^{\rm lo}(x)$ of the adiabatic primordial fluctuations to 
be spatially linearly modulated according to
\begin{equation}
\widetilde{\cal{R}}^{\rm lo}(\vec{x})
 = {\cal{R}}^{\rm lo}(\vec{x})\left(1 + A\frac{\vec{x}\cdot{\vec{\hat d}}}{r_{\rm LS}}\right),
\end{equation}
where ${\cal{R}}^{\rm lo}(\vec{x})$ is statistically isotropic with power spectrum 
${{\cal P}_{\cal{R}}^{\rm lo}}(k)$, $A \le 1$ and $\vec{\hat d}$ are the 
amplitude and direction of modulation, respectively, and $r_{\rm LS}$ is the 
comoving radius to last scattering.  This leads, to a good approximation, to 
the total temperature or polarization multipole covariance
\begin{align}
C_{\ell m\ell'm'} &\equiv \langle a_{\ell m}a_{\ell'm'}^*\rangle\\
   &= C_\ell\delta_{\ell\ell'}\delta_{mm'} +
      \frac{\delta C_{\ell\ell'}}{2}\sum_M \Delta X_M\xi^M_{\ell m\ell'm'},
\label{eq:asymcovar}
\end{align}
to first order in $A$.  Here $C_\ell$ is the usual $\Lambda$CDM anisotropy 
power spectrum; $\delta C_{\ell\ell'} \equiv 2(C_\ell^{\rm lo} + 
C_{\ell'}^{\rm lo})$, where $C_\ell^{\rm lo}$ is the power spectrum calculated 
in the usual way from ${{\cal P}_{\cal{R}}^{\rm lo}}(k)$; $\Delta X_M$ is the 
multipole decomposition of $A\vec{\hat n}\cdot\vec{\hat d}$; and the 
$\xi^M_{\ell m\ell'm'}$ coefficients couple $\ell$ to $\ell \pm 1$ via
\begin{equation}
\xi^M_{\ell m\ell'm'} \equiv \sqrt{\frac{4\pi}{3}}
   \int Y_{\ell'm'}(\vec{\hat n})Y_{1M}(\vec{\hat n})
        Y_{\ell m}^*(\vec{\hat n})d\Omega.
\end{equation}

   In principle the scale dependence of the asymmetry spectrum 
${{\cal P}_{\cal{R}}^{\rm lo}}(k)$ is completely free, but here we take 
three phenomenological forms which are capable of producing a large-scale 
asymmetry with a small number of parameters.  First, we consider a simple 
power-law modulation,
\begin{equation}
\mathcal{P}^{\rm lo}_{\mathcal{R}}(k)
   = \mathcal{P}^0_\mathcal{R}\left(k_0^{\rm lo}\right)
     \left(\frac{k}{k_0^{\rm lo}}\right)^{n^{\rm lo}_{\rm s} - 1},
\end{equation}
where $\mathcal{P}^0_\mathcal{R}(k)$ is the usual 
$\Lambda$CDM spectrum, and $n^{\rm lo}_{\rm s}$ and $k_0^{\rm lo}$ are the 
tilt and pivot scale of the modulation.  We consider only red asymmetry tilts 
with $n^{\rm lo}_{\rm s} \le n_{\rm s}$, and choose $k_0^{\rm lo} =
1.5\times10^{-4}\,{\rm Mpc}^{-1}$.  We also consider a $\tanh$ model, 
defined according to
\begin{equation}
\mathcal{P}^{\rm lo}_{\mathcal{R}}(k)
  = \frac{1}{2}\mathcal{P}^0_\mathcal{R}(k)
    \left[1 - \tanh{\left(\frac{\ln k - \ln k_{\rm c}}{\Delta\ln k}\right)}\right].
\end{equation}
This spectrum approaches that of $\Lambda$CDM on scales larger than 
$k_{\rm c}$, with a width determined by $\Delta\ln k$.  That is, scales 
well above the cutoff $k_{\rm c}$ will be modulated with amplitude $A$, and 
scales below will be unmodulated.  Finally, we consider a model with a linear 
gradient in the scalar tilt, $n_{\rm s}$, across our volume.  In this case 
the asymmetry spectrum can be written as
\begin{equation}
C_\ell^{\rm lo}
   = -\frac{\Delta n_{\rm s}}{2}\frac{dC_\ell}{dn_{\rm s}},
\end{equation}
with modulation amplitude $\Delta n_{\rm s}$.  There will be an implicit 
dependence on the pivot scale $k_*$ for this model.

   Given the multipole covariance, Eq.~(\ref{eq:asymcovar}), we can construct 
a maximum likelihood estimator for the modulation, $\Delta X_M$.  In the 
noise-free, full-sky case this takes the form~\citep{mszb11,planck2014-a18}
\begin{equation}
\Delta\hat{X}_M = \frac{1}{4}\sigma_X^2 \sum_{\ell m \ell' m'} \frac{\delta
C_{\ell\ell'}}{C_\ell C_{\ell'}} \xi^M_{\ell m\ell'm'}a_{\ell m}^*a_{\ell'm'},
\end{equation}
where the cosmic variance of the estimator is given by
\begin{equation}
\sigma_X^2 = 12\left(\sum_\ell(\ell + 1)
             \frac{\delta C_{\ell\ell + 1}^2}{C_\ell C_{\ell + 1}}\right)^{-1}.
\end{equation}
The modifications we use to deal with realistic skies are described in 
detail in~\citet{planck2014-a18} and \citet{czsbg17}.

\begin{figure*}
\begin{center}
\includegraphics{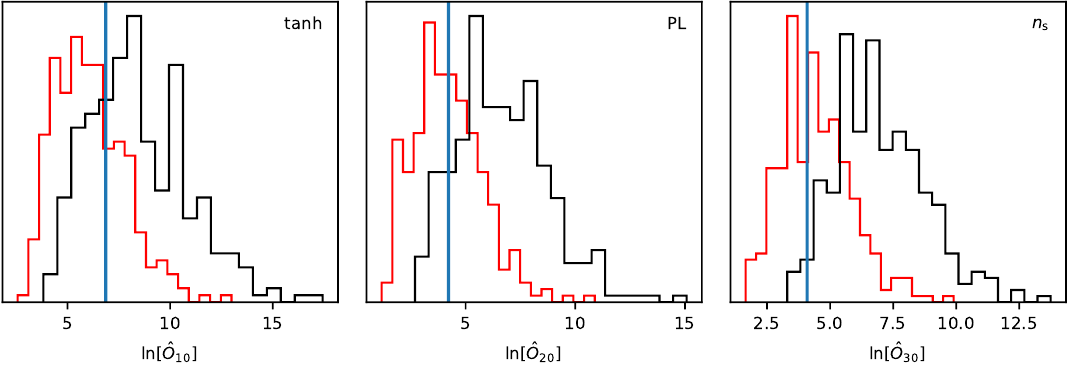}
\end{center}
\caption{Histograms of the quantity $\hat{O}_{j0}$ for the $\tanh$, 
power-law, and $n_{\rm s}$ gradient modulation models using \Planck\ 
temperature data combined with 300 statistically isotropic 
polarization simulations (red outlines) or 300 polarization 
simulations modulated according to the best-fit parameters from the 
temperature data (black).  The blue lines indicate the values for 
the actual \texttt{SMICA} polarization data.  A large value relative 
to the isotropic (red) simulations would indicate that the 
modulation model is preferred over $\Lambda$CDM.}
\label{fig:Oj0hist_PL}
\end{figure*}

   To decide whether the polarization data support the modulation model 
or not, we consider the quantity $\hat{O}_{j0}$, which is the ratio of the 
maximum likelihood for modulation model $j$ to that of 
$\Lambda$CDM~\citep{czsbg17}.  In Fig.~\ref{fig:Oj0hist_PL} we plot for the 
three adiabatic models histograms of $\hat{O}_{j0}$ calculated for 300 
statistically isotropic polarization simulations (sharing the required 
$TE$ correlation with the real $T$ data) added to the \Planck\ 
temperature data (red outlines).  This indicates our expectation for 
$\hat{O}_{j0}$ for the scenario that the 
temperature asymmetry is due to a statistical fluctuation and not to a 
physical modulation.  We also plot in Fig.~\ref{fig:Oj0hist_PL} 
histograms for 300 polarization 
simulations modulated with the best-fit parameters from the \Planck\ temperature 
data (black outlines), to represent the scenario that the asymmetry 
is due to a physical modulation.  In both cases the polarization 
simulations contain realistic levels of noise for \Planck.  By comparing 
the isotropic and modulated histograms, we can see 
that the quantity $\hat{O}_{j0}$ can serve to distinguish the two 
scenarios, but only relatively weakly for \Planck\ noise~\citep{czsbg17}.  
The blue lines indicate the values using the actual \texttt{SMICA} polarization 
data (the results for the other component-separation methods are similar).  
We see that for these models the data do not help to decide whether we 
have a physical modulation or not, with $p$-values of $43\,\%$, 
$30\,\%$, and $57\,\%$ for the power-law, $\tanh$, and $n_{\rm s}$ gradient 
models, respectively, relative to the isotropic simulations.

\subsection{Quadrupolar asymmetry}
\label{sec:quadrupoleasym}

   We will next explore models that predict a quadrupolar 
direction dependence in the primordial power spectrum.  In 
\citetalias{planck2014-a24} we found no evidence for such a modulation, but 
several inflationary models have been constructed which predict this 
effect \citep{Ackerman:2007,Soda:2012,Tsujikawa:2014rta}.  
Therefore it is important to extend those results with the 
improved polarization data.  We now attempt to reduce the effect of unresolved 
point sources using the bias-hardened estimator approach 
of \citet{planck2014-a17}.  In \citetalias{planck2014-a24} we pointed out 
that some models of quadrupolar asymmetry 
predict a modification to the angular power spectra as well.  Here we will 
account for such modifications in our analysis, increasing the constraining 
power of temperature data, in particular for tilted models with 
non-scale-invariant modulation spectral index.  Note that independent searches 
relaxing our approximation of power-law spectra have also been carried out 
\citep{Durakovic:2017prf}. 

We assume a modulation of the primordial comoving curvature power spectrum 
of the form
\begin{equation}
\mathcal{P}_{\mathcal{R}}({\vec k}) = \mathcal{P}_{\mathcal{R}}^0(k)\left[1 + g(k)
   \left(\vec{\hat k}\cdot\vec{\hat{d}}\right)^2\right],
\label{Pkgk}
\end{equation}
which can be rewritten as
\begin{equation}
\mathcal{P}_{\mathcal{R}}({\vec k}) = \mathcal{P}_{\mathcal{R}}^0(k)
   \left[1 + \frac{1}{3}g(k) + \sum_{m}g_{2m}(k)\,Y_{2m}(\vec{\hat k})\right].
\label{prim_power_glm}
\end{equation}
Here
\begin{equation}
g_{2m}(k) \equiv \frac{8\pi}{15}g(k)\,Y_{2m}^*(\vec{\hat{d}}),
\end{equation}
with $g_{2m}(k)$ satisfying $g_{2,-m}(k)=(-1)^m\,g^*_{2m}(k)$.  We parameterize 
the scale dependence of the modulation as $g(k) = g_*(k/k_*)^q$, with 
pivot scale $k_* = 0.05\,\mathrm{Mpc}^{-1}$.  For $q \ne 0$, in addition 
to producing a quadrupolar modulation of the anisotropies, this model 
affects the CMB isotropic power spectra via the term $g(k)/3$ in 
Eq.~(\ref{prim_power_glm}).  We 
therefore consider a joint constraint with the isotropic power spectra 
likelihood to improve constraints over the modulation alone.

\begin{figure*}
\begin{center}
\includegraphics[width=\columnwidth]{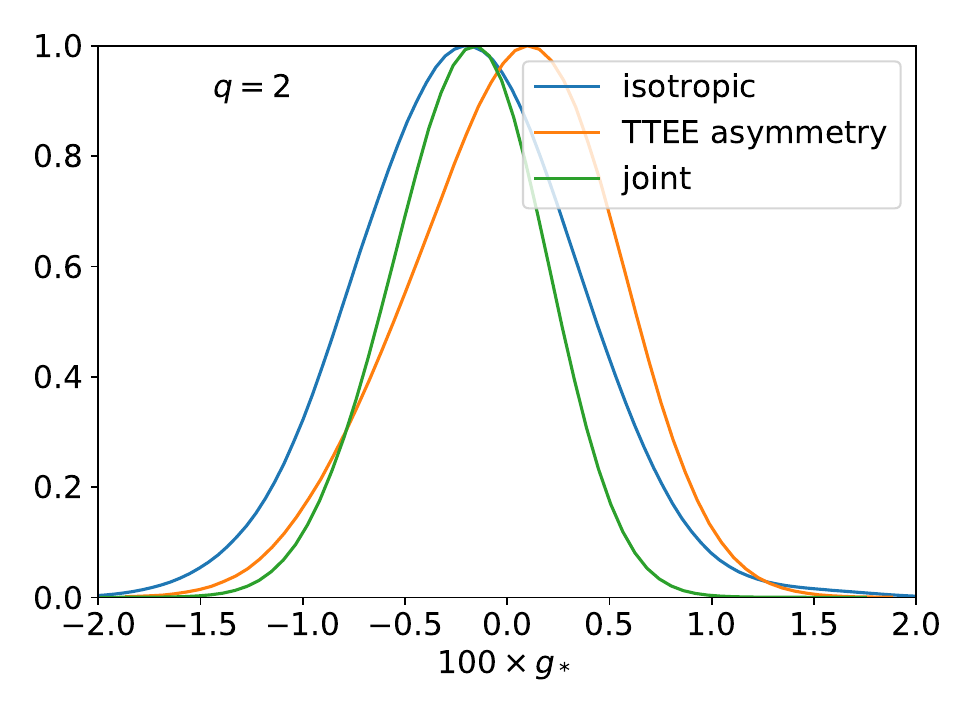}
\includegraphics[width=\columnwidth]{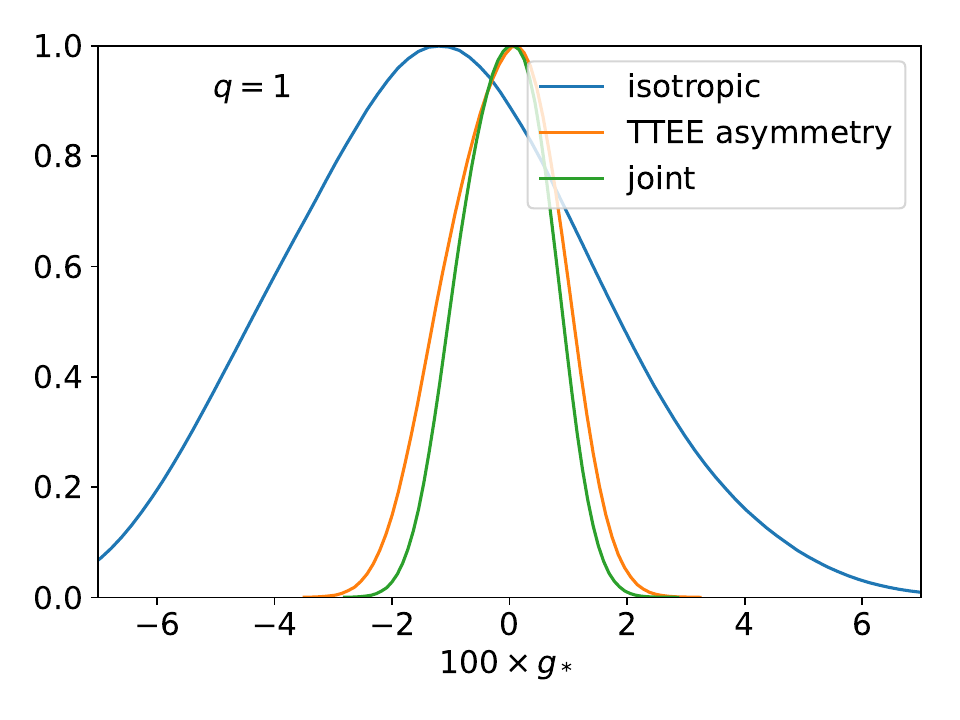}
\includegraphics[width=\columnwidth]{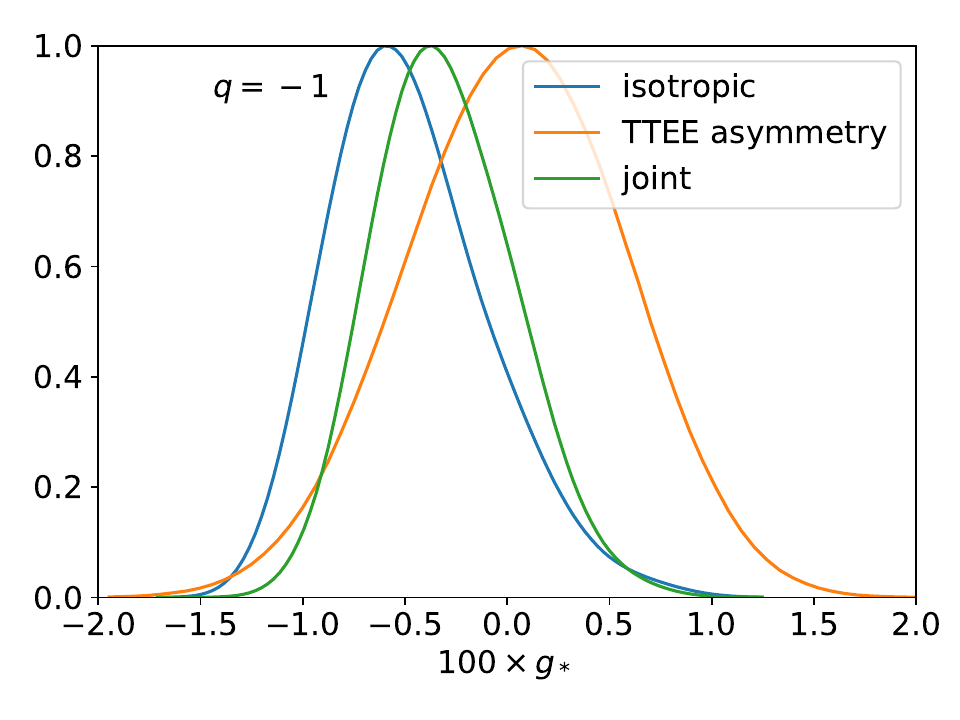}
\includegraphics[width=\columnwidth]{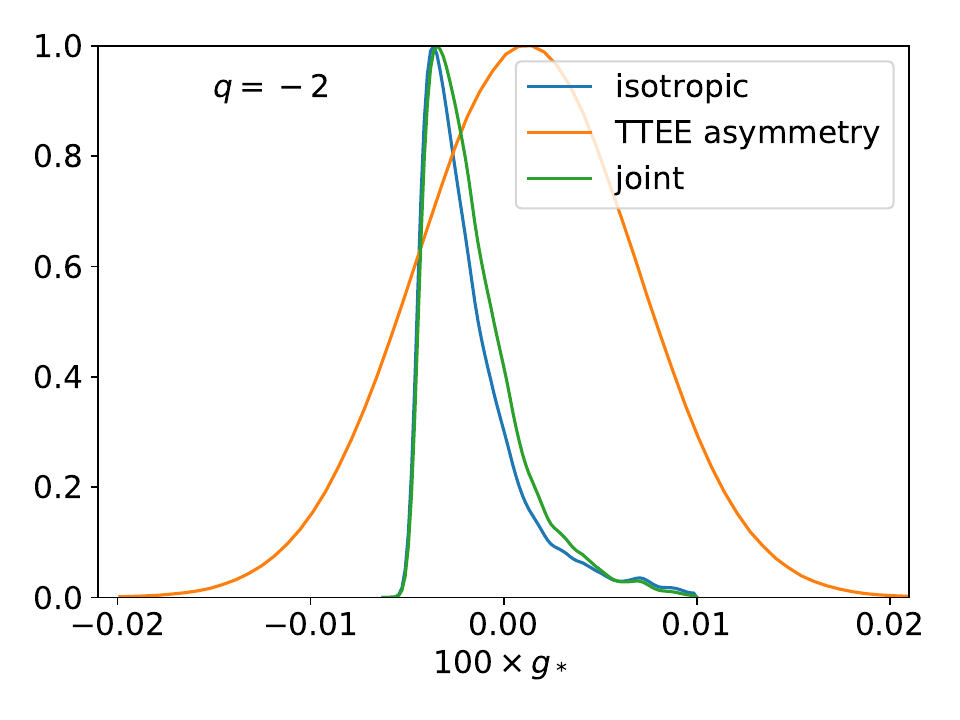}
\end{center}
\caption{Marginalized posteriors for quadrupolar modulation parameter 
$g_*$, using \texttt{SMICA} data for the $TT$+$EE$ asymmetry 
constraints (orange curves) and \Planck\ TT,TE,EE+lowE for the isotropic 
constraints (blue curves), which probe the modification to the power 
spectrum via Eq.~(\ref{prim_power_glm}).  {\it Top:} 
constraints for $q = 2$ and 1 (left and right, respectively).  {\it Bottom:} 
constraints for $q = -1$ and $-2$ (left and right, respectively).  
Strongly negative $g_*$ is suppressed for $q < 0$, due to the 
unphysical prediction of negative power.} \label{fig:gstar_qs}
\end{figure*}

\begin{table*}[ht!]
\begingroup
\newdimen\tblskip \tblskip=5pt
\caption{
Minimum-$\chi^2$ $g_*$ values for quadrupolar modulation,
determined from the
\texttt{SMICA} foreground-cleaned maps.  Also given are $p$ values,
defined as the fraction of isotropic simulations with larger $|g_*|$
than the data.  The $TT$ results use $\ell_{\rm min} = 2$ and
$\ell_{\rm max} = 1200$, while $EE$ uses $\ell_{\rm min} = 2$ and
$\ell_{\rm max} = 850$.  These results indicate that the data are
consistent with cosmic variance in statistically isotropic skies.}
\label{tab:gstar}                            
\nointerlineskip
\vskip -3mm
\footnotesize
\setbox\tablebox=\vbox{
   \newdimen\digitwidth
   \setbox0=\hbox{\rm 0}
   \digitwidth=\wd0
   \catcode`!=\active
   \def!{\kern\digitwidth}
   \newdimen\signwidth
   \setbox0=\hbox{+}
   \signwidth=\wd0
   \catcode`?=\active
   \def?{\kern\signwidth}
\halign{\hbox to 0.7cm{#\leaderfil}\tabskip 1em&
\hfil#\hfil&\tabskip 0.8em&
\hfil#\hfil&
\hfil#\hfil&
\hfil#\hfil&
\hfil#\hfil\tabskip 0pt\cr
\noalign{\doubleline}
\omit&\multispan2 \hfil$TT$\hfil & \multispan2 \hfil$EE$\hfil
     &\multispan2 \hfil$TT + EE$\hfil\cr
\noalign{\vskip -5pt}
\omit&\multispan6\hrulefill\cr
\omit\hfil$q$\hfil
   &\omit\hfil $g_*$ \hfil&\omit\hfil $p$ value [\%]\hfil
   &\omit\hfil $g_*$ \hfil&\omit\hfil $p$ value [\%]\hfil
   &\omit\hfil $g_*$ \hfil&\omit\hfil $p$ value [\%]\hfil\cr
\noalign{\vskip 4pt\hrule\vskip 6pt}
\quad$-2$ & $-6.83\times10^{-5}$ & 75.7 & $?1.23\times10^{-4}$ & 54.7 & $-6.90\times10^{-5}$ & 75.0 \cr
\noalign{\vskip 3pt}
\quad$-1$ & $-8.56\times10^{-3}$ & 64.7 & $?1.44\times10^{-2}$ & 30.0 & $-6.15\times10^{-3}$ & 86.0 \cr
\noalign{\vskip 3pt}
\quad$?0$ & $?1.08\times10^{-2}$ & 82.7 & $?3.17\times10^{-2}$ & 55.3 & $?1.07\times10^{-2}$ & 83.0 \cr
\noalign{\vskip 3pt}
\quad$?1$ & $?7.77\times10^{-3}$ & 82.7 & $?5.09\times10^{-2}$ & 24.0 & $?7.75\times10^{-3}$ & 82.3 \cr
\noalign{\vskip 3pt}
\quad$?2$ & $?4.92\times10^{-3}$ & 78.3 & $?5.62\times10^{-2}$ & 17.0 & $?4.92\times10^{-3}$ & 78.7 \cr
\noalign{\vskip 3pt\hrule\vskip 4pt}}}
\endPlancktable                    
\endgroup
\end{table*}

\begin{table*}[ht!]
\begingroup
\newdimen\tblskip \tblskip=5pt
\caption{
As for Table~\ref{tab:gstar}, but for the quantity $g_2 \equiv
\sqrt{\sum_m |g_{2m}|^2/5}$ for a completely general quadrupolar modulation.}
\label{tab:g2}                            
\nointerlineskip
\vskip -3mm
\footnotesize
\setbox\tablebox=\vbox{
   \newdimen\digitwidth
   \setbox0=\hbox{\rm 0}
   \digitwidth=\wd0
   \catcode`!=\active
   \def!{\kern\digitwidth}
   \newdimen\signwidth
   \setbox0=\hbox{+}
   \signwidth=\wd0
   \catcode`?=\active
   \def?{\kern\signwidth}
\halign{\hbox to 0.7cm{#\leaderfil}\tabskip 1em&
\hfil#\hfil&\tabskip 0.8em&
\hfil#\hfil&
\hfil#\hfil&
\hfil#\hfil&
\hfil#\hfil\tabskip 0pt\cr
\noalign{\doubleline}
\omit&\multispan2 \hfil$TT$\hfil & \multispan2 \hfil$EE$\hfil
     &\multispan2 \hfil$TT + EE$\hfil\cr
\noalign{\vskip -5pt}
\omit&\multispan6\hrulefill\cr
\omit\hfil$q$\hfil
   &\omit\hfil $g_2$ \hfil&\omit\hfil $p$ value [\%]\hfil
   &\omit\hfil $g_2$ \hfil&\omit\hfil $p$ value [\%]\hfil
   &\omit\hfil $g_2$ \hfil&\omit\hfil $p$ value [\%]\hfil\cr
\noalign{\vskip 4pt\hrule\vskip 6pt}
\quad$-2$ & $3.30\times10^{-5}$ & 82.3 & $9.00\times10^{-5}$ & 27.0 & $3.32\times10^{-5}$ & 81.0 \cr
\noalign{\vskip 3pt}
\quad$-1$ & $4.34\times10^{-3}$ & 66.0 & $6.81\times10^{-3}$ & 40.0 & $3.24\times10^{-3}$ & 87.0 \cr
\noalign{\vskip 3pt}
\quad$?0$ & $7.65\times10^{-3}$ & 51.7 & $1.79\times10^{-2}$ & 45.7 & $7.62\times10^{-3}$ & 51.7 \cr
\noalign{\vskip 3pt}
\quad$?1$ & $5.39\times10^{-3}$ & 58.0 & $2.95\times10^{-2}$ & 12.7 & $5.38\times10^{-3}$ & 58.3 \cr
\noalign{\vskip 3pt}
\quad$?2$ & $3.15\times10^{-3}$ & 58.7 & $3.39\times10^{-2}$ & !6.0 & $3.15\times10^{-3}$ & 58.7 \cr
\noalign{\vskip 3pt\hrule\vskip 4pt}}}
\endPlancktable                    
\endgroup
\end{table*}

As in \citetalias{planck2014-a24}, we obtain constraints on the modulation 
parameters by forming 
quadratic maximum-likelihood estimates, $\hat{g}_{2m}$, for the data and 
simulations. For this we use the component-separated data and 300 simulations 
provided by \texttt{NILC}, \texttt{SEVEM}, \texttt{SMICA}, and 
\texttt{COMMANDER}. For brevity we only show the \texttt{SMICA} results.  We 
can then compute a covariance $G$ and likelihood as
\begin{equation}
{\cal L} \propto |G|^{-1/2}\label{g2Mlike} \exp\left[-\frac{1}{2}
              M^{\textsf T}G^{-1}M\right],\nonumber
\end{equation}
where
\begin{equation}
M \equiv \hat{g}_{2m} - g_{2m}(g_*,\vec{\hat{d}}).
\end{equation}
We then evaluate the marginalized (over the angles) posterior for $g_*$. For the 
isotropic constraints we simply include the modulation parameters in a 
\texttt{CosmoMC} run using \Planck\ TT,TE,EE+lowE data, and then evaluate the 
marginalized (over all other \LCDM\ parameters) posterior for $g_*$.

   For $q > 0$ the effect of $g_*$ on the isotropic spectra occurs mainly at 
high $\ell$, and is highly degenerate with $n_{\rm s}$.  This degeneracy leads to 
slightly less stringent 
constraints than what one would achieve with a fixed $n_{\rm s}$.  We show the 
marginalized posteriors for this case in the top panels of 
Fig.~\ref{fig:gstar_qs}, where we see that the isotropic constraints are 
roughly comparable in strength to (and fully consistent with) the constraints 
from the asymmetry data.

For $q < 0$  the isotropic constraints are much more constraining than the 
modulation constraints, as seen in the bottom panels of Fig.~\ref{fig:gstar_qs}.  
This is because for large scales the factor $k_*/k$ can become large and a 
negative $g_*$ will decrease isotropic power on those scales, which is 
compensated for by increasing $A_{\rm s}$ and $\tau$.  Strongly negative $g_*$ values 
are disallowed by predicting unphysical negative power spectra at low 
$\ell$.  Note that even the parameter ranges in which the power spectra 
are reduced to close to zero are likely beyond the perturbative regime for 
the models in question, and so should be approached with caution.  
The isotropic constraints still prefer a slightly negative $g_*$, 
likely due to being able to fit the power deficit at large scales.  The joint 
constraint in this case is then greatly improved by the isotropic data.

   Minimum-$\chi^2$ and $p$ values (relative to isotropic simulations) for 
$g_*$ are presented in Table~\ref{tab:gstar}.  The addition of polarization 
does not affect the temperature results greatly.

   Finally, when allowing the completely general form of quadrupolar 
modulation, i.e.,
\begin{equation}
\mathcal{P}_{\mathcal{R}}({\vec k}) = \mathcal{P}_{\mathcal{R}}^0(k)
   \left[1 + \sum_{m}g_{2m}\left(k/k_*\right)^q\,Y_{2m}(\vec{\hat k})\right],
\end{equation}
with no restriction on the $g_{2m}$, we present results for the quantity 
$g_2 \equiv \sqrt{\sum_m |g_{2m}|^2/5}$ in Table~\ref{tab:g2}.  In all 
cases there is no significant detection of quadrupolar modulation, as 
quantified by the $p$ values.

\label{modulationSection}

\newpage

\section{Conclusions \label{sec:conclusions}}


This paper summarizes the status of cosmic inflation in light of the \Planck\ 2018 release. 
The main improvements are in the \Planck\ polarization likelihoods. The 2018 release now 
includes a low-$\ell$ HFI polarization likelihood based on the 100- and 143-GHz channels. 
This likelihood is now the baseline, whereas the \Planck\ 2015 likelihood was based only 
on the LFI 70-GHz channel data, which also have been updated in this release. Corrections 
for beam-leakage effects, which had been flagged in the 2015 release as the main limitation 
of the $TE$ and $EE$ data at that time, have improved the accuracy of the high-$\ell$ polarization likelihoods. 
Our analyses focus on the results obtained using the \Planck\ baseline likelihoods alone, but results supplemented 
by the BK15 likelihood (when tensors are included) and a compilation of BAO likelihoods are also given in order to help break 
cosmological parameter degeneracies. We summarize the main results of this paper in the form of responses to a number of key questions.
\begin{enumerate}
\item
\textit{What is the value of the scalar tilt?}

\noindent Using a characterization of polarization anisotropy better at all multipoles in this release, 
we find that $n_\mathrm{s}=0.9649\pm 0.0042$ at 68\,\% CL, including the full information provided by \Planck\ (TT,TE,EE+lowE+lensing). The 2018 uncertainty is approximately 2/3 of that obtained with the \Planck\ 2015 baseline likelihood. Importantly, this determination rules out perfect scale invariance (i.e., $n_\mathrm{s}=1$) at $8.4\sigma$.
From an inflationary perspective, this result is consistent with slow-roll inflation evolving towards a natural exit.

\item
\textit{Does $n_\mathrm{s}$ depend on the wavelength?}

We investigated the possibility of a running spectral index, as well as a running of the running [i.e., the next two (subleading) terms in a power series expansion of $\ln (\mathcal{P}_\mathcal{R})$ in $\ln (k)$], corresponding to non-negligible third- and fourth-order derivatives of the inflationary potential. Starting with its first 2013 cosmological release, \Planck\ has removed any hint of a running spectral index, which had been suggested by pre-\Planck\ data and would have pointed to inflationary models beyond the slow-roll approximation. \Planck\ 2018 sets $d n_\mathrm{s}/d \ln k = -0.005 \pm 0.013$ as the tightest 95\,\% CL constraint, when $d^2 n_\mathrm{s}/d \ln k^2 = 0$. No hints of further extensions, such as running of the running, are found with \Planck\ 2018 data. These results are consistent with the simplest slow-roll dynamics for the inflaton. A detection of running at the level predicted by slow-roll models will require a combination of future ambitious CMB anisotropy experiments and galaxy surveys.

\item
\textit{Is the Universe spatially flat?}

Most simple models of inflation predict a spatially flat universe, although inflationary models with a minimum degree of fine tuning producing a hyperbolic universe have been constructed. 
\Planck\ has been the first experiment to constrain the spatial curvature at the percent level without any external information, thanks to the CMB lensing likelihood. Although negative values of $\Omega_K \sim -0.01$ provide a non-statistically significant improvement to the fit of \Planck\ temperature and polarization data (compared to the minimal $\Lambda$CDM model), \Planck\ 2018 data including lensing constrain $\Omega_K = -0.011^{+0.013}_{-0.012}$ at 95\,\% CL. Combining with BAO data further tightens the uncertainty, constraining $\Omega _K$ to lie within 0.4\,\% of a flat spatial geometry (at $95\,\%$ CL).

\item
\textit{Are tensor modes required?}

Inflationary models predict that tensor modes were also excited during the nearly 
exponential expansion, with a power spectrum amplitude proportional to the energy 
scale of inflation. Using the measurement of CMB temperature and $E$-mode polarization 
anisotropies from the quadrupole into the acoustic peak region, \Planck\ has reduced the 
degeneracy between the tensor-to-scalar ratio $r$ and $n_{\rm s},$ establishing the bound 
$r_{0.002} < 0.10$ at 95\,\% CL, assuming $n_{\rm t} = - r/8$ as predicted by the simplest inflationary models.
When the \Planck\ likelihood is combined with the $B$-mode polarization likelihood of 
the BICEP2-Keck Array experiment, a tight 95\,\% CL upper limit of $r_{0.002}<0.056$ is obtained, 
corresponding to a 95\,\% CL bound on the energy scale of inflation of $V_*^{1/4} < 1.6 \times 10^{16}\,{\mathrm{GeV}}$. 
\Planck\ 2018 and BK15 data also set tight bounds on gravitational waves generated in the early Universe when $r$ and $n_{\rm t}$ 
are varied independently, complementary to the results obtained by the direct-detection interferometers LIGO and VIRGO at much higher frequencies.

\item
\textit{Which inflationary models are best able to account for the data?}

Starting with the 2013 release using only a part of the data, \Planck\ has substantially 
tightened the constraints on slow-roll inflationary models, ruling out hybrid models 
with $n_{\rm s} >1$ and power-law inflation \citepalias{planck2013-p17}. In combination with the BK15 data, \Planck\ 2018 now strongly disfavours
monomial models with $V(\phi) \propto \phi^p$ and $p > 1$, natural inflation, and low-scale SUSY models. 
Within the representative cases studied in this paper, inflationary models such as $R^2$, T and E $\alpha$-attractor models, D-brane inflation, and those 
having a potential with exponential tails provide  
good fits to \Planck\ and BK15 data.
We used 
two methods to reconstruct the inflaton potential beyond the slow-roll approximation: by Taylor expanding the inflaton potential or Hubble parameter in the observable region; and through a free-form reconstruction of the potential with cubic splines. No statistically significant detection beyond the second derivative of the potential was found, suggesting that the slow-roll approximation is adequate for the \Planck\ 2018 likelihood in combination with the BK15 data.

\item
\textit{What model-independent constraints can be placed on the primordial power spectrum?}

We reported on three different methods for the non-parametric reconstruction of the primordial power spectrum (penalized likelihood,
a Bayesian spline reconstruction.  
and a method based on cubic splines).  
All three methods give broadly consistent results.
In no case is any statistically significant evidence for a deviation 
from a pure power law found. The constraints on the deviations are at the few-percent level for wavenumbers 
in the range $0.005\,\mathrm{Mpc}^{-1} \lesssim k \lesssim 0.2$\,Mpc$^{-1}$ probed by the CMB, the precise constraint depending on the level of smoothing allowed.

\item
\textit{Is there evidence for features in the primordial power spectrum?}

We explored several classes of theoretically motivated parametric models with strong departures from a power law for the primordial power spectra and tested their predictions using combinations of \Planck\ temperature and polarization power spectra. We also carried out an analysis using bispectrum data as well. 
No statistically significant evidence for features was found. 

\item
\textit{Were the primordial cosmological perturbations solely adiabatic?}

A key question is whether the primordial cosmological fluctuations consisted exclusively of adiabatic growing-mode perturbations or whether isocurvature perturbations, possibly correlated with the adiabatic mode and with each other, were also excited. The new polarization data has helped to sharpen constraints on the allowed isocurvature fraction compared to the \Planck\ 2015 results. In correlated mixed adiabatic and isocurvature models, the 95\,\% CL upper bound for the non-adiabatic contribution to the observed CMB temperature variance is $|\alpha_{\text{non-adi}}|<$ 1.3\,\%, 1.7\,\%, and 1.7\,\% for CDM, neutrino density, and neutrino velocity isocurvature, respectively. For this 
release we also report constraints on a scale-invariant compensated baryon-CDM isocurvature mode, which is uncorrelated with the adiabatic mode. This mode would cause an additional lensing-like smoothing at high $\ell$ and modify the lensing potential at $\ell \lesssim 40$. By using the temperature, polarization, and lensing data, we obtain the constraint $\dms = 0.0037^{+0.0016}_{-0.0021}$ at 68\,\% CL for the variance of the baryon isocurvature density perturbation. A detection of isocurvature modes would suggest the need for a theory beyond single-field inflation, which is able to excite only one mode.

\item
\textit{Were the primordial fluctuations statistically isotropic?}

The \Planck\ analysis has confirmed evidence at low statistical significance of anomalies in the CMB temperature anisotropies on large angular scales that are not alleviated in models with nontrivial topology or an anisotropic expansion \citep{planck2014-a20}. This motivates an exploration of inflation-based models giving such violation of statistical isotropy.  We have found no statistically significant evidence in favour of a curvaton model for dipolar asymmetry (compared to the base-$\Lambda$CDM model), nor any evidence for a quadrupolar asymmetry in the temperature or polarization anistropies. Theoretical models producing the observed temperature dipolar asymmetry make a prediction for the polarization dipolar asymmetry. We tested whether the fit to the temperature dipolar asymmetry gives a prediction for the polarization asymmetry consistent with the data. We found no statistically significant evidence that the pattern seen in temperature is repeated in polarization. However, the discriminating power of this test is weak, due to the low polarization signal-to-noise ratio on large angular scales.

\end{enumerate}
The \Planck\ 2013, 2015, and 2018 releases have substantially improved the constraints on the space of inflationary models, 
as described above. Future CMB polarization data will be crucial for further constraining those inflationary models that currently 
provide an adequate fit to \Planck\ and other data. Forthcoming $E$-mode polarization data will be decisive for determining whether 
the intriguing features in the temperature power spectrum, such as the deficit at $\ell \simeq 20$--30, the smaller average amplitude 
at $\ell \lesssim 40$, and other anomalies at higher multipoles require new physics or whether these features are simply the result of 
statistical fluctuations plus instrumental noise. Improved measurements of the $B$ modes promise to constrain inflation even more tightly and it will be 
interesting to see how the search for $B$ modes evolves. One possibility would be a convincing detection of inflationary gravitational waves, 
but a tighter upper limit of $r \lesssim 10^{-3}$ is also an achievable outcome. Either case would substantially advance our understanding of 
inflation and the constraints on the physics of the very early Universe.


\begin{acknowledgements}
We are grateful to Jan Hamann and Jim Zibin for extensive help with the final editing of this
manuscript.
The Planck Collaboration acknowledges the support of: ESA; CNES and 
CNRS/INSU-IN2P3-INP (France); ASI, CNR, and INAF (Italy); NASA and 
DoE (USA); STFC and UKSA (UK); CSIC, MINECO, JA, and RES (Spain); 
Tekes, AoF, and CSC (Finland); DLR and MPG (Germany); CSA (Canada); 
DTU Space (Denmark); SER/SSO (Switzerland); RCN (Norway); SFI 
(Ireland); FCT/MCTES (Portugal); ERC and PRACE (EU). A description 
of the Planck Collaboration and a list of its members, indicating 
which technical or scientific activities they have been involved in, 
can be found at 
\href{http://www.cosmos.esa.int/web/planck/planck-collaboration}{\texttt{http://www.cosmos.esa.int/web/planck/planck-collaboration}}. 
\end{acknowledgements}

\bibliographystyle{aat}
\bibliography{Planck_bib,Inflation_references}

\end{document}